\documentclass[preprint,12pt]{elsarticle}

\usepackage{amssymb}
\usepackage{graphicx}
\usepackage{bm}
\usepackage{braket}
\usepackage{placeins}
\usepackage{flafter}
\usepackage{caption}
\usepackage[skip=0.5ex]{subcaption}
\usepackage{fourier}
\usepackage{lineno}
\usepackage{color}
\usepackage[bb=ams, cal=cm, scr=boondox, frak=euler]{mathalpha}
\let\amsmathbb\mathbb
\AtBeginDocument{%
    \let\mathbb\relax
    \newcommand{\mathbb}[1]{\amsmathbb{#1}}
}
\newcommand\myeq{\mathrel{\stackrel{\makebox[0pt]{\mbox{\normalfont\tiny Eq. (11)}}}{=}}}
\newcommand\myeqq{\mathrel{\stackrel{\makebox[0pt]{\mbox{\normalfont\tiny Eq. (10)}}}{=}}}
\newcommand\myeqqq{\mathrel{\stackrel{\makebox[0pt]{\mbox{\normalfont\tiny Eq. (178)}}}{=}}}
\newcommand\myeqqqq{\mathrel{\stackrel{\makebox[0pt]{\mbox{\normalfont\tiny Eq. (181)}}}{=}}}
\journal{Surface Science Reports}

\begin{document}

\begin{frontmatter}



\title{Advancements in Secondary and Backscattered Electron Energy Spectra and Yields Analysis: from Theory to Applications}

\author[inst1,inst2]{Simone Taioli}

\affiliation[inst1]{organization={European Centre for Theoretical Studies in Nuclear Physics and Related Areas, Fondazione Bruno Kessler},
        city={Trento},
            postcode={38123}, 
            country={Italy}}

\affiliation[inst2]{organization={Trento Institute for Fundamental Physics and Applications (TIFPA), Istituto Nazionale di Fisica Nucleare}, 
            city={Trento},
            postcode={38123}, 
            country={Italy}}

\author[inst1,inst2]{Maurizio Dapor}

\begin{abstract}
Over the past decade, experimental microscopy and spectroscopy have made significant progress in the study of the morphological, optical, electronic and transport properties of materials. These developments include higher spatial resolution, shorter acquisition times, more efficient monochromators and electron analysers, improved contrast imaging and advancements in sample preparation techniques. These advances have driven the need for more accurate theoretical descriptions and predictions of material properties. Computer simulations based on first principles and Monte Carlo methods have emerged as a rapidly growing field for modeling the interaction of charged particles, such as electron, proton and ion beams, with various systems, such as slabs, nanostructures and crystals.
This report delves into the theoretical and computational approaches to modeling the physico-chemical mechanisms that occur when charged beams interact with a medium.
These mechanisms encompass single and collective electronic excitation, ionization of the target atoms and the generation of a secondary electron cascade that deposits energy into the irradiated material.
We show that the combined application of ab initio methods, which are able to model the dynamics of interacting many-fermion systems, and Monte Carlo methods, which capture statistical fluctuations in energy loss mechanisms by random sampling, proves to be an optimal strategy for the accurate description of charge transport in solids.
This joint quantitative approach enables the theoretical interpretation of excitation, loss and secondary electron spectra, the analysis of the chemical composition and dielectric properties of solids and contributes to our understanding of irradiation-induced damage in materials, including those of biological significance.
\end{abstract}




\end{frontmatter}

%\linenumbers

\section{Introduction}\label{Introduction}

The study of the interaction of beams of charged particles, which can consist of electrons, protons or ions, with matter is relevant both for fundamental physics and for technological applications. 
Streams of charged particles are effectively used to probe the optical, dielectric \cite{werner2023,li2024}, electronic \cite{taioli2009electronic,umari2012communication}, and structural \cite{pedrielli2022search} properties of materials. They are also routinely used in the manufacture of customised devices, in plasma processing and in electron lithography, microscopy and spectroscopy \cite{ding1996,joy1996,ding2001,Vos2001,Orosz2004,Yubero2008,ding2008,larciprete2013,salvatpuyol2013,bellissimo2020,tougaard2022,polak2021,khan2023,pauly2023,dapor2023editorial}. 
In addition, they play a role in areas as diverse as the assessment of damage and potential failure of materials, for example on the walls of fusion reactors or during re-entry of spacecraft exposed to ionised plasma or on board space stations \cite{durante2011}, or to test the charging processes of test masses due to cosmic radiation in the upcoming Laser Interferometer Space Antenna (LISA) for the detection of gravitational waves \cite{taioli2023role}.
They also play an important role in recent developments in cancer therapy such as hadron therapy \cite{Solovyov2009,Surdutovich2014,Surdutovich2015,Nikjoo2016,Solovyov2017,conte2017,conte2018,Friedrich2018,Surdutovich2019,taioli2020relative,de2022energy,de2022simulating,daporEPJD2015,PhysRevB.96.064113}, in which the shower of low-energy ($< 50$ eV) secondary electrons generated by the impact of charged ions on the biological medium \cite{gorfinkiel2005electron,taioli2006waterwaves} can cause lethal damage to diseased cells through ionisation, excitation and dissociation, while minimising the effects of irradiation on healthy tissue.

This review describes both state-of-the-art and established theoretical and computational methods that enable robust predictive analysis and accurate interpretation of experimental measurements, e.g. backscattering coefficients, energy loss spectra, secondary electron energy distribution and yields.
Our analysis covers the main mechanisms triggered by the collision of charged particles in matter, with a focus on electrons. In addition, we show the application of these methods to a variety of test cases and discuss the results that can be obtained with different levels of theory in terms of their accuracy.

These methods are based on the principles and laws of quantum mechanics that govern at the microscopic scale. Their application to the calculation of material properties, discovery and design of novel materials is generally referred to as first-principles (or ab initio) simulation. 
However, the accurate representation of the electron-matter interaction using quantum mechanical approaches comes at the price of a high computational cost, which becomes prohibitive with increasing system size. Therefore, numerous semi-empirical approaches have been proposed to describe this interaction, which can also provide a quantitative and at the same time computationally feasible modelling of these phenomena.

All these methods for modelling collisions between charged particles and target objects are based on the concept of the scattering cross-section. This measurable quantity (or derived quantities such as the mean free path and the stopping power) encodes the accessible information about the various interactions that can occur when particles pass through materials. It implicitly includes the microscopic properties of the material, such as the energy-momentum dispersion and the constitutive equations that can be calculated from first-principles by taking into account the quantum many-body correlation, e.g. the exchange and spin-orbit interactions, excitonic and polaronic effects.
In particular, access to the total cross-section allows to characterise the contribution to the motion of electron beams penetrating materials in terms of single or collective excitation, ionisation, Auger decay, electron-phonon scattering and trapping phenomena \cite{taioli2010electron,taioli2015computational}.

Due to the enormous number of particles involved in these collision processes (e.g. the electron density of metals, such as gold and copper, at the Fermi level is $\approx 10^{28}-10^{29}$ electrons/m$^3$, while the incident electron beam flux is $\approx 10^{16}$ electrons/(mm$^2$sec)),
a statistical approach to modelling charge transport in solid targets can be useful.
Transport Monte Carlo (TMC) is such a statistical approach based on the concept of classical trajectory \cite{Shimizu_1992,joy1995monte,Dapor2003book,Dapor2023}. This theoretical and numerical framework greatly simplifies the modelling of charge transport in solids and significantly reduces the computational effort. In TMC, electrons (or ions in general, even considering the dynamic change of their charge as they move in matter) are treated as point-like classical particles endowed with momentum and energy, and their trajectories are driven, modified and extinguished by elastic and inelastic scattering mechanisms that occur with different probabilities.
We emphasise at this point that the use of the concept of trajectory to represent the motion of electrons, especially when their kinetic energy becomes very low due to multiple inelastic scattering processes, is questionable due to the de Broglie wavelength, which is comparable to typical interatomic distances. 
Due to the wave nature of the particles, the trajectory can essentially be defined with the uncertainty $\Delta x \approx \hbar/p$, where $p$ is the momentum of the lightest of the colliding particles (be it the projectile or the target). Therefore, the concept of trajectory, whose resolution cannot be lower than $\approx \Delta x$, loses its meaning for impact parameters smaller than such a value. According to Liljequist, however, the wave character and diffraction effects of low-energy electrons up to a few eV can be safely neglected in liquid water and amorphous media \cite{5Liljequist2008,5Liljequist2013,LILJEQUIST201445}.

The bridge between the microscopic (or quantum mechanical) description and the macroscopic (or classical) representation of charge transport in solids is the dielectric 
function $\varepsilon({\bm r},{\bm r'},t-t')$, which describes the linear response of a material at the space-time point $({\bm r},t)$ to an external electrical perturbation at the space-time point $({\bm r'},t')$. 
For a periodic solid characterised by translation invariance, only the space-time interval $({\bm r}-{\bm r'}, t-t')$ is considered to maintain causality, which is transformed in Fourier space into the wave vector ${\bm k}$ and the frequency $\omega$, thus $\varepsilon({\bm k},\omega)$. 
In general, the variables ${\bm k}$ and $\omega$ in the dielectric function  are the wave vector and the frequency of the external electromagnetic field that perturbs the system and correspond to the crystal momentum ${\bm k}$, i.e. the momentum-like vector associated with the electrons in a crystal lattice and the electron energy (in a solid, $\omega=\omega({\bm k})$ determines the dispersion law or, what is the same, the electronic band structure). Only in the case of collision events must these quantities be interpreted as momentum transfer and energy loss. We note that in this report we use $\bm{k}~(k)$ to denote the crystal momentum (modulus) of the electrons in a solid or the wave vectors of the incident electrons (since they are in principle indistinguishable), $\bm{k'}~(k')$ for the wave vector (modulus) of the electron in the final state after inelastic scattering, while $\bm{q}=\bm{k}-\bm{k'}~(q)$ denotes the momentum (modulus) transferred from the impinging charged particles to the target material.
The physical interpretation as crystalline momentum or momentum transfer is explicitly stated if this is not clear from the context.

By having access to the behaviour of $\varepsilon({\bm{k},\omega})$ in the entire momentum and energy range for a given material, together with the type of incident particles (be they electrons, protons or ions), one can model the various physical phenomena that are triggered by the collisions with the target atoms, in the form of the inelastic scattering cross-section (ISCS), the inelastic mean free path (IMFP) and the stopping power (SP) of the material \cite{1Ritchie57}. This is the information required in input, along with the elastic mean free path (EMFP) and the cumulative probabilities of elastic and inelastic scattering, for running TMC simulations. 

In general, ab initio methods are able to provide a high degree of accuracy \cite{taioli2009electronic,umari2012communication,pedrielli2022search,pedrielli2021electronic,azzolini2017monte,azzolini2018anisotropic}
in the determination of such inputs for the description of charge transport in solids.
However, they suffer from an inherently high computational cost, even though they generally start from different levels of approximation, such as the decoupling of electronic and nuclear motions (i.e. the Born--Oppenheimer approximation) and the pseudopotential scheme, where you can represent the interaction of the valence electrons with the atomic cores by a weak effective ``pseudopotential''. Therefore, one usually resorts to less accurate, faster, analytical or semi-empirical formulations, keeping in mind that these approaches are only applicable in certain energy ranges and for certain targets. 

An example of this are the elastic processes between electron and target. These collisions can be interpreted in the general theoretical framework of the formal theory of scattering \cite{triggiani2023elastic,Newton,TENNYSON201029} by calculating the matrix elements of the transition operator from first principles \cite{taioli2010electron}.
However, even if the general theory of elastic scattering is largely developed, there are many relevant difficulties in its specific application to solids and molecules and in some cases also to atomic problems, such as the approximations involved in the calculation of the scattering potential, the choice of the basis set or simply the computational scaling, which until recently have hindered the production of high quality theoretical spectra.
In this context, one can resort to the Mott differential elastic scattering cross-section \cite{1Mott}, which can be calculated by computing the direct and spin-flip scattering amplitudes in the framework of the relativistic partial wave expansion method.
Mott's theory of the elastic scattering cross section (ESCS) was developed specifically for scattering on spherically symmetric potentials, i.e. atoms. To speed up the simulation, such potentials are typically tabulated or fitted to ab initio data \cite{Salvat1987}. Within this approach, however, the ESCS for molecular or solid systems neglects or poorly includes the effects of the close presence of multiple atomic centres when calculating the scattering potential, such as dipole moment and multiple scattering. 
The calculations can be further simplified if the conditions of the first Born approximation are met, i.e. high (non-relativistic, i.e. $E << m_e$, where $m_e$ is the rest mass of the electron) energy of the incident particle and low atomic number of the target system. In this case, the relativistic Mott cross-section can be approximated by the non-relativistic screened Rutherford formula \cite{Dapor2023}.

The accuracy ladder can also be climbed down when evaluating the IMFP. In this respect, the most commonly used scheme is the dielectric theory \cite{1Ritchie57}. This approach is valid within the first Born approximation and is based on the knowledge of the energy loss function (ELF) in the entire momentum and energy phase space, the so-called Bethe surface. The ELF encodes the information about the excitation spectrum \cite{RevModPhys.74.601} of the target material and thus about the underlying electronic structure. It can be shown that it is given by the imaginary part of the inverse dielectric function. It is noteworthy that the ELF is independent of the beam and can therefore be (indirectly) determined by probing the system with photons or electrons.
In optical experiments, the ELF is usually only determined for the vanishing momentum transfer (the so-called optical ELF), due to the small wavelength of light typically used
($5\cdot10^{-2}$nm$^{-1}$). The question therefore arises as to how the energy-momentum dispersion of the respective material can be included in the evaluation of the ELF, as this is necessary for an exact determination of the IMFP.
This problem can be addressed either with a semi-empirical approach, where a fit to the experimental optical ELF is made using Lorentz functions (Drude-Lorentz approach) and its extension beyond the optical limit is typically achieved by introducing a polynomial dependence on momentum transfer that guarantees the correct limits \cite{Ashley1988,Ashley1990,TPP1993}, or with an analytical continuation into the momentum axis via the Mermin model (known as the Mermin energy loss function-generalised oscillator strength (MELF-GOS) approach \cite{GarciaMolina2012,EAbril}). Conversely, if the experimental optical ELF is not available, one can resort to first principles simulations, in which the dispersion is determined ``automatically'' and treated at the same theoretical level as the optical ELF.
Using ab initio simulations to calculate the ELF as a function of momentum and energy transfer, improved results can be obtained with respect to measured data  such as reflection energy loss spectra (REELS), secondary electron spectra (SES) and secondary (SEY) or total (TEY) electron yields, especially in the low energy range (around the mean excitation energy of the target atoms) \cite{taioli2023role,taioli2020relative,azzolini2017monte}.

An even simpler method of modelling energy loss is the Bethe-Bloch formula \cite{1Egerton}, which gives the stopping power of the target material if the atomic number and kinetic energy of the electron beam are known. Here we introduce the Bethe-Bloch formula by using the field equations in linear response from which we can establish a link to the dielectric model and find its relativistic extension as well as approximations for high and low density and momentum.
The Bethe-Bloch approach is based on the same assumptions as the Drude model of electrical conduction in solids. In this framework, the bound electrons are modelled as driven damped harmonic oscillators, which are jointly described by a mean excitation energy. 

Other important mechanisms of energy loss during electron transport in solids occur at kinetic energies of the electrons that are lower than a multiple of the fundamental gap (typically below 20 eV) and are: (i) the generation (and annihilation) of phonons due to the interaction of the electron with the lattice and (ii) trapping phenomena due to polarisation effects in insulating materials and defects in metals and semiconductors. To describe the former processes, the electron-phonon cross-section can be introduced either with the semi-empirical Fr\"ohlich theory \cite{1Frohlich} or with ab initio simulations \cite{ZHOU2021107970}. Trapping phenomena can also be treated either with an empirical relation to determine their contribution to the total IMFP \cite{1Ganachaud} or with ab initio methods \cite{Franchini2021}. 
Finally, knowing the EMFP and IMFP for each of the energy loss mechanisms, one can perform a TMC simulation \cite{Shimizu_1992,joy1995monte,Dapor2003book,Dapor2023}. We emphasise that accurate modelling of both elastic and inelastic scattering events is crucial to obtain a realistic description of particle trajectories and to model the transport of electrons. In this respect, the fully quantum mechanical approach – which treats a system at the level of individual electrons and nuclei – is the most accurate option for calculating the ELF. To demonstrate this, we give here some examples of their application to the ELF of metals (gold), semiconductors (cerium oxide) and insulators (tantalum oxides) by comparison with semi-empirical approaches.

Please note that TMC simulations for electron energies higher than 10 keV provide quite good results compared to experimental data when using the Rutherford differential elastic scattering cross-section and the Bethe-Bloch formula for the stopping power. In this range, the TMC method is used in conjunction with the continuous-slowing-down approximation (CSDA). In this approximation, the electrons experience a continuous loss of energy. This completely neglects the fact that the electrons undergo several inelastic collisions, the energy losses fluctuate statistically (energy straggling) and in some cases the electrons can also lose their entire energy in a single collision. CSDA can be used to a certain extent for the calculation of the backscattering coefficient. In general, the most accurate strategy is to simulate the energy straggling of the scattered particles, which amounts to treating in detail all the mechanisms - quasi-elastic scattering by phonons, polaronic effects and inelastic collisions with bound electrons or plasmons - triggered by the passage of charged particles in solid and liquid materials, be they insulators, semiconductors or metallic samples.

In the next sections, we will discuss the theoretical and computational models that deal with elastic scattering from atomic nuclei and also describe the energy loss and trajectory deviation of energetic particles travelling in materials.

\section{The excitation processes initiated by the interaction of charged particle beams with matter}

\subsection{Why electron energy loss spectroscopy?}

Electron spectroscopy consists of a relatively large number of experimental techniques based on analysing the energy and angular pattern of electrons scattered back or transmitted through a sample surface \cite{taioli2010electron,Dapor2023}.

Here the discussion focuses on electrons acting both as perturbing probes and as carriers of information. The extension to a beam of charged particles, such as ions and protons, follows the same reasoning.
\begin{figure}[hbt!]
    \centering
    \includegraphics[width = 1.0 \textwidth, angle = 0]{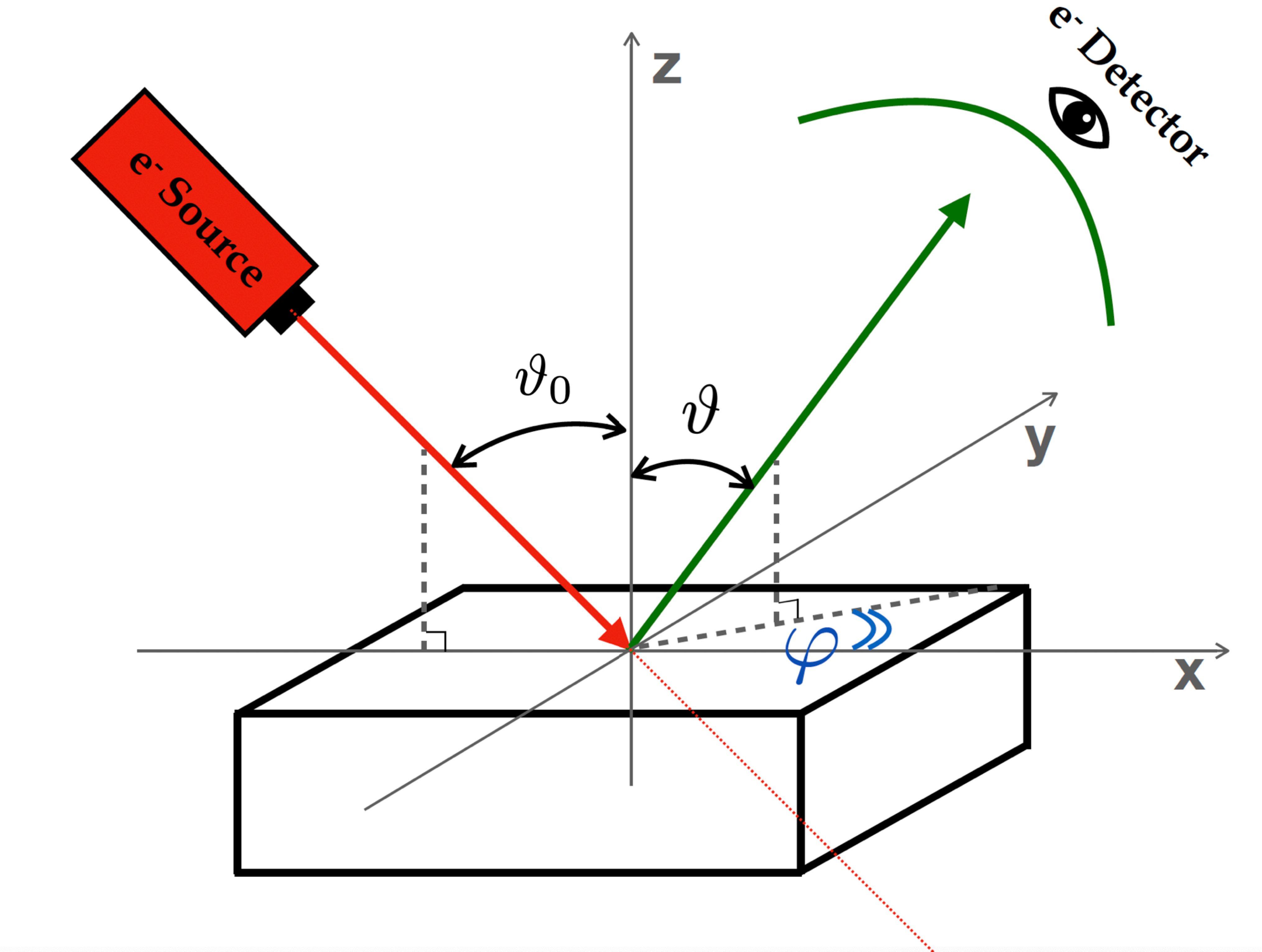}
    \caption{Layout of a reflection energy loss experiment using an electron beam. Source: Reprinted from \cite{AZZOLINI2020109420}.}
    \label{fig:scattexp}
\end{figure}
The geometry of the experiment that we have in mind in this review concerns an electron beam that impinges on a target material with variable kinetic energy at an adjustable angle with respect to the surface normal (see Fig. \ref{fig:scattexp}).
Due to their charge, electrons can be handled by electromagnetic fields and can therefore be easily detected, counted and analysed with regard to their energy and angular distribution patterns.

\begin{figure}[hbt!]
    \centering
    \includegraphics[width=1.0\linewidth]{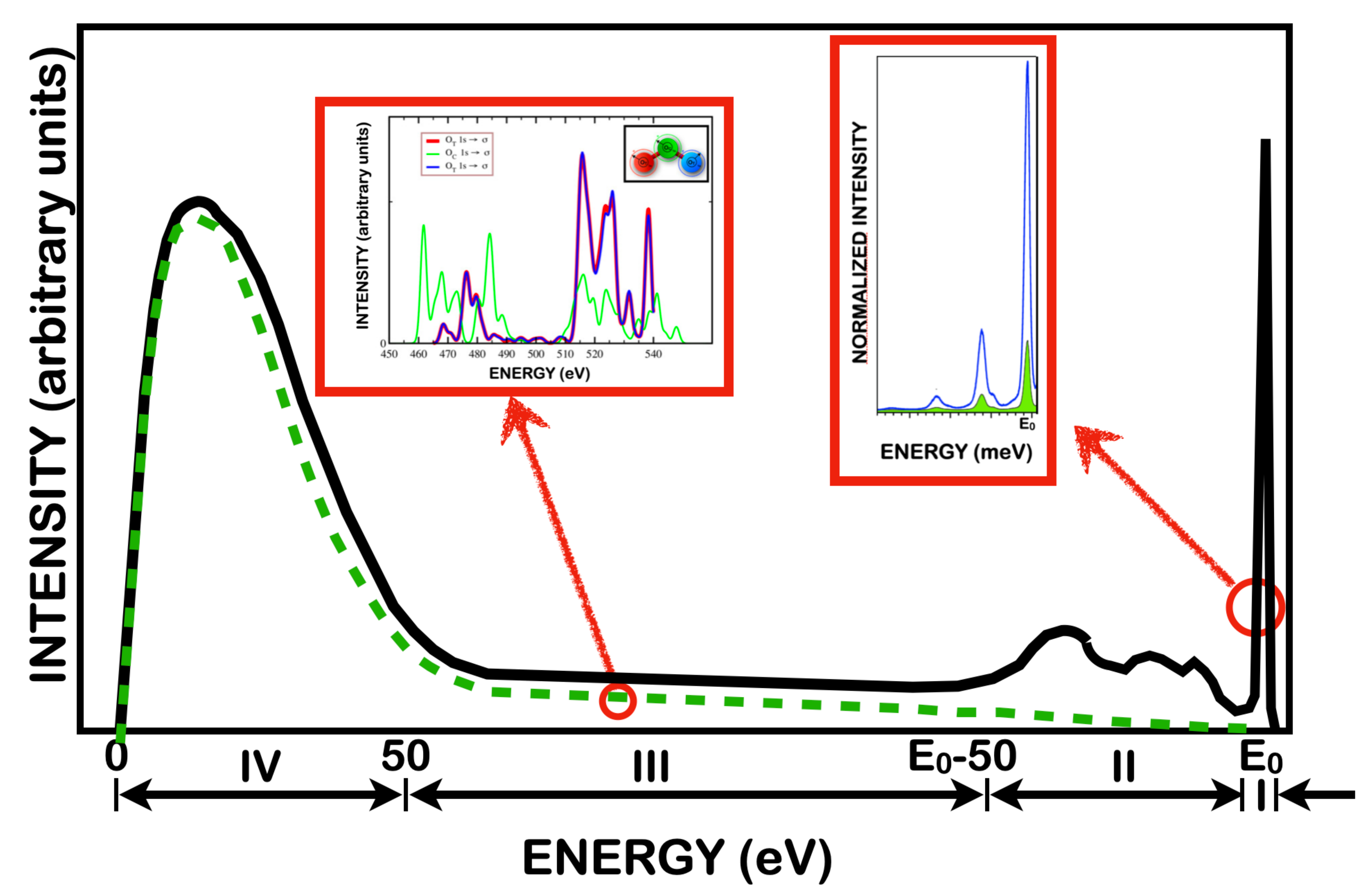}
    \caption{Electron energy distribution spectrum, which includes both primary (solid line) and secondary (dashed line) electrons, showing from right to left: I) the elastic peak and the phonon losses (zoomed in the inset); II) the surface and bulk plasmon excitation; III) Auger transitions (zoomed in the inset) and excitation and emission of core electrons; IV) the peak of the low-energy secondary electrons. Source: adapted from \cite{ibach1977}.}
    \label{fig:spectrumex}.
\end{figure}

The resulting electron signals recorded by the electron analyser contain contributions from various processes involving individual and collective charges, such as electron excitation and emission, Auger decay, plasmons, phonons, trapping and charge density waves. The electronic transitions generated by the external perturbation of the charged beam lead to a well-resolved line shape, the so-called electron energy spectrum, and thus to different types of spectroscopy. A general energy loss spectrum in reflection geometry, which represents the number (intensity) of electrons emerging from a solid surface as a function of their emission energy after they have undergone single or multiple collisions with the atomic constituents, is shown in Fig. \ref{fig:spectrumex} \cite{ibach1977}.
This plot reflects the electrical, magnetic, optical, thermal and structural properties of the analysed material. By analysing this spectral line shape, information can be obtained, e.g. about the atomic components, about the presence of impurities and defects in a test sample and about the dielectric properties, such as the refractive index, the extinction and absorption coefficients.

In the following subsections we describe the
collision processes that can be activated within a medium by an external perturbation and the corresponding spectroscopies.

\subsection{Elastic and quasi-elastic collisions}\label{elas_section}

The elastic peak (peak I in Fig. \ref{fig:spectrumex}) collects all backscattered electrons (or electrons transmitted through a thin film) that have suffered negligible energy losses at the detector (see Fig. \ref{fig:scattexp} for an experimental set-up with reflection geometry). In this definition we include the electrons that have undergone quasi-elastic collisions with phonons, where the energy transferred is often too low to be resolved experimentally (of the order of a few meV). We find that the energy position of the elastic peak may be slightly shifted due to the recoil energy absorbed by the target nuclei. Elastic peak electron spectroscopy (EPES) is an analytical technique that analyses the shape of the elastic peak line \cite{1Gergely,1JablonskiII} and in particular the energy of the elastic peaks of different chemical species in a sample. Due to the low mass, this method is particularly effective in detecting hydrogen (for incident electron energies in the range of 1000-2000 eV, the difference in recoil energy is 2-4 eV).The elastic peak typically has a full width at half maximum (FWHM) of the order of 1 eV, which is due to a combination of Doppler effect, analyser resolution and energy distribution of the primary beam.

\subsection{Plasmon energy loss}

The broad region at 20-50 eV from the elastic peak (energy range II in Fig. \ref{fig:spectrumex}, to the left of the zero-loss peak) collects the primary electrons that are backscattered (or transmitted through a thin film) after they have suffered energy losses due to collective plasmon excitation, inter- and intraband transitions with the valence electrons of the target components.
This phenomenon is more pronounced in electrons of the outer shell. According to Drude theory, these electrons can be modelled as harmonic oscillators with a characteristic angular frequency $\omega_0$. In this context, the time period $\tau$ in which the electronic interaction between a charged beam and an irradiated solid can act is approximately proportional to the ratio of the impact parameter $b$ and the beam velocity $v$, $\tau\approx b/v$, resulting in $b\approx v/\omega_0$. For kinetic energies of the electron beam that are increasingly higher than $\hbar\omega_0$, which is the case for the outer electrons, the range of the effective interaction $b$ increases, while the oscillator still has enough time to react to the external disturbance, resulting in a collective excitation.

The focus of electron energy loss spectroscopy is on analysing the plasmon peak to obtain information about the density of states of valence and conduction electrons (DOS), dielectric properties, composition, structure and type of bonding. This energy loss process can occur multiple times; therefore, the peak repeats with lower intensity (due to the lower probability of multiple independent inelastic scattering events) at equidistant energies up to 100-200 eV from the elastic peak. The interaction of the primary electron beam with finite samples can also lead to the excitation of surface plasmons \cite{1Ritchie57}, which are longitudinal charge density waves oscillating along the sample-air interface.
The plasmon peak associated with this delocalised excitation is approximately $1/{\sqrt 2}$ (see section \ref{surpla}) of the energy of the bulk plasmon peak and can be used, for example, to detect molecular adsorption on surfaces.

\subsection{Localised excitation and Auger transitions}

Towards higher energy losses (a few hundred to a few thousand eV away from the elastic peak, in the energy region III of Fig. \ref{fig:spectrumex}), the electron energy spectrum shows several sharp peaks of relatively low intensity (with respect to the plasmon losses) corresponding to the excitation of localised inner shell electrons. The energy losses of these transitions appear as sharp steps in the energy spectrum corresponding to the ionisation thresholds. In addition, the spectral line shape in this energy range is enriched by the presence of resonant autoinisation and Auger lines, \cite{taioli2010electron,taioli2009surprises,taioli2021resonant,colle2004abc,colle2004auger,colle2004ab}, which are due to the non-radiative decay of excited or singly ionised atoms (see the well-resolved peaks shown in the left inset in energy region III of Fig. \ref{fig:spectrumex})). In particular, Auger decay involves the emission of a secondary electron after the initial ionisation of core-levels, which provides useful information about many-body Coulomb interactions within the system, core-level shift for chemical analysis and the excitonic properties of materials.

\subsection{Secondary emission}

Finally, on the far left of the energy loss spectrum (see peak IV in Fig. \ref{fig:spectrumex}) is the pronounced tail of electrons with energies of less than 50 eV, which, however, is sufficient to escape from the solid target, the so-called secondary electron peak.
This energy range mainly concerns the secondary electrons, which are generated either by the ionisation of core-levels of the target atoms or by the emission of valence electrons or by the decay of plasmons. Of course, particles of the primary beam, although typically a negligible fraction (unless the primary energy is also very small), can also be part of this peak and are indistinguishable from the secondary electrons.
To determine the yield of secondary electrons, conventional methods therefore integrate the spectrum over the range $[0,50]$ eV for a given kinetic energy of the primary beam, including the tail of the low-energy backscattered electrons. 
The TEY (i.e. the sum of the secondary electron yield and the backscattering coefficient), on the other hand, is obtained by integrating the entire spectrum from 0 to the kinetic energy of the primary electrons E$_0$.

In scanning electron microscopy (SEM), for example, the emission of secondary electrons plays an important role in the formation of contrasts in the image of a sample \cite{1DaporetalSE2010}. Conversely, secondary electrons can lead to unwanted charging of devices \cite{taioli2023role}.

\subsection{A test case for the ELF: the material structure}

Electron emission from the surface of materials irradiated with particle beams has proven to be a unique tool for characterising the chemical and structural properties of solids, especially in the near-surface layers.

To demonstrate the sensitivity of these electron-based analytical techniques, we use the ELF to solve a central problem in condensed matter physics, namely the identification of the atomic arrangement of solids. Structure is indeed closely related to the properties and functionalities of materials.

\begin{figure}[hbt!]
    \centering
\includegraphics[width=1.0\linewidth]{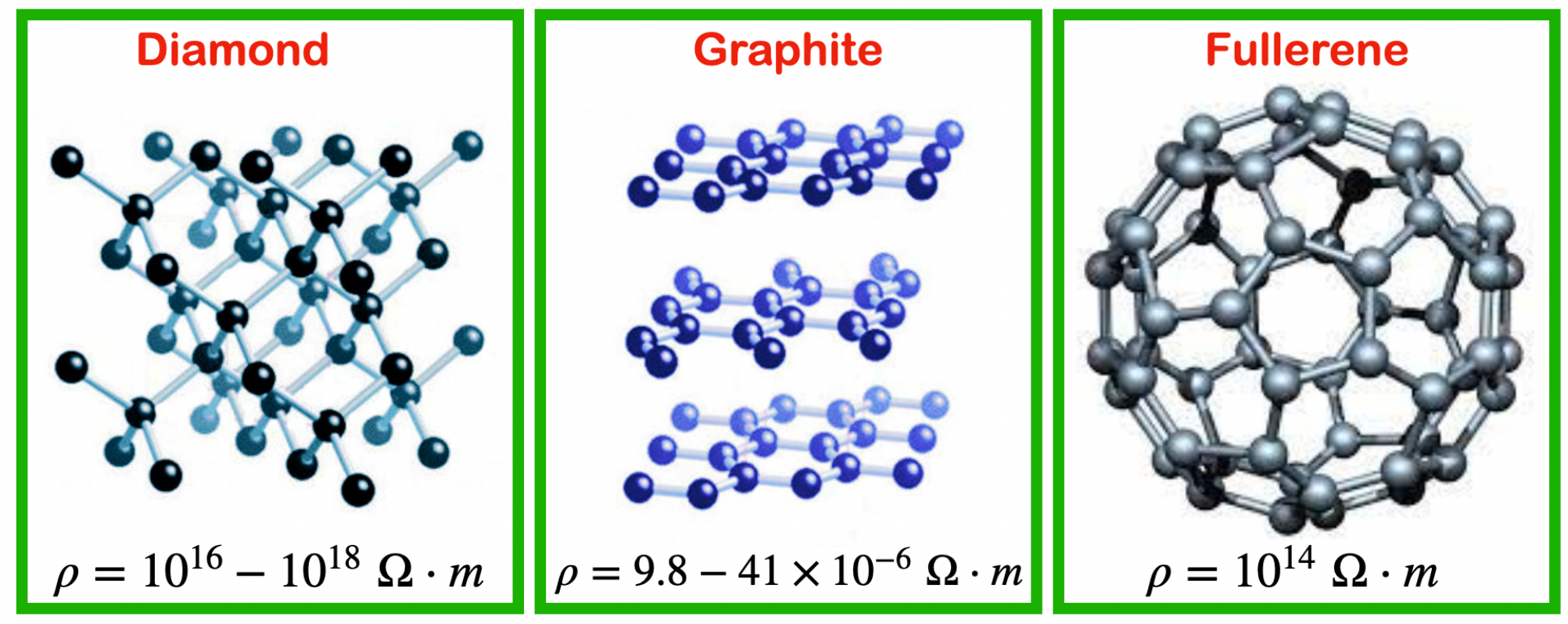}
    \caption{The atomic structure of diamond (left panel), graphite (middle panel) and fullerene (right panel). The corresponding resistivity of these structures at room temperature is also indicated in the insets of the figure. For graphite the range refers to planar and non-planar values, for fullerene to a thin film.}
    \label{fig:struct}
\end{figure}

We find that even solids consisting of a single element of the periodic table in different configurations can have properties that span an extremely wide range, which can extend over many orders of magnitude.
Carbon, for example, is a very versatile chemical element due to its tetravalent nature, which can combine under different conditions in multiple bonding schemes such as $sp^1$, $sp^2$ and $sp^3$.
The number of known chemical compounds of carbon (which by mass is only the 4th most abundant element in the universe after H, He and O and only the 15th most abundant in the earth's crust) is far greater than the sum of all others (over 10 million species).

In Fig. \ref{fig:struct} we show three carbon-based allotropes, namely diamond, graphite and fullerene. Diamond is a wide-gap insulator at room conditions, characterised by high thermal conductivity, charge mobility and mechanical robustness (due to $sp^3$ hybridisation).
Graphite is the most stable allotrope of carbon and is a semi-metal.
It is a $sp^2$ layered material with an in-plane strength similar to diamond and weak van der Waals out-of-plane bonds, which makes it soft and mouldable and ensures optimal thermal and electrical conductivity. C$_{60}$-fullerene is a $sp^2$ semiconductor with a gap of 1.6 eV at room conditions, characterised by aromaticity. Due to its dome shape, C$_{60}$ is mechanically and chemically stable, but not completely unreactive (with the aim of releasing the elastic energy stored in it).

Interestingly, the modulus of elasticity, which characterises the mechanical response of a solid to mechanical stress, is $\approx 1000$ GPa for diamond, while it is $\approx 2$ GPa for fullerene and $\approx 4$ GPa for graphite, differing by three orders of magnitude despite being made of the same element. In this review, we discuss in particular the ability of solids to conduct electricity, which is measured by their electrical resistance.
The room temperature resistivity of the carbon allotropes reported in the insets of Fig. \ref{fig:struct} also differ by many orders of magnitude.

If we exclude the possibility that the origin of the large variation in the properties of these structures is related to the nature and concentration of the atoms, which is comparable, it turns out to be due to the specific way in which the valence electrons of the constituent atoms interact when they arrange themselves in arrays with a step distance of a few angstroms, i.e. the bond length of solids.

\begin{figure}[hbt!]
    \centering
    \includegraphics[width=1.0\linewidth]{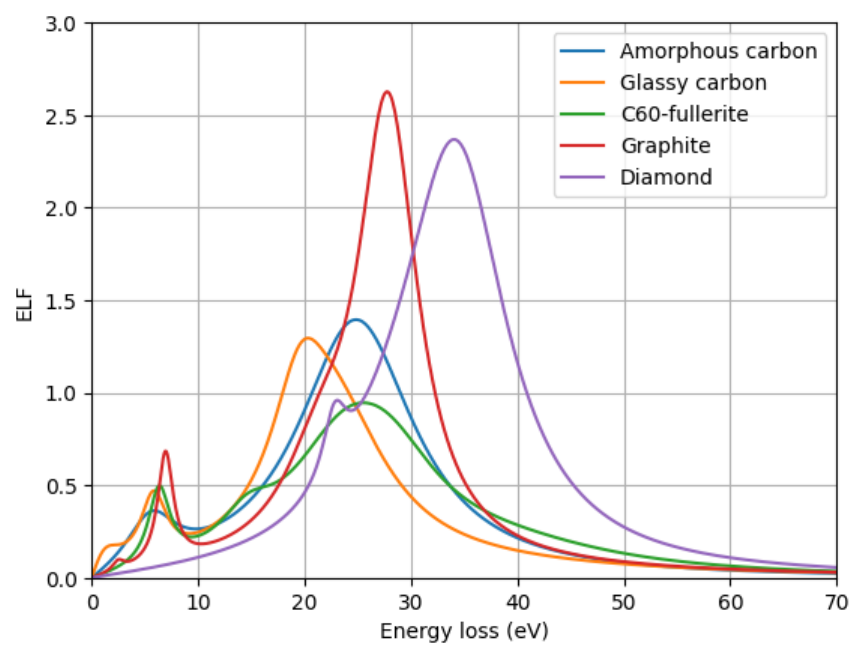}
    \caption{ELF of different allotropic forms of carbon as a function of energy transfer, calculated with the best-fit parameters of Garcia-Molina et al. \cite{garciamolina2006}.}
    \label{fig:spectrumex2}
\end{figure}

The question we ask ourselves is: Can ELF and the associated energy loss signals distinguish between these different allotropes? The curves in Fig. \ref{fig:spectrumex2} show the ELF line shapes (which, as we will see, are directly related to the energy loss spectrum) of various carbon allotropes, including those analysed above. It shows how sensitive the energy loss signals are to the structural arrangement of the materials. Different structures show peaks that are clearly split, energetically shifted and characterised by different intensities.

\section{Theory: analytical models for calculating the backscattering coefficient and the secondary energy spectrum}\label{Theory}

\subsection{The backscattering coefficient}

The electron backscattering coefficient is usually defined by the ratio of the primary electrons of the beam, which emerge from the surface of an irradiated solid with a kinetic energy of more than 50 eV, to the total number of primary electrons incident on the target.
In this context, the possible generation of secondary electrons is therefore neglected and secondary electrons do not appear in this definition. We emphasise that there are both secondary electrons with higher energy than any predefined limit and backscattered electrons with lower energy than 50 eV. However, these numbers are negligible compared to the relevant quantities.
Several studies have been carried out to calculate the backscattering coefficient, based either on semi-empirical assumptions \cite{6Everhart,Archard} or more accurate numerical simulations \cite{Shimizu_1992}.
However, we note that some of the earlier analyses used the Rutherford cross section, which is typically limited to energies greater than 5-10 keV and an atomic number less than 45. In Ref. \cite{PhysRevB.46.618} the range of validity of such an approximation was discussed in detail. In the next section, we will briefly explain some analytical methods for estimating the backscattering coefficient. These approaches are based on crude assumptions to simplify the model, such as the consideration of only one scattering event per particle. Nevertheless, they are used for a faster comparison with the experimental data to identify trends with respect to more accurate but more computationally intensive numerical TMC simulations. The backscattering coefficient is used, for example, in the evaluation of the layer thickness \cite{1DaporetalBSE2011} and for chemical characterisation.

\subsection{Backscattering Coefficient: Everhart Model, Archard Model, Vicanek and Urbassek Model}

\subsubsection{Everhart model}

A simple analytical formula for calculating the backscattering coefficient at normal incidence of electron beams characterised by a kinetic energy of more than $\sim 10$ keV and a target atomic number of $Z\leq 45$ was proposed by T.E. Everhart \cite{6Everhart} as follows:

\begin{equation}
\eta\,=\,\frac{a-1+0.5^a}{a+1}\;,
\label{EverhartFormula}
\end{equation}
where $a=0.045Z$. 

\begin{figure}[hbt!]
    \centering    \includegraphics[width=1.0\linewidth]{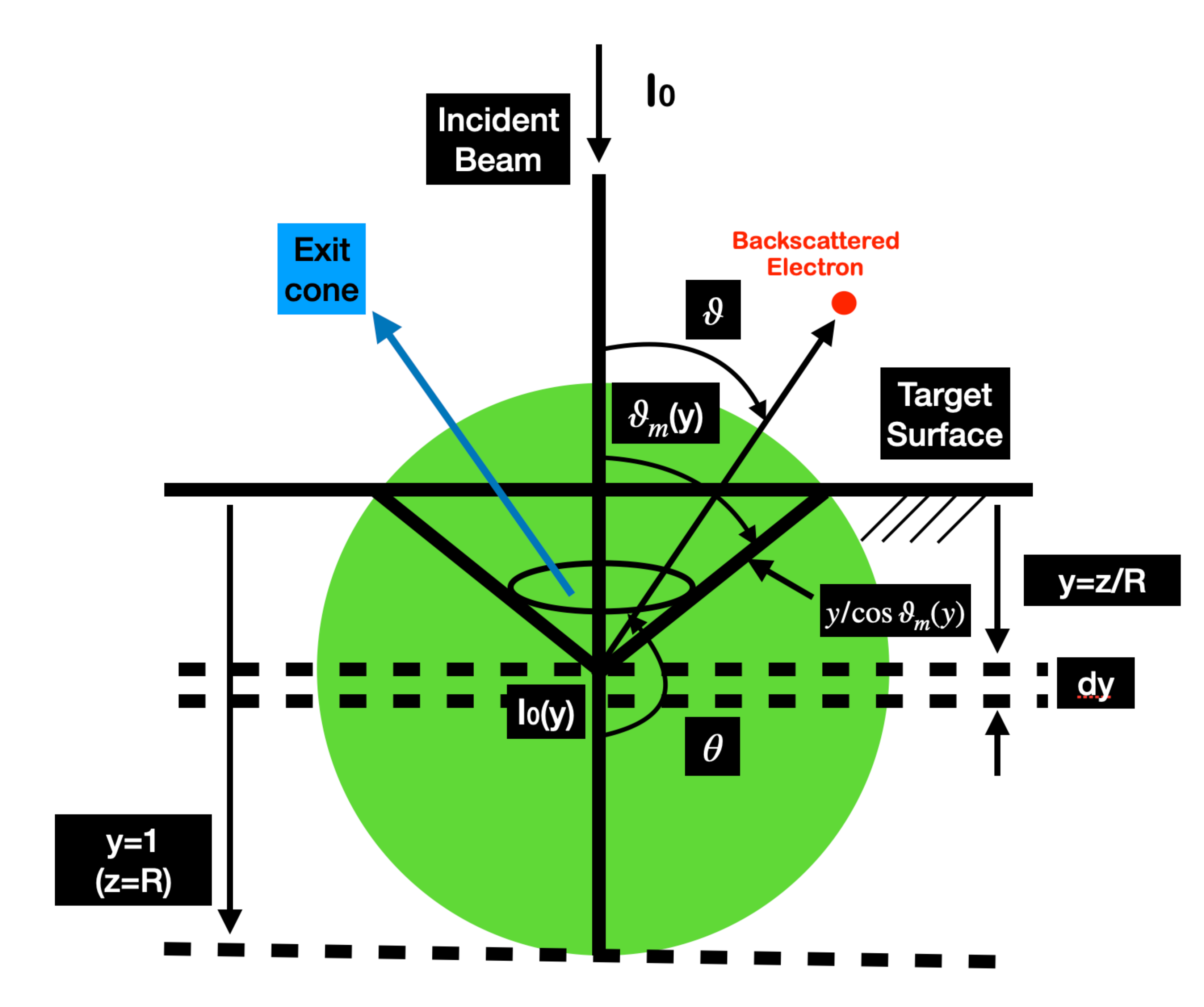}
    \caption{Geometry of Everhart's backscattering model.}
 \label{fig:everhart}
\end{figure}

This formula can be deduced by applying the
Thomson--Whiddington semi-empirical relation for the continuous energy loss \cite{6Thomson,Whiddington}, which in turn is based on the Bethe-Bloch model \cite{Dapor2003book,1Egerton} for the inelastic interaction (see section \ref{bethe_block} below):
\begin{equation}\label{BBW}
v^4-v_0^4= \frac{4}{m_e^2}(E^2-E_0^2) = - c_t\rho z,
\end{equation}
where $v_0$ ($E_0$), $v$ ($E$), are the electron velocities (energies) of the primary beam outside the sample or at a depth $z$ inside the material; $\rho$ is the mass density of the sample, $c_t$ is a constant and $m_e$ is the electron mass. For $v=0$, we obtain the maximum range that a beam of normally incident electrons can reach if they move in a straight line within the material and continuously lose energy through inelastic scattering processes characterised by negligible angular scattering, i.e:

\begin{equation}\label{zmax}
R=\frac{v_0^4}{c_t\rho}=\frac{4E_0^2}{c_t\rho m_e^2}.
\end{equation}
$R$ represents the maximum penetration depth, i.e. the longest path that the electrons can travel without being elastically scattered out of the primary beam (see Fig. \ref{fig:everhart}).
This definition implicitly assumes that the electrons are only removed from the primary beam by elastic scattering events.

By defining the supplementary scattering angle $\vartheta=\pi-\theta(y)$ (see Fig. \ref{fig:everhart}), elastic collisions in the high-energy range can be described by the following Rutherford scattering cross-section for a bare Coulomb interaction 

\begin{equation}\label{ruthil}
d\sigma_\mathrm{el}\,=\,\frac{Z^2 e^4}{c_t\rho m_e^2}\frac{1}{R-z}\frac{2\pi\sin\vartheta\,d\vartheta}
{(1+\cos\vartheta)^2}=\frac{Z^2 e^4}{c_t\rho m_e^2R}\left(\frac{1}{1-y}\right)\frac{2\pi\sin\vartheta\,d\vartheta}
{(1+\cos\vartheta)^2}\;,
\end{equation}
where $e$ is the electron charge, $Z$ is the atomic number of the target material and, in the last equation, we have used $E^2=c_t\rho m_e^2(R-z)/4$ from Eqs. (\ref{BBW}) and (\ref{zmax}). 
The depth $z$ that the electrons reach before they are elastically scattered is described in Eq. (\ref{ruthil}) by introducing the reduced fraction $y=z/R$ ($0 < y <1$). 
 
The contribution to the backscattering coefficient comes only from the particles that can escape through the sample surface. These are the electrons scattered into the ``exit cone'', which is characterised by an opening half-angle $\vartheta_{\rm m}(y)$ determined according to the following relation (see Fig. \ref{fig:everhart}):

\begin{equation}\label{travdist}
y+\frac{y}{\cos(\vartheta_{\rm m}(y))}=1,
\end{equation}
where the left-hand side represents the total distance the electron travels in the solid, and the right-hand side represents the maximum distance the electron can travel within the solid target, which corresponds to the maximum penetration depth. The electron flux at a fractional depth $y$ within the material, $I_0(y)$, can be calculated by assuming charge conservation between incident and elastically scattered electrons and the validity of Eq. (\ref{ruthil}) as

\begin{equation}
I_0(y)=I_0 -\frac{\pi Z^2 e^4 N_A}{c_tm_e^2M}\int_0^y \frac{I_0(y')}{1-y'}dy'=I_0 -a\int_0^y \frac{I_0(y')}{1-y'}dy',
\label{EverhartFormula2}
\end{equation}

\noindent where $N_A$ is the Avogadro number, $M$ is the molecular mass of the sample, $I_0$ is the primary radiation flux incident on the sample surface ($y=0$), and we have also defined Everhart's single scattering coefficient
\begin{equation}\label{aaasss}
a=\frac{\pi Z^2 e^4 N_A}{c_tm_e^2M}.
\end{equation}
Eq.~(\ref{EverhartFormula2}) can be easily integrated and returns
\begin{equation}\label{EverhartFormula3}
I_0(y)=I_0(1-y)^a.
\end{equation}
The double differential backscattering coefficient of electrons at a fractional depth $y$ is given by:

\begin{eqnarray}\label{aaasssurdo}
d^2\eta &=&\frac{d^2I_0(y)}{I_0}=\frac{I_0(y)}{I_0}d\sigma_\mathrm{el}\frac{N_A\rho}{M}dz=\frac{Z^2 e^4 N_A}{c_tm_e^2M}\frac{I_0(y)dy}{I_0(1-y)}\frac{2\pi\sin{\vartheta}d\vartheta}{(1+\cos\vartheta)^2}=\nonumber \\
&=&\frac{Z^2 e^4 N_A}{4c_tm_e^2M}\frac{I_0(y)dy}{I_0(1-y)}\frac{2\pi\sin{\vartheta}d\vartheta}{\cos^4(\vartheta/2)}=\frac{a}{4\pi}\frac{I_0(y)dy}{I_0(1-y)}\frac{d\Omega}{\cos^4(\vartheta/2)}=\nonumber\\
&=&a(1-y)^{a-1}dy\frac{1}{4\pi}\frac{d\Omega}{\cos^4(\vartheta/2)},
\end{eqnarray}
where in the last line we have used Eq. (\ref{EverhartFormula3}).
Eq. (\ref{aaasssurdo}) can be integrated over both the supplementary scattering angle $\vartheta$ (between 0 and $\vartheta_{\rm m}(y)$), which can be calculated with Eq. (\ref{travdist}) and over the fractional depth $y$ (between 0 and $1/2$ ($z = R/2$), see Fig. \ref{fig:everhart}) to obtain the Everhart formula, Eq.~(\ref{EverhartFormula}), for backscattering from a bulk target at normal incidence. This relation can be extended to thin films and oblique incidence \cite{Niedrig2020AnalyticalMI}.
However, due to the neglect of the velocity distribution of electrons penetrating the target, the angular distribution of primary electrons caused by inelastic collisions, and the existence of multiple elastic large-angle collisions, the semi-empirical coefficient $a$ must be modified to reproduce the experimental data \cite{6Everhart}, limiting the ability of the Everhart approach to predict the backscattering coefficients of materials. 

\subsubsection{Archard model}

The Archard model of backscattering \cite{Archard} assumes that the primary beam travels straight upwards up to a distance $D$ within the target, the so-called diffusion depth, and then diffuses isotropically so that the path of the electrons is equal to the maximum penetration depth $R$. According to Archard, the diffusion depth is proportional to $R$:
\begin{equation}\label{Archard1}
D \approx\frac{40R}{7Z}.
\end{equation}

\begin{figure}[hbt!]
    \centering
\includegraphics[width=1.0\linewidth]{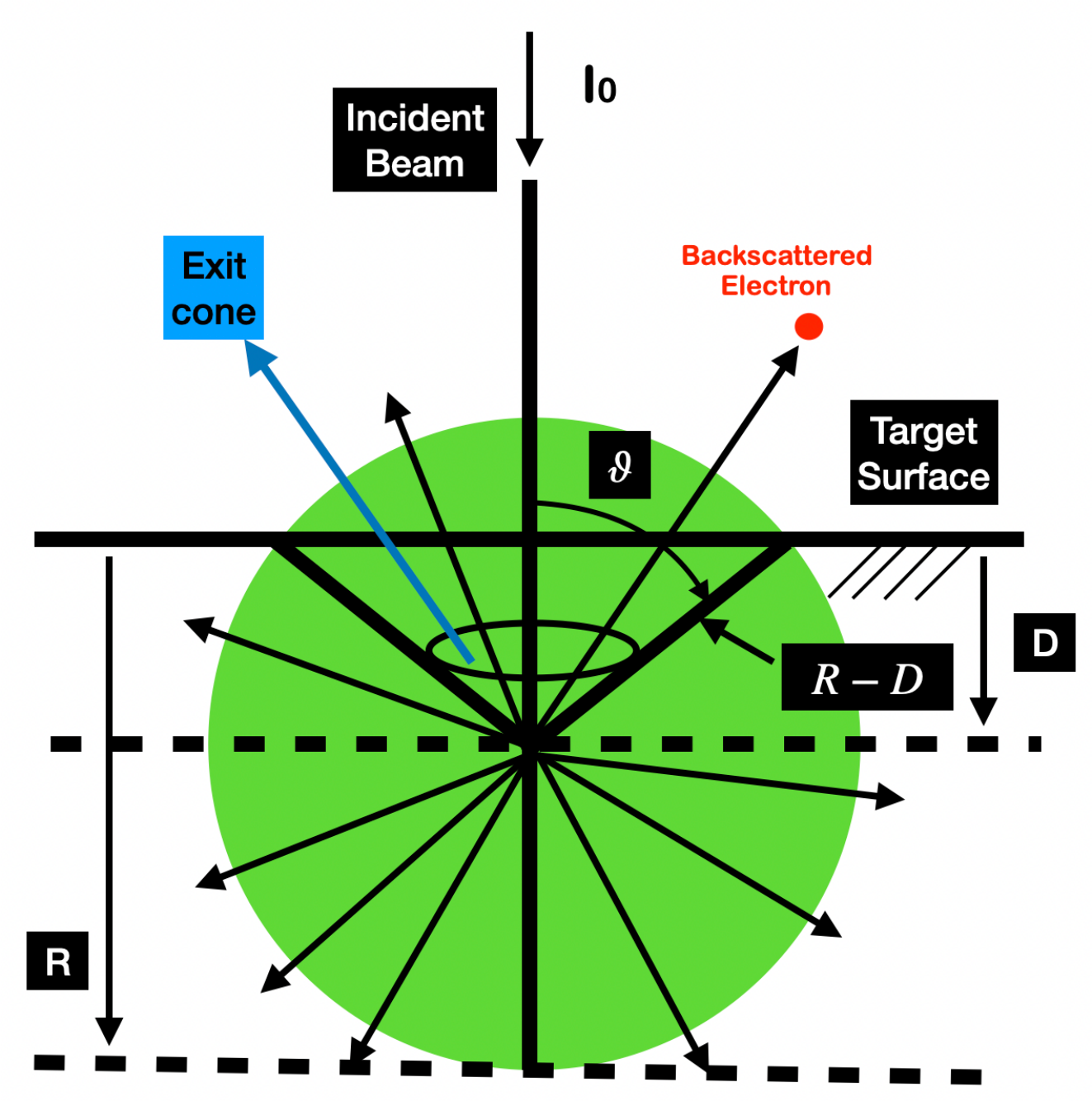}
    \caption{Geometry of Archard's backscattering model.}
 \label{fig:archard}
\end{figure}
The contribution to the backscattering coefficient comes from those electrons that fulfil the following relation (see Fig. \ref{fig:archard}):
\begin{equation}\label{Archard2}
\cos\vartheta\,=\,\frac{D}{R-D},
\end{equation}
which gives, using Eqs. (\ref{Archard1}) and (\ref{Archard2}),
\begin{equation}\label{Archard3}
\eta=\frac{1}{4\pi}\int_0^{\vartheta}d\Omega= \frac{1}{4\pi}\int_0^{\vartheta}{2\pi\,\sin x\,dx}~~\myeq~~\frac{1}{2}\left(1-\frac{D}{R-D}\right)~~\myeqq~~\frac{7Z-80}{14Z-80}.
\end{equation}
\begin{figure}[hbt!]
    \centering
    \includegraphics[width=1.0\linewidth]{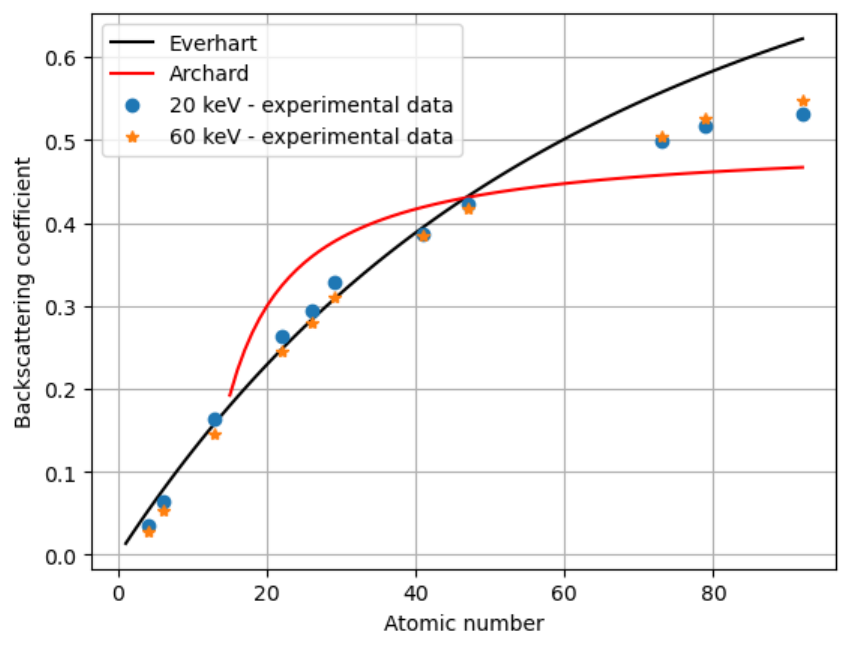}
    \caption{Backscattering coefficient calculated according to Everhart (Eq.~(\ref{EverhartFormula}), black line) and Archard (Eq.~(\ref{Archard3}), red line)
 compared with the experimental data from Neubert and Rogaschewski \cite{6NeubertRogaschewski}
(full colored circles: 20 keV, stars: 60 keV).}
    \label{fig:archard-everhart-comparison}
\end{figure}

In Fig. \ref{fig:archard-everhart-comparison} we show a comparison of the backscattering coefficients of bulk solids calculated with the theoretical models of Everhart (black curve) and Archard (red curve) with experimental data (symbols). 

\subsubsection{Vicanek and Urbassek Model}

Another analytical model for calculating the backscattering coefficient for a wide range of incident beam kinetic energies was proposed by Vicanek and Urbassek \cite{6VicanekUrbassek1991}, who calculated the so-called mean number of wide-angle collisions as follows:

\begin{equation}
n_{\rm wa}\,=\,N\,R\,\sigma_{\mathrm{tr}}\,,
\label{mnwac}
\end{equation}

\noindent where $R$ is the maximum range of penetration, $N$ is the number of target atoms per unit volume and $\sigma_{\mathrm{tr}}$ is the transport cross-section, given by

\begin{equation}
\sigma_{\mathrm{tr}}\,=\,2 \pi \int_0^{\pi}(1-\cos\vartheta)\,\frac{d\sigma_{\mathrm{el}}}{d\Omega}\sin\vartheta\,d\vartheta\,.
\end{equation}

Note that the quantity $n_{\rm wa}$ defined by Eq. (\ref{mnwac}) provides an approximate estimate of the number of wide-angle collisions that the electrons undergo inside the solid target before they come to a complete stop due to inelastic collisions.

Vicanek and Urbassek \cite{6VicanekUrbassek1991} have shown that the backscattering coefficient for light ions can be calculated as follows:

\begin{equation}
\eta\,=\,\frac{1}{\sqrt{1+a_1\frac{\mu_0}{\sqrt{n_{\rm wa}}}+a_2\frac{\mu_0^2}{n_{\rm wa}}
+a_3\frac{\mu_0^3}{n_{\rm wa}^{3/2}}+a^4\frac{\mu_0^4}{n_{\rm wa}^2}}}\,,
\label{vu91}
\end{equation}
where $a_1= 3.39$, $a_2= 8.59$, $a_3= 4.16$, $a_4= 135.9$, $\mu_0=\cos\vartheta_0$ and $\vartheta_0$ is the polar angle of incidence of the beam.

The formula of Vicanek and Urbassek, Eq.~(\ref{vu91}), provides results that are in reasonable agreement with the available experimental data for light ions \cite{6VicanekUrbassek1991}, electrons and positrons \cite{6Colemanetal1992,6DaporJAP1996}. In the Tables \ref{VUbck1keVpositrons} and \ref{VUbck3keVpositrons} the backscattering coefficients of 1000 and 3000 eV positron beams on Al, Cu, Ag and Au, calculated according to Eq. (\ref{vu91}) \cite{6DaporJAP1996}, together with a comparison with the experimental data of Coleman et al. \cite{6Colemanetal1992} are presented.

\begin{table}[hbt!]
\begin{center}
\begin{tabular}{c|c|c}
&&\\
Atomic number&Eq.~(\ref{vu91}) \cite{6VicanekUrbassek1991}&Experimental data \cite{6Colemanetal1992}\\
&&\\
\hline
\hline
&&\\
13&0.086&0.069\\
&&\\
\hline
&&\\
29&0.110&0.135\\
&&\\
\hline
&&\\
47&0.100&0.106\\
&&\\
\hline
&&\\
79&0.111&0.123\\
&&\\
\hline
\end{tabular}
\caption[]{Comparison between the backscattering coefficients of 1000 eV positrons impinging on Al ($Z=13$), Cu ($Z=29$), Ag ($Z=47$) and Au ($Z=79$) \cite{6DaporJAP1996} obtained using the Vicanek's and Urbassek's model  \cite{6VicanekUrbassek1991}, Eq.~(\ref{vu91}), with the experimental data by Coleman et al. \cite{6Colemanetal1992}.}
\label{VUbck1keVpositrons}
\end{center}
\end{table}

\begin{table}[hbt!]
\begin{center}
\begin{tabular}{c|c|c}
&&\\
Atomic number&Eq.~(\ref{vu91}) \cite{6VicanekUrbassek1991}&Experimental data \cite{6Colemanetal1992}\\
&&\\
\hline
\hline
&&\\
13&0.107&0.086\\
&&\\
\hline
&&\\
29&0.159&0.177\\
&&\\
\hline
&&\\
47&0.156&0.168\\
&&\\
\hline
&&\\
79&0.186&0.186\\
&&\\
\hline
\end{tabular}
\caption[]{Comparison between the backscattering coefficients of 3000 eV positrons impinging on Al ($Z=13$), Cu ($Z=29$), Ag ($Z=47$) and Au ($Z=79$) \cite{6DaporJAP1996} obtained using the Vicanek's and Urbassek's model  \cite{6VicanekUrbassek1991}, Eq.~(\ref{vu91}), with the experimental data by Coleman et al. \cite{6Colemanetal1992}.}
\label{VUbck3keVpositrons}
\end{center}
\end{table}

\subsection{Secondary electron energy distribution: Wolff model, Amelio model, Chung and Everhart model}

Analytical models have also been proposed to describe the energy distribution of secondary electrons. Although the MC method (see section \ref{MonteCarlo}) is the most suitable approach to calculate these energy distributions in terms of accuracy and general applicability, purely analytical models are still useful to get an idea of the general behaviour of secondary electron spectra.
In this respect, a model based on the Boltzmann diffusion equation was developed by Wolff
\cite{Wolff,PhysRev.95.1415},
in which secondary electron current can be described by the following equation
\begin{equation}\label{je}
j(E)\,\propto \,P(E)\,\frac{\lambda(E)}{E^{x(E)}}\,,
\end{equation}
where $\lambda(E)$ is the electron mean free path and $P(E)$ refers to the probability that a charge reaching the surface with energy $E$ has a normal velocity sufficient to overcome the potential barrier $\Phi_{\rm W}$ between the vacuum level and the bottom of the conduction band (we note that $\Phi_{\rm W}$ is usually referred to as {\em work function} for metals and {\em electron affinity} for semiconductors and insulators). In this approach to the calculation of the secondary electron spectra for $E$ greater than 2$E_{\rm F}$ ($E_{\rm F}$ = Fermi energy), $x(E)$ in Eq. (\ref{je}) becomes almost constant ($x(E)~\approx~2.1$) and \cite{Wolff,PhysRev.95.1415}:

\begin{equation}
P(E)\,=\,1\,-\,\sqrt{\frac{\Phi_{\rm W}+E_F}{E}}.
\end{equation}

Amelio \cite{10.1116/1.1315884} reformulated the expression of current density on the basis of the Streitwolf equation \cite{https://doi.org/10.1002/andp.19594580308} as follows:
\begin{eqnarray}
&&j(E')\,=\,\frac{E'+W}{4}\times \nonumber\\
&& \left\{\frac{E'}{E'+W}\psi_0(E'+W)\,-\,\frac{5}{2}\psi_2(E'+W)\left[\frac{3}{2}\left(1-\frac{W^2}{(E'+W)^2}\right)\,-\,\frac{E'}{E'+W}\right]\right\}, \nonumber\\
\end{eqnarray}
where $W\,=\,E_{\rm F}+\Phi_{\rm W}$, $E'\,=\,E-W$ and $\psi_{0,2}$ are solutions of the diffusion equation as in Refs. \cite{Wolff,PhysRev.95.1415,10.1116/1.1315884} while Chung and Everhart \cite{10.1063/1.1663306} proposed the following expression:

\begin{equation}
j(E)\,\propto\,\frac{E-E_F-\Phi_{\rm W}}{(E-E_F)^4}\,.
\label{CandE}
\end{equation}

\begin{figure}[hbt!]
    \centering
    \includegraphics[width=1.0\linewidth]{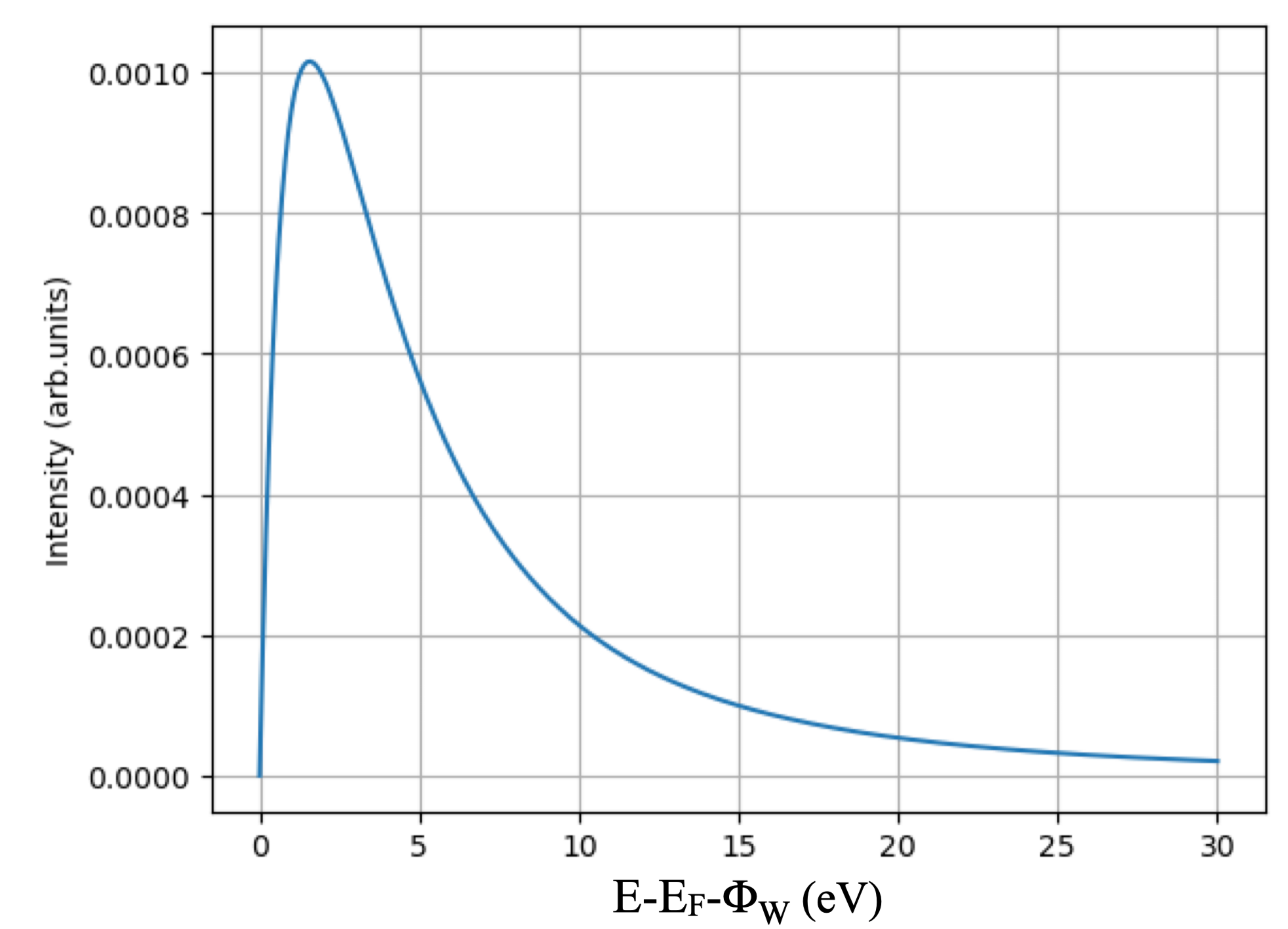}
    \caption{Shape of the energy distribution of the secondary electrons of Cu, calculated with the formula of Chung and Everhart, Eq. (\ref{CandE}), represented as a function of $E-E_{\rm F}-\Phi_{\rm W}$, where $E$ is the energy of the emerging electron, $E_{\rm F}$ = 7.0 eV is the Fermi energy and $\Phi_{\rm W}$ = 4.7 eV is the work function.}
    \label{fig:SESCuCandE}
\end{figure}

In this respect, Fig. \ref{fig:SESCuCandE} shows the secondary electron spectrum emerging from copper modelled using the theory of Chung and Everhart.

\section{The Monte Carlo method}\label{MonteCarlo}

Monte Carlo (MC) is a powerful numerical approach that uses random numbers to evaluate multiple integrals and, compared to quadrature methods, achieves favourable scaling for high-dimensional space at a given accuracy.
Its use has spread very effectively in various areas of condensed matter physics. The implementations of the MC framework range from classical physics for modelling the interaction of particles in solids (e.g. electrons colliding with a target material \cite{Dapor2023}) to the simulation of the quantum properties of the electron gas \cite{Ceperley}.
In this respect, the Quantum Monte Carlo method (QMC) is an extremely accurate tool for determining the electronic properties of many-body systems \cite{becca2017quantum}. However, it still suffers from a large scaling prefactor, preventing its systematic use in the calculation of the properties of medium and large systems.
This section therefore focuses on classical trajectory MC (TMC).

TMC uses elastic and inelastic mean free paths to simulate the interaction processes by averaging over a large number of single-particle trajectories obtained by comparison with random numbers.
In order to provide an accurate interpretation of the experimental data and to increase the predictive power, TMC requires above all accurate values for the elastic and inelastic scattering cross-sections.
If these are available, TMC can achieve statistically significant results in a few CPU hours on current hardware. In addition, the TMC methods and routines are embarrassingly parallel and are perfectly suited to the latest high-performance computers.

\subsection{The Monte Carlo strategies: continuous-slowing-down approximation vs. energy straggling}

In TMC, two main strategies can be devised for the simulation of electron transport in solid targets.

The {\it continuous-slowing-down approximation} is based on the assumption that the electrons continuously lose energy on their trajectory within the solid, while the direction is mainly changed by elastic collision processes. This approach has proven useful, for example, in the calculation of backscattering or transmission coefficients, for which it is not necessary to track the generation of secondary electrons and to know the details of the statistical fluctuations of electron energy losses.

In contrast to this view, if the description of inelastic collision events along the electron trajectory must take such statistical fluctuations into account, a second approach known as {\it energy straggling} must be adopted. In this framework, the electrons are followed in their motion within the target material and all possible elastic (change of direction) and inelastic (energy loss) scattering events are taken into account. This second strategy is crucial, for example, when simulating the energy spectrum of secondary electrons escaping from a solid surface or the plasmon loss peaks.
 
%In the next sections, we will provide a general overview of these two strategies as well as the description of the assessment of the elastic and inelastic mean free paths. The application of the Monte Carlo method to a specific set of test cases  will be instead outlined at the end of our review.

Throughout our discussion, we will use spherical coordinates and assume that the target surface lies in the plane $z = 0$, which is hit by an electron beam with an angle of incidence $\theta_0$, measured from the surface normal ($z$-axis) and given kinetic energy (see Fig. \ref{fig:scattexp}).

\subsubsection{The continuous-slowing-down approximation}

In this approach, the collisions are stochastic processes that follow an exponential (Poisson) law (see section \ref{generpseudo} for the mathematical details), characterised by the step length
\begin{equation}
\Delta s=-\lambda_{\mathrm{el}} \ln(\mu_1),
\label{ds}
\end{equation}
where $\mu_1$ is a random number that is sampled uniformly in the range [0,~1] and $\lambda_{\mathrm{el}}$ is the EMFP:

\begin{equation}\label{lambda_el}
\lambda_{\mathrm{el}}=\frac{1}{N \sigma_{\mathrm{el}}}\,.
\end{equation}

In Eq. (\ref{lambda_el}) $N$ is the number of atoms per unit volume in the solid and $\sigma_{\mathrm{el}}$ is the total elastic scattering cross-section obtained by integrating the differential elastic cross-section over the total solid angle as follows:

\begin{equation}
\sigma_{\mathrm{el}}(E)\,=\,\int\,\frac{d\sigma_{\mathrm{el}}}{d\Omega}\,d\Omega\,=\,\int_0^{2\pi}d\phi\int_0^{\pi}d\vartheta\frac{d\sigma_{\mathrm{el}}}{d\Omega}\sin{\vartheta}\,.
\end{equation}

In this approximation, only the elastic events dictate the average step length along the electron trajectories.
The particle that undergoes an elastic collision suffers a change in trajectory characterised by the elevation (or scattering) angle $\theta$ and the azimuth angle $\varphi$ relative to the last direction before the impact.
Assuming cylindrical symmetry, the angle $\varphi$ can be chosen by drawing a random number from the interval [0,~$2\pi$].
The scattering angle $\theta$ can be calculated by equating the cumulative elastic scattering probability in an angular range [0,~$\theta$] with a random number $\mu_{2}$ sampled uniformly in the range [0,~1]:

\begin{equation}
\mu_2\,=P_{\mathrm{el}}(\theta,E)\,=\,\frac{2\,\pi}{\sigma_{\mathrm{el}}}\,\int_0^{\theta}\frac{d\sigma_{\mathrm{el}}}{d\Omega}\,\sin{\vartheta}\,d\vartheta.
\label{Pelheta}
\end{equation}

\begin{figure}[hbt!]
    \centering
    \includegraphics[width=1.0\linewidth]{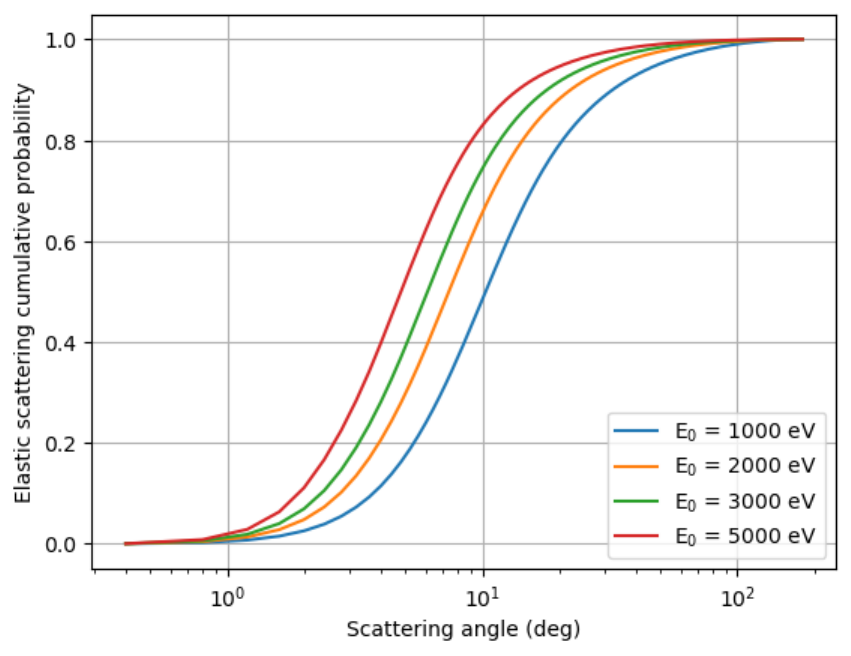}
\caption{Cumulative probability of elastic scattering for C in the angular range $[0,~\theta<\pi]$ and in the primary energy range [1000,~5000] eV, determined by the Mott cross-section (see Eq. (\ref{decs})).}
\label{Pel_Si}
\end{figure}

A typical example of such a cumulative elastic scattering probability is shown in Fig. \ref{Pel_Si} for a carbon atom target, where the Mott approach is used to calculate the ESCS (see the following subsection \ref{elasmott} for details). 

Knowing the scattering angles $\theta$ and $\varphi$ (angles refer to the direction of motion before the collision) the direction of motion of the electron after scattering can be estimated. Finally, the angular coordinates of the electron's reference system can be related to the laboratory system using suitable transformations \cite{Shimizu_1992,joy1995monte,Dapor2003book,Dapor2023}.

The most important component of the TMC method based on the continuous-slowing-down approximation is the stopping power, which is used to calculate the continuous energy loss along the electron trajectory.
In this model, the energy lost by the electron along a section $\Delta z$ is given by
\begin{equation}
\Delta E = \frac{dE}{dz} \Delta z\,
\end{equation}
where $-dE/dz$ is the stopping power of the electrons.

The trajectory of each electron is tracked until an energy threshold is reached (e.g. 50~eV for the evaluation of the backscattering coefficient) or until it leaves the target surface.
To calculate the stopping power several empirical and analytical formulas \cite{joyluo} and statistical models \cite{TUNG1979427} have been developed; to this end, also the Bethe formula (see section \ref{bethe_block}) can be used.
However, we emphasise that the latter approach is based on the first order Born approximation, which fails at low energy. Therefore, the stopping power based on these models is inaccurate.

It is worth noting that a high number of trajectories, typically in the order of 10$^5$ to 10$^6$ depending on the specific problem, is necessary to obtain statistically significant results and to improve the signal-to-noise ratio.
However, we stress that this approach completely neglects the statistical fluctuations of the energy loss mechanisms, i.e. the so-called energy straggling. In this respect, TMC simulations based on the continuous-slowing-down approximation cannot provide detailed information on energy loss patterns. Therefore, its use can only be recommended for the calculation of backscattering and transmission coefficients of primary electrons, while it must be avoided for the evaluation of electron energy distributions.

\subsubsection{The energy straggling approach}

The energy straggling approach differs from the continuous-slowing-down approximation because it requires a thorough knowledge of all inelastic scattering processes that lead to electron energy loss, in particular electron-electron and electron-phonon interactions as well as electron trapping phenomena. The relative contribution of these energy loss mechanisms to the spectrum varies with the kinetic energy of the electrons and the target material, e.g. at low energy electron-phonon scattering is likely to be found in insulators \cite{1Frohlich}.
The assumption about the step length (see section \ref{generpseudo} for the mathematical details) in the energy straggling strategy is similar to the exponential law used in the continuous-slowing-down approximation:
\begin{equation}\label{steplength_es}
\Delta s = -\lambda\,\ln(\mu_1)\,,
\end{equation} 
where $\mu_1$ is a uniformly distributed random number in the range [0,~1], which is generally different from $\mu_1$ in Eq. (\ref{ds}).
The elastic mean free path $\lambda_{\mathrm{el}}$ given in Eq. (\ref{lambda_el}) of the continuous-slowing-down approximation is replaced in Eq. (\ref{steplength_es}) by the total mean free path:
\begin{equation}\label{lambdatotal}
\lambda=\frac{1}{N\,(\sigma_{\mathrm{el}}+\sigma_{\mathrm{inel}})},
\end{equation}
which now contains all relevant elastic and inelastic scattering phenomena represented by the elastic scattering cross-section $\sigma_{\mathrm{el}}$ and the inelastic scattering cross-section $\sigma_{\mathrm{inel}}$. The total IMFP refers to the Coulomb repulsion of the incident electron with the core and valence electrons of the target atoms as well as to the quasi-elastic electron-phonon scattering and electron trapping (polaronic effects) in insulator materials.
These individual contributions add up to the total inelastic scattering cross-section $\sigma_{\mathrm{inel}}$.

\subsection{The Monte Carlo algorithm}

TMC models the process of charge diffusion within the target material by allowing particles to take random walks controlled by individual collisions selected by a probabilistic algorithm. We describe such an algorithm in the context of the energy straggling approach.

\subsubsection{Pseudo-random numbers distributed according to the exponential law}
\label{generpseudo}

The TMC algorithm proceeds according to the theorems of probability theory by generating random variables $\xi$ in the range $[a,b]$, which are distributed according to a given probability density $p(s)$. $\xi$ can be obtained by solving the following implicit relation:
\begin{equation}
\int_a^{\xi}\!\! p(s) \, ds= \mu,
\label{generate}
\end{equation}
where $\mu$ represents a random number that is uniformly distributed in the range [0,1].
To generate a random variable $\xi$ that is distributed according to the exponential law in the range $[0, \infty)$, for example, the Eq. (\ref{generate}) must be solved with the exponential probability density:
\begin{equation}
p_{\xi}(s)\,=\,\frac{1}{\bar{\lambda}}\exp\left(-\frac{s}{\bar{\lambda}}\right),
\label{poissonlambdaconstant}
\end{equation}
where $\bar{\lambda}$ is the mean value of the distribution (for TMC this is the mean free path). According to the probability distribution of Eq. (\ref{poissonlambdaconstant}), the step lengths of Eqs. (\ref{ds}) or (\ref{steplength_es}) for the electron jumps within the material can be obtained. 

\subsubsection{Elastic and inelastic collisions} 

The probability that the electron is inelastically scattered within the medium can be calculated by:

\begin{equation}
{\Pi}_{\mathrm{inel}}=\frac{\sigma_{\mathrm{inel}}}{\sigma_{\mathrm{inel}}+\sigma_{\mathrm{el}}}\,=\,\frac{\lambda}{\lambda_{\mathrm{inel}}},
\end{equation}

\noindent while the probability of elastic scattering is naturally

\begin{equation}
{\Pi}_{\mathrm{el}}=1-{\Pi}_{\mathrm{inel}}.
\end{equation}
The TMC algorithm extracts a random number $\mu_2$, which is sampled uniformly in the interval [0,~1] and compares it with the probability of elastic scattering ${\Pi}_{\mathrm{el}}$: If $\mu_2 < {\Pi}_{\mathrm{el}}$, then the collision is elastic, otherwise it is inelastic.

\begin{figure}[hbt!]
\centering
\includegraphics[width=0.7\linewidth,angle=-90]{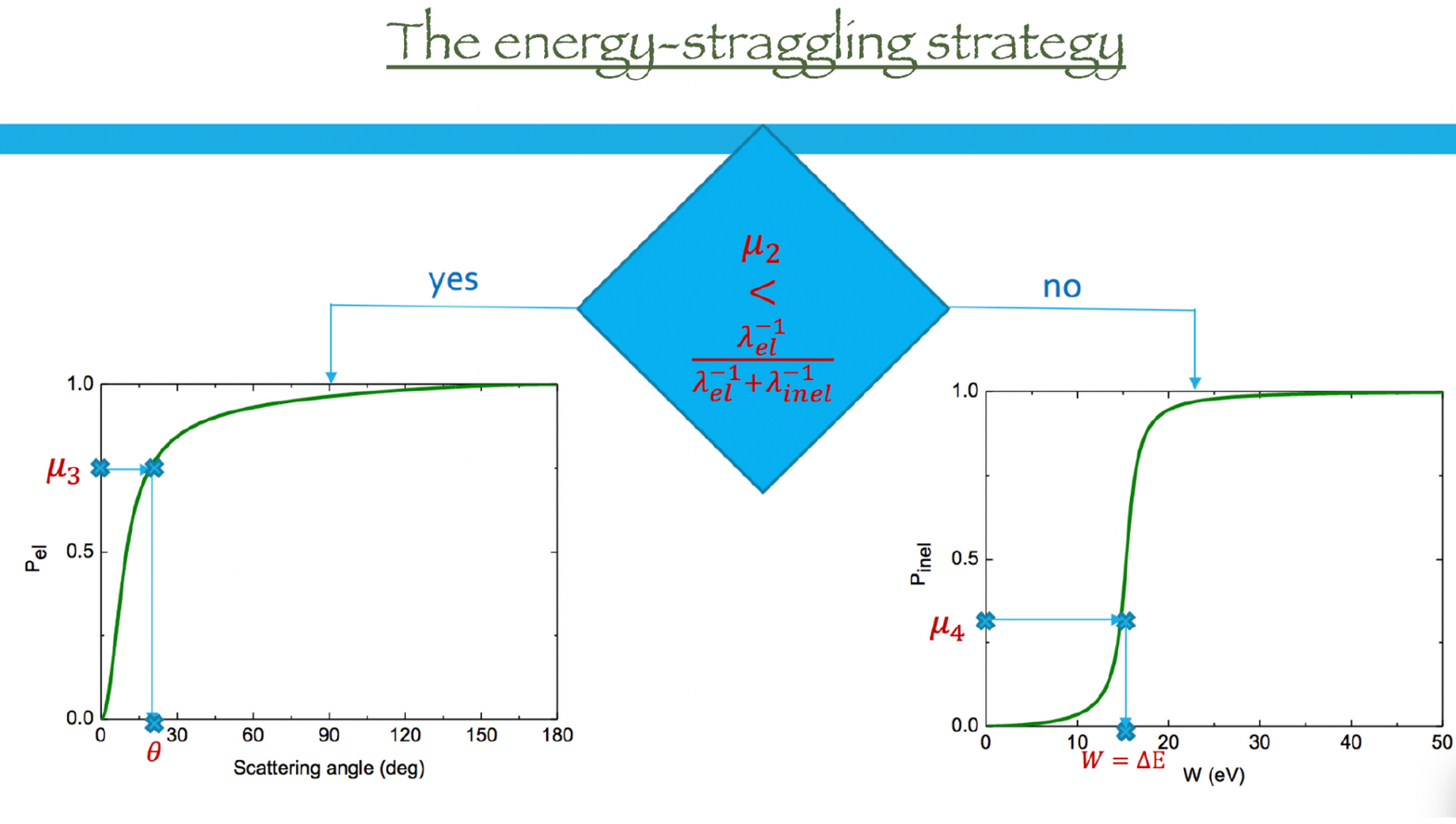}
\caption{The Transport Monte Carlo workflow using the energy straggling strategy. The elastic cumulative probability function appearing in the left panel is defined in Eq. (\ref{P_ELASTIC}) and the inelastic cumulative probability function appearing in the right panel is defined in Eq. (\ref{P_INELASTIC}).}
\label{MC_flow}
\end{figure}

A workflow of the TMC algorithm is shown in Fig. \ref{MC_flow}, where the elastic cumulative probability function appearing in the left side of the figure is defined in Eq. (\ref{P_ELASTIC}) and the inelastic cumulative probability function appearing in the right side of the figure is defined in Eq. (\ref{P_INELASTIC}).

\subsubsection{Polar scattering angles for elastic collisions}

Elastic scattering collisions of electrons on target atoms mainly lead to a change in the direction of motion due to the large mass and charge (even if it is reduced by effective electron screening) of the nuclei. This change in the trajectory can be estimated by calculating the polar scattering angle using the following relation (see also Eq. (\ref{Pelheta})):

\begin{equation}
\label{P_ELASTIC}
\mu_3=P_{\mathrm{el}}(\theta,E)=\frac{1}{\sigma_{el}}\int_0^{\theta}\frac{d\sigma_{el}}{d\Omega}\,2\pi\,\sin\vartheta\,d\vartheta,
\end{equation}
where the cumulative probability of elastic scattering $P_{\mathrm{el}}$ in an angular range $[0,~\theta]$ is equated with a random number $\mu_3$, which is uniformly sampled in the range $[0,~1]$.
An example of such a cumulative elastic scattering probability is shown in Fig. \ref{Pel_Si} for a carbon atom target, where the Mott approach is used to calculate the ESCS (see the following subsection \ref{elasmott} for details). 

\subsubsection{Energy loss and polar scattering angle in inelastic collisions}

Conversely, the energy lost by electrons in inelastic scattering events as they pass through a target can be determined by generating a random number sampled uniformly in the range $[0,~1]$ and equating it to the cumulative probability of inelastic scattering $P_{\mathrm{inel}}$ in a range $[0,~W]$, as follows:

\begin{equation}
\label{P_INELASTIC}
\mu_4=P_{\mathrm{inel}}(W,E)=\frac{1}{\sigma_{\mathrm{inel}}}\int_0^W \frac{d\sigma_{\mathrm{inel}}}{dw}dw\,.
\end{equation}

\begin{figure}[hbt!]
\centering
\includegraphics[width=1.0\linewidth]{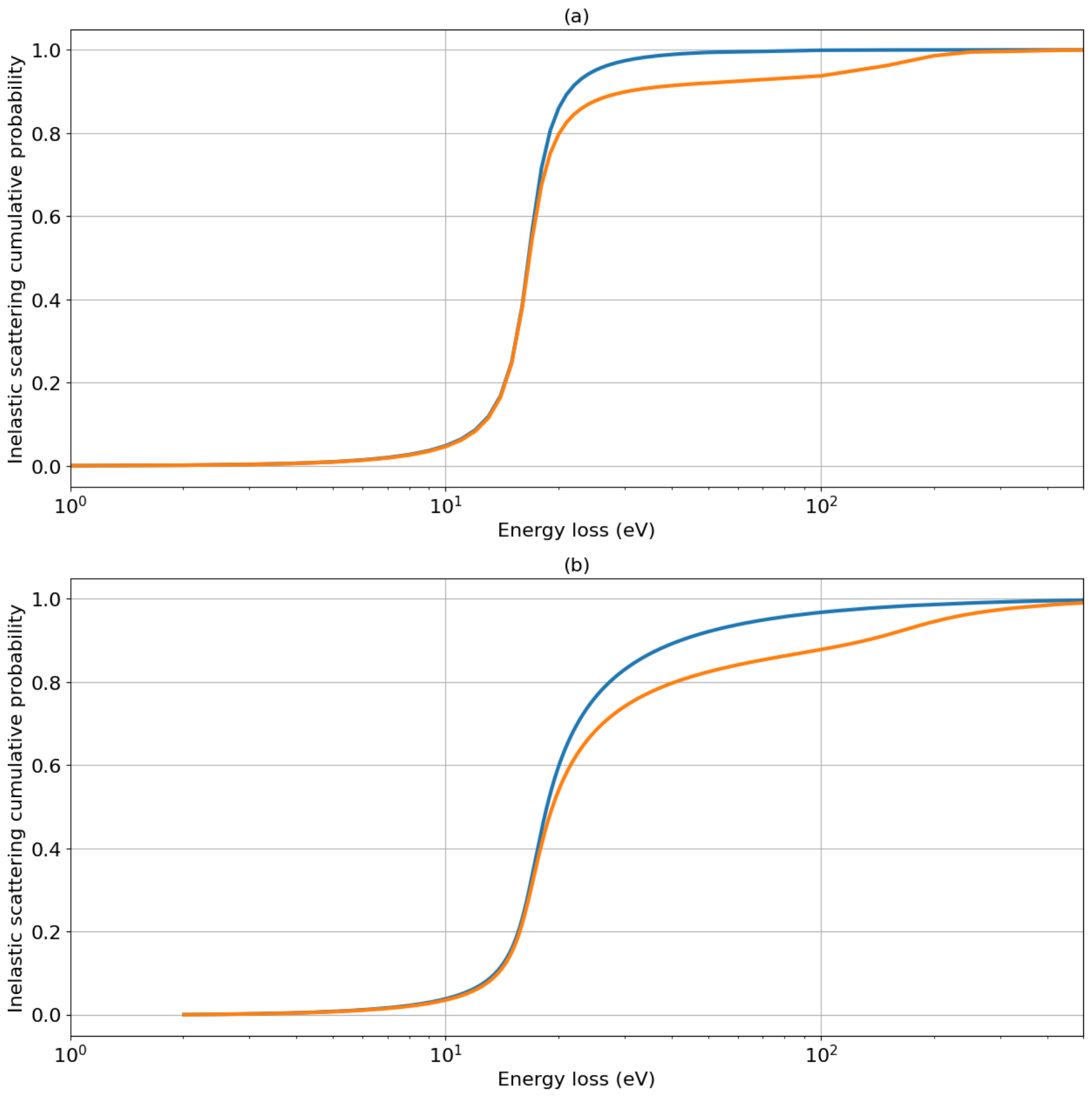}
\caption{Cumulative probability for the sampling of the energy loss of electrons undergoing inelastic collisions in Si calculated according to Ritchie's dielectric theory (see section \ref{Ritchie}). The primary kinetic energy of the electrons is 1000 eV. Blue lines: only valence electrons. Orange lines: all atomic electrons. Panel (a): Ashley model \cite{Ashley1988,Ashley1990}. Panel (b): Yubero and Tougaard model \cite{yuberotougaard}.}
\label{PincINS_Si}
\end{figure}

In Fig. \ref{PincINS_Si} we give the cumulative probability with which we calculate the energy loss of a 1000 eV electron beam impinging on bulk silicon according to Ritchie's dielectric theory (see section \ref{Ritchie}).
This cumulative function provides the proportion of electrons that lose less than or equal to $W$ energy. The curves were generated with two models to extend the ELF beyond the optical limit (zero momentum transfer). In particular, the Ashley model \cite{Ashley1988,Ashley1990} (panel (a) of Fig. \ref{PincINS_Si}) and the model of Yubero and Tougaard \cite{yuberotougaard} (panel (b) of Fig. \ref{PincINS_Si}) are used for this purpose. In addition, the ELFs within the optical limit were calculated using two different methods. In one scenario (blue lines) only the four valence electrons were considered, while in the other scenario (orange lines) all fourteen atomic electrons were considered. In the first case, the electron beam only excites the plasmon at $\approx$~16.6 eV. In the second case, however, electrons from the $L$-shell and the $K$-shell are also involved. This more comprehensive description of the process made it possible to account for energy losses due to the excitation of these oscillators. The effect of this improved accuracy can be seen in the behaviour shown by the orange curves in Fig. \ref{PincINS_Si} in the energy loss range from about 30 eV to 250 eV with respect to the electrons of the $L$-shell.

If the kinetic energy of the electron is higher than the ionisation energy of the atomic constituents of the sample, a secondary electron can be ejected with a kinetic energy of $E=W-B$, where $B$ is the binding energy of the electron. Conversely, if the kinetic energy of the electron is not sufficient to ionise the atomic centres ($W \leq B$),
electron-electron scattering leads to excitation without the generation of secondary electrons.

We point out that, depending on the energy, other mechanisms of energy loss can occur, such as electron-lattice interaction and trapping.
In particular, electron-phonon scattering can lead to energy loss (gain), where the scattered electron loses (gains) an amount of energy equal to the energy of the generated (annihilated) phonon, typically fractions of eV. This scattering mechanism can be described using either Fr\"ohlich theory \cite{1Frohlich} or first-principles methods \cite{qe,elk}. 

The Fr\"ohlich theory is suitable for dealing with very low-energy electrons where the probability of electron-phonon interaction is considerable.
With this model, the interaction of the electrons with the longitudinal optical phonons can be included in the TMC scheme by simply adding the reciprocal of the mean free path of the electron-phonon interaction,

\begin{equation}\label{ephon}
\lambda_{\mathrm{phonon\index{Phonon}}}^{-1}=\frac{1}{a_0}\,\frac{\varepsilon_0\,-\,\varepsilon_{\infty}}{\varepsilon_0\,\,\varepsilon_{\infty}}\,\frac{W_{\mathrm{ph}}}{E}\,\frac{n(T)\,+\,1}{2}\,\ln\left[\frac{1\,+\,\sqrt{1\,-\,W_{\mathrm{ph}}/E}}{1\,-\,\sqrt{1\,-\,W_{\mathrm{ph}}/E}}\right]\,, 
\end{equation}
to the reciprocal of the IMFP \cite{1Ganachaud}, where $a_0$ is the Bohr radius, $W_{\mathrm{ph}}\,=\,\hbar\omega$ is the energy of the generated phonon, $E$ is the energy of the incident electron, $n(T)$ is the average number of phonons at temperature $T$ given by the Bose occupation distribution ($k_{\mathrm{B}}$ is the Boltzmann constant):

\begin{equation}
n(T)\,=\,\frac{1}{\exp{(\hbar\,\omega/k_{\mathrm{B}} T})\,-\,1}\,,
\end{equation}  
and $\varepsilon_{\mathrm{\infty}},\varepsilon_0$ are the high and zero frequency dielectric constants, respectively.
In principle, a more general equation using a specific phonon mode $\nu$ and frequency $\omega_{\bm{q},\nu}$ should be used to account for the dispersion with momentum transfer $\bm{q}$ instead of Eq. (\ref{ephon}). However, due to the complexity of this calculation and the importance of the electron-phonon interaction only at very low energy, reasonable results can also be obtained with a mean phonon excitation energy $\hbar\omega$, i.e. assuming almost flat dispersion modes.

In the ab initio method, the electron-phonon coupling strength is \cite{ZHOU2021107970}:
\begin{equation}\label{elphon}
t^{\rm el-ph}_{\bm{q},\nu}=\frac{1}{N(E_\mathrm{F})\omega_{\bm{q},\nu}}\sum_{mn}\int_{1\mathrm{BZ}}\frac{d{\bm k}}{\Omega_{\mathrm{BZ}}}|g_{mn,\nu}(\bm{k},\bm{q})|^2
\delta({\varepsilon_{n{\bm{k}}}}-E_{\mathrm{F}})
\delta({\varepsilon_{n{\bm{k}}+{\bm{q}}}}-E_{\mathrm{F}}),
\end{equation}
where $N(E_\mathrm{F})$ is the density of states per spin at the Fermi level $E_\mathrm{F}$, $\Omega_{\mathrm{BZ}}$ is the cell volume of the first Brillouin zone (1BZ) and $\omega_{\bm{q},\nu}$ are the phonon frequencies.
Eq. (\ref{elphon}) represents the response to a perturbation characterised by a wave vector ${\bm q}$ and a mode $\nu$, which acts on an electron in the quantum state $n$ with energy $\varepsilon_{n{\bm{k}}}$ and crystalline momentum $\bm{k}$ and leads to the band state $(m,\bm{k}+\bm{q})$ with energy $\varepsilon_{m{\bm{k}}+{\bm{q}}}$.
The electron-phonon matrix elements can be determined from first-principles simulations \cite{qe,elk} as follows:
\begin{equation}
g_{mn,\nu}(\bm{k},\bm{q})=\frac{1}{\sqrt{2\omega_{\bm{q},\nu}}}\bra{\psi_{m,{\bm k}+{\bm q}}({\bm r})}\partial_{{\bm q},\nu}\bm{V}^\mathrm{scf}({\bm r})\ket{\psi_{n,{\bm k}}({\bm r})},
\end{equation}
where $\ket{\psi_{n,{\bm k}}}$ and $\ket{\psi_{m,{\bm k}+{\bm q}}}$ represent the electronic wave functions that can be obtained as a result of a self-consistent density functional calculation and $\partial_{{\bm q},\nu}\bm{V}^\mathrm{scf}({\bm r})$ is the derivative of the self-consistent deformation potential associated with a phonon of wavevector ${\bm q}$, branch index $\nu$, and frequency $\omega_{\bm{q},\nu}$
\cite{baroni_book}. Recall that the $k$-point lattice must be dense enough to achieve good self-consistency and accurate phonon calculation, which may lead to high computational cost for this calculation.

Similarly, low-energy electrons can be trapped in the lattice of polarisable materials by exciting a quasiparticle, the so-called polaron. While the characterisation and modelling of polarons in solids based on first-principles simulations is feasible and implemented in several computational codes \cite{Franchini2021}, an effective semi-empirical formula to treat polaronic effects can be used as follows \cite{1Ganachaud}:

\begin{equation}
\lambda^{-1}_{\mathrm{pol}}\,=\,C\,e^{-\gamma\,E},
\label{pol}
\end{equation}
where $C$ and $\gamma$ are constants that depend on the specific dielectric material.

Inelastic electron-electron collisions can lead not only to energy losses but also to a change in the polar scattering angle. Using the classical model for binary collisions \cite{Dapor2023}, which is based on momentum and energy conservation, it can be shown that (i) the polar scattering angles of the primary electron ($\theta$) resulting from the collision (electrons can be distinguished in classical physics) and the scattered secondary electron ($\theta_s$) are complementary $\theta+\theta_s=\pi/2$ or, which is the same, $\sin\theta_s=\cos\theta$; (ii) the scattering angle $\theta$ can be derived by $W/E=\sin^2\theta$ i.e. $\theta =\arcsin\sqrt{W/E}$, where $W$ is the energy loss and $E$ is the kinetic energy of the incident electron before the inelastic collision. In the small-angle approximation, 
the value of the maximum momentum transfer corresponds to a cutoff of the polar scattering angle of $\sqrt{W/E}$ (Bethe ridge) \cite{EGERTON2017115}. 
We note that if you have access to the dispersion in ${\bm q}$, i.e., the ${\bm q}$-dependence of energy on both the modulus and the direction of momentum transfer (for example, using ab initio simulations), you could determine the inelastic scattering angle by directly using the differential inelastic scattering cross section in the momentum transfer ${\bm q}$ to go beyond the classical binary collision model. 
The ${\bm q}$-dependence can then be used directly for the calculation of the IMFP (and thus generalises Eq. (\ref{lineldiff}), which is written for a spherically symmetric potential). 

\subsubsection{Secondary electron emission}

In the energy straggling approach, the trajectory of the electron is tracked either until its kinetic energy falls below a threshold value that depends on the specific physical problem to be modelled, or until it is trapped in the solid or emerges from the surface (in the latter case, the electron is counted as part of the emission spectrum). In Electron Energy Loss  Spectroscopy (EELS), for example, the kinetic energy cut-off can be $(E_0-150)$~eV, since plasmons typically occur in the range [$E_0-150,E_0]$~eV ($E_0$ is the energy of the elastic peak). Conversely, to simulate the energy distribution and the yield of the secondary electrons, the electron trajectories must be followed until their kinetic energy is in the order of magnitude of the work function (electron affinity) of the analysed material.

\begin{figure}[hbt!]
\centering
\includegraphics[width=1.0\linewidth]{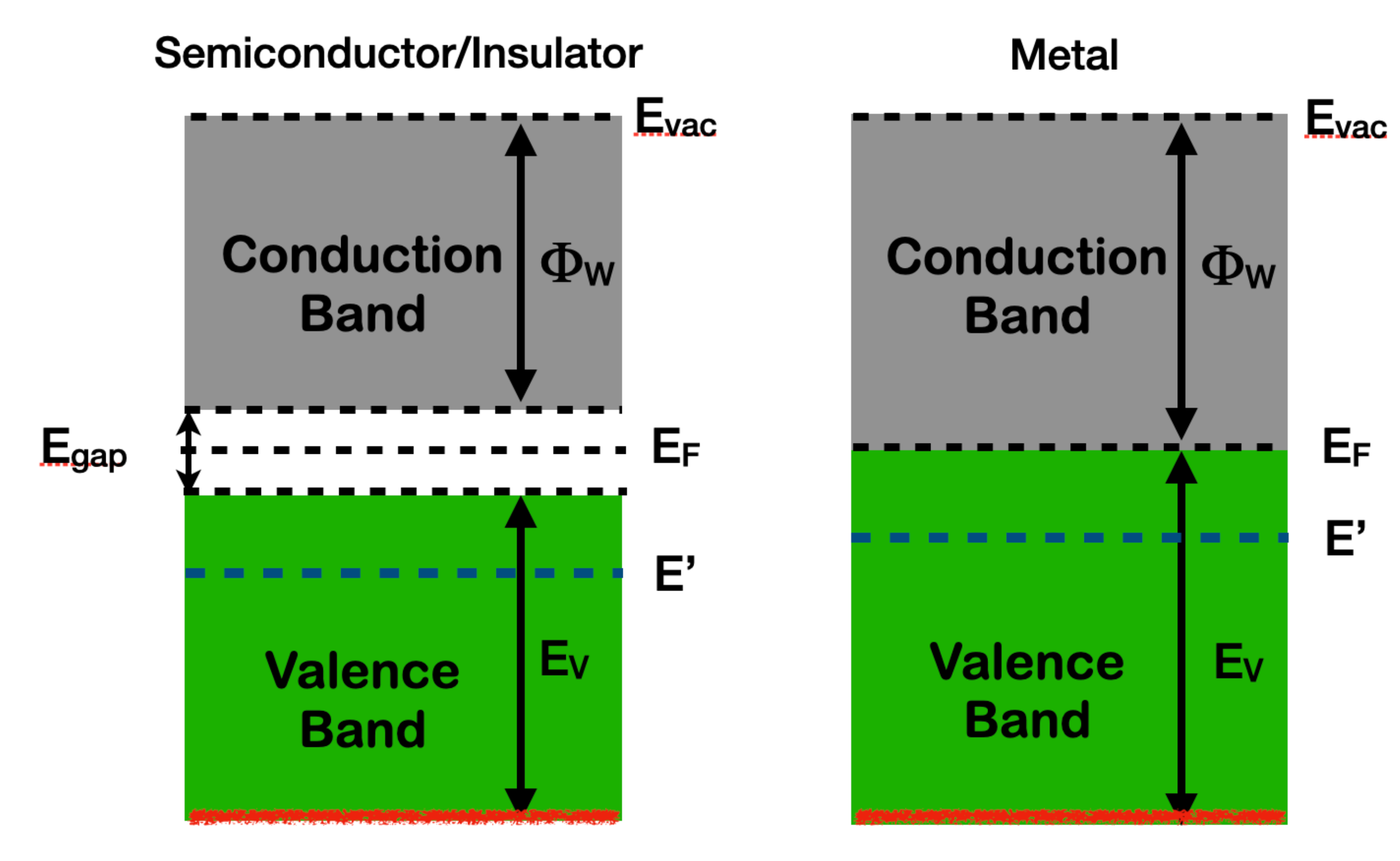}
\caption{Band structure, where $E_{\rm vac}$=vacuum level, $E_{\rm gap}$=band gap,$E_{\rm F}$=Fermi level, $\Phi_{\rm W}$= work function/electron affinity, $E_{\rm V}$=valence bandwidth. Adapted from \cite{KHAN2023112257}.}
\label{valcond}
\end{figure}

In order to escape from the surface of the sample, the electrons must overcome the barrier at the solid-vacuum interface, which can be modelled as a step potential. The transmission coefficient can be calculated as follows \cite{Shimizu_1992,Dapor2023}: 
\begin{equation}\label{transmission}
T = \frac{4\sqrt{1 - \Phi_{\rm W}/(E\cos^2 \vartheta)}}{\left[1 + \sqrt{1 - \Phi_{\rm W}/(E\cos^2 \vartheta)} \right]^2}.
\end{equation}
According to Eq. (\ref{transmission}) the electrons must obey certain angular energy conditions, i.e:
\begin{equation}\label{eq:condsecond}
E \cos^2 \vartheta \ge \Phi_{\rm W},
\end{equation}
where $\vartheta$ is the angle formed between the surface normal and the intercept of the electron path on the surface, $\Phi_{\rm W}$ is the work function (metals) or the electron affinity (semiconductors/insulators) of the material and $E$ is the residual kinetic energy of the electrons.
The conditions for the emission of electrons at the sample surface can be fulfilled according to our statistical scheme if $\mu_5 < T$, where $\mu_5$
is a random number that is sampled uniformly in the interval $[0,1]$. We find that the value of $\Phi_{\rm W}$ (see Fig. \ref{valcond}) is of utmost importance when modelling secondary emission and yield. In particular, a higher work function (electron affinity) leads to smaller calculated values of a secondary electron yield of a material.

We would like to point out that the electron energies in the solid in this paper refers to the bottom of the conduction band (the Fermi level $E_{\rm F}$ for a metal, see Fig. \ref{valcond}).
The secondary electron thus emerges from the inelastic collision with an energy $E_s=W-(E_{\rm gap}+E_{\rm v}-E')$, where $E'$ is the energy of the electron in the valence band \cite{KHAN2023112257}, whose bandwidth is $E_{\rm V}$.

In conclusion, we find that the typical number of electron trajectories $\cal N$ should be between 10$^7$ and 10$^{10}$ to achieve statistical significance when using the energy straggling strategy to simulate energy distribution spectra. This number depends crucially on the material, the initial energy and the desired signal-to-noise ratio, which increases with $\sqrt{\cal N}$.

Having discussed the MC workflow, we are now in a better position to describe the different approaches to calculate the elastic and inelastic scattering cross-sections.

\section{Elastic and inelastic scattering mechanisms: semi-empirical vs. ab initio methods}

When modelling charge transport in solids and biomaterials, which are of interest e.g. for hadron therapy to cure cancer, we are interested in all inelastic, elastic and absorption scattering mechanisms.
\begin{figure}[hbt!]
\centering
\includegraphics[width=1.0\linewidth]{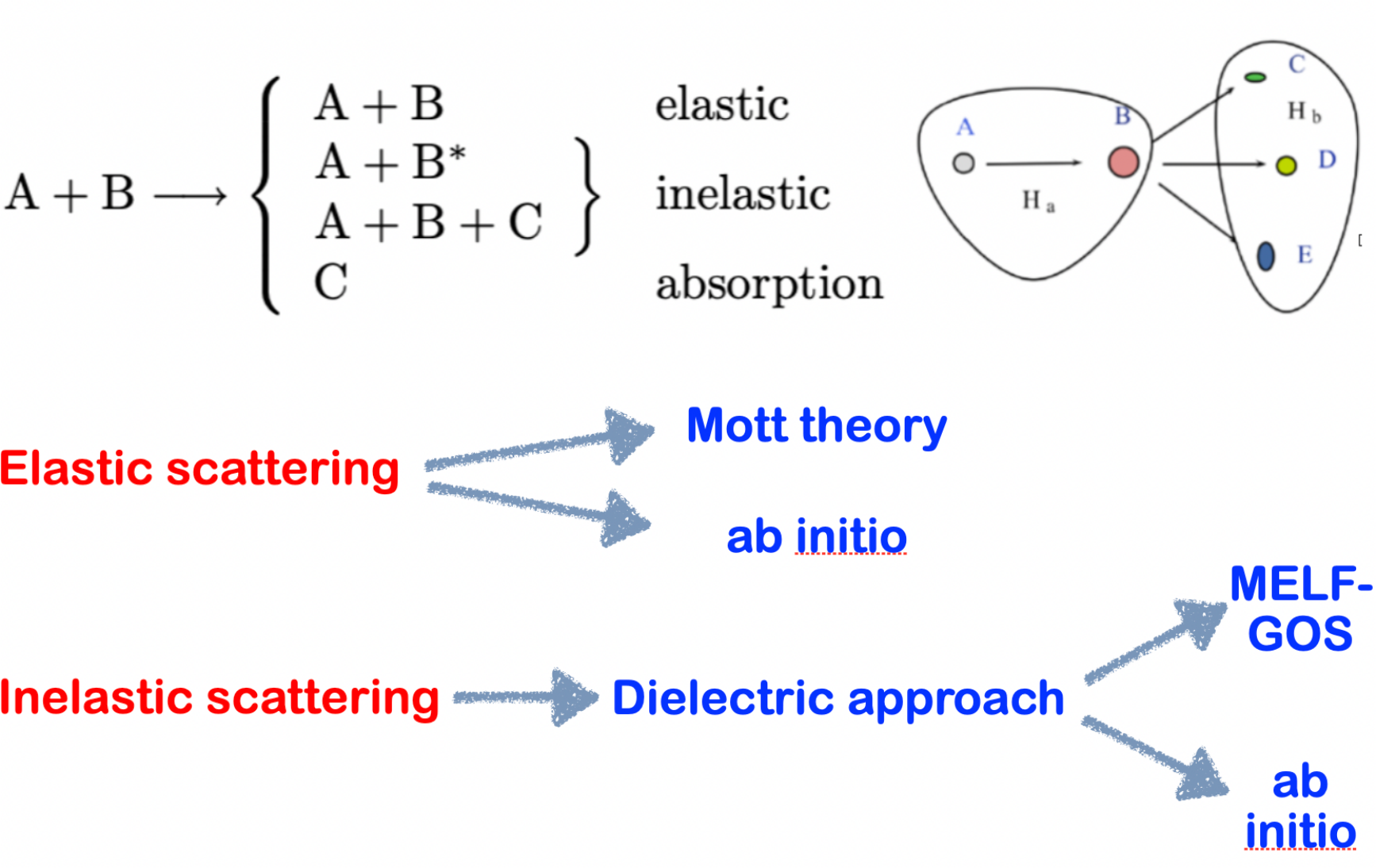}
\caption[]{Idealized scattering experiment}
\label{sketch_scatt}
\end{figure}

In Fig. \ref{sketch_scatt} we sketch an idealised scattering experiment in which an electron (A) -- or generally an ion -- moves towards a target (B). The outcomes of the collision can be summarised as elastic, inelastic and absorption scattering, including the possibility that the number of products (described by the free Hamiltonian $H_b$) may differ from that of the reactants (described by the free Hamiltonian $H_a$).

In elastic scattering, the number of particles remains the same before and after the collision. To a first approximation, the momentum modulus and the kinetic energy of the incident electron are also conserved.\footnote{Strictly speaking, in an elastic collision only the sum of the kinetic energies of the reactants, i.e. in this case the electron and the atom, and the products before and after the collision is conserved. Even if the mass of the electron is much smaller than the mass of the atom, the electrons emerge from the elastic collisions with a slightly lower kinetic energy due to the recoil. Such an effect is the basis of elastic peak electron spectroscopy (EPES) or electron Compton scattering (ECS). Since the mean recoil energy is inversely proportional to the mass of the target atom, the elastic peak of hydrogen exhibits the strongest shifts. Therefore, EPES has been proposed as a technique for the detection of hydrogen in polymers and hydrogenated carbon-based materials. See section \ref{elas_section} for a discussion.} Thus, the only result of elastic scattering between electron and target is the deflection of the electron trajectory within the system. In general, the effective screening of the electron cloud leads to a small angular deviation of the electron trajectory. When the kinetic energy transfer is very small, of the order of fractions of eV, electron scattering is often considered quasi-elastic (e.g. in electron-phonon scattering or when the atomic displacement energy is extremely small).

The elastic scattering event is modelled by calculating the differential elastic scattering cross-section (DESCS), which measures the probability of scattering into a solid angle.
A key difference in the calculation of the DESCS is whether the target is an atom, a molecule or a solid.
This is essentially related to the symmetry of the scattering potential, which is spherical in the case of an electron-atom collision.

In the latter case, one can indeed obtain an analytical formula for the DESCS in the framework of quantum mechanics by applying the partial wave expansion method, which is valid in both the relativistic and non-relativistic regimes \cite{3KesslerBook,3BurkeJoachainBook,3ELSEPA,Dapor2022}.
In particular, we will present the calculation of the DESCS in a relativistic quantum mechanical framework \cite{1Mott}, based on the solution of the Dirac equation in a central field (Mott theory). The spherical symmetry of the scattering potential further simplifies this solution, as the scattering wave function can be written as a product of angular and radial functions. Mott's theory generally provides a more accurate description of the elastic scattering of electrons impinging on atoms than the Rutherford formula, which is derived as a non-relativistic limit within the first Born approximation and can therefore only be safely used when the kinetic energy of the electrons is high and the scattering potential is weak (which typically corresponds to a low atomic number of the scattering centres). For accomplished reviews on this topic, the interested reader can take a look at the following reports \cite{Dapor2023,1Egerton,3KesslerBook,3BurkeJoachainBook,3ELSEPA,Dapor2022,3Rileyetal,3Sigmund,3EgertonII,3Jablonski,MayolSalvat1997}.

In this context, we will also describe a novel ab initio method \cite{triggiani2023elastic,taioli2021relativistic} for calculating the elastic scattering cross-section, which is based on the formal theory of scattering and the use of localised basis sets to implement the Dirac-Hartree-Fock mean-field method (see section \ref{abinito}). Compared to Mott's theory, this approach has the advantage that it can be extended to non-spherical symmetric scattering potentials, such as molecules and solids and works well in the low energy range. However, its application in the high energy range is cumbersome due to the difficulty of representing strongly oscillating scattering wave functions with the localised Gaussian functions used in this approach. In principle, this problem could be solved by expanding the scattering wave function with Hermite Gaussian functions modulated by plane waves \cite{Colle1988HermiteGF}.

Inelastic scattering leads to a loss of energy and momentum as well as a deflection of the electron trajectory. For example, impinging electrons can be slowed down by exciting either collective charges, such as plasmons, or individual orbital electrons of the atomic constituents of the solid target.

The theoretical and computational modelling of the energy losses of electrons in materials can be carried out using the Bethe-Bloch stopping power formula \cite{3Bethe} (see section \ref{bethe_block}), semi-empirical approaches \cite{3Lane,3Kanaya} and the dielectric theory of Ritchie \cite{1Ritchie57} (see section \ref{Ritchie}). These methods work for relatively high kinetic energies of the electrons. In this energy range, the trajectory of the point-like charged particle is driven by electromagnetic fields, which are determined by Maxwell's equations. 

The Bethe-Bloch formula is based on the assumptions underlying the Drude model of electrical conduction in a solid. This model considers the bound electrons in an atom as driven, damped harmonic oscillators and fails at low kinetic energies of the electrons, as it is based on the first-order Born approximation (see section \ref{bethe_block}).

Conversely, the dielectric or Ritchie's approach is based on the fundamental idea that the dielectric function (actually, the inverse imaginary part of the dielectric function) determines the spectra of electron energy losses.
Within this theoretical framework, which is also valid at relatively low kinetic energy (i.e. below the $K$-shell binding energy of the constituents of the sample), the theory comes in different flavours, which differ in the way the dielectric function is evaluated as a function of energy and momentum transfer. The latter can be calculated either with empirical parameterisations of measured ELF data (typically recorded for vanishing momentum transfer), such as the Drude-Lorentz model (see section \ref{drudino}), which can be extended to finite momentum by including semi-empirical (see  Sections \ref{drudino} and \ref{pennino}) or analytical (see section \ref{ELF_exp2}) dispersion laws, or by first-principles simulations (see section \ref{ELF_ab}).

We emphasise here that the contributions to the energy loss can come from all microscopic inelastic scattering mechanisms, such as excitation and ionisation processes, electron-polaron, electron-phonon and electron-plasmon interactions, and merge into a measurable macroscopic quantity, namely the IMFP. The latter quantity can be measured and, together with the EMFP, represents an essential input for TMC.

\subsection{Elastic scattering: Mott's theory}\label{elasmott}

The DESCS can be calculated using the Mott's theory \cite{1Mott}. This approach is based on either the analytical or the numerical solution of the Dirac equation
\begin{eqnarray}\label{rotinv}
H\Psi(\bm{r})&=&E_{\gamma,\kappa}\Psi(\bm{r}),\label{rotinv11}
\end{eqnarray}
where ${\gamma,\kappa}$ are the energy and the relativistic quantum number that identify the state. The relativistic Hamiltonian for a charge in an electromagnetic field is ($c=1$):
\begin{equation}\label{Dirac_eq}
H\,=\,e\varphi+{\bm \alpha}\cdot({\bm p}-e{\bm A})+\beta m_e\;.
\end{equation}
In Eq. (\ref{Dirac_eq}), ${\bm{\alpha}}$ and $\beta$ are the $4 \times 4$ Dirac matrices \cite{Dapor2022,taioli2021relativistic},
$(\varphi,{\bm A})$ stands for the electromagnetic four-potential, $m_e$ for the electron mass and ${\bm p}$ for the momentum.

This method is particularly suitable for the treatment of electron-atom collisions due to the rotational invariance of the atomic potential $\varphi(r)=-Ze/r$ (we recall that this expression applies to one-electron atoms, which include hydrogen, singly or multiply ionised elements. For many-electron atoms we have to take the electron screening into account by multiplying $\varphi(r)$ by a screening function such as a Yukawa-type function \cite{Salvat1987}, or considering it as a mean-field Dirac-Hartree-Fock potential).

In relativistic quantum mechanics, the total angular momentum $\bm{J}={\bm{L}}+(1/2)\bm{S}$ of a rotationally invariant system commutes with the Dirac-Hamiltonian, Eq. (\ref{Dirac_eq}). Here $\bm{L}$ is the orbital angular momentum, $\bm{S}=\rm{diag}(\bm{\sigma},\bm{\sigma})$ is the $4\times 4$ spin-angular momentum matrix and $\bm{\sigma}=(\sigma_x, \sigma_y,\sigma_z)$ is the vector whose components are the $2\times 2$ Pauli matrices.

The eigenvalues of ${\bm J}, J_z$ together with the energy thus represent good quantum numbers to label the quantum states (and not $\bm{L}, L_z$ as in Schr{\"o}dinger's mechanics).
For a central potential, the solutions of the Dirac equation (\ref{rotinv}) can be factorised into the product of radial and angular functions \cite{Greiner} as follows:

\begin{eqnarray}\label{rotinv2}
\Psi_{\gamma,\kappa,\mu}(\bm{r})&=& \left(\begin{array}{c}
\psi_{\gamma,\kappa,{\rm L}}\left(r\right)\chi_{\kappa,\mu}\left(\vartheta,\varphi\right)\\
\psi_{\gamma,\kappa,{\rm S}}\left(r\right)\chi_{-\kappa,\mu}\left(\vartheta,\varphi\right)
\end{array}\right)=
\left(\begin{array}{ll} \frac{u_{\kappa}(r)}{r} \  \chi_{\kappa,\mu} (\Omega) \\ i \frac{v_{\kappa}(r)}{r} \  \chi_{-\kappa,\mu}(\Omega) \end{array}\right),\label{rotinv1}
\end{eqnarray}
\noindent where $\chi_{\kappa,\mu}$ is the tensor product of orbital and spin spherical harmonics, each characterised by the Dirac quantum numbers $(\kappa,\mu)$ and ${\rm L,S}$ are the large and small bi-component parts of the Dirac spinor.
We recall that, for $j=l+1/2$ (spin up),
\begin{equation}
\kappa=-\left(j+\frac{1}{2}\right)=-(l+1)
\end{equation}
and, for $j=l-1/2$ (spin down), 
\begin{equation}
\kappa=\left(j+\frac{1}{2}\right)=l\;.
\end{equation}
In Eq. (\ref{rotinv1}) the angular eigenfunctions fulfill the following completeness relation
\begin{equation}
\sum_{m}\frac{4\pi}{2l+1}\int\chi_{\kappa,m}^{+}\left(\vartheta^{\prime},\varphi^{\prime}\right)\chi_{\kappa,m}\left(\vartheta^{\prime},\varphi^{\prime}\right)Y_{l,0}^{*}\left(\vartheta^{\prime},\varphi^{\prime}\right)d\Omega^{\prime}=2\left|\kappa\right|\sqrt{4\pi}\delta_{l0}\,.
\end{equation}
By using the factorization (\ref{rotinv2}) and considering for the sake of simplicity a system of many-fermions interacting via a central potential (e.g. the spherically symmetric Dirac-Hartree-Fock mean-field potential) the Dirac equation (\ref{rotinv}) in matrix form reads

\begin{equation}
\frac{\partial}{\partial r}\left(
\begin{array}{c}
u_{\kappa}(r)\\
v_{\kappa}(r)
\end{array}\right)=\left(\begin{array}{cc}
-\frac{\kappa}{r} & \frac{E}{c}-\frac{V(r)}{c}+c\\
\frac{V(r)}{c}+c-\frac{E}{c} & \frac{\kappa}{r}
\end{array}\right)\left(\begin{array}{c}
u_{\kappa}(r)\\
v_{\kappa}(r)
\end{array}\right)\label{DHF}
\end{equation}

\noindent with the total potential energy:  

\begin{equation}\label{potdhf}
V(r)=-\frac{Z}{r}+\int \frac{\rho(r')}{|r-r'|}d^3 r'-V_{\mathrm{ex}}(r,r'),
\end{equation}
which is given by the sum of the nuclear potential energy ($Z$ is the atomic number), the Hartree potential ($\rho(r)$ is the electron density at $r$) and the 
exchange-correlation potential $V_{\mathrm{ex}}(r,r')$. In principle, the last term is a non-local Fock exchange potential that can be approximated by a local density approximation (LDA) for the electron gas \cite{Salvat1987,Slater1951}, which reads:

\begin{equation}\label{rho13}
    V_{\mathrm{ex}}=\frac{9}{4} \left[ \frac{3}{\pi} \rho (r) \right]^{1/3}.
\end{equation}

We emphasise that the Dirac equation (\ref{DHF}) applies in general to any other centrally symmetric potential, such as the bare Coulomb potential, the Yukawa potential (which can represent the screened Coulomb interaction) and the Wood-Saxon potential \cite{taioli2021relativistic,Woodsaxon} (which can approximate the screened nucleon-nucleon interaction).

The radial Eq.~(\ref{DHF}) with the local interaction of Eq.~(\ref{rho13}) can be solved self-consistently, and the eigenvalues are the electron binding energies.
To numerically integrate Eq. (\ref{DHF}), Runge-Kutta method or the method of collocations \cite{taioli2021relativistic,morresi2018nuclear} can be used.
As a result of the integration of Eq.~(\ref{DHF}), the four-component spinor given by Eq.~(\ref{rotinv2}) and the atomic potential of Eq.~(\ref{potdhf}), which also represents the scattering potential for an impinging particle,  are known. Usually a radial grid [0,$R$] is used, on which the scattering potential can be fitted to a sum of Yukawa potentials \cite{3ELSEPA}.
In scattering problems, it is common to analyse the behaviour of the scattering wavefuntions at an observation point far from the scattering centre to assess the effect of the potential after it has decayed (we assume that it is a well-behaved potential that dies off faster than $\approx 1/r^2$). Therefore, the endpoint $R$ must be large enough to represent the asymptotic behaviour of the scattering wave functions. 

From the formal theory of scattering, the asymptotic behaviour of the Dirac spinor at a point far outside the range of the central potential (e.g. the screened Dirac-Hartree-Fock atomic potential) is as follows:

\begin{equation}\label{asympt}
\Psi_i
\begin{array}[t]{c}
\sim \\
{\scriptstyle r \rightarrow \infty}
\end{array}
a_i \exp(iKz)+b_i(\theta,\varphi)\frac{\exp(iKr)}{r}\;,
\end{equation}
where $K\,=\,m_ev/\hbar$ is the electron momentum resulting from the relativistic dispersion relation $K^2=E^2-m_e^2$ and, for simplicity, the incident direction points in the $z$ axis. Recall that $\Psi_i~( i=1,2,3,4)$ in relativistic quantum mechanics is a four-component spinor, with bi-dimensional large and small components. Therefore, the asymptotic behaviour of Eq.~(\ref{asympt}) characterises all four components of the scattered wave.
If we restrict the discussion to the large part of the Dirac spinor ($i=1,2$), we can obtain the asymptotic condition for the spin parallel to the direction of incidence (spin up ) by setting
$a_1\,=\,1$, $a_2\,=\,0$, $b_1\,=\,f^+(\theta,\varphi)$, $b_2\,=\,g^+(\theta,\varphi)$, where $f^+$ and $g^+$ are the direct and spin-flip scattering amplitudes, which can be determined by the relativistic partial wave expansion method \cite{3ELSEPA,Dapor2022}. The asymptotic behaviour in Eq. (\ref{asympt}) is given, in this particular case, by

\begin{equation}\label{psi1_sc}
\Psi_1
\begin{array}[t]{c}
\sim \\
{\scriptstyle r \rightarrow \infty}
\end{array}
\exp(iKz)+
f^+(\theta,\varphi)\frac{\exp(iKr)}{r}\;,
\end{equation}

\begin{equation}\label{psi2_sc}
\Psi_2
\begin{array}[t]{c}
\sim \\
{\scriptstyle r \rightarrow \infty}
\end{array}
g^+(\theta,\varphi) \frac{\exp(iKr)}{r}\;.
\end{equation}

The solution of Dirac's equation (\ref{DHF}) in a central field must correspond to the asymptotic behaviour given by Eqs. (\ref{psi1_sc}) and (\ref{psi2_sc}) at a distance large enough for the potential, Eq.~ (\ref{potdhf}), to be negligible.
Due to the central symmetry of the atomic potential, such a solution can be obtained by expanding the relativistic scattering wave functions $\Psi_1$ and $\Psi_2$ in spherical harmonics as follows:

\begin{equation}\label{psi1_g}
\Psi_1=\sum_{l=0}^{\infty}[A_lG_l^++B_lG_l^-]P_l(\cos\theta)\;,
\end{equation} 

\begin{equation}\label{psi2_g}
\Psi_2=\sum_{l=1}^{\infty}[C_lG_l^++D_lG_l^-]P_l^1(\cos\theta)\exp(i\varphi)\;,
\end{equation}
where $P_l^1(\cos\theta)$ are the associated Legendre polynomials
\begin{equation}
P_l^1(x)\,=\,(1\,-\,x^2)^{1/2}\frac{dP_l(x)}{dx},
\end{equation}
and
\begin{equation}\label{asymptotic_1}
G_l^{\pm}
\begin{array}[t]{c}
\sim \\
{\scriptstyle r \rightarrow \infty}
\end{array}
j_l(Kr) \cos \delta_l^{\pm} - n_l(Kr)\sin \delta_l^{\pm}\;.
\label{asyG} 
\end{equation}
$\delta_l^{\pm}$ are the so-called phase shifts and encode the effect of the scattering potential;
$j_l$ and $n_l$ are the regular and irregular spherical Bessel functions whose asymptotic behaviour is

\begin{equation}
j_l(Kr) 
\begin{array}[t]{c}
\sim \\
{\scriptstyle r \rightarrow \infty}
\end{array}
\frac{1}{Kr}\sin\left(Kr-\frac{l\pi}{2}\right)\;,
\label{assbessel}
\end{equation}
\begin{equation}\label{asymptotic_2}
n_l(Kr) 
\begin{array}[t]{c}
\sim \\
{\scriptstyle r \rightarrow \infty}
\end{array}
-\frac{1}{Kr}\cos\left(Kr-\frac{l\pi}{2}\right)\;,
\end{equation}
that is, from Eq. (\ref{asymptotic_1}),

\begin{equation}
G_l^+
\begin{array}[t]{c}
\sim \\
{\scriptstyle r \rightarrow \infty}
\end{array}
\frac{1}{Kr}\sin\left(Kr-\frac{l \pi}{2}+\delta_l^+\right)\;,
\label{gipiu}
\end{equation}
and
\begin{equation}
G_l^-
\begin{array}[t]{c}
\sim \\
{\scriptstyle r \rightarrow \infty}
\end{array}
\frac{1}{Kr}\sin\left(Kr-\frac{l \pi}{2}+\delta_l^-\right).
\label{gimeno}
\end{equation}

\noindent We note that such an expansion is similar to the partial wave expansion method used in non-relativistic quantum mechanics for the treatment of scattering from a central potential, with the notable differences that: (i) in the non-relativistic case, the angular momentum ${\bm l}$ is a good quantum number (while the total angular momentum ${\bm j}={\bm l}+{\bm s}$ is conserved in the relativistic approach); (ii) the wave function can be represented by a two-component vector (while it is a four-component spinor in the relativistic approach).

The phase shifts $\delta_l^{\pm}$ represent the effect of the scattering potential of Eq. (\ref{potdhf}) on the outgoing waves after elastic collisions, which corresponds to a phase change.
The coefficients $A_l$, $B_l$, $C_l$ and $D_l$ can be determined by fitting the asymptotic behaviour of Eqs. (\ref{psi1_sc}) and (\ref{psi2_sc}) to Eqs. (\ref{psi1_g}) and (\ref{psi2_g}), which results in
\begin{eqnarray}
A_l=(l+1)i^l\exp(i\delta_l^+)\;, \label{mio} \\
B_l=l i^l\exp(i\delta_l^-)\;,\label{tuo} \\
C_l=-i^l\exp(i\delta_l^+)\;, \label{suo}\\
D_l=i^l\exp(i\delta_l^-)\;\label{loro}.
\end{eqnarray}
Finally, substituting Eqs. (\ref{mio}), (\ref{tuo}), (\ref{suo}) and (\ref{loro}) in Eqs. (\ref{psi1_g}) and (\ref{psi2_g}) one obtains:
\begin{equation}\label{ggg}
\Psi_1=\sum_{l=0}^{\infty}[(l+1)\exp(i\delta_l^+)G_l^+
+l\exp(i\delta_l^-)G_l^-]i^lP_l(\cos\theta)\;,
\end{equation}
\begin{equation}\label{gggg}
\Psi_2=\sum_{l=1}^{\infty}[\exp(i\delta_l^-)G_l^--\exp(i\delta_l^+)G_l^+]i^lP_l^1(\cos\theta)
\exp(i\varphi)\;.
\end{equation}
By comparison to the Eqs. (\ref{psi1_sc}) and (\ref{psi2_sc}), the Eqs. (\ref{ggg}) and (\ref{gggg}) give the following expressions for the scattering amplitudes of electrons that hit atomic targets with a spin parallel to the direction of motion:
\begin{eqnarray}
f^+(\theta,\varphi)
&&=f^+(\theta)\nonumber\\
&&=\frac{1}{2iK}\sum_{l=0}^{\infty}
\{(l+1)[\exp(2i\delta_l^+)-1]\nonumber\\
&&+l[\exp(2i\delta_l^-)-1]\}P_l(\cos\theta)\;,
\end{eqnarray}
\begin{equation}
g^+(\theta,\varphi)=\frac{1}{2iK}\sum_{l=1}^{\infty}
[\exp(2i\delta_l^-)-\exp(2i\delta_l^+)]P_l^1(\cos\theta)\exp(i\varphi)\;.
\end{equation}
With this result, the differential elastic scattering cross-section can be determined as the fraction between the number of particles scattered in $d\Omega$ per unit time (the second term of Eq. (\ref{asympt})) and the incident flux \cite{Dapor2023}:

\begin{equation}\label{differ}
\frac{d\sigma_{\mathrm{el}}}{d\Omega}\,=\,\frac{|b_1|^2+|b_2|^2}{|a_1|^2+|a_2|^2}=|f^+|^2+|g^+|^2\,.
\end{equation}
\noindent We note that Mott's approach neglects the recoil of the proton. Similar expressions for the differential elastic scattering cross-section can also be obtained for completely unpolarised electron beams

\begin{equation}
\frac{d\sigma_{\mathrm{el}}}{d\Omega}=|f|^2+|g|^2\,,
\label{decs}
\end{equation}
where $f=f^+$ and $g=g^+\exp(-i\varphi)$ \cite{3KesslerBook,Dapor2022}.
{\footnote {Please note that, in general, the differential elastic scattering cross-section is given by $$
\frac{d\sigma^{\rm el}}{d\Omega}=(|f|^2+|g|^2)
[1+S(\vartheta){\bm P} \cdot \hat{\bm n}]\,,
$$
where 
$$
S(\vartheta)=i\frac{fg^*-f^*g}{|f|^2+|g|^2}
$$
is the Sherman function, ${\bm P}$ is the spin-polarisation and $\hat{\bm n}$ is the unit vector normal to the plane of scattering. If the beam is completely unpolarised, then ${\bm P}$=0 and we obtain Eq. (\ref{decs}) \cite{3ELSEPA,Dapor2022}.
}
The problem of determining the DESCS by Eq. (\ref{decs}) is thus reduced to calculate the phase shifts $\delta_l$. Normally, they are obtained via the asymptotic form of the wave function, which is calculated by solving the Schr{\"o}dinger equation (or Dirac equation) within the scattering region with specific boundary conditions.

Finally, the total ESCS ($\sigma_{\mathrm{el}}$) and the transport cross-section ($\sigma_{\mathrm{tr}}$) can be determined by numerical integration as follows:
\begin{equation}
\sigma_{\mathrm{el}}=2\,\pi\,\int_0^{\pi}\frac{d\sigma_{\mathrm{el}}}{d\Omega}\sin\theta\,d\theta\;,
\label{sitot}
\end{equation} 
\begin{equation}
\sigma_{\mathrm{tr}}=2\,\pi\,\int_0^{\pi}(1-\cos\theta)\frac{d\sigma_{\mathrm{el}}}{d\Omega}\sin\theta\,d\theta\;.
\label{sitr}
\end{equation}

Mott's formula, Eq.~\ref{decs}, has known mathematical limits when the electron is non-relativistic (the electron rest mass $m_e >> E$) and the first Born approximation for the scattering amplitude can still be applied (high energy of the incident particle and a low atomic number of the target system).
This limit value of Eq. (\ref{decs}) is known as the screened Rutherford cross-section. It can be derived by considering the scattering of a non-relativistic particle in the static Coulomb potential of a proton without considering the interaction due to the intrinsic magnetic moments of the electron or proton, i.e.
only the interaction between the electric charges of the particles counts.

In particular by modelling the electron scattering from a screened Coulomb atomic potential via the following Yukawa potential:
\begin{equation}\label{Yukawa}
V(r)=\frac{V_0}{\mu r}e^{-\mu r},
\end{equation}
where $V_0$ and $1/\mu$ refer to the strength and range of the potential ($V\rightarrow 0$ for $r>> 1/\mu$), in the framework of the first Born approximation, which is valid for
$E\,\gg\,e^2Z^2/(2 a_0)$ \cite{3Bohm} ($a_0$ is the Bohr radius), the DESCS reads:

\begin{equation}
\frac{d\sigma_{\mathrm{el}}}{d\Omega}\,=\,\frac{Z^2 e^4}{4 E^2}\,\frac{1}{(1\,-\,\cos\theta\,+\,\alpha_s)^2}\,,
\label{Rutherford}
\end{equation}
where
\begin{equation}
\alpha_s\,=\,\frac{m_e e^4 \pi^2}{h^2} \frac{Z^{2/3}}{E}
\end{equation}
is the correction factor for the screening, $h$ is the Planck's constant, $e$ is the electron charge and $m_e$ is the electron mass.
We note that the bare Coulomb potential can be recovered by taking the limit $\mu \rightarrow 0$, provided that $V_0/\mu=-Ze^2$ is a fixed constant ($Z$ stands for the target atomic number and $e$ for the electron charge). Eq. (\ref{Yukawa}) corresponds to an exponential attenuation of the nuclear potential as a function of the distance $r$ from the centre of mass of the nucleus and can therefore represent a model of the electron-atom interaction in the presence of electronic screening.
The analytic closed form of Eq.~(\ref{Rutherford}) corrects the formula established by Rutherford for the scattering cross-section of charged hard spheres, which diverges at very small angles.

We note that there are three relevant problems when using Eq.~(\ref{sitot}) for the calculation of the total elastic scattering cross-section starting from the Mott expression (\ref{decs}).

The first problem when dealing with elastic scattering in condensed matter systems is the excessively high values of the Mott elastic scattering cross-section found at very low kinetic energy of the beam, despite the presence of the screened Coulomb potential of Eq. (\ref{potdhf}). This corresponds to small EMFPs (often smaller than the interatomic distance in solid targets). To correct this behaviour, one can use a truncation function, according to Ganachaud and Mokrani \cite{1Ganachaud}, to multiply the ESCS:
\begin{equation}
R_c(E)\,=\,\tanh[\alpha_c(E^2)]\,,
\label{cutoff1}
\end{equation}
where $\alpha_c$ is a parameter that determines the slope of the cut-off (e.g. $\alpha_c=0.003$ eV$^{-2}$ for graphite \cite{azzolini2018anisotropic}). Such a cut-off function can be written for a gapped solid \cite{1Ganachaud}:
\begin{equation}
R_c(E)\,=\,\tanh[\alpha_c(E/E_g)^2]\,.
\label{cutoff2}
\end{equation}

In this way, the ESCS is forced to become negligible at low kinetic energy of the electrons \cite{Dapor2023,1Ganachaud,DAPOR201895} (we recall that in this energy range the quasi-elastic electron-phonon scattering cross-section increases).

A second problem is the polarisation of the electron cloud of the target atom due to the close passage of a charge. 
Within the Mott approach, this effect can be treated semi-empirically by using an approximate form of the polarisation potential~\cite{3ELSEPA}. Furthermore, if the projectile is an electron beam, one must also consider exchange correlation effects \cite{3ELSEPA}.

A third problem is the fact that the relativistic Mott theory is a genuinely atomic approach. In fact, the form of the Dirac equation (\ref{DHF}) is valid for a central symmetric potential (as in Eq. (\ref{potdhf})). Nevertheless, one often has to model electron- or ion-molecule scattering where radial symmetry is lost, e.g. for determining the dose delivered by ion beams in liquid water at energies relevant for hadron therapy
\cite{taioli2020relative,de2022simulating,triggiani2023elastic,taioli2021relativistic}.
In this respect, Eq. (\ref{decs}) must be extended to take into account the DESCS of electrons impinging on individual molecules by using the following formula \cite{3ELSEPA}:

\begin{equation}\label{molsca}
\frac{d\sigma_{\mathrm{el}}(E,\theta)}{d\Omega}\,=\,\sum_{m,n}\,\exp{(i {\bm q} \cdot {\bm r}_{mn})}\,[f_m(\theta) f^*_n(\theta)\,+\,g_m(\theta) g^*_n(\theta)]\,,
\end{equation}
where $f_{m,n}(\theta)$ and $g_{m,n}(\theta)$ are the direct and spin-flip scattering amplitudes obtained by solving the Dirac equation for the atomic constituents $m,n$ of the molecule (or solid), ${\bm r}_{mn}\,=\,{\bm r}_m\,-\,{\bm r}_n$ is the $m$ to $n$ bond length and ${\bm q}$ and $\theta$ are the momentum transfer and scattering angle.
Eq. (\ref{molsca}) generalises Eq. (\ref{decs}) to the molecular case by using the sum of the scattering amplitudes of the individual atomic species, weighted by a phase term, instead of the scattering probabilities to account for interference and intramolecular scattering. However, we emphasise that in this framework the scattering potential is represented by a mere superposition of independent atomic centres as if they were not bound in a molecule, thus neglecting the effect of the overlap between the long-range tail of the atomic Coulomb interaction. In this respect, the effects of multiple scattering (or rescattering) due to the presence of surrounding molecules, e.g. in liquid water, can only be approximated by averaging over all possible spatial orientations \cite{taioli2020relative,de2022simulating,3ELSEPA,Dapor2022}.
Furthermore, the presence of a permanent dipole moment for some molecular systems, e.g. in water, which makes a strong contribution to the DESCS, is not automatically included in the generalised Mott's approach, Eq. (\ref{molsca}). The dipole interaction is usually added by semi-empirical corrections \cite{3ELSEPA}. The contribution of this long-range interaction to the elastic scattering cross-section is also an open problem in electron-atom scattering \cite{McEachran2003}.

To overcome these shortcomings, one must rely on a complete ab initio treatment of the intermolecular and intramolecular scattering interaction.

\subsection{Elastic scattering: ab initio methods}\label{abinito}

In the context of the formal theory of scattering, the DESCS can be written in the form of the on-shell $T-$operator as follows \cite{taioli2010electron,taioli2021relativistic}: 
\begin{equation}\label{eq:diff_elastic_cross_sect}
\frac{d\sigma_{el}}{d\Omega}=\frac{m_e^2}{4\pi^2}|\langle{\phi_{k\hat{ \bm n}}}|T^+(E)|\phi_{\bm{k}}\rangle|^2=\frac{m_e^2}{4\pi^2}|\langle{\phi_{{k\hat {\bm n}}}}|V|{\psi^+_{\bm{k}}}\rangle|^2,
\end{equation} 
where $m_e$ is the mass of the impinging electron, $\phi_{k{\hat{\bm n}}}({\bm r})=e^{ik{\hat{\bm n}}\cdot{\bm{r}}}$ and $\phi_{\bm{k}}({\bm{r}})=e^{i{\bm{k}}\cdot{\bm r}}$ are respectively the free outgoing and incoming wave functions in the position basis of an electron, which is far outside the range of the scattering potential $V$ along the direction defined by the unit vectors ${\hat{\bm n}}$ after the scattering (at time $t=+\infty$) and $\bm{k}$ long before the scattering (at time $t=-\infty$); $\psi^{+}_{\bm{k}}=\psi^{+}_{\bm{k}}(E)$ is the scattering wave function at energy $E=\hbar^2k^2/2m_e$, which is formally connected to the free wave by the following relation:
\begin{equation}\label{moller}
\psi^{+}_{\bm{k}}(E)=\Omega_+(E)\ket{\phi^{+}_{\bm{k}}(E)},
\end{equation}
where $\Omega_+(E)=\lim_{t\rightarrow -\infty}e^{iHt}e^{-iH_0t}=I+G_+(E)V$ is the M{\"o}ller operator and $G_+(E)=1/(E-H)=\lim_{\eta\rightarrow 0}1/(E-H+i\eta)$ is the Green's function or the resolvent of the complete Hamiltonian $H=H_0+V=\hbar^2k^2/2m_e+V$ and $\eta$ is a positive infinitesimal number.
The M{\"o}ller operator maps the free asymptotic eigenstates of $H_0$ into the scattering states of $H$ at the same positive energy $E$ (elastic scattering processes are defined by energy conservation) and fulfils the relation \cite{Newton}:
\begin{equation}\label{moller2}
H\Omega_+=\Omega_+H_0.
\end{equation}
The operator equivalence given in Eq.~(\ref{moller2}) leads to the Schr{\"odinger equation when applied to the free wave $\ket{\phi_{\bm{k}}}$ on both sides:
\begin{equation}\label{moller3}
H\ket{\psi^+_{\bm{k}}(E)}=E\ket{\psi^+_{\bm{k}}(E)},
\end{equation}
which shows that the scattering wave functions are eigenstates of the full Hamiltonian $H$.
Eqs. (\ref{moller}) and (\ref{moller3}) are thus equivalent formulations of the problem of elastic scattering.
Eq. (\ref{moller}) can be written in configuration space as:
\begin{equation}\label{green_90}
\psi^+_{\bm k}({\bm r})=e^{i{\bm{k}}\cdot{\bm r}}+\int d{\bm r'} G_0(E,{\bm r},{\bm r'})V({\bm r'})\psi^+_{\bm k}({\bm r'}),
\end{equation}
where $G_0(E,{\bm r},{\bm r'})=\bra{\bm r}\frac{1}{E-H_0+i\eta}\ket{\bm r'}=-\frac{m_e}{2\pi}\frac{e^{ik|{\bm r}-{\bm r'}|}}{|{\bm r}-{\bm r'}|}$ is the Green's function or propagator associated with the free Hamiltonian in the position basis. The first-order Born approximation, which is valid at high kinetic energy and weak scattering potential (e.g. low atomic number of target components), consists in the approximation of $V\ket{\psi^+_{\bm k}}$ by $V\ket{\phi^+_{\bm k}}$. From the so-called Lippmann--Schwinger equation (LS) for the scattering wave function Eq. (\ref{green_90}), the asymptotic behaviour of Eq. (\ref{asympt}), i.e. a free wave plus an outgoing spherical wave, can be obtained provided that the potential is of short range. The effect of the potential is contained in the coefficient in front of the second term in Eq. (\ref{asympt}), the square of which is the DESCS.

We note that knowledge of the transition operator for a given energy is crucial, as it is directly related to the ESCS (see Eq. (\ref{eq:diff_elastic_cross_sect})), which is a measurable observable.
The on-shell transition operator that appears in Eq. (\ref{eq:diff_elastic_cross_sect}), can also be written in the form of the M{\"o}ller operator as $T(E)=V\Omega_+$.
Thus, the scattering problem can be approached either by calculating the Green's function of the full Hamiltonian, through which we can calculate the M{\"o}ller operator and then the transition matrix element given in Eq. (\ref{eq:diff_elastic_cross_sect}), or by calculating the scattering wave function via the numerical solution of Eq. (\ref{moller3}) for positive eigenvalues $E$, where $H$ is either the many-body Schr{\"o}dinger or Dirac Hamiltonian (see Eq. (\ref{DHF})).

The solution of Eq. (\ref{moller3}) can be obtained using either the Hartree-Fock (HF) or the Dirac-Hartree-Fock (DHF) method \cite{taioli2021relativistic,morresi2018nuclear}, which approximates the complex many-body Coulomb interaction by an average potential acting on each electron of the system, including the incoming particle, and generated by all electrons and the nucleus. The self-consistent solution of the HF or DHF equations on a radial grid of points provides both the electronic levels of the system and the scattering potential $V$ (see Eqs. (\ref{potdhf}), (\ref{rho13})). Finally, the continuum wave function of Eq. (\ref{green_90}) is used to determine the DESCS in Eq. (\ref{eq:diff_elastic_cross_sect}).

This approach is particularly suitable for modelling the scattering of a non-relativistic particle with no internal degrees of freedom that hits a target represented by a spin-independent potential $V$ that decays faster than $1/r^2$.
However, this is essentially a toy model that cannot represent, for example, the three-dimensional problem of an electron elastically scattered by an atomic or molecular target.
Such a three-dimensional elastic electron-atom collision can be mapped to a one-dimensional, non-interacting multichannel scattering problem \cite{taioli2010electron,taioli2015computational}. In the position basis, each partial wave is the product of a radial function, represented in a grid covering the range $[0,R]$ large enough to reproduce the scattering wave function, and a spherical harmonic characterised by the angular momentum $l$ and its projection onto a chosen axis $m$, identifying each independent channel (here we describe the elastic process using non-relativistic quantum theory, hence $l,m$ are good quantum numbers).
For a rotationally invariant potential, the partial wave method in the non-relativistic regime, in which the Hamiltonian commutes with both $\bm{L}^2$ and $L_z$ (see Eq. (\ref{rotinv1}) for the relativistic theory), yields for the scattering wave function

\begin{equation}\label{pwave}
\Psi({\bm r})=\sum_{lm}\frac{\phi_{lm}(r)}{r}Y^l_m(\vartheta,\varphi).  
\end{equation}

In this respect, scattering can be schematised by a particle with internal degrees of freedom whose motion is bound to the semi-infinite line $[0,\infty[$. The internal degrees of freedom span a finite Hilbert space ${\cal H}^n$ in which each channel $e_{\alpha}$ is a basis vector. A physical state of the system therefore belongs to the Hilbert space spanned by the square-integrable functions $\phi[0:\infty[ \rightarrow {\cal H}^n$, which cancel at 0. In general, a state in ${\cal H}^n$ can be represented by the expansion
\begin{equation}
\phi(r)=\sum_{\alpha=1}^{n}\phi_{\alpha}(r)e_{\alpha},
\end{equation}
where the wave function $\phi_{\alpha}(r)$ is the projection of the state into the single channel identified by $\alpha=(l,m)$. Since each channel is independent, the Green's functions in the multichannel scattering framework are the same as in the single channel, with the caveat that a dependence of the energy on a threshold is added, characterised by the index $\alpha$,
so that the perturbed channel-dependent Hamiltonian is given by
\begin{equation}\label{hammer}
H\phi_{\alpha}(r)=\left(-\frac{1}{2m}\frac{d^2}{d^2r}+E_{\alpha}\right)\phi_{\alpha}(r)+\sum_{\beta} V_{\alpha,\beta}(r)\phi_{\beta}(r),
\end{equation}
where $E_{\alpha}$ is the energy of the channel $\alpha$ and $V_{\alpha,\beta}(r)$ is the channel-dependent scattering potential.
We note that in this description the phase shifts are also channel-dependent. In this respect, the LS equation (\ref{green_90}), which is calculated using the Hamiltonian of Eq. (\ref{hammer}), can be generalised in coordinate space:
\begin{eqnarray}
\braket{\beta,r|\psi^+_{\alpha,\bm k_{\alpha}}}&=&\psi^+_{\alpha,\bm k_{\alpha}}(\beta, {\bm r})=\delta_{\alpha,\beta}e^{i{\bm{k_{\beta}}}\cdot{\bm r}}\nonumber \\ &+&\sum_{\gamma}\int d{\bm r'} G_0(E-E_{\beta},{\bm r},{\bm r'})V_{\beta,\gamma}({\bm r'})\psi^+_{\alpha,\bm k_{\alpha}}({\bm r'}),\label{hammer_2}
\end{eqnarray}
from which the elements of the $T$-matrix can be derived as $t_{\beta,\alpha}^+(E)=\bra{\phi_{\beta,{\bm k}_{\beta}}}V\ket{\psi^+_{{\alpha,{\bm k}_{\alpha}}}}$.

Eq. (\ref{hammer}) can be solved either with Runge-Kutta on a radial grid or with the finite difference method.
Such grid-based approaches, which can also be used to find the self-consistent solutions of the Dirac or Schr{\"o}dinger equation for systems with many electrons (see Eqs. (\ref{DHF}) and (\ref{moller3}), respectively), are suitable for investigating scattering processes in spherically symmetric systems, such as electron-atom collisions \cite{taioli2021relativistic,morresi2018nuclear}. These radial mesh methods have recently been extended and generalised to the electroweak Hamiltonian together with Eq. (\ref{eq:diff_elastic_cross_sect}) to also study electron-atom collisions and $\beta$ decay processes in astrophysical scenarios, with the aim of modelling the abundance of chemical elements in solar-like and evolved stars as a result of nucleosynthesis processes \cite{morresi2018nuclear,palmerini2016lithium,vescovi2019effects,taioli2022theoretical,palmerini2021presolar,mascali2022novel,mascali2023new,palmerini2023presolar,agodi2023nuclear}. 

In order to study non-spherical, aperiodic, polycentric systems, as in the case of collisions of electrons with molecules or clusters \cite{triggiani2023elastic}, methods based on the expansion of the wave function on Gaussian basis functions (GBFs) are the most suitable. Grid-based methods can be actually regarded as a limiting case of the Gaussian-based approaches when GBFs with infinitesimally small widths are used \cite{taioli2010electron,taioli2015computational,taioli2009surprises,taioli2021resonant}. In this method, the non-kinetic terms of the Hamiltonian (essentially the Coulomb potential in a non-relativistic approach) and the continuum orbital are projected onto a functional space spanned by GBFs.
The GBFs are expressed as follows:
\begin{equation}\label{gaussian}
g(\bm{r})=g(u,v,w;a,\bm{R};\bm{r})=C{\partial^{u+v+w}\over
\partial X^u \partial Y^v\partial Z^w}\left({2a\over\pi}\right)^{3/4}
\exp[-a (\bm{r}-\bm{R})^2],
\end{equation}
where $\bm{R}\equiv(X,Y,Z)$ represents the centre of the GBF, $a$ determines the GBF width and ($u,v,w$) indicates the order of the derivative, which is related to the symmetry of the GBF (e.g. zero-order derivative has symmetry $s$, the first derivative has symmetry as an orbital of type $p$, the second derivative an orbital of type $d$). The GBFs are normalised by the coefficient $C$.

The GBFs can be centred on the nuclei of the constituents (or in principle on any other point in space) and therefore this method is suitable for polycentric molecular systems such as water, water clusters and molecular water aggregates (see Refs. \cite{triggiani2023elastic,taioli2021relativistic} for more details on this method).
GBFs are used to project the scattering potential into a functional space of finite size, while the kinetic energy (in Dirac theory $H_0=c{\bm \alpha} \cdot {\bm p} +\beta mc^2$) is not projected to ``recover'' the continuum.

To evaluate the mono- and bielectronic integrals that appear in the (relativistic) many-body Hamiltonian, several GBFs characterised by different derivative orders and centres can be included \cite{taioli2010electron,taioli2021relativistic}. Depending on the scattering potential, such integrals can be analytical \cite{taioli2021relativistic}, e.g. in the case of the Yukawa potential of Eq. (\ref{Yukawa}). 

The projected LS equation (\ref{hammer_2}) for the scattering wave function can be expressed as follows:
\begin{equation}\label{LSp1}
\psi_{\gamma {\bm k}}^+({\bm r})=e^{i \bm{k} \cdot \bm{r}}+
  G_0^+(\varepsilon) V_\gamma^t(\bm {r})\psi_{\gamma \bm{k}}^+(\bm{r}),
 \end{equation}
where $V_\gamma^t=\sum_{\lambda\mu\nu\tau}
  |\lambda\rangle S^{-1}_{\lambda\mu}\langle\mu|\hat V_\gamma|\nu\rangle S^{-1}_{\nu\tau}
  \langle\tau|,\quad S_{\lambda\mu}=\langle\lambda|\mu\rangle$
is the projected (HF or DHF) self-consistent potential of the molecular or solid state system and can be obtained by numerically solving the Schr{\"o}dinger or Dirac equation with wavefunctions projected onto the same set of GBF \cite {taioli2021relativistic,morresi2018nuclear}. The elements of the GBF basis set are selected in such a way that the difference between the true potential and the projected potential is minimised within the scattering region. The advantage of this method over Mott's approach is that the scattering potential in the LS equation (\ref{LSp1}) contains all the important contributions from the interaction between the electrons of the system and with the incoming electron, including the exchange if the projectile is an electron, as well as other terms such as the permanent dipole moment in polar molecules. 
Normally, the static exchange approximation is used in the calculation of the potential, where the effects of the continuum orbital on the bound orbitals are ignored. In section \ref{DESCS_section} we show an application of this ab initio method to the calculation of the total ESCS of electrons scattered by liquid water.
However, in order to accurately account for the dynamic polarisation of the electron cloud, one should go beyond mean-field approaches and include further orbitals and Slater determinants using multi-configuration methods such as CI.  
We emphasise that the theoretical framework and the relevant equations we have described apply to both non-relativistic and relativistic approaches, even though the wave functions in the latter framework are 4-component spinors
\cite{taioli2021relativistic}.

Another method for modelling elastic electron-atom scattering according to first principles was developed a few years ago \cite{PhysRevLett.96.073202}.
This approach aims to solve the problem of incorporating electron correlation into the continuum wave function with the same level of accuracy as bound-state calculations, a challenge for computer simulations based on conventional scattering methods. Theoretical and computational methods for computing bound states have indeed reached a quite advanced stage, with several codes implementing correlated methods, such as Many-Body Perturbation Theory (MBPT) \cite{taioli2009electronic,umari2012communication},
configuration interaction (CI) \cite{taioli2010electron,taioli2021resonant}, multiconfiguration
Dirac-Hartree-Fock (MCHF) \cite{jonsson2022introduction} and time-dependent DFT (TDDFT) methods \cite{elk,PhysRevLett.99.043005}.
However, scattering states are more difficult to treat numerically due to their (in principle infinite) spatial extent. In this framework, the problem of elastic scattering can be treated with the same computational effort as the problem of bound states.
This approach was applied to rotationally invariant potentials \cite{PhysRevLett.96.073202,PhysRevLett.99.043005}, such as the calculation of the non-relativistic elastic scattering cross-sections of electron-(Ar,Kr,He) collisions, for which one can apply partial wave analysis and assume a functional dependence of the scattering wave function as in Eq. (\ref{pwave}).
From the rotational invariance in conjunction with the conservation of probability it follows that the elastic scattering process can only change the phase of the scattering wave function solution of Eq. (\ref{moller3}) \cite{Dapor2023}, which can thus be represented by a free wave shifted by a phase (phase shift).
The details of the scattering potential determine such a phase shift and the associated observables, such as the ESCS, which can be written:
\begin{equation}
\sigma_{\mathrm{el}}=\frac{4\pi}{k^2}\sum_{l}(2l+1)\sin^2{\delta_l}.    
\end{equation}
This is also the conceptual framework used in the $R$-matrix method \cite{TENNYSON201029}, in which electron (or positron) scattering is modelled by dividing the system into an asymptotic region, in which the non-interacting wave function is analytically known, and a scattering region in which the scattering wave function is typically represented by a partial wave expansion and the electron-molecule interaction is taken into account by highly accurate quantum mechanical methods such as CI or coupled cluster (CC). The fitting of the wave function and its derivative at the boundary between the inner and outer solution provides important scattering information, such as the phase shift and the transmission probability. This approach follows the same philosophy as the $R$-matrix method, i.e. it is based on the idea of dividing the space by cutting out a spherical cavity. However, in contrast to the $R$-matrix, specific boundary conditions are introduced at the edges of the cavity (e.g. the scattering wave functions are set to zero) to convert the continuum problem of electron scattering into the problem of searching for discrete bound-state energies, which can be performed using existing efficient numerical codes implementing mean-field (HF, DHF, DFT) or correlated methods (CI, MBPT, TDDFT). The radius of the cavity can also be easily modified to span the energy spectrum for scattering states.
The asymptotic behaviour of the continuum wave functions beyond the range of the scattering potential is given in Eqs. 
(\ref{assbessel}) and (\ref{asymptotic_2}), which provide at the cavity boundary \cite{PhysRevLett.96.073202}:
\begin{equation}\label{scatt_shift}
\tan(\delta_{l}(k))=-\frac{j_l(kR_c)}{n_l(kR_c)}\simeq -\sqrt{2E_n}R_c +r_{nl}, 
\end{equation}
where $R_c$ is the cavity radius, $k=\sqrt{2E_n}$, $r_{nl}$ is the $n$th zero of the spherical Bessel function $j_l(r)$ for a given angular momentum $l$ and $E_n$ is the $n$th (excited) state energy, which must be used in conjunction with the $n$th zero of the spherical Bessel function. By increasing the radius of the cavity, we can calculate the phase shifts at all energies. Eq. (\ref{scatt_shift}) shows that the phase shift of the continuum problem is related to discrete energies of the same potential \cite{PhysRevLett.96.073202,PhysRevLett.99.043005}. The latter can be obtained either by direct diagonalisation of the Hamiltonian matrix or by methods for bound states, such as TDDFT or CI, which accurately account for the screening of the Coulomb interaction. This method is therefore suitable for calculating phase shifts in $(s,p,d)$-wave scattering of electrons and related quantities, such as the scattering length, at an affordable computational cost. 

\subsection{Inelastic scattering}\label{Ritchie_theory}

The Ritchie theory enables the calculation of the IMFP (and thus also derived quantities such as the stopping power) by establishing a relationship between the dielectric function and the energy loss in a medium \cite{dimitris2012}.
It is a macroscopic theory derived from Maxwell's equations that can be used to evaluate optical macroscopic observables such as absorption, extinction and refraction coefficients and reflectivity, which can be measured by optical spectroscopy experiments.
However, it is based on knowledge of the dielectric function, which is expressed by the band structure of a solid and is therefore a microscopic quantity.
As such, it can be calculated using first-principles approaches such as density functional theory (DFT) and also beyond DFT \cite{taioli2009electronic,umari2012communication,RevModPhys.74.601}, which are able to accurately account for many-body effects such as electron correlation, but also defects, impurities and lattice vibrations. In the next sections, we will discuss the connection between the microscopic dielectric function based on simulations of the electronic quantum structure and the macroscopic description of the electron-matter interaction based on Maxwell's equations, which is the key to quantitatively analyse the optical and transport properties of solids.

\subsubsection{Complex dielectric function and absorption coefficient}

The complex dielectric function $\varepsilon$
is introduced in connection with the constitutive equations of materials; in particular, if non-linear effects are neglected, these are as follows
\begin{eqnarray}\label{constitutive}
\bm{D}(\bm{r}, t) = \int d\bm{r'}\int dt' \varepsilon(\bm{r},\bm{r'},t-t')\bm{E}(\bm{r'}, t'),\\
\bm{j}(\bm{r}, t) = \int d\bm{r'}\int dt' \sigma(\bm{r},\bm{r'},t-t')\bm{E}(\bm{r'}, t'),\label{currden}
\end{eqnarray}

\noindent where $\bm{E}$ and $\bm{D}$ are the electric field and the electric displacement, and $\sigma$ and $\bm{j}$ are the conductivity and the current density, respectively.
In the frequency domain Eq. (\ref{constitutive}) and (\ref{currden}) are given by:
\begin{equation}\label{constitutive2}
\bm{D}(\omega) =  \bar\varepsilon(\omega)\bm{E}(\omega),
\end{equation}
\begin{equation}\label{constitutive90} 
\bm{j}(\omega) =  \sigma(\omega)\bm{E}(\omega),
\end{equation}
assuming that the spatial average (symbolised by the bar symbol in Eq.~(\ref{constitutive2})) was performed over the coordinates $(\boldsymbol{r},\boldsymbol{r}')$. Eq. (\ref{constitutive90}) is also known as Ohm's law.\\
\noindent For a polarisable material, such as a dielectric solid, the polarisation $ {\bm P}(\omega)=\varepsilon_0\bar\chi(\omega) \bm{E}(\omega)$ must be added, where $\bar\chi$ is the macroscopic electrical susceptibility and Eq. (\ref{constitutive2}) reads:
\begin{equation}\label{constitutive3}
\bm{D}(\omega) =  \varepsilon_0\bm{E}(\omega)+4\pi \varepsilon_0{\bm P}=\varepsilon_0(1+4\pi\bar\chi)\bm{E}(\omega),
\end{equation}
\noindent where $\varepsilon_{0}\approx 1$ is the  permittivity of free space, which defines:
\begin{equation}\label{constitutive4}
\bar\varepsilon(\omega)=1+4\pi\bar\chi(\omega).
\end{equation}

\noindent Using the wave equation for an electric field in a medium

\begin{equation}\label{wave_eq}
\nabla^2\bm{E}=\frac{\mu\bar\varepsilon}{c^2}\frac{\partial^2 \bm{E}}{\partial^2 t},
\end{equation}

\noindent where $\mu$ is the magnetic permeability, the solution for a wave propagating along the $x$-direction is a dumped wave

\begin{equation}\label{wave_eq1}
\bm{E}(x,t)=\bm{E_0}e^{i\omega x{\mathscr{N}}/c}e^{-i\omega t},
\end{equation}

\noindent where we have introduced the complex refractive index:

\begin{equation}\label{wave_eq2}
{\mathscr{N}}=\sqrt{\bar\varepsilon}=\nu+i\kappa.
\end{equation}

In Eq. (\ref{wave_eq2}), the real and imaginary parts of ${\mathscr{N}}$ are the refractive index and the extinction coefficient, respectively, which are related to the real ($\bar\varepsilon_1$) and imaginary ($\bar\varepsilon_2$) parts of the macroscopic dielectric function as follows:
\begin{eqnarray}\label{wave_eq3}
\bar\varepsilon_1=\nu^2-\kappa^2, \\
\bar\varepsilon_2=2\nu\kappa.
\end{eqnarray}

\noindent Eq. (\ref{wave_eq1}) can be written:

\begin{equation}\label{wave_eq4}
\bm{E}=\bm{E_0}e^{i\frac{\omega}{c} \nu x}e^{-\frac{\omega}{c}\kappa x}e^{-i\omega t}=\bm{E_0}e^{i\frac{\omega}{c} \nu x}e^{-\frac{x}{\delta}}e^{-i\omega t},
\end{equation}

\noindent where we have defined the optical skin depth $\delta$ (related to the absorption coefficient $\alpha$): 

\begin{eqnarray}
\delta=\frac{c}{\omega \kappa}, \label{wave_eq5} \\  \alpha=\frac{2\omega\kappa}{c}=\frac{\omega\bar\varepsilon_2}{\nu c},\label{wave_eq6}
\end{eqnarray}
as the (inverse) distance at which the amplitude (intensity) of the field is reduced by $1/e$.
Eq. (\ref{wave_eq6}) provides a relation between a macroscopic quantity such as the (frequency-dependent) absorption coefficient, which can be measured in optical reflectivity experiments, and the imaginary part of the dielectric function, which can be obtained by microscopically analysing the band structure of solids and performing a macroscopic average in space.
In section \ref{micro_to_macro} we discuss exactly what is meant by macroscopic average.

\subsubsection{Electron energy loss: the Ritchie theory}\label{Ritchie}

The response of conduction band electrons to an external electromagnetic perturbation generated by a moving electron can be described by the complex dielectric function $\bar{\varepsilon}(\bm{q},\omega)$, where $\bm{q}$ is the wave vector and $\omega$ is the frequency of the field generated by a point charge travelling through a solid \cite{taioli2010electron,Dapor2023,1Ritchie57,1Egerton,3Sigmund,3EgertonII,CRaether}. In this context, $\bm{q}$ and $\hbar\omega$ must therefore also be interpreted as the momentum transfer and the energy lost by the incident charged particle, which the field can transfer to the bound electrons of the atoms of the sample. The energy loss experiment we are modelling is shown in Fig. \ref{fig:scattexp}. 

We point out that in this section we deal with the collision of fast charged particles impinging on materials in different states and phases, focussing on the exchange of energy, momentum and the change in their direction of motion. In order to develop a theoretical model of these collisions between charged particles, a number of assumptions must be made. First, the speed of the incident particles is much higher than the typical speed of electrons orbiting in atoms, so that the first-order Born approximation can be applied (even if this approach can also be applied to slow particles with some caveats), while it is not relativistic (see section \ref{bethe_block} for an extension to relativistic regimes). In this context, the bound electron can be considered at rest. This assumption allows us to neglect (i) the effect of the magnetic field generated by the moving particles on the atomic electrons; (ii) the longitudinal field along the direction of motion of the impinging particle, so that the bound electron at rest sees a transverse field, which can thus be equated to a normal electromagnetic radiation pulse (see Ref. \cite{jackson}). In addition, the incident charged particle interacts electromagnetically with both the nuclei and the surrounding electronic cloud. However, the nuclei (as a whole) are characterised by a much greater mass than the electrons, so that they only absorb a very small amount of energy compared to the latter. The energy loss is therefore only due to the interaction of the incoming charge with the atomic electrons, while the collision with the nuclei in the solids is essentially elastic, also due to their higher charge, and leads to a change in the direction of incidence. If the projectiles are particles with a large mass (e.g. $\mu,\pi$ mesons or ions), the deviation from the trajectory is also smaller than in the case of electrons, whose motion, in contrast, is almost diffusive and characterised by large-angle scattering. Furthermore, although the system is treated as a quantum object, which enters in the calculation of the dielectric function, the effects of quantum electrodynamics are disregarded in the treatment of electromagnetic fields. Finally, since the velocity of the bound electron is much lower than that of the projectile, the energy and momentum transfer from the colliding particle to the system can be modelled by calculating only the part that is due to the electric field of the incident particle at the instantaneous position of the bound electron. 

In this context, the external perturbation density

\begin{equation}\label{densityr}
\rho_{\mathrm{ext}}(\bm{r},t)\,=\,-\,e\,\delta(\bm{r}-\bm{v}t)\,
\end{equation}

\noindent represents an elementary charge $-e$ moving with the (non-relativistic) uniform velocity $\bm{v}$ at time $t$ and at position $\bm{r}$. We point out that by external perturbation we mean both the incoming electrons of the primary beam, which may lose energy and therefore reduce their velocity compared to the initial value when entering the solid, and the secondary electrons generated by ionization events due to electron-electron collisions within the target. 

In Fourier space, Eq. (\ref{densityr}) can be written:
\begin{eqnarray}\label{rho_1}
&&\rho_{\rm ext}(\bm{q},\omega)\,=\frac{1}{(2\pi)^4}\,\int\,d^3r\int_{-\infty}^{+\infty}\,dt\,\exp[-i(\bm{q}\cdot\bm{r}-\omega\,t)]\,\rho(\bm{r},t)\,=\nonumber\\
&&=\frac{1}{(2\pi)^4}\,\int\,d^3r\int_{-\infty}^{+\infty}\,dt\,\exp[-i(\bm{q}\cdot\bm{r}-\omega\,t)]\,[e\,\delta(\bm{r}-\bm{v}t)]\,=\nonumber\\
&&=\,2\pi e\frac{1}{(2\pi)^4}\int_{-\infty}^{+\infty}\,dt\,\exp[-i(\bm{q}\cdot\bm{v}-\omega) t]\,=\,\nonumber\\
&&=\,\frac{1}{(2\pi)^3} e\,\delta(-\bm{q}\cdot\bm{v}+\omega)\,,
\end{eqnarray}
so as

\begin{equation}
\rho_{\rm ext}(\bm{r},t)\,=\,\int\,d^3q\int_{-\infty}^{+\infty}\,d\omega\exp[i(\bm{q}\cdot\bm{r}-\omega\,t)]\,[ e\,\delta(-\bm{q}\cdot\bm{v}+\omega)]\,.
\label{rhoFourier}
\end{equation}

The integration limits of the frequency variable are purely mathematical and have no physical meaning, since the frequency is a positively defined observable quantity. These limits are defined by the conservation of energy within the Dirac $\delta$ function. The integration limits of the momentum are determined by the energy-momentum conservation laws (see Eq. (\ref{kpm})). In general, integration in ordinary space $\bm{r}$ is carried out on the volume of interest, which is the unit cell for a periodic solid, for example. The electric potential generated by the external charge density of Eq. (\ref{rho_1}) can be determined using the Poisson equation, which reads in Fourier space:

\begin{equation}
q^2V_{\rm{ext}}(\bm{q},\omega)=\,4\pi\,\rho_{\rm ext}(\bm{q},\omega)\,.
\label{PoissonEq}
\end{equation}

The dielectric medium, which is perturbed by the external field $V_{\rm ext}$, reacts by inducing a counteracting charge density $\rho_{\rm ind}$, which in turn generates such a potential:
\begin{equation}
q^2V_{\rm ind}(\bm{q},\omega)=\,4\pi\,\rho_{\rm ind}(\bm{q},\omega)\,=\,4\pi\,\chi(\bm{q},\omega) V_{\rm ext}(\bm{q},\omega),
\label{PoissonEq_ind}
\end{equation}

\noindent where we have assumed that the response of the (for simplicity's sake homogeneous) system is linear with the external perturbation via the polarisability response function $\chi(\bm{q},\omega)$ (see also Eq. (\ref{constitutive3})).
The total effective potential acting inside the medium is given by the sum of the induced and external potentials:

\begin{eqnarray}\label{PoissonEq_tot}
q^2V_{\rm tot}(\bm{q},\omega)&=&\,4\pi\,[\rho_{\rm ext}(\bm{q},\omega)+\rho_{\rm ind}(\bm{q},\omega)] \nonumber \\ &=&\,4\pi\left[\frac{q^2}{4\pi}V_{\rm ext}(\bm{q},\omega)+\chi(\bm{q},\omega)V_{\rm ext}(\bm{q},\omega)\right],
\end{eqnarray}
that is:
{\begin{eqnarray}\label{PoissonEq_tot_2}
V_{\rm tot}(\bm{q},\omega)&=&\left[1+\frac{4\pi}{q^2}\chi(\bm{q},\omega)\right]V_{\rm ext}(\bm{q},\omega)=
\bar\varepsilon^{-1}(\bm{q},\omega)V_{\rm ext}(\bm{q},\omega)\nonumber =\\ &=& \frac{4\pi}{q^2}\bar{\varepsilon}^{-1}(\bm{q},\omega)\rho_{\rm ext}(\bm{q},\omega)=\frac{e}{2\pi^2q^2}\bar{\varepsilon}^{-1}
(\bm{q},\omega)\delta{(\omega-\bm{q} \cdot \bm{v})},
\end{eqnarray}
where the last equality is obtained by applying Eq. (\ref{rho_1}). 

If we assume that the perturbing electron can be treated as a classical point-like particle, we can derive the electrostatic field from Maxwell's equations, which is in Fourier space (recall that in real space $\bm{E}_{\rm tot}(\bm{r},t)=-\nabla_{\bm{r}}V_{\rm tot}(\bm{r},t)$):

\begin{equation}\label{PoissonEq_tot_3}
\bm{E}_{\rm tot}(\bm{q},\omega)=-i\bm{q}V_{{\rm tot}}(\bm{q},\omega)=-\frac{ie}{2\pi^2q^2}\bar{\varepsilon}^{-1}
(\bm{q},\omega)\delta{(\omega-\bm{q} \cdot \bm{v})}\bm{q}.
\end{equation}

The energy lost by an electron due to its interaction with the total electric field $\bm{E}_{\mathrm{tot}}$ generated during its passage through the solid can be written:
\begin{equation}\label{work_el}
dW\,=\,\bm{{\cal F}}\,\cdot\,d\bm{r}=-e\bm{E}_{\rm tot}\cdot\,d\bm{r},
\end{equation}
where $dW$ is the work performed by the electrical force $\bm{{\cal F}}$ acting on the unit charge.
At the spatial position $\bm{r}=\bm{v}t$ the electric potential is
\begin{equation}
\bm{E}_{\rm tot}\cdot d\bm{r}\,=\,\bm{E}_{\rm tot}\cdot\frac{d\bm{r}}{dt}\,dt\,=\,\bm{v}\cdot\bm{E}_{\rm tot}\,{dt}.
\end{equation}
The energy loss rate of the electrons inside the solid can also be interpreted as Joule heating caused by the interaction between the charge carriers and the atomically bound electrons, which are typically assumed to be dumped harmonic oscillators as in the Drude approach (see section \ref{drudino}). Using the Eq. (\ref{work_el}), the Joule heating per unit volume can be written as follows
\begin{equation}\label{enloss}
\frac{dW}{dt}\,=-e\bm{v}\,\cdot\,{\bm E}_{\rm tot}\,=-\int{d\bm{r}e\bm{v}\delta(\bm{r}-\bm{v}t)\cdot\,{\bm E}_{\rm tot}\,}=\int{d\bm{r}\,{\bm j_{\rm ext}}\cdot\,{\bm E}_{{\rm tot}}\,},
\end{equation}
where ${\bm j_{\rm ext}}=-e\bm{v}\delta(\bm{r}-\bm{v}t)$ is the current density of the unbound electrons moving in the sample at the velocity $\bm v$.
Since
\begin{equation}
{\bm E}_{\rm tot}(\bm{r},t)=\int d\bm{q}\int_{-\infty}^{+\infty} d\omega e^{i(\bm{q}\cdot \bm{r}-\omega t)}\bm{E}_{\rm tot}(\bm{q},\omega),
\end{equation}
one obtains:
\begin{eqnarray}\label{dieletcric_formalism}
\frac{dW}{dt}=\nonumber \\ =\int d\bm{r}e\bm{v}\delta(\bm{r}-\bm{v}t)\cdot\int d\bm{q}\int_{-\infty}^{+\infty} d\omega ~e^{i(\bm{q}\cdot \bm{r}-\omega t)}\frac{ie}{2\pi q^2}\bm{q}\bar\varepsilon^{-1}(\bm{q},\omega)\delta(\omega-\bm{q}\cdot \bm{v})\nonumber \\
= \frac{ie}{2\pi}\int d\bm{q} \int d\omega ~\bm{q}\cdot \bm{v}\frac{1}{q^2}e^{i(\bm{q}\cdot \bm{v}-\omega)t}\bar\varepsilon^{-1}(\bm{q},\omega)\delta(\omega-\bm{q}\cdot \bm{v})\nonumber \\
=-\frac{e^2}{\pi^2}\int d\bm{q} \frac{1}{q^2} {\rm Im}\left( \frac{\omega}{\bar\varepsilon(\bm{q},\omega)}\right).
\end{eqnarray}

Eq. (\ref{dieletcric_formalism}), in which $\hbar q$ and $\hbar\omega$ must be interpreted as the momentum and energy lost by the fast (non-relativistic) electron in its interaction with matter, represents a fundamental result for the study of the interaction of electrons with solid targets (but can easily be extended to ions and charged particles such as protons in general) and is the basis of the dielectric formalism. It relates the energy loss of electrons passing through matter to the imaginary part of the inverse dielectric function ${\rm Im}\left[ \frac{1}{\varepsilon(\bm{q},\omega)}\right]$. This quantity, which can be determined indirectly from experiments using analytical techniques such as EELS, is known as the energy loss function (ELF). As such, it can be directly compared to dielectric function calculations based on either analytical/semi-empirical models or first principles. The shape of the ELF spectrum as a function of momentum and energy transfer is a characteristic feature of the material, regardless of the type (e.g. mass, charge) of the probe particles used. The details of the projectile actually only occur when calculating the cross-section. 
We note that the ELF is a definite positive quantity, so the sign of $\mbox{Im}\left[\frac{1}{\bar\varepsilon(q,W)}\right]$ must be chosen accordingly.

From the knowledge of the ELF we can obtain information about inelastic scattering observables, such as the differential inverse inelastic mean free path as follows
\begin{equation}
\frac{d\lambda_{\mathrm{inel}}^{-1}}{dW}\,=\,\frac{1}{\pi a_0 E}\,\int_{q_-}^{q_+}\frac{dq}{q}\mbox{Im}\left[\frac{1}{\bar\varepsilon(q,W)}\right]\,,
\label{lineldiff}
\end{equation}
where $a_0$ is the Bohr radius, $E$ is the kinetic energy of the impacting electron and $\hbar q, W$ are the momentum transfer and the energy loss. 
By further integration over the energy loss, the inverse inelastic mean free path can be calculated.
In Eq. (\ref{lineldiff}) the integration limits can be deducted by the energy-momentum conservation laws as
\begin{equation}
\hbar\,q_{\pm}\,=\,\sqrt{2\,m_e\,E}\,\pm\,\sqrt{2\,m_e\,(E\,-\,W)}.
\label{kpm}
\end{equation}

\subsection{From micro to macro}\label{micro_to_macro}

We recall that the transition rate (transition probability per unit time) when a photon of frequency $\omega$ is absorbed from an initial state $\ket{i}$ to a final state $\ket{f}$, which are eigenstates of the unperturbed Hamiltonian $H_0=\sum_j{\bm p}_j^2/2m_e$ with eigenvalues $E_i$ and $E_f$ respectively, is given by Fermi's Golden rule:
\begin{equation}\label{Fgoldrule}
P_{i \rightarrow f}=2\pi |\hat{\bm{e}}\cdot{\bm M}_{if}|^2\delta({E_f-E_i-\omega}), 
\end{equation}
where ${\bm M}_{if}=\bra{i}e^{i{\bm q}\cdot{\bm r}} \cdot {\bm p}\ket{f}$
is the matrix element of the perturbed Hamiltonian $\hat{H}_1=-{\bm A}({\bm r}_j,t)\cdot {\bm p}_j$ and ${\bm A}({\bm r}_j,t)=A_0 \hat{\bm{e}}e^{i({\bm q}\cdot{\bm r}-\omega t)}+ c.c.$ is the vector potential of the electromagnetic field with $\hat{\bm{e}}$ the polarisation vector and ${\bm q}$ the radiation wave vector.

The absorption coefficient is defined as the fraction of the rate of energy absorbed per unit volume over all possible initial and final states, i.e. $\omega \sum_{if}P_{i\rightarrow f}$, divided by the incident energy flux, which is given by $u(c/\nu)=\frac{\nu^2A_0^2\omega^2}{2\pi c^2}(c/\nu)$, where $\nu$ is the refractive index of the material, namely
\begin{equation}
\alpha=\frac{4\pi^2}{\nu\omega c} \sum_{if}|\hat{\bm{e}}\cdot{\bm M}_{if}|^2\delta({E_f-E_i-\omega}).    
\end{equation}
This result can be transferred to a crystalline solid by replacing the initial and final levels ($i,f$) by the indices $(v,{\bm k})$ and $(c,{\bm k+ \bm q})$, which represent the occupied valence and empty conduction band levels (characterised by the wave vectors ${\bm k},{\bm k+\bm q}$, which lie in the 1BZ if the photon momentum transfer $\bm q$ is small (we remind that e.g. UV photons have momentum of 3$\times$10$^{-27}$ J$\cdot$s/m; however high-energy photons, such as X-rays, can impart significant momentum). 
Finally, for small momentum transfer ($q\rightarrow 0$) the imaginary part of the complex dielectric function is given by Eq. (\ref{wave_eq6}):
\begin{equation}\label{micromacro}
\bar \varepsilon_2=2\frac{4\pi^2}{\Omega}\lim_{{\bm q} \rightarrow 0} \frac{1}{q^2} \sum_{vc,{\bm k}}|\bra{c,{\bm k+\bm q}}e^{i{\bm q}\cdot{\bm r}}\ket{v,{\bm k}}|^2\delta({\varepsilon_{c,{\bm k+\bm q}}-\varepsilon_{v,{\bm k}}-\omega}).
\end{equation}
The expression for the real part can be determined using the Kramers-Kronig relations \cite{Dapor2023}.

We note that in Eq. (\ref{micromacro}) the spatial average is performed on the exponential factor times the single-particle wave functions $\ket{v,{\bm k}},\ket{c,{\bm k+\bm q}}$. This equation bridges the gap between the microscopic picture of the solid and the macroscopic absorption index measured in experiments. The sum over the band indices $v,c$ and the crystal momentum ${\boldsymbol k}$ in the right-hand side of Eq.~(\ref{micromacro}) implies that the definition of the macroscopic dielectric function includes all electrons and allowed transitions and retains only the information about the momentum transfer, i.e. the wavenumber of the perturbing electromagnetic field. In this respect, the dielectric function is the fundamental quantity to link the microscopic and macroscopic description of the dielectric properties in solids.

Finally, we note that the microscopic dielectric function for a perfect crystal in reciprocal space can be represented by the matrix $\varepsilon_{{\bm G},{\bm G'}}({\bm q},\omega)$, where ${\bm q}$ is the momentum in the 1BZ. It is tempting to write the average in real space as $\bar\varepsilon(W)=\lim_{{\bm q}\rightarrow 0}\int d{\bm r} \int d{\bm r'}e^{i{\bm q}({\bm r}-{\bm r'})}\varepsilon({\bm r},{\bm r'},W)$ 
However, the total electric field $\bm E_{\rm tot}$ is connected to the external field $\bm D$ via Eq. (\ref{constitutive2}); therefore, to calculate the macroscopic dielectric function, the physically correct average over $\varepsilon({\bm q},\omega)^{-1}$ should be carried out as follows \cite{Wiser,Adler}:
\begin{equation}\label{eq:eM}
\bar{\varepsilon}( \bm{q}, \omega) =  \left [ \varepsilon^{-1}(\bm{q}, \omega)  \right ]^{-1}_{\bm{G}=0,\bm{G'}=0}.
\end{equation}
It is worth mentioning that the inversion of the complete dielectric matrix is performed before taking the head of the matrix corresponding to the element ($\bm{G}=0,\bm{G'}=0$). With this method
%, instead of $\bar{\varepsilon}'( \bm{q}, W) ={\varepsilon}( \bm{q}, \omega)_{\bm{G}=0,\bm{G'}=0}$, 
all off-diagonal entries are actually included. In this way, the so-called crystalline local field effects \cite{doi:10.1119/1.12734}, which are related to the local anisotropy of the material, can be taken into account. These effects can be important for systems such as Au \cite{taioli2023role,GURTUBAY2001123,PhysRevB.88.195124}.
Finally, we note that the averaged dielectric matrix given by Eq. (\ref{eq:eM}) can now be compared with the measured absorption index and the EEL spectra as given by Eqs. (\ref{abs1}) and (\ref{abs2}).

\subsection{Energy loss function: Mermin model, Drude-Lorentz theory, Peen model and ab initio methods}\label{ELF_exp}

In section \ref{Ritchie} we have determined how the dielectric function is related to the energy loss of electrons that hit a solid target and move within it. In the next three subsections, we will discuss the Drude-Lorentz, Mermin and Penn models \cite{EAbril,EMermin,EPlanes,PhysRevB.35.482}. These methods propose analytical representations of the ELF to make the dielectric formalism a viable approach for simulating charge transport within a medium.

\subsubsection{Mermin model}\label{ELF_exp2}

The analytical representation of the macroscopic Mermin dielectric function \cite{EMermin} is as follows:

\begin{equation}\label{merm}
\varepsilon_{\mathrm M}({\bm q},\omega)\,=\,1+\frac{(1+i/\omega\tau)[\varepsilon^0({\bm q},\omega+i/\tau)-1]}{1+(i/\omega\tau)[\varepsilon^0({\bm q},\omega+i/\tau)-1]/[\varepsilon^0({\bm q},0)-1]}\,,
\end{equation}
where ${\bm q}$ is the momentum transfer, $\omega$ the energy loss and $\tau$ the relaxation time.
In order to take the momentum dispersion of the electron energy into account, a dispersion law must be established for the dielectric function. This is necessary to extend the optical ELF to finite momentum transfer.
In the Mermin approach, such a dispersion law is given by the Lindhard dielectric function \cite{ELindhard}:

\begin{equation}
\varepsilon^0({\bm q},\omega)\,=\,1+\frac{4\pi^2q^2}{e^2}B({\bm q},\omega)\,, \label{Lindhard1}
\end{equation}

\begin{equation}
B({\bm q},\omega)\,=\,\int \frac{d{\bm p}}{4\pi^3}\frac{f_{{\bm p}+{\bm q}/2}-f_{{\bm p}-{\bm q}/2}}{\omega-(\varepsilon_{{\bm p}+{\bm q}/2}-\varepsilon_{{\bm p}-{\bm q}/2})/\hbar}, \label{Lindhard2}
\end{equation}
where $f_{{\bm p}}$ is the Fermi-Dirac distribution and $\varepsilon_{{\bm p}}=p^2/2m_e$ is the energy of a free electron in a non-interacting Fermi gas. 
Assuming that we are dealing with a homogeneous, isotropic and infinite medium (we can omit in that case the dependence on the momentum vector and retain only that of $|{\bm{q}}|= q$),
Eqs. (\ref{Lindhard1}) and (\ref{Lindhard2}) can be integrated numerically or solved analytically as follows \cite{EAbril,3Sigmund,EPlanes}:
\begin{equation}
\varepsilon^0(q,\omega)\,=\,1+\frac{\chi_{\mathrm F}^2}{z^2}[f_1(u,z)\,+\,i\,f_2(u,z)]\,,
\end{equation}
where $u=\omega/(q v_{\mathrm F})$, $z=q/(2q_{\mathrm F})$, $\chi_{\mathrm F}^2=e^2/(\pi\,\hbar\,v_{\mathrm F})$, $v_{\mathrm F}$ is the Fermi velocity and $q_{\mathrm F}=m_ev_{\mathrm F}/\hbar$. The functions $f_1(u,z)$ and $f_2(u,z)$ are given by

\begin{equation}
f_1(u,z)\,=\,\frac{1}{2}\,+\,\frac{1}{8z}\,[g(z-u)\,+\,g(z+u)]\,,
\end{equation}

\begin{equation}
f_2(u,z)=\left\{\begin{array}{lll}
&& \frac{\pi}{2}u\,,\,\,\,\,\,\,\,\,\,\,\,\,\,\,\,\,\,\,\,\,\,\,\,\,\,\,\,\,\,\,\,\,\,\,\,\,\,\,\,\, z+u<1 \\
&& \frac{\pi}{8z}[1-(z-u)^2]\,,\,\,\,\,\,\, |z-u|<1<z+u \\
&& 0\,,\,\,\,\,\,\,\,\,\,\,\,\,\,\,\,\,\,\,\,\,\,\,\,\,\,\,\,\,\,\,\,\,\,\,\,\,\,\,\,\,\,\,\,\, |z-u|>1\,,
\end{array}
\right.
\end{equation}
where

\begin{equation}
g(x)\,=\,(1-x^2)\,\ln\left|\frac{1+x}{1-x}\right|\,.
\end{equation}

\subsubsection{The Drude-Lorentz and the Generalized Oscillator Strength methods}\label{drudino}

The Drude--Lorentz (DL) \cite{Dapor2023} and the Mermin Energy Loss Function Generalized Oscillator Strength (MELF-GOS) \cite{GarciaMolina2012,EAbril,EPlanes,PhysRevA.72.052902,https://doi.org/10.1002/sia.5947}
represent other effective and accurate numerical approaches to calculate the ELF over the entire momentum and energy surface. These models can account for the contribution of valence and inner-shell electron excitations to the ELF.

The DL model of electrical conductivity in metals considers electrons as oscillators characterised by an elastic constant $m_e\omega^2_0$, a natural frequency $\omega_0$ and a frictional damping constant $\gamma_0$ (which is related to the relaxation time $\tau_0=1/\gamma_0$).
If both free and bound electrons are included in the DL model,
the dielectric function reads \cite{jackson}:
\begin{equation}\label{druso}
\bar\varepsilon(\omega)=1
-\frac{4\pi e^2N Z_V}{m_e}
\sum_i \frac{f_i}{\omega^2-\omega^2_i-i\omega\gamma_i},
\end{equation}

\noindent where $n=NZ_V$ is the number density of valence electrons in a target with $N$ atoms per unit volume and
$Z_V$ outer shell electrons, $\gamma_i$ are frictional damping coefficients and $f_i = Z_i/Z_V$ is the fraction of free ($f_0,\omega_0$) or bound ($f_i,\omega_i$) excited electrons (which of course obey the sum rule $\sum_i f_i =1$) characterised by an excitation energy $\omega_i$ ($\omega_0=0$ for free electrons).
By neglecting the bound electrons ($f_0=1$) one obtains:
\begin{equation}\label{DLmio}
\bar\varepsilon(\omega)=1
- \frac{\omega^2_P}{\omega^2-i\omega\gamma_0},
\end{equation}
where $\omega^2_P=\frac{4\pi e^2N Z_V}{m_e}$, which gives for the real and imaginary part of the DL dielectric function:
\begin{eqnarray}
\bar\varepsilon_1(\omega)=1
- \frac{\omega^2_P}{\omega^2+\gamma^2_0},\label{real_eps}\\
\bar\varepsilon_2(\omega)=
- \frac{\gamma_0}{\omega}\frac{\omega^2_P}{\omega^2+\gamma^2_0}.
\end{eqnarray}
We remember that the condition $\bar\varepsilon_1=0$ must be fulfilled for plasmon excitation to take place (for frictionless plasmons we get $\omega=\omega_P$, i.e. the bulk plasma frequency).

In both methods, the optical or long-wavelength (${\bm q}\rightarrow 0$) limit of the ELF of the target material, typically obtained by optical or electron energy loss experiments, is fitted to an analytical form given for the valence electrons (or outer electrons) by a linear combination of DL functions

\begin{equation}\label{Drude2}
{\rm Im}\left[ \frac{1}{\varepsilon_{\rm M}(A_i,\omega_i,\gamma_i;{\bm q} =0,\omega)} \right] = \frac{A_i \gamma_i \omega}{(\omega_i^2-\omega^2)^2+(\gamma_i  \omega)^2}, 
\end{equation}

\noindent as follows:

\begin{eqnarray}\label{mermin} 
&&{\rm Im}\left[ \frac{1}{\bar{\varepsilon}({\bm q}=0,\omega)}\right]_{\rm outer} = \nonumber \\ && \sum_i F(\omega-\omega_{\mathrm{th},i}) 
%\frac{A_i}{W_i^2} 
{\rm Im}\left[ \frac{1}{\varepsilon_{\rm M}(A_i,\omega_i,\gamma_i;{{\bm q}=0},\omega)} \right],
\end{eqnarray} 

where 
\begin{equation}\label{smooth}
F(\omega-\omega_{\mathrm{th},i})=\frac{1}{1+e^{-\Delta_i
(\omega-\omega_{\mathrm{th},i})}}
\end{equation}
is introduced to smear the onset of the electronic excitations of the outer shell at the threshold energies $\omega_{\mathrm{th},i}$.
In Eq. (\ref{mermin}) $\bar{\varepsilon}({{\bm q}=0},\omega)$ is the macroscopic optical dielectric function (see discussion in section \ref{micro_to_macro}).
$A_i$, $\omega_i$, $\gamma_i$ and $\Delta_i$
are fitting parameters that are determined by a best-fit. They correspond to the relative weight, position and width of the peaks observed in the experimental optical ELF, while $\Delta_i$ is added to more accurately reproduce the onset of electronic excitation from the inner shells.

We note that the ELF obtained by using the Mermin dielectric function in Eq. (\ref{merm}) reduces to the DL ELF of Eq. (\ref{Drude2}) in the optical limit, so that the Mermin and DL oscillators provide the same values for the fitting parameters in the optical limit \cite{EdelaCruzandYubero}.

Inner shell electrons are included in the representation of the ELF using atomic generalised oscillator strengths (GOS) as follows \cite{GarciaMolina2012}:

\begin{equation}\label{mermin2}
\mathrm{Im}\left[ \frac{1}{\bar{\varepsilon}({\bm q}=0,\omega)}\right]_{\mathrm{inner}} = \frac{2\pi^2 {N}}{\omega}\sum_j \alpha_j \sum_{nl} \frac{df_{nl}^j({\bm q}, \omega)}{d \omega} \Theta(\omega - \omega_{{\rm th},nl}^{j}),
\end{equation}

\noindent where $\frac{df_{nl}^j({\bm q}, W)}{d W}$ are the atomic GOS hydrogen wave functions obtained using an effective nuclear charge for each inner orbital $(n,l)$ of the $j$th atomic constituent of the target material with stoichiometric weight $\alpha_j$ and ionisation energy $\omega_{{\rm th},nl}^{j}$. In Eq. (\ref{mermin2}), ${N}$ is the atomic or molecular number density of the target material.

The accuracy of the fit in both the DL and MELF-GOS methods can be verified by checking the fulfilment (i) of the $f$-sum rule, according to which the integral of the ELF multiplied by the energy loss gives an effective number of electrons per atom (typically the number of valence or total electrons):

\begin{equation}\label{fsumrule}
Z_{\mathrm{eff}} (\omega_\mathrm{max}) =\frac{2}{\pi\Omega_P^2}\int_0^{\omega_\mathrm{max}} \omega \times \mathrm{Im}\left[ \frac{1}{\bar\varepsilon(\omega,q\rightarrow 0)} \right]\,d\omega\,,
\end{equation}
where $\omega_P^2=\Omega_P^2Z_V$
(please note that $Z_{\mathrm{eff}}(\omega_\mathrm{max})\rightarrow Z$ is to be understood as $\omega_\mathrm{max}\rightarrow \infty$, where $Z$ is the target atomic number); ii) of the perfect-screening sum-rule, where the integral:
\begin{equation}\label{psumrule}
P_{\mathrm{eff}} (\omega_\mathrm{max}) =\frac{2}{\pi}\int_0^{\omega_\mathrm{max}} \frac{1}{\omega} \times \mathrm{Im}\left[ \frac{1}{\bar\varepsilon(\omega,q\rightarrow 0)} \right]\,d\omega,
\end{equation}
is an increasing monotonic function that asymptotically goes to 1 when $\omega_\mathrm{max}\rightarrow \infty$.

Once the best-fit parameters have been determined using the experimental data for the optical ELF and carefully checked using the sum rules, the DL and MELF-GOS methods differ in the way they deal with the extension to finite momentum transfer necessary to account for the dispersion of the excitation spectrum in the evaluation of energy losses (see the ${\boldsymbol q}$-dependence of $\bar\varepsilon$ in Eq. (\ref{dieletcric_formalism})).

In the context of the MELF-GOS, it is assumed that the energy-momentum dispersion law is given by the Mermin function of Eqs. (\ref{Lindhard1}) and (\ref{Lindhard2}).

Conversely, the extension of the ELF to the finite momentum transfer in the DL approach is achieved by introducing a functional dependence of the energy on the modulus of the electron momentum $\omega=\omega(|{\bm{q}}|)=\omega(q)$ in Eq. (\ref{Drude2}) \cite{GarciaMolina2012,10.1063/1.3626460}, assuming, of course, that the solid is homogeneous and isotropic.
In these conditions, the most general dispersion law that can be derived considering the correct limit forms for the momentum transfer $q \rightarrow 0$ (Mermin or Drude formula) and $q \rightarrow \infty$ (single-particle dispersion) is the following \cite{10.1063/1.3626460}:
\begin{equation}\label{fullRPA}
\hbar\omega_n(q)=  \hbar\omega_{n}+ \alpha_{\mathrm{RPA}}\frac{\hbar^2q^2}{2m_e},
\end{equation}
where $\omega_{n}$ is the $n$ peak plasmon frequency for $q=0$, $\alpha_{\mathrm{RPA}}=6E_\mathrm{F}/(5\omega_{P})$ ($E_\mathrm{F}$ is the Fermi energy of the material), $\omega_P$ is the nominal bulk plasmon frequency, where $e$, $\rho$, $m_e$ are the electron charge, density and mass, respectively. We note that for $q \rightarrow 0$ this expression approaches the value $\omega_{n}$, which corresponds to the plasma frequency in a metallic bulk system; for finite $q$ it approaches the quadratic energy-momentum dispersion of the random-phase approximation (RPA) or, what is the same, Lindhard's dielectric function (see Eq. (\ref{Lindhard1})), which corresponds to the $n$-independent single-particle excitation for a 3D electron gas. In the most general case, the damping coefficients $\gamma_i$ in the DL relation (\ref{Drude2}) can also be varied over the $q$-space.

\subsubsection{Penn model}\label{pennino}

A third possibility for determining the optical ELF and its extension to finite momentum transfer is the Penn model \cite{TPP1993,PhysRevB.35.482}, also known as the Full Penn Algorithm (FPA). In this approach, the ELF is represented by a convolution of the imaginary part of the inverse Lindhard dielectric function (instead of using the Mermin dielectric function as in the MELF-GOS approach) with a spectral density function obtained from experimental optical data.

In particular, Penn assumes that the dielectric properties of a solid are described by the Lindhard dielectric function \cite{EMermin,ELindhard} ${\varepsilon_{\mathrm{L}}(q,\omega;\omega_P)}$ (see Eqs. (\ref{Lindhard1}), (\ref{Lindhard2})), as for a gas of free electrons (no exchange-correlation interaction is accounted for).
In this model the ELF can be written

\begin{eqnarray}
\mathrm{Im}\left[\frac{1}{\bar\varepsilon(q,\omega)}\,\right]&=&\,\int_0^\infty d\omega_P\,g(\omega_P)\,\mathrm{Im}\left[\frac{1}{\varepsilon_{\mathrm{L}}(q,\omega;\omega_P)}\right],\label{penn1}
\\
g(\omega)\,&=&\,\frac{2}{\pi\omega}\mathrm{Im}\left[\frac{1}{\bar\varepsilon(\omega)}\right]\label{penn2},
\end{eqnarray}

\noindent where in Eq. (\ref{penn1}) $\omega_P(r)=\left[ \frac{4\pi e^2}{m_e}n_P(r)\right]^{1/2}$ generalizes the definition of the volume plasmon frequency of a free electron-gas with constant density $\rho$ ($\omega_P=\left[ \frac{4\pi e^2}{m_e}n\right]^{1/2}$) to the corresponding long-wavelength limit of a free-electron-like material with spherically-symmetric charge density $n_P(r)$. 
The fictitious charge density $\rho_P(r)$ is determined by fulfilling the following implicit equation in $\omega_P(r)$
\begin{equation}\label{penn5}
\mathrm{Im}\left[\frac{1}{\bar\varepsilon(0,\omega)}\,\right]=\mathrm{Im}\left[\frac{1}{\bar\varepsilon(\omega)}\,\right],
\end{equation}
where the analytically known left-hand side, given in Eq. (\ref{penn1}), can be specified for momentum transfer $q=0$ and the right-hand side is the optical ELF, which can be obtained from optical or electron energy loss experiments. Therefore, no fitting procedure is required for this model as for the MELF-GOS.
A further simplification can be achieved by using the plasmon pole approximation for the dielectric function, i.e:
\begin{equation}\label{penn3}
\mathrm{Im}\left[\frac{1}{\varepsilon_{\mathrm{L}}(q,\omega;\omega_P)}\right]\,\approx\,-\frac{\pi}{2}\frac{\omega^2_p}{\omega(q)}\delta(\omega-\omega(q)),
\end{equation}
with the following dispersion law for the plasmon frequency

\begin{equation}\label{penn4}
\omega^2_q(\omega_P)\,=\,\omega_P^2\,+\,\frac{1}{3}[v_{\mathrm F}(\omega_P) q]^2\,+\,(\hbar q^2/2m_e)^2,
\end{equation}

\noindent where $v_{\mathrm F}(\omega_P)$ is the Fermi velocity. Using Eqs. (\ref{penn3}) and (\ref{penn4}), the ELF in Eq. (\ref{penn1}) becomes \cite{DaBo,d504rm44w}:
\begin{equation}\label{penn6}
\mathrm{Im}\left[\frac{1}{\bar\varepsilon(q,\omega)}\right]\,=\,\mathrm{Im}\left[\frac{1}{\bar\varepsilon(\omega_0)}\right]\Big/\left[1+\frac{\pi q^2}{6k_{\mathrm{F}}(\omega_0)}\right],
\end{equation}
where $\omega_0$ is the solution of the equation:
\begin{equation}\label{penn7}
\omega-\omega_q(\omega_P)=0.
\end{equation}
Using Eq. (\ref{penn6}), you can therefore obtain the ELF extended to finite momentum transfer if you know the optical ELF.
If the dispersion relation of Eq.(\ref{penn3}) is replaced by a quadratic law in $q$, the ELF can be written:
\begin{equation}\label{penn8}
\mathrm{Im}\left[\frac{1}{\bar\varepsilon(q,\omega)}\right]\,=\,\frac{\omega-q^2/2}{\omega}\mathrm{Im}\left[\frac{1}{\bar\varepsilon(\omega-q^2/2)}\right].
\end{equation}
We emphasise that an application of this approach to calculate the IMFP of several metals, compared to the ab initio results, is shown in Fig. \ref{fig:lambda_inel}.

\subsubsection{Surface and interface excitation}\label{surpla}

The surface acts as an interface between the material and the vacuum. The dielectric function varies with the material, so that the excitation energy of the plasmons at the interface should differ from that of the bulk plasmons. We note that the electric field at the interface between two materials must remain continuous, i.e:
\begin{equation}
\bar\varepsilon_a(\omega)+\bar\varepsilon_b(\omega) = 0,
\end{equation}
where $\bar\varepsilon_a$ and $\bar\varepsilon_b$ are the macroscopic dielectric functions of the material $a$ and $b$ respectively.
If the condition $\bar\varepsilon_1 =0$ is fulfilled at the interface, a plasmon excitation is expected to be observed.
In particular, at the interface between metal and dielectric, if the dielectric medium is the vacuum (in this case $\bar\varepsilon_b =1$), this condition, neglecting frictional damping ($\gamma_0=0$), provides the surface plasmon frequency (see Eq. (\ref{real_eps})) 
\begin{equation}
1-\frac{\omega^2_P}{\omega^2}+1=0,
\end{equation}
that is $\omega=\omega_{\rm sp}=\frac{\omega_P}{\sqrt{2}}$ (for aluminum this is $\approx 11.2$ eV, see Fig. \ref{fig:prova13}).

In the MC simulation, the modelling of the surface plasmon excitation in the REEL spectrum can be achieved by introducing the contribution of the surface inelastic scattering. This can be obtained by using the surface ELF which can be described as follows
\cite{PhysRevB.29.4878,PhysRevB.39.8209,YUBERO1990173}:
\begin{equation}\label{elf_surf}
{\mathrm{ELF}_s} = \mathrm{Im}\left(\frac{1}{\bar\varepsilon({\bf q}, \omega)+1}\right).
\end{equation}
It has been observed that the scattering zone of the surface extends into the solid and the vacuum \cite{PhysRevB.66.085411} and that the decay length of surface excitations is approximately $t_s = v/\omega_s$, where $v$ is the electron velocity and $\omega_s=\omega_P/2$ is the surface plasmon frequency \cite{PhysRevB.74.075421}. In TMC simulations, it must be assumed that the thickness at which surface plasmons are excited is $v/2\omega_s$ both in the solid state and in a vacuum \cite{6VicanekUrbassek1991}. In vacuum, $\lambda_{\mathrm{el}}^{-1}= 0$, so that for $-v/2\omega_s \le z \le 0$ (assuming that the $z$-axis is oriented from the surface into the bulk), $\lambda = \lambda_{\mathrm{inel}}$, which must be calculated using the surface ELF of Eq. \ref{elf_surf}. A similar approach must be used to calculate the cumulative probabilities. For $-v/2\omega_s \le z \le v/2\omega_s$ the inelastic cumulative probability and IMFP must be calculated using the surface ELF of Eq. \ref{elf_surf}. For $z \ge v/2\omega_s$, the inelastic cumulative probability and IMFP must be calculated using the bulk ELF of Eq. \ref{lineldiff}. The cumulative probability of elastic scattering has to be calculated only when electrons are in the bulk. In this way we are able to determine the depth dependent surface excitations.
In Figs. \ref{fig:prova13} and \ref{fig:prova14} we show a calculation of the REEL spectrum of aluminum and silicon respectively using this approach, where a good agreement is achieved for the absolute values of the plasmon peaks owing to the inclusion of the  inner shell electrons in the calculation of the ELF. Further theoretical models for dealing with surface plasmon excitation can be found in Refs. \cite{1Ritchie57,WERNER2001L461,CHEN1996131}.
A second possibility is to perform an ab-initio simulation of the ELF. However, the presence of a surface (or interface) that must be included in the ELF model leads to a simulation cell that is much larger than the bulk, which typically consists of a few periodically repeating atoms. 

We point out that the dispersion law is also influenced by surface or interface effects.
In particular, for semi-infinite or finite systems, such as slabs, the occurrence of surface plasmon excitation requires an additional term of the plasmon frequency, which is linear with the momentum transfer $\hbar q$, as follows \cite{doi:10.1002/sia.5878}:
\begin{equation}\label{SB}
\hbar^2\omega^2=  \hbar^2\omega_{P}^2+\hbar^2\alpha q+\beta\frac{\hbar^2q^2}{m}+  \frac{\hbar^4q^4}{4m^2}\,,
\end{equation}
where
$\alpha=\sqrt{3E_{\mathrm F}/(5m)}\omega_P$ and $\beta=6E_{\mathrm F}/5$.
In this respect, the equal inclusion of bulk and surface plasmons is important to obtain a correct interpretation of the measured REEL spectra, which contain both contributions.
In Fig. \ref{fig:lambda_inel} in the Results section we show the effects of the different dispersion models with respect to the calculation of the IMFPs \cite{azzolini2018anisotropic,AZZOLINI2020109420,10.1063/1.3626460,azzolini2018secondary} also including the surface plasmon with its relevant dispersion.
}

\subsubsection{Semi-empirical approach to beam-target correlation}

The extension to finite momentum transfer using RPA (see Eq.~\ref{fullRPA}) is a source of error, as it is based on the Fermi electron gas theory, in which it is assumed that the particles are independent or interact via a mean field. In particular, the short-range exchange and correlation effects, which are expected to become relatively important at low energy, i.e. for scattering at large angles, which tends to be characterised by a finite $q$, are poorly treated in the RPA.
Such many-body interactions, related to the Coulomb repulsion between the electrons in the systems and between the incoming and atomic electrons and the Pauli exclusion principle, can be easily accounted for in the DL and MELF-GOS models in a phenomenological way.
In particular, the exchange interaction resulting from the indistinguishability of the incident, target and ejected electrons is typically treated via the Born--Ochkur approximation \cite{3BornOchkur,3FernandezVarea,3deVera}.
In this model, the ELF is modified by rescaling the direct scattering in the calculation of the differential inverse inelastic mean free path in Eq. (\ref{lineldiff}) as follows

\begin{equation}
\frac{d\lambda_{\mathrm{inel}}^{-1}}{dW}\,=\,\frac{1}{\pi a_0 E}\,\int_{q_-}^{q_+}\frac{dq}{q}\,[1\,+\,f_{\mathrm{ex}}(q)]\,\mbox{Im}\left[\frac{1}{\bar\varepsilon(q,W)}\right]\,,
\label{exchange}
\end{equation}
where the exchange function is 

\begin{equation}
f_{\mathrm{ex}}(q)\,=\,\left(\frac{\hbar q}{m_ev}\right)^4\,-\,\left(\frac{\hbar q}{m_ev}\right)^2\,,
\label{BornOchkur}
\end{equation}

\noindent $v$ is the velocity of the electrons, $\hbar q$ is the momentum transfer, $m_e$ is the electron mass and $a_0$ is the Bohr radius. 
This approximation may require corrections for incident energies below 200 eV, where exchange effects are important and significantly affect the inelastic scattering cross section. This correction can be implemented by using an interference phase \cite{3Bourke} to account for the reduced scattering rates due to the Pauli exclusion principle. We note that the inclusion of exchange effects in the ELF is generally particularly important for insulators and semiconductors beyond the optical limit, while they have less dramatic effects for metals and semimetals due to more effective screening. 

We emphasise that all these methods for calculating the ELF suffer from at least three common shortcomings, namely: (i) the need for experimental optical data, which are only available for a few materials; (ii) the implicit use of Fermi electron-gas theory to treat the electron-electron interaction, neglecting or crudely approximating the exchange-correlation effects; (iii) the need for an analytical (see Eqs. (\ref{fullRPA}) and (\ref{SB})) or a built-in (see Eq. (\ref{Lindhard1})) relation for modelling the dispersion law at finite momentum.

The ab initio methods offer a practicable solution to these disadvantages. Although these approaches are computationally more expensive than the DL and MELF-GOS models, they can in principle provide the most accurate results with regard to the ELF \cite{taioli2023role,azzolini2017monte}.
These methods are discussed in the next section.

\subsubsection{First principles simulations of the ELF}\label{ELF_ab}

For the description of electronic motion, the ab initio methods rely only on the laws of quantum mechanics, without resorting to experiments. In particular, the calculation of the ELF can be performed at different levels of theory, with increasing accuracy from the RPA via the Many-Body Perturbation Theory (MBPT) \cite{taioli2009electronic,umari2012communication} and the Time-Dependent Density Functional Theory (TDDFT), \cite{azzolini2017monte,doi:10.1002/sia.5878,segatta2017quantum} over the entire momentum dispersion with the same computational effort as the ${\bm q}\rightarrow 0$ calculations.
These models are based on a different description of the correlated motion of electrons. However, they are all based on numerical or analytical models of the Coulomb repulsion between electrons. 
In this regard, various exchange-correlation functionals have been developed over the years to treat the Coulomb interaction  using known constraints in the low and high density regions, leading to different approximating potentials \cite{medvedev2017density}. While the search for universal and all-purpose functionals remains a challenge, artificial intelligence-based approaches have also been successfully applied in this regard recently \cite{Borlido,Ryabov}.

In general, the ab initio methods within the dielectric formalism provide a more rigorous treatment of screening effects in the evaluation of the ELF, especially for finite momentum transfers ${\bm q}$, than the semi-empirical approaches implemented in DL and MELF-GOS \cite{10.1667RR3281,doi:10.1063/1.4824541}. They can treat excitonic effects, polarons and electron-phonon interactions at the same level of theory.
It is therefore to be expected that they have a higher accuracy than DL and MELF-GOS, especially for finite ${\bm q}$. In fact, DL and MELF-GOS start with the fitting of experimental data and thus inherit the finite resolution of the measurement apparatus. Ab initio ELFs show a more complex line shape characterised by many peaks \cite{taioli2023role}. Most importantly, experimental results must be available to elaborate the DL and MELF GOS methods. In this context, a discussion on the effects of different dielectric response models on the calculation of REEL spectra, IMFP and stopping power of graphite and diamond can be found in Ref. \cite{azzolini2017monte}. There it is shown that the use of ab initio calculated ELF leads to a better agreement between calculated and experimental data, in the low energy range ($<$ 100 eV) compared to the widely used DL and MELF-GOS models.

The most important ingredient for accessing the ELF is the macroscopic dielectric function. The latter can be obtained by calculating the microscopic polarisation function (or density-density response) $\chi({\bm{r},\bm{r'}},\omega)$, which is defined as follows:
\begin{equation}\label{chidefine}
\rho^{\mathrm{ind}}({\bm{r}},\omega)=\int d{\bm{r'}}\chi({\bm{r},\bm{r'}},\omega)V^{\rm{ext}}({\bm{r'}},\omega) \, \mbox{,}
\end{equation}
where $\rho^{\mathrm{ind}}({\bm{r}},\omega)$ is the electron density induced by the external Coulomb potential $V^{\mathrm{ext}}$ (see Eq. (\ref{PoissonEq_ind})) and $\omega$ is the electromagnetic field frequency.
$\chi({\bm{r},\bm{r'}},\omega)$ is associated with the electronic band structure of the target material. Assuming periodic boundary conditions, one can take advantage of the Bloch theorem for pristine crystals, according to which one-particle states can be conveniently expanded using a plane wave basis as follows:
\begin{equation}
\phi_{{\bm q},n}({\bm r})=\frac{1}{\sqrt{V}}\sum_{\bm G} u_{{\bm q},n}({\bm G})\exp^{i({\bm q}+{\bm G})\cdot {\bm r}},
\end{equation}
where $V$ is the volume of the simulation cell, $\hbar{\bm q}$ is the electron momentum transfer located in the 1BZ, ${\bm G}$ is a reciprocal lattice vector, the quantum number $n$ characterises the band and the function $u_{{\bm q},n}$ has the same periodicity as the crystal lattice.
The polarisability in Eq. (\ref{chidefine}) can also be conveniently expressed for periodic crystals by using the Fourier transform from real to reciprocal space as a matrix $\chi_{\bm{G},\bm{G'}}({\bm q},\omega)$ in which the reciprocal lattice points $\bm{G},\bm{G'}$ identify the rows and columns for each momentum and energy transfer, respectively. The full polarisability $\chi_{\bm{G},\bm{G'}}({\bm q},\omega)$ of a periodic solid in reciprocal space can then be obtained by solving the following Dyson-like screening equation \cite{RevModPhys.74.601, Weissker2010}:
\begin{eqnarray}
\chi_{\bm{G},\bm{G'}}({\bm q},\omega) = \chi^0_{\bm{G},\bm{G'}}({\bm q},\omega)+ \sum_{\bm{G''},\bm{G'''}}\chi^0_{\bm{G},\bm{G''}}({\bm q},\omega) \nonumber \\ \times \Big[ v_{\bm{G''}}({\bm q})\delta_{\bm{G''},\bm{G'''}}+ K^{xc}_{\bm{G''},\bm{G'''}}({\bm q},\omega) \Big]\chi_{\bm{G'''},\bm{G'}}({\bm q},\omega),\label{dyson}
\end{eqnarray}
which relates the non-interacting density-density response function $\chi^0_{\bm{G},\bm{G'}}({\bm q},\omega)$ for independent particles to the full response. The single-particle states appearing in $\chi^0_{\bm{G},\bm{G'}}({\bm q},\omega)$ are usually derived by solving the mean-field Hartree-Fock equations or using the Kohn-Sham DFT (KS-DFT) scheme.
In the KS-DFT scheme, the polarisability of the independent particles is written as a second order transition:
\begin{equation}\label{chi_ind}
\chi^0_{\bm{G},\bm{G'}}({\bm q},\omega)=2\sum_{vc}\frac{\bra{v}e^{-i({\bm{q}+\bm{G}})\cdot\bm{r}}\ket{c}\bra{c}e^{i({\bm{q}+\bm{G'}})\cdot\bm{r'}}\ket{v}}{\omega-(\varepsilon_c-\varepsilon_v)+i\eta},
\end{equation}
where $\ket{v},\ket{c}$ and $\varepsilon_v,\varepsilon_c$ are the KS eigenfunctions and eigenvalues corresponding to the valence and conduction bands of the crystal and $\eta$ is an positive infinitesimal number. Using the expression (\ref{chi_ind}) instead of the Dyson Eq. (\ref{dyson}), also known as RPA, can lead to results that agree well with the experimental data (see section \ref{applic_inel}), even though the single-particle KS polarisability is generally not satisfactory for the description of electronic spectra.

In Eq. (\ref{dyson}) the exchange-correlation kernel $K^{xc}_{\bm{G''},\bm{G'''}}({\bm q},\omega)=\left\{ \frac{d}{d\rho}v_{\rm xc}[\rho]\right\}_{\rho=\rho(\bm{r},t)}$ encodes all non-trivial many-body effects, while $v_{\bm{G'}}({\bm q})=4\pi/|{\bm q}+{\bm G'}|^2$ is the Fourier-transformed bare Coulomb interaction.
The kernel $K^{xc}_{\bm{G''},\bm{G'''}}({\bm q},\omega)$ can be modelled using the adiabatic local density approximation (ALDA) \cite{doi:10.1146/annurev.physchem.55.091602.094449}, by which the time dependence is local over the instantaneous DFT ground state density
\begin{equation}\label{adia}
v_{\rm  xc}[\rho]({\bm r},t)=v_{\rm  xc}^{\rm adia}[\rho]({\bm r},t)=v_{\rm  xc}^{\rm GS}[\rho]({\bm r})|_{\rho({\bm r'},t)=\rho(\bm r')_{GS}},
\end{equation}
where $v_{\rm xc}^{\rm GS}[\rho]({\bm r})|_{\rho({\bm r'},t)=\rho(\bm r')_{GS}}$ is the exchange-correlation potential for the ground-state density $\rho(\bm r')_{GS}$. $v_{\rm xc}^{\rm GS}[\rho]({\bm r})$ can be calculated with any variant of the standard exchange correlation functionals, such as LDA, GGA or hybrid functionals to account for exact exchange \cite{doi:10.1146/annurev.physchem.55.091602.094449}.
Recently, other functionals with memory (non-local in time) have also been developed \cite{PhysRevLett.79.1905}.

The full polarisability ${\chi_{\bm{G},\bm{G'}}}({\bm q},\omega)$ is related to the microscopic dielectric function as follows:
\begin{equation}\label{chichi}
\varepsilon_{\bm{G},\bm{G'}}({\bm q},\omega)=1-v_{\bm G'}({\bm q})\chi_{\bm{G},\bm{G'}}({\bm q},\omega) \, \mbox{,}
\end{equation}
where $1=\delta_{\bm{G},\bm{G'}}$ is the identity operator. Eq. (\ref{chichi}) shows that the microscopic $\varepsilon_{\bm{G},\bm{G'}}$ in the entire energy-momentum surface can ultimately be obtained from electronic structure calculations, i.e. from the knowledge of the electronic wave functions. We note that macroscopic observables are typically obtained from experiments, such as the absorption coefficient in Eq. (\ref{wave_eq6}) or the EEL spectra. In order to directly compare computer simulations and experiments, it is therefore necessary to average the variables in Eq. (\ref{chichi}) (see discussion in section \ref{micro_to_macro}).
In this context, we recall that the macroscopic $\bar\varepsilon(\omega)$ can be obtained from Eq.~(\ref{chichi}) by using the expression in Eq.~(\ref{eq:eM}).
Finally, the absorption coefficient according to Eq. (\ref{wave_eq6}) is proportional to ${\rm Im}(\bar\varepsilon)$:
\begin{equation}\label{abs1}
\alpha \propto {\rm Im}(\bar\varepsilon)=-v\times {\rm Im}(\bar\chi)\,,
\end{equation}
while the electron energy loss from Eq. (\ref{dieletcric_formalism}) is proportional to the imaginary part of the reciprocal of the macroscopic dielectric function:
\begin{equation}\label{abs2}
{\mathrm {ELF}} \propto {\rm Im}\left(-\frac{1}{\bar\varepsilon}\right)=-v\times {\rm Im}\left(\frac{\bar\chi}{1-v\bar\chi}\right).
\end{equation}
where the macroscopic polarizability $\bar\chi$ is the function for which $\bar\varepsilon=1-v\bar\chi_{\bm{G}=\bm{G'}=0}({\bm q},W)$.

\subsection{Inelastic scattering: the Bethe-Bloch formula}\label{bethe_block}

The dielectric theory describes energy loss as a polarisation phenomenon that can be modelled macroscopically by Maxwell's electromagnetic field equations and microscopically by the quantum response of a target to an external perturbation, such as the collision of an electron beam. Previously, Bethe \cite{3Bethe,Bloch1,Bloch2} had proposed a different approach to calculating stopping power. This approach was based on the idea that electromagnetic interactions triggered by the impact of a charged projectile on a target system can be described by the classical response of a collection of independent harmonic oscillators with given energy spacing and dipole strength. In particular, the excitation cross-section for such an electron-atom collision is \cite{3Sigmund}:
\begin{equation}\label{cs_bethe_block}
\sigma_j=\frac{2\pi e^4}{m_ev^2}\int \frac{dW}{W^2}|F_{j0}(\bm{q})|^2,    
\end{equation}
where $\bm{q}$ and $W=\hbar^2q^2/2m_e$ are the momentum and energy transfer and 
\begin{equation}\label{almo_1}
F_{n0}(\bm{q})=\bra{n} \sum_{\nu=1}^{Z}e^{i\bm{q}\cdot{\bm{r}_\nu}}\ket{0}
\end{equation}
\noindent is the form factor, the square of which is the generalised atomic oscillator strength, for a transition from the ground state $0$ to the excited state $n$ of the system of $Z$ charges in the positions $\bm{r}_\nu$. The presence of $F_{n0}(\bm{q})$ in the following equations suggests the interpretation of the variable ${\bm q}$ as momentum transfer.
This cross-section can be derived using semiclassical arguments, where bound electrons can be considered as quantised harmonic oscillators \cite{Dapor2023}; however, we will discuss the rigorous full quantum mechanical derivation, which can also be more easily extended to relativistic regimes \cite{3Sigmund}.

In this context, we assume that a homogeneous, isotropic and infinite medium is perturbed by an external point charge, as in Eq. (\ref{densityr}). From Eq. (\ref{enloss})
the stopping power results as follows:
\begin{equation}\label{stopping}
-\frac{dW}{dx} = -\frac{e}{v}{\bm v}\cdot{\bm E_{\mathrm{tot}}}({\bm r=\bm v t},t),
\end{equation}
as the stopping force is opposite to the velocity of the particles.
By using the scalar and vector potentials
($\Phi({\bm r},t),{\bm A}({\bm r},t))$ to represent the electromagnetic fields and assuming Coulomb gauge (${\bm \nabla}\cdot{\bm A}=0$), the field equations can be written in Fourier space (see Eq. (\ref{PoissonEq_tot})):
\begin{eqnarray}
q^2\Phi({\bm q},\omega)=4\pi\rho_{\rm tot}({\bm q},\omega),
\label{fieldeq_trasf_phy}\\
\left(q^2-\frac{\omega^2}{c^2}\right){\bm A}({\bm q},\omega) -\frac{\omega}{c}{\bm q}\Phi({\bm q},\omega)=\frac{4\pi}{c}{\bm j}_{\rm tot}({\bm q},\omega).\label{fieldeq_trasf_A}
\end{eqnarray}

In a polarisable medium, the total charge and current densities can be divided into induced and external contributions as follows (see Eqs. (\ref{PoissonEq}), (\ref{PoissonEq_ind})):
\begin{eqnarray}    
\label{fieldeq_trasf77}
\rho_{\rm tot}=\rho_{\rm ind}+\rho_{\rm ext},\\
{\bm j}_{\rm tot}={\bm j}_{\rm ind}+{\bm j}_{\rm ext}=\sigma {\bm E_{\rm tot}},
\label{fieldeq_trasf78}
\end{eqnarray}
where the induced components represent the (linear) response of the system to the external perturbation of Eq. (\ref{densityr}). Eq. (\ref{fieldeq_trasf78}) is the Ohm's law (see also Eq. (\ref{constitutive90})).
Furthermore, the vector potential and the wave vector are orthogonal in the Coulomb gauge, so that by projecting Eq. (\ref{fieldeq_trasf_A}) into transverse (orthogonal to ${\bm q}$) and longitudinal (parallel to ${\bm q}$) components we obtain
\begin{eqnarray}
\frac{\omega}{c}{\bm q}\left(1+\frac{4\pi i \sigma_{\rm long}({ q},\omega)}{\omega} \right)\Phi({\bm q},\omega)=\frac{4\pi}{c}{\bm j}_{\rm ext, long}({\bm q},\omega)\label{fieldeq_trasf23},\\
\label{fieldeq_trasf2}
\left(q^2-\frac{\omega^2}{c^2}\right)\left( 1+\frac{4\pi i \sigma_{\rm trans}({q},\omega)}{\omega}\right){\bm A}({\bm q},\omega) =\frac{4\pi}{c}{\bm j}_{\rm ext, trans}({\bm q},\omega),
\end{eqnarray}
\noindent 
where $\sigma_{\rm trans}, \sigma_{\rm long}$ and ${\bm j}_{\rm ext, trans}, {\bm j}_{\rm ext, long}$ are the conductivity and external current density along the transverse and longitudinal directions, respectively.

\noindent Using the continuity equation:
\begin{equation}
{\bm q}\cdot{\bm j}_{\rm ext}({\bm q},\omega)=\omega\rho_{\rm ext}({\bm q},\omega),
\end{equation}
one can rewrite Eq. (\ref{fieldeq_trasf23}) as:
\begin{equation}
q^2\left(1+\frac{4\pi i \sigma_{\rm long}({ q},\omega)}{\omega} \right)\Phi({\bm q},\omega)=4\pi\rho_{\rm ext}({\bm q},\omega)=q^2\varepsilon_{\rm long}\Phi({\bm q},\omega),\label{fieldeq_trasf231}
\end{equation}
and, analogously, Eq. (\ref{fieldeq_trasf2}) as:
\begin{equation}\label{scnd_eq}
\left(q^2-\frac{\omega^2}{c^2}\right) 
\varepsilon_{\rm trans}{\bm A}({\bm q},\omega) =\frac{4\pi}{c}{\bm j}_{\rm ext, trans}({\bm q},\omega),
\end{equation}
where we have defined the dielectric functions in the longitudinal and transverse directions $\varepsilon_{\rm long}=\left(1+\frac{4\pi i \sigma_{\rm long}({q},\omega)}{\omega} \right)$ and $\varepsilon_{\rm trans}=\left(1+\frac{4\pi i \sigma_{\rm trans}({q},\omega)}{\omega} \right)$, respectively.
By inserting Eqs. (\ref{fieldeq_trasf231}), (\ref{scnd_eq}) and (\ref{rho_1}) in Eq. (\ref{stopping})
one obtains \cite{3Sigmund}:
\begin{eqnarray}\label{almosta}
-\frac{dW}{dx}=\frac{ie^2}{\pi v^2}\int_0^\infty \frac{dq}{q}\int_{-qv}^{qv}d\omega \omega\left(\frac{1}{\varepsilon_{\rm long
}({q},\omega)}-\frac{v^2}{c^2}\frac{q^2-\omega^2/v^2}{q^2-\varepsilon_{\rm trans
}({q},\omega)\omega^2/c^2}\right).\nonumber \\
\end{eqnarray}
In general, the second term in Eq. (\ref{almosta}), which represents the contribution of the transverse excitation to the stopping force, can be neglected in the non-relativistic (up to the quasi-relativistic, typically for $\beta=v/c < 0.5$) regime, i.e. $dW/dx\approx dW/dx|_{\rm long}$.  

To simplify the derivation of the Bethe-Bloch stopping formula, Eq. (\ref{cs_bethe_block}), we assume that the target system is represented by a one-electron atom at position ${\bm R}$ and the impinging projectile is a non-relativistic electronic charge unit, so that only the first term in Eq. (\ref{almosta}) can be retained. For materials, this hypothesis is equivalent to the assumption that the electron clouds of the atomic components do not overlap significantly. The generalised oscillator strength $|F_{j0}(\bm{q})|^2$ of Eq. (\ref{almo_1}) for this problem can be determined using first-order perturbation theory. The interaction between the perturbing electromagnetic field generated by the incoming charged particle and the one-electron target atom is described by the potential energy
\begin{equation}\label{almost_field}
V({\bm r},t)=-e\Phi({\bm r}-{\bm R},t)=\frac{-e^2}{|{\bm r}-{\bm b}(t)|},
\end{equation}
where $\Phi({\bm r}-{\bm R},t)$ is the potential generated by the atomic electron at the position $({\bm r}-{\bm R})$ and ${\bm c}(t)={\bm b}_0+{\bm v}t-{\bm R}={\bm b}(t)-{\bm R}$ is the trajectory of the impacting projectile characterised by the impact parameter ${\bm b}_0$.
Within the framework of the first-order perturbative approach, the wave function of the bound electron can be expanded in the eigenstates of the unperturbed Hamiltonian $H_0\ket{n}=\varepsilon_n\ket{n}$ of the target as follows:
\begin{equation}\label{almost_field_1}
\Psi({\bm r},t)=\sum_{n}c_n(t)e^{-i\varepsilon_n t/\hbar|}\ket{n},
\end{equation}
where
\begin{equation}\label{almost_field_2}
c_n(t)\simeq \delta_{n0}+c_n^{(1)}=\delta_{n0}+\frac{1}{i\hbar}\int_{-\infty}^{t}dt' e^{i\omega_{n0}t'}\bra{n}V({\bm r},t')\ket{0}.
\end{equation}
In Eq. (\ref{almost_field_2})
$\delta_{j0}$ is the Kronecker symbol and $\hbar\omega_{n0}=\varepsilon_n-\varepsilon_0$ is the energy jump between the ground state and the $n$th excited state of the system. 
In Fourier space, the matrix element in Eq. (\ref{almost_field_2}) is given using Eq. (\ref{almost_field}) by
\begin{eqnarray}\label{almost_field_21}
\bra{n}V({\bm r},t)\ket{0}&=&-e\int d{\bm q}\int d\omega \int d{\bm r}\Phi({\bm q},\omega)\psi^*_n({\bm r}-{\bm R})\psi_0({\bm r}-{\bm R})\rangle e^{-i\omega t}e^{i{\bm q}\cdot {\bm r}}\nonumber \\&=&-e\int d{\bm q}\int d\omega \Phi({\bm q},\omega)e^{i{\bm q}\cdot {\bm R}}F_{n0}({\bm q})e^{-i\omega t},
\end{eqnarray} 
where $F_{n0}({\bm q})$ is given in Eq. (\ref{almo_1}) and $\psi_{n,0}({\bm r}-{\bm R})$ are the electron wavefunctions corresponding to the $n,0$ states.
We emphasise that the presence of $F_{n0}(\bm{q})$ implies the interpretation of $\hbar\bm{q}$ as the momentum transfer associated with the electronic excitation.

At time $t=\infty$, which is asymptotically far away from the scattering centre, where the potential has a negligible effect on the target, we thus obtain from Eq. (\ref{almost_field_2}): 
\begin{equation}\label{almost_field_3}
c_n^{(1)}(\infty)=-\frac{e}{\hbar}\int d{\bm q} \int d\omega \Phi({\bm q},\omega)e^{i{\bm q}\cdot {\bm R}} F_{n0}({\bm q})\frac{e^{i(\omega_{n0}-\omega)t}}{\omega_{n0}-\omega-i\Gamma},
\end{equation} 
where the damping constant $\Gamma$ was added to force the convergence of the integral in the complex plane near the singular points.

\noindent From Eqs. (\ref{fieldeq_trasf_phy}) and (\ref{fieldeq_trasf231}), we can also write in Fourier space the total induced charge  for the bulk solid as:
\begin{equation}\label{almost_field_69}
\rho_{\rm ind}({\bm q},{\omega})=-\frac{q^2}{4\pi}[\varepsilon_{\rm long}-1]\Phi({\bm q},\omega).
%=\int d{\bm q} \int d\omega e^{i{\bm q}\cdot {\bm r}}e^{-i\omega t}\rho_{\rm ind}({\bm r},t).
\end{equation}
\noindent 
%in the left hand side of Eq. \ref{almost_field_69} can be defined in terms of the generalized oscillator strength 
For a medium with a number density $N$ consisting of one-electron atoms, the total induced charge density to first order in $\Phi$ can be estimated by perturbation theory (see Eq. (\ref{almost_field_1})) from the induced charge density for a single electron $\rho_{\rm pol}$ as follows:
\begin{eqnarray}\label{piopio}
\rho_{\rm ind}({\bm r},t)&=&N\int d{\bm R}~\rho_{\rm pol}({\bm r},t)= -Ne\int d{\bm R}~|\Psi({\bm r},t)|^2=\nonumber \\&\myeqqq&-Ne\int d{\bm R}~(\psi^{(0)*}({\bm r},t)\psi^{(1)}({\bm r},t)+\psi^{(0)}({\bm r},t)\psi^{(1)*}({\bm r},t))=
\nonumber \\&\myeqqqq&-\frac{Ne^2}{\hbar}\int d{\bm R}~\Big[\sum_n \Big(\psi_0^*({\bm{r-R}},t)\psi_n({\bm {r-R}},t) \times \nonumber \\&\times&\int d{\bm q}
\int d\omega \Phi({\bm q},\omega)e^{i{\bm q}\cdot {\bm R}} F_{n0}({\bm q})\frac{e^{-i\omega t}}{\omega_{n0}-\omega-i\Gamma}\Big) {\rm + c.c.}\Big]=\nonumber\\
&=& \int d{\bm q}
\int d\omega\rho_{\rm ind}({\bm q},{\omega})e^{i{\bm q}\cdot {\bm r}} e^{-i{\bm \omega} {\bm t}}. 
\end{eqnarray}

\noindent Noticing that $\int{d{\bm R}~\psi_0^*({\bm{r-R}},t)\psi_n({\bm {r-R}},t)e^{i{\bm q}\cdot {\bm R}}}=e^{i{\bm q}\cdot {\bm r}}F_{0n}(-\bm q)$ and $\Phi^*({\bm q},\omega)=\Phi({-\bm q},-\omega)$ we obtain from Eq. (\ref{piopio}):

\begin{equation}
\rho_{\rm ind}({\bm q},{\omega})=-\frac{Ne^2}{\hbar}\Phi({\bm q},\omega)\Bigg(\sum_n \frac{F_{0n}({-\bm q})F_{n0}({\bm q})}{\omega_{n0}-\omega-i\Gamma}+\sum_n \frac{F_{0n}({\bm q})F_{n0}({-\bm q})}{\omega_{n0}+\omega+i\Gamma}\Bigg).\label{almost_field_70}
\end{equation}
\noindent Finally, if we insert Eq. (\ref{almost_field_69}) into the left-hand side of Eq. (\ref{almost_field_70}) and take into account the isotropy of the medium, by which $F_{0n}({\bm q})=F_{n0}({-\bm q})$, we obtain the longitudinal dielectric function as \cite{3Sigmund}:
\begin{equation}\label{almost_field_71}
\varepsilon_{\rm long}(q,\omega)=1+\frac{m_e\omega^2_P}{\hbar q^2}\sum_n F_{n0}^2({\bm q}) 
\times \left(\frac{1}{\omega_{n0}-\omega-i\Gamma}+\frac{1}{\omega_{n0}+\omega+i\Gamma}\right),
\end{equation}
where $\omega^2_P=4\pi Ne^2/m_e$ .

We note that for a non-relativistic Fermi gas of free electrons, the wave function is represented by a plane wave, i.e:
\begin{equation}
\ket{n}=\frac{1}{\sqrt{V}}e^{i{\bm k}_n\cdot{\bm r}},   
\end{equation}
where $V$ is the volume occupied by the gas. 
Eq. (\ref{almo_1}) then results in $F_{n0}({\bm q})=\delta_{{\bm k}_n,{\bm q}}$, which used in Eq. (\ref{almost_field_71}) yields:
\begin{equation}\label{viva_1}
\varepsilon_{\rm long}(q,\omega)=1+\frac{\omega^2_P}{\omega_q^2-(\omega+i\Gamma)^2}, 
\end{equation}
which is the DL dielectric function given in Eq. (\ref{DLmio}) to fit the experimental optical ELF with $\omega_q=\hbar q^2/2m_e$.
From this expression one can derive
\begin{equation}\label{viva_2}
\frac{i}{\varepsilon_{\rm long}(q,\omega)}=-\frac{i\omega^2_P}{\omega_q^2+\omega_P^2-(\omega+i\Gamma)^2},
\end{equation}
which can be inserted into the non-relativistic limit of Eq. (\ref{almosta}) by neglecting the second term to obtain :
\begin{equation}\label{exp_suml_5}
-\frac{dW}{dx}\approx-\frac{dW}{dx}{\Bigg |}_{\mathrm {long}}=\frac{e^2\omega_P^2}{v^2}\int \frac{dq}{q}=\frac{e^2\omega_P^2}{2v^2}\ln\left( \frac{\zeta+\sqrt{\zeta^2-1}}{\zeta-\sqrt{\zeta^2-1}}\right),
\end{equation}
where $\zeta=mv^2/\hbar\omega_P$.
For $\zeta>> 1$, i.e. for projectiles with kinetic energies much greater than the mean excitation energy, the asymptotic expansion of Eq. (\ref{exp_suml_5}) can be written as follows:
\begin{equation}\label{exp_suml_6}
-\frac{dW}{dx}=\frac{4\pi Ne^4}{mv^2}\Bigg(\ln \left( \frac{2mv^2}{\hbar\omega_P}\right) - \left( \frac{\hbar\omega_P}{2mv^2}\right)^2 -\frac{3}{2}\left( \frac{\hbar\omega_P}{2mv^2}\right)^4 +...\Bigg),
\end{equation}
which is Bethe's asymptotic formula for a non-relativistic charge that hits a target at high speed ($mv^2>>\hbar \omega_P$), which can be approximated as a Fermi gas with free electrons.
As an aside, it should be noted that if Cauchy's main partial equation
\begin{equation}\label{Cauchy}
\frac{1}{\omega-\omega_{n0}+i\Gamma}={\cal P}\frac{1}{\omega-\omega_{n0}}- i\pi \delta({\omega-\omega_{n0}})
\end{equation} 
is used in Eq. (\ref{viva_2}), we obtain (the stopping power in Eq. (\ref{almosta}) must be a real quantity):
\begin{equation}\label{viva_5}
{\rm Re}\Bigg(\frac{i}{\varepsilon_{\rm long}(q,\omega)}\Bigg)=-\frac{\pi\omega^2_p}{2\alpha_q}\Big( \delta(\omega-\alpha_q)-\delta(\omega+\alpha_q)\Big),
\end{equation}
where $\alpha_q^2=\omega^2=\omega_q^2+\omega_P^2=(\hbar q^2/2m_e)^2+\omega_P^2$, which reveals the quadratic dependence of the energy-momentum transfer relation for a Fermi gas (the so-called RPA, see Eq. (\ref{fullRPA})). Due to the presence of the Dirac functions, this expression shows well-resolved resonances and is therefore referred to as the plasmon pole approximation.

The Bethe stopping formula can also be obtained in the limiting case of low target density by using the non-relativistic limit of Eq. (\ref{almosta}) and inserting the expansion of Eq. (\ref{almost_field_71}) up to first order in $N$:
\begin{equation}\label{exp_suml}
\frac{1}{\varepsilon_{\rm long}}\approx 1 - \frac{m_e\omega^2_P}{\hbar q^2}\sum_n|F_{n0}({\bm q})|^2 \times \left(\frac{1}{\omega_{n0}-\omega-i\Gamma} + \frac{1}{\omega_{n0}+\omega+i\Gamma} \right).
\end{equation}
Eq. (\ref{almosta}) can be integrated over $\omega$ by using Eq. (\ref{Cauchy})
to obtain the Bethe-Bloch formula for the stopping power of a non-relativistic target with low density:
\begin{equation}\label{exp_suml_2}
-\frac{dW}{dx}=\frac{2\pi e^4N}{m_ev^2}\sum_n \hbar \omega_{n0}\int \frac{dW}{W^2}|F_{n0}({\bm q})|^2,
\end{equation}
which is equivalent to Eq.  (\ref{cs_bethe_block}) by interpreting the integrand as a differential cross-section.

In the relativistic regime (typically for $\beta=v/c> 0.9$), we are reminded that (i) the transverse contribution to the stopping power (second term in Eq. (\ref{almosta})) cannot be neglected; (ii) the wave functions are characterised by four-component spinors as in Eq. (\ref{rotinv1}).
Assuming that we are dealing with a gas of free relativistic electrons, the derivation of the longitudinal dielectric function for relativistic particles follows the same reasoning that we have presented to obtain Eq. (\ref{almost_field_71}), which in this case leads to the following expression \cite{3Sigmund}:
\begin{eqnarray}\label{relat_field_13}
\varepsilon_{\rm long}(q,\omega)&=&1+\frac{m\omega^2_P}{\hbar q^2}B_q^2\Bigg(\frac{1}{\omega_{q}^+-\omega-i\Gamma}+\frac{1}{\omega_{q}^+ +\omega+i\Gamma}+
\nonumber \\ &+& \frac{(b_q q)^2}{\omega_{q}^- -\omega-i\Gamma}+\frac{(b_q q)^2}{\omega_{q}^- +\omega+i\Gamma}\Bigg)=\nonumber\\
&=& 1+\frac{\omega^2_P}{\omega_q^2-(1+\hbar\omega_q/mc^2)(\omega+i\Gamma)^2+\hbar^2(\omega+i\Gamma)^4/(2mc^2)^2}, \nonumber \\
\end{eqnarray}
where $\hbar\omega^{\pm}_q=\pm E_q-mc^2$ represents the excitation energy for states with positive (+) or negative (-) energy, $B_q^2=(E_q+mc^2)/(2E_q)$ and $b_q=\hbar c/(E_q+mc^2)$. We note that Eq. (\ref{relat_field_13}) is far more complex than Eq. (\ref{almost_field_71}) valid for the non-relativistic case; however, it is reduced to the first term of the expansion in Eq. (\ref{exp_suml_6}) under the assumption of $\hbar\omega_P/(2mc^2)<< 1$.

The contribution of the transverse excitation to the stopping power can be rewritten as follows
\begin{eqnarray}
-\frac{dW}{dx}{\Bigg |}_{{\rm trans}}&=&\frac{2\pi e^4N}{mv^2}\sum_n \hbar\omega_{n0}\int \frac{dW}{W^2}\times |F_{n0}({\bm q})|^2 \frac{2mv^2W/(\hbar\omega_{n0})^2-1}{[2mc^2W/(\hbar\omega_{n0})^2-1]^2}\nonumber \\
&=&\frac{2\pi e^4NZ}{mv^2}\sum_n \int \frac{dW}{W}\times |f_{n0}({W})|^2 \frac{2mv^2W/(\hbar\omega_{n0})^2-1}{[2mc^2W/(\hbar\omega_{n0})^2-1]^2},\nonumber\\
\label{rel_5}
\end{eqnarray}
for $W>\hbar \omega_{n0}$,
where
\begin{equation}\label{rel_6}
f_{n0}(W)=\frac{1}{Z}\frac{\hbar\omega_{n0}}{W}|F_{n0}({\bm q})|^2
\end{equation}
are the generalised oscillator strengths that can be approximated by a Poisson distribution $f_{n0}(W)\propto (W/\hbar \omega_0)^{(n-1)}e^{-W/(\hbar \omega_0)}$ in a semi-classical theory in which the bound electron is modelled as a harmonic oscillator.

Given the velocity-dependent weight factor in the integrand, one can conveniently express Eq. (\ref{rel_5}) in the limits of low (essentially within the validity of the dipole approximation, a regime typical of weak or distant interactions in real space), medium (where the non-relativistic description of particle dynamics is still valid) and high (relativistic, typical of strong or close collisions in real space) momentum transfers \cite{3Sigmund}.
Summing up all these contributions we derive the relativistic Bethe formula for a relativistic charged projectile:

\begin{equation}
-\frac{dW}{dx}
=\frac{4\pi e^4NZ}{mv^2}\sum_n f_{n0}\Bigg[\ln\left(\frac{2mv^2}{\hbar\omega_{n0}}\right)-\ln\left(1-\frac{v^2}{c^2}\right)- \frac{v^2}{c^2}\Bigg] 
\label{rel_9}
\end{equation}

A comparison with Eq. (\ref{exp_suml_6}) shows that the second and third terms in Eq. (\ref{rel_9}) come from relativistic corrections.
We also note that for $v<< c$, i.e. far away from the relativistic limit, the expansion in Eq. (\ref{rel_9}) consists only of the first term, which can be rewritten as
\begin{equation}
-\frac{dW}{dx}
=\frac{4\pi e^4NZ}{mv^2}\ln \frac{2mv^2}{I},
\label{rel_11}
\end{equation}
where $\ln I=\sum_nf_{n0}\ln(\varepsilon_n-\varepsilon_0)$ is commonly referred to as the mean logarithmic excitation energy. \footnote{Please note that if we take into account the indistinguishability of the scattered electrons from the ejected ones, Eq. (\ref{rel_11}) becomes
$$-\frac{dW}{dx}
=\frac{4\pi e^4NZ}{m_ev^2}\ln \sqrt{\frac{\rm{e}}
{2}}\,\frac{m_ev^2}{2I}=\frac{2\pi e^4NZ}{E}\ln\frac{1.166 E}{I}$$
where $E=m_ev^2/2$ is the electron kinetic energy and $\rm{e}$ is Euler's number \cite{Dapor2023,Ashley1988,Ashley1990,3Sigmund}.}

Further details on the derivation of the relativistic Bethe relation with an in-depth analysis of the different momentum transfer regimes can be found in Ref. \cite{3Sigmund}. The Bethe-Bloch stopping power formula can be used in conjunction with the continuous-slowing-down approximation to model charge transport in solids as part of an MC approach.

\subsection{Inelastic scattering: the Auger decay}\label{Auger_inles}

Depending on the kinetic energy of the impinging projectile, inelastic electron scattering can lead to excitation and ionisation of the atomic constituents of a solid.
Especially if the excitation or ionisation affects inner shells, Auger states can be generated. These are quasi-bound states that can decay through the emission of a secondary electron, the Auger electron, or through radiation. The decaying system is in a singly or doubly ionised state, depending on whether the initial excitation is neutral (autoionising state) or ionising (Auger state), i.e. whether the projectile has enough energy to promote an electron into an empty bound orbital or into the continuum.
Often the primary excitation process can lead to simultaneous ionisation
(shake-off) or excitation (shake-up) of a valence electron, resulting in satellite lines in the spectrum (typically these processes account for 10\% of the total spectral intensity). In this section, dedicated to Auger decay, we use atomic units to simplify the equations.

In general, we consider ionisation and Auger emissions as resonance processes in which the excited system decays to the higher charge state. Auger decay in particular is typically treated as a two-step process, where the first ionisation is followed by the non-radiative decay. In order to determine the contribution of these processes to the line shape (see Fig. \ref{fig:spectrumex}), the corresponding autoionisation or Auger cross-sections must be evaluated. This task can be fulfilled within the framework of Fano's theory of decay processes \cite{PhysRev.124.1866} in conjunction with the projected potential approach developed in section \ref{abinito} for elastic scattering.
In the framework of Fano, the scattering wave function, which is an eigenstate of the full Hamiltonian of the system, can be represented by the following linear combination of configurations
\begin{equation}
|\Psi^-_{\alpha,{\bm k}}\rangle=\sum_j^{m_d} a_{j\alpha}({\bm k})|\Phi_j\rangle+
\sum_{\beta=1}^{m_c}\int\limits_0^\infty\frac{d{\bm p}}{(2\pi)^3}\ket{\chi^-_{\beta,{\bm p}}} b_{\beta,
\alpha}({\bm p},{\bm k}), \label{TF1}
\end{equation}
where $|\Phi_j\rangle$ is a discrete bound resonance state generated by the primary collision and degenerate with several continua $\{|\chi_{\beta,{\bm p}}^-\rangle \}$ that asymptotically represent the decay channels of the ionised target identified by the quantum number $\beta$ (e.g. the angular momentum and its projection along an axis in spherically symmetric systems) and the momentum $\bm{p}$ of the emitted electron \cite{1982HD469A}. $|\Psi^-_{\alpha{\bm k}}\rangle$ is a stationary scattering state at energy $E=E_\alpha+{k^2\over2}$, which is asymptotically defined by the momentum ${\bm k}$ of the emitted electron and the state $|\alpha\rangle$ of the singly ionised target (for an autoionisation process) or doubly ionised target (for an Auger process).
Eq. (\ref{TF1}) is of course written under the assumption that the Born--Oppenheimer approximation is valid, i.e. that the typical time scale of electronic excitation/de-excitation is much faster than that of the nuclei. Therefore, the discrete intermediate states can represent, for example, electronic bound states with the corresponding roto-vibrational levels of the target molecule.
The continuum wave functions $\{|\chi_{\alpha,{\bm k}}^-\rangle\}$ are characterised by the incoming wave boundary conditions to represent a free electron asymptotically, which is ejected with the kinetic energy ${\epsilon=k^2/2}$ in the direction ${\bm k}$ and a system in the ionised state $\alpha$ with the energy $E_{\alpha}$, so that $E=E_{\alpha}+\epsilon$.
The coefficients of Eq.~(\ref{TF1}) are obtained by solving the following equations
\begin{equation}\langle \Phi_l|H-E|\Psi^-_{\alpha,{\bm k}}\rangle =\langle \chi^-_{\gamma,{\bm \tau}}
|H-E|\Psi^-_{\alpha,{\bm k}}\rangle =0 ~~~~\forall ~l,\gamma,{\bm{\tau}},\label{TF5}
\end{equation}
which explicitly read:
\begin{equation}(E_l-E)a_{l\alpha}({ \bm k})+\sum_{\beta=1}^{m_c}\int\limits_0^\infty\frac{d{\bm p}}{(2\pi)^3}
M_{l\beta}(E,{\bm p})b_{\beta\alpha}({\bm p},{\bm k})=0\,,\label{TF6}\end{equation}
\begin{equation} \sum_j^{m_d}a_{j\alpha}({\bm k})M_{\gamma j}^-(\tau,E)^*+(\epsilon+
E_\gamma-E)b_{\beta\alpha}({\bm{\tau}},{\bm k})=0\,, \label{TF7}\end{equation}
where the elements of the Hamiltonian matrix constructed using discrete and continuum
functions are the following
\begin{eqnarray}
\langle \Phi_l|H-E|\chi_{\beta,{\bm p}}^-\rangle=M_{l\beta}(E,{\bm p}), \label{TF8}\\
\bra{\Phi_l} H-E |\Phi_j\rangle =(E_l-E)\delta_{lj},\label{TF88} \\
\bra{\chi^-_{\gamma,{\bm \tau}}}H-E\ket{\chi^-_{\beta,{\bm p}}}=(E_{\gamma}+\frac{\tau^2}{2}-E)\delta_{\gamma\beta}(2\pi)^3\delta({\bm{\tau}}-{\bm p}).
\end{eqnarray}
The matrix element $M_{l\beta}$ in Eq. (\ref{TF8}) couples the discrete electronic state with the continuum state and from Eq. (\ref{TF88}) ${E_l}$ is the energy of the discrete level corresponding to the quasi-bound states ${|\Phi_l\rangle}$.
The states $\ket{\chi^-_{\beta,\tau}}$ in Eq.~(\ref{TF1}) are continuum states obtained by diagonalising the full Hamiltonian of the system
\begin{eqnarray}\langle \chi_{\beta,\epsilon}|H-E|\chi_{\gamma,\epsilon^\prime}\rangle =
(E_\beta+\epsilon-E)\delta(E_\beta+\epsilon-E_\gamma-\epsilon^\prime)
+V_{\beta\gamma}(\epsilon,\epsilon^\prime;E),\label{TF3}\nonumber \\
\end{eqnarray}
where $\{|\chi_{\beta,\epsilon}\rangle \}$
are interacting continuum states.

To avoid singularities in the coefficients
$\{a_\alpha,b_{\beta\alpha}\}$, Eqs.~(\ref{TF6}) and (\ref{TF7}) are shifted into the complex plane, where we obtain the following expression for the scattering wave function
\begin{eqnarray}
|\Psi^-_{\alpha,{\bm k}}\rangle = |\chi^-_{\alpha,{\bm k}}\rangle + \sum_{jl}
{M_{l\alpha}(E,{\bm k})\over (E-E_{l})\delta_{lj}-\Delta_{lj}-i{\Gamma_{lj}\over 2}} \times\nonumber \\
\left[|\Phi_j\rangle +\lim_{\nu\rightarrow 0}\sum_{\beta}^{m_c}
\int\limits_0^\infty\frac{d\bm p}{2\pi^3}{|\chi^-_{\beta,{\bm p}}\rangle M^*_{\beta j}({\bm p},E)\over
E-E_\beta-\frac{p^2}{2}-i\nu}\right],\label{TF9}
\end{eqnarray}
where

\begin{equation}\Gamma_{lj}=2\pi\sum_{\beta}^{m_c}
\int\limits_0^\infty \frac{d\bm p}{2\pi^3}M_{l\beta}(E,{\bm p})M^*_{j\beta}(E,{\bm p})\delta(E-E_{\beta}-\frac{p^2}{2}),
\label{TF10}\end{equation}
\begin{equation}
\Delta_{lj}=\sum_{\beta}^{m_c}
{\cal P}\int\limits_0^\infty \frac{d\bm p}{2\pi^3} \frac{M_{l\beta}(E,{\bm p})M^*_{j\beta}(E,{\bm p})}{
E-E_\beta-\frac{p^2}{2}},\label{TF11}
\end{equation}
represent the width and energy shift of the transition. 

Finally, if we know for each quantum number $\alpha$ and each momentum ${\bm k}$, which asymptotically define the remaining ionised system and the emitted electron, the stationary states $\{|\Psi^-_{\alpha,{\bm k}}\rangle \}$, we can derive the cross-section for the autoionisation (or Auger) process. In this process, thus, the initial collision perturbs the system by promoting an electron from an inner-shell occupied orbital of the ground state $\ket{0}$ to an empty resonance state $|\Phi_j\rangle$, which decays non-radiatively into the channel $|\chi^-_{\alpha,{\bm k}}\rangle$ representing a single (double) ionised atom and an electron in the continuum.
In particular, the autoionisation cross-section is proportional to the square element of the following on-shell $T$ matrix ($H_{1}$ is the transition dipole operator that occurs in Eq. (\ref{Fgoldrule})):
\begin{equation}
{\bm T}^+_{0\rightarrow \alpha}({E})=\langle \chi^-_{\alpha,{\bm k}}
|H_1|0\rangle  +{\langle \chi^-_{\alpha,{\bm k}}|H-E|\Phi_j\rangle 
\langle \Phi_j^-|H_1|0\rangle 
\over {E}-{E_l}\delta_{lj}-\Delta_{lj}-i{\Gamma_{lj}\over 2}},\label{TF11d}
\end{equation}
which can be obtained using Eqs.~(\ref{TF8}), (\ref{TF10}) and (\ref{TF11}) with the definition
\begin{equation}
\ket{\Phi_j^-}=\left[|\Phi_j\rangle +\lim_{\nu\rightarrow 0}\sum_{\beta}^{m_c}
\int\limits_0^\infty\frac{d\bm p}{2\pi^3}{|\chi^-_{\beta,{\bm p}}\rangle M^*_{\beta j}({\bm p},E)\over
E-E_\beta-\frac{p^2}{2}-i\nu}\right],
\end{equation}
which represents the state $\ket{\Phi_j}$ modified by a mixture of continuum states.
In Eq. (\ref{TF11d}), the first term represents the direct ionisation of the atom in the solid (it represents a direct double ionisation in the Auger process). In the second term, the transition operator connects the initial ground state $|0\rangle$ with the final continuum state $|\chi_{\alpha,\bm k}^- \rangle$ via the resonance state $\ket{\Phi_j}$.
Finally, the cross-section of the resonant autoionisation process is
\begin{eqnarray}
&&{\partial\sigma_{0\rightarrow\alpha}\over\partial{\bm k}}
({\bm k};\omega)=\left({4\pi^2\omega\over c}\right)
|<0|H_{1}|\chi^-_{\alpha\vec k}>\nonumber\\
&+&\sum^{m_c}_{\beta=1}\lim_{\nu\rightarrow\infty}\int{d{\bm p}\over(2\pi)^3}
{<0|H_{1}|\chi^-_{\beta{\bm p}}>
{\overline\Gamma}_{\beta\alpha}
({\bm p},{\bm k})\over E-E_{\beta}-{p^2\over 2}-i\nu}\nonumber\\
&+&\sum^{m_d}_{jl=1}<0| H_{1}|\Phi_j>\Lambda_{jl}^{-1}
<\Phi_l|H-E|\chi^-_{\alpha{\bm k}}>|^2 
\delta[(E_0+\omega)-(E_{\alpha}+{k^2\over 2})],\nonumber\\ \label{TF11e}
\end{eqnarray}
where
\begin{equation}
\Lambda_{lj}=(E-E_l)\delta_{lj}-\Delta_{lj}-i{\Gamma_{lj}\over
2},
\end{equation}
and 
\begin{eqnarray}
{\overline \Gamma}_{\beta\alpha}(\vec k,\vec p)
&=&\sum^{m_d}_{jl}M^+_{\beta j}(\vec p,E)\Lambda_{jl}^{-1}M_{l\alpha}
(E,{\bm k})=\nonumber \\
&=&\sum^{m_d}_{jl}<\chi^-_{\beta{\bm p}}|H -E|\Phi_j>\Lambda_{jl}^{-1}
<\Phi_l|H-E|\chi^-_{\alpha{\bm k}}>.
\end{eqnarray}

The first two terms on the right-hand side of Eq. (\ref{TF11e}) describe the direct ionisation process from the ground state $|0\rangle$ to the final decay state
$|\chi^-_{\alpha\vec k}\rangle$. The third term, on the other hand, provides the resonance contribution, i.e. the contribution that can be attributed to the decay process via bound intermediate states $\{|\Phi_j\rangle\}$. 

An expression similar to that in Eq. (\ref{TF11e}) with $\bra{0}H_{1}\ket{\chi^-_{\alpha\vec k}}$
replaced by \\$\bra{0}H_{1}\ket{\chi^-_{\alpha{\bm k}}\eta_{\bm k_1}}$, and furthermore, $\bra{0}H_{1}\ket{\Phi_j}$ substituted by
$\bra{0}H_{1}\ket{\Phi_j\eta_{\bm k_1}}$ and $\delta[(E_0+\omega)-(E_{\alpha}+{k^2\over 2})]$ by
$\delta[(E_0+\omega)-(E_{\alpha}+{k^2+k_1^2\over 2m})]$ gives the detailed expression of the cross-section for an Auger process.
In the case of an Auger process, the direct process is a double ionisation process and therefore the first two terms on the right-hand side of Eq. (\ref{TF11e}) contain matrix elements connecting a bound state (the ground state) to the continuum states with two unbound electrons. The third term, on the other hand, is characterised by matrix elements between discrete and continuum states with only one unbound electron. The interference between these different terms leads to asymmetric Fano profiles in the electron spectrum, which are specific to the transition. Depending on the kinetic energy of the projectile, the first two terms are usually smaller than the third term, which can usually be neglected. Within this approximation, Eq.~(\ref{TF11e}) shows a Lorentzian profile (rather than a Fano profile) characterised by the decay rate $\Gamma_\alpha=\sum_{\beta}\Gamma_{\beta\alpha}$ into the channel $|\alpha\rangle$.

In an Auger process, the global $N$-electron stationary state with the primary electron at the energy $\epsilon_1=
{{k_1}^2\over 2}$ and momentum
$\bm k_1$, can be represented by the following antisymmetric product:
\begin{equation}
|\Theta_{\alpha}({\bm k},{\bm k_1})\rangle=\hat {\cal A}\left[
|\Psi^-_{\alpha{\bm k}}\rangle|\eta_{\bm k_1}\rangle\right]\label{ME17}
\end{equation}
where the wave function $|\eta_{\bm k_1}\rangle$ describes the primary electron, which only interacts very weakly with the secondary Auger electron when the kinetic energy of the projectile is much higher than the ionisation energy of the inner-shell electrons.

In the same framework, the derivation of the cross-section can be generalised to a molecular system in which an isolated discrete electronic state supporting a range of vibrational levels is coupled to several electronic continua (open channels) with their vibrational levels.
In fact, in order to be compared with the experimental data, the theoretical cross-section, Eq.~(\ref{TF11e}), must take into account the intrinsic properties of the target, the finite resolution of the electron spectrometer and the broadening of the autoionisation (or Auger) lines due to the energy losses of the emitted electrons on their way out of the solid before they leave the material surface \cite{taioli2010electron}. In the latter case, the ab initio spectrum can be used as input for an MC scheme \cite{taioli2009surprises,taioli2009mixed}.
An application of this idea to calculate the Auger spectrum of silica, including the energy loss mechanisms, is developed in the Results section.

The cross section in Eq. (\ref{TF11e}) can be determined by calculating the resonance state and the continuum wave functions.
To this end, we use the static exchange approximation, which amounts to calculating separately
the bound intermediate states using the methods of quantum chemistry (e.g. HF, CI, MBPT) \cite{taioli2010electron} or condensed matter physics (e.g. DFT) \cite{carvalho2021computational}) and the scattering wave function using the projected potential approach.

The interacting decay channels are represented by the antisymmetric normalised product
\begin{equation}
\chi_{\alpha{\bm k}}(1,..N-1,N)=\hat {\cal A}
[\Theta_{\alpha}(1,...,N-1)\psi_{\alpha{\bm k}}(N)]\,,\label{TH5}
\end{equation}
where $\psi_{\alpha{\bm k}}(N)$ is the spin orbital of the emitted electron and $\Theta_{\alpha}(1,...,N-1)$ is the Slater determinant representing the bound state of the remaining $(N-1)$ system ($(N-2)$ for Auger processes), which is the solution of the following secular problem
\begin{equation}
\langle \Theta_{\alpha}|H^{N-1}|\Theta_{\beta}\rangle =
E_{\alpha}\delta_{\alpha\beta}\,,\label{TH6}
\end{equation}
in a functional space spanned by the one-particle orbitals $|\theta_j\rangle,~j=1,2,...,M$ with $M>>N-1$, which are expanded in a GBF basis set (see Eq. (\ref{gaussian})). 

To construct the continuum orbital in Eq.~(\ref{TH5}), which represents a particle emitted in the effective field generated by the bound electrons of the system, we use the method of projected potentials developed in section \ref{abinito}. Specifically, we define a model Hamiltonian for an $N$-electron system as follows:
\begin{equation}
H(1,..,N)=\sum^N_{i=1}[ T(i)+V_{en}^{\pi}(i)]
+{1\over 2}\sum^N_{i\neq j} v^{\pi}(i,j),\label{TH1}
\end{equation}
where
\begin{equation}
T(i)=-{1\over 2}\nabla^2_i~,~~~
V_{en}^{\pi}(i)=\hat\pi(i)V_{en}(i)\hat\pi(i),\label{TH2}\end{equation}
\begin{equation}v^{\pi}(i,j)=\hat\pi(i)\hat\pi(j) v(i,j)\hat\pi(i)\hat\pi(j),
\label{TH3}
\end{equation}
and $\hat\pi(i)=\sum_{l=1}^{m}~|g_l(i)\rangle \langle g_l(i)|$ is a projector into an $m$-dimensional space spanned by the Gaussian functions and $V^{en}(i)=-\sum_{\mu}{Z_{\mu}\over|{\bm r}_i-{\bm R}_{\mu}|}$ and $v(i,j)=
{1\over|{\bm r}_i-{\bm r}_j|}$ are the electro-nuclear and electron-electron Coulomb interaction potentials.

Instead of directly solving the secular problem of the Hamiltonian (\ref{TH1}), the continuum eigenfunctions can be found (as we have demonstrated in the case of the elastic scattering in section \ref{abinito}) by using the LS with the appropriate boundary conditions (see Eq. (\ref{LSp1}), valid for the elastic collisions):
\begin{equation}
\psi^-_{\alpha {\bm k}}(\bm r)=e^{i\bm k\cdot\bm r}+
G_0^-(\epsilon)\hat T_\alpha(\epsilon)e^{i\bm k\cdot\bm r},
\label{ML11}\end{equation}
where $G_0^-(\epsilon=\frac{k^2}{2})=\frac{1}{\epsilon-\frac{\hat k^2}{2}-i\eta}$ is the Green's function of the free particle, $\eta$ is an infinitesimal positive number and $\hat T_\alpha$ is the transition operator given by the equation:
\begin{equation}
\hat T_\alpha=V^{\pi}_\alpha+ V^\pi_\alpha\hat G_0^-(\epsilon)
\hat T_\alpha.\label{ML12}
\end{equation}
The interacting channels defined in Eq. (\ref{TH5}) are coupled by the following interchannel interaction
\begin{eqnarray}
&&\langle \chi_{\alpha{\bm k}}|H^N|\chi_{\beta{\bm p}}\rangle =
(2\pi)^3\delta({\bm k} - {\bm p})\delta_{\alpha\beta}\left({k^2\over2}+E_{\alpha}\right)+
\langle \psi_{\alpha{\bm k}}|V^{en}_{\pi}\delta_{\alpha\beta}+
W_{\pi}^{\alpha\beta}|\psi_{\beta{\bm p}}\rangle,\nonumber\\ &&\label{TH11}\end{eqnarray} 
where
\begin{equation}
W_{\pi}^{\alpha\beta}(1)=\sum_{j=2}^N
\langle \Theta_{\alpha}(2,.j.,N)|v_{\pi}(1,j)(I- \hat {\cal P}_{1,j})|
\Theta_{\beta}(2,.j.,N)\rangle ,\label{TH12}\end{equation}
and $\hat {\cal P}_{1,j}$ is the operator that determines the exchange of the
$(1,j)$ variables and $I$ is the identity operator. In Eq. (\ref{TH12}) $\Theta_{\alpha}$ are the solutions of the bound-state problem of Eq. (\ref{TH6}). The matrix elements defined in Eq.~(\ref{TH11}), through which the channels are mixed via the effective interchannel interaction, can be viewed as a representation of an effective one-particle Hamiltonian over a set of basis vectors \hbox{$\{|\varphi_{\alpha{\bm k}}\rangle \}$} with internal degrees of freedom, which moves in an effective potential depending on the internal states $\{|\alpha\rangle \}$ of the particle. The diagonalisation of Eq. (\ref{TH11}), projected onto the GBFs, yields the wave functions for the different channels that are open to decay and can be used in Eq. (\ref{TF11e}) to determine the autoionization or Auger cross-sections. Finally, the spectrum of autoionisation (Auger) can be plotted as a function of the kinetic energy of the emitted electrons (see Fig. \ref{QMMC}).

\section{Results and Discussion} \label{Results}

In this section, we present a number of applications of the previously discussed methods for modelling elastic and inelastic collisions as well as MC calculations of energy losses and secondary spectra.

\subsection{Differential and total elastic scattering cross-sections}\label{DESCS_section}

To test the accuracy of Mott's cross-section (see section \ref{elasmott}) in reproducing the experimental results, we present the DESCS obtained using the Eq. (\ref{decs}) for an unpolarised electron beam with a kinetic energy of 1000 eV scattered by bulk aluminum (Fig. \ref{fig:prova1}) and gold (Fig. \ref{fig:prova2}) as a function of the scattering angle measured from the surface normal (see Fig. \ref{fig:scattexp} for the experimental setup).
In Fig. \ref{fig:prova3} we show the same calculation for mercury at lower kinetic energy (300 eV). The agreement with experimental data and other calculations shown by dots and stars in the figures is excellent at all energies and scattering angles. 

\begin{figure}[hbt!]
\centering
\includegraphics[width=1.0\textwidth]{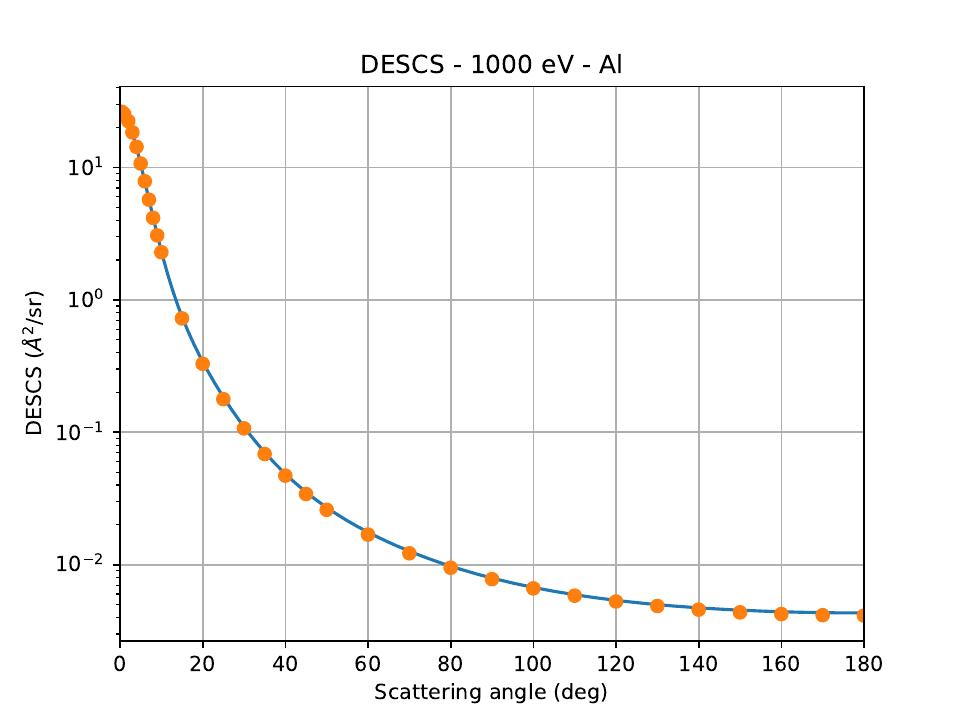}
\caption{Differential elastic scattering cross section calculated with Mott's formula of Eq. (\ref{decs}) (solid line) for an electron beam with a kinetic energy of 1000 eV scattered by bulk aluminum as a function of the scattering angle measured from the surface normal (see Fig. \ref{fig:scattexp} for the experimental setup). The dots represent the calculations from Ref. \cite{3Rileyetal}.}
\label{fig:prova1}
\end{figure}
 
\begin{figure}[hbt!]
\centering
\includegraphics[width=1.0\textwidth]{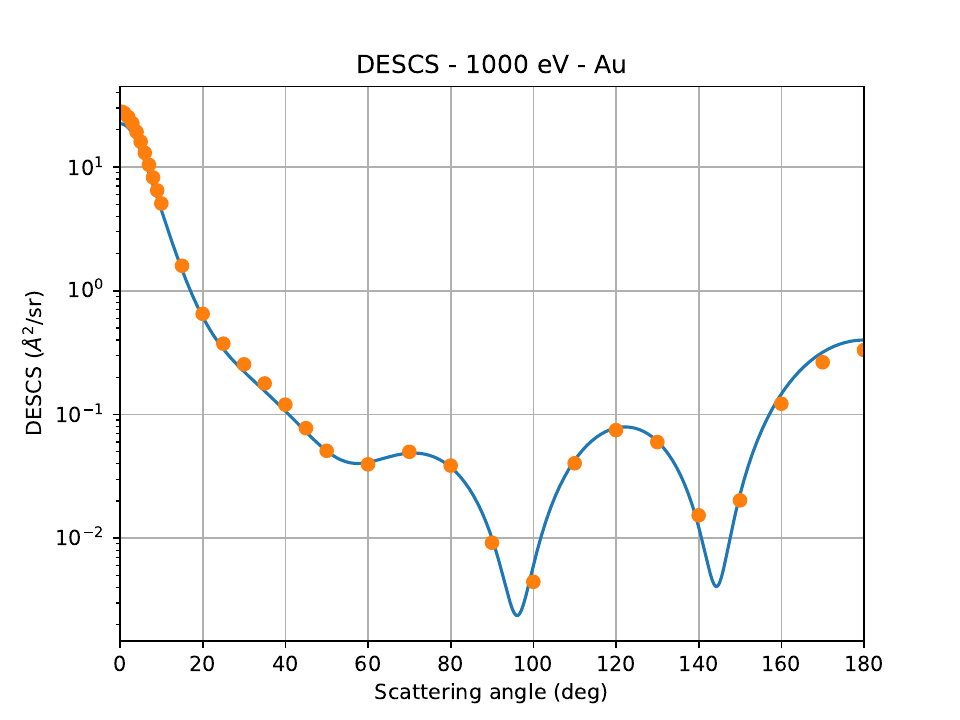}
\caption{Differential elastic scattering cross-section calculated with Mott's formula of Eq. (\ref{decs}) (solid line) for an electron beam with a kinetic energy of 1000 eV scattered by bulk gold as a function of the scattering angle measured from the surface normal (see Fig. \ref{fig:scattexp} for the setup of the experiment). The dots represent the calculations from Ref. \cite{3Rileyetal}.}
\label{fig:prova2}
\end{figure}

\begin{figure}[hbt!]
\centering
\includegraphics[width=1.0\textwidth]{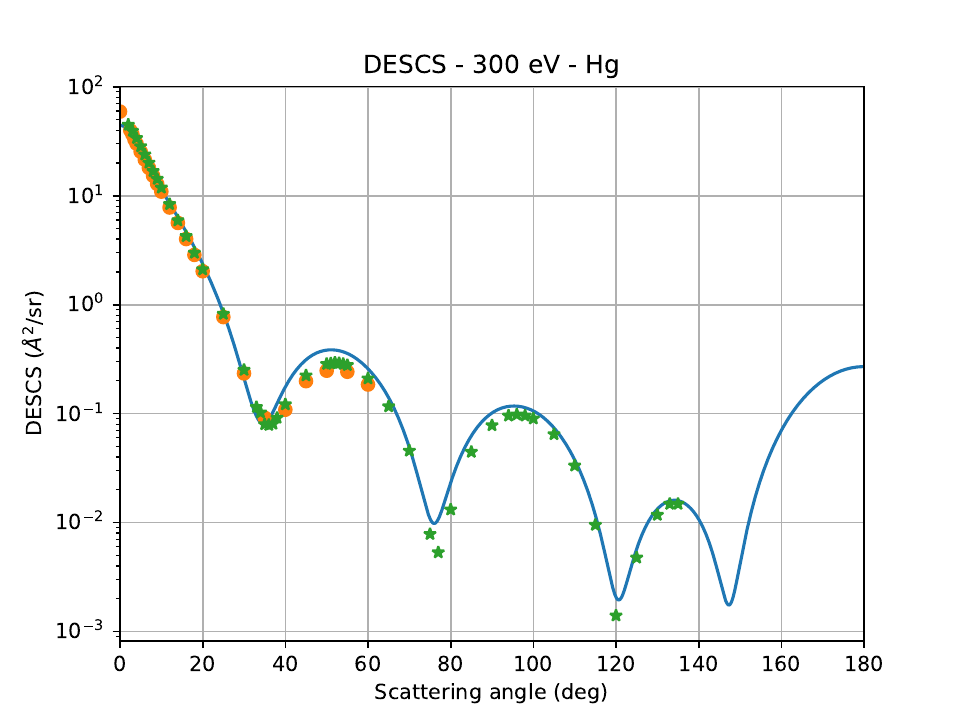}
\caption{Differential elastic scattering cross-section, calculated with Mott's formula of Eq. (\ref{decs}) (solid line) for an electron beam with a kinetic energy of 300 eV scattered by bulk mercury as a function of the scattering angle measured from the surface normal (see Fig. \ref{fig:scattexp} for the setup of the experiment). The dots \cite{10.1063/1.1672634} and stars \cite{Holtkamp_1987} represent the experimental data.}
\label{fig:prova3}
\end{figure}

\begin{figure}[htb!]
\centering
\includegraphics[width=1.0\textwidth]{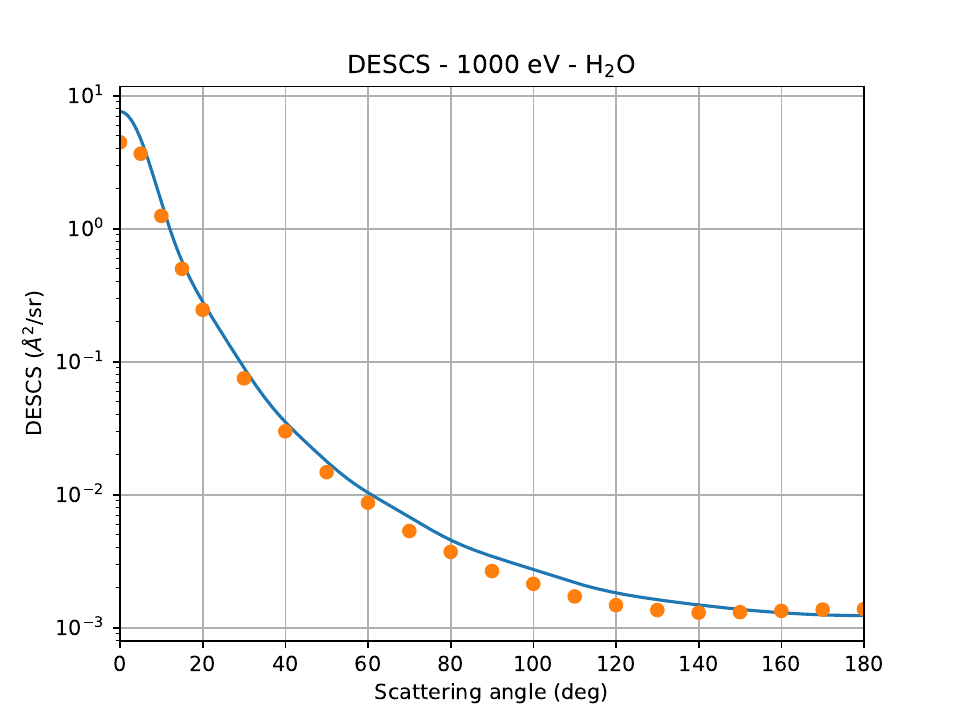}
\caption{Differential elastic scattering cross-section calculated with Mott's formula of Eq. (\ref{molsca}) (solid line) for an electron beam with a kinetic energy of 1000 eV scattered by a water molecule as a function of the scattering angle measured from the surface normal (see Fig. \ref{fig:scattexp} for a plot of the experimental setup). The dots represent experimental data from Ref. \cite{Katase_1986}.}
\label{fig:prova4}
\end{figure}

\begin{figure}[htb!]
\centering
\includegraphics[width=1.0\textwidth]{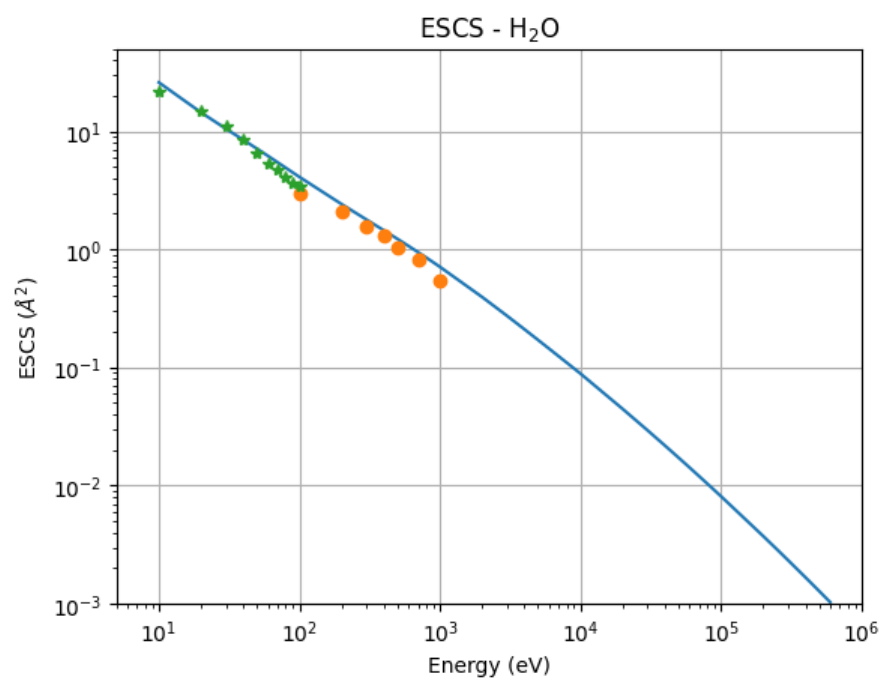}
\caption{Total elastic scattering cross-section calculated using Mott's theory and the formula of Eq. (\ref{sitot}) (solid line) of electrons scattered by a water molecule as a function of the electron kinetic energy. The dots \cite{Katase_1986} and stars \cite{10.1063/1.1799251} represent experimental data.}
\label{fig:prova5}
\end{figure}

Furthermore, with the help of Eq. (\ref{molsca}), which generalises Mott's atomic formula to molecular systems, we can also plot in Fig. \ref{fig:prova4} the DESCS of a single water molecule.
Finally, in Fig. \ref{fig:prova5} the Mott's ESCS (solid line) in a large kinetic energy range, which is obtained by integrating the DESCS over the scattering angle $\vartheta \in [0,\pi]$ and the azimuth angle $\varphi \in [0,2\pi]$ (see Eq. (\ref{sitot}) for a centre-symmetric potential) at each energy is compared with the experimental data (dots and stars). The ESCS is still calculated for a single water molecule.

Nevertheless, it may be important to increase the accuracy of the ESCS calculation by including solid-state effects, i.e. the effects of the environment surrounding the atom, e.g. bound in a solid or in a molecule in the gas or liquid phase.
This task can be accomplished by using the first-principles relativistic approach developed in section \ref{abinito}.
In this context, we plot in Figs. \ref{fig:single_molecule}, \ref{fig:cs_water_mol} and \ref{fig:cs_zundel} (see Ref. \cite{triggiani2023elastic}) the calculation of the ESCS of water for a single molecule, a cluster of three molecules and the Zundel cation (protonated water).

\begin{figure}[htb!]
\begin{center}
\includegraphics[width=1.0\textwidth,trim={0.6cm 0.6cm 0cm 0cm},clip=true]{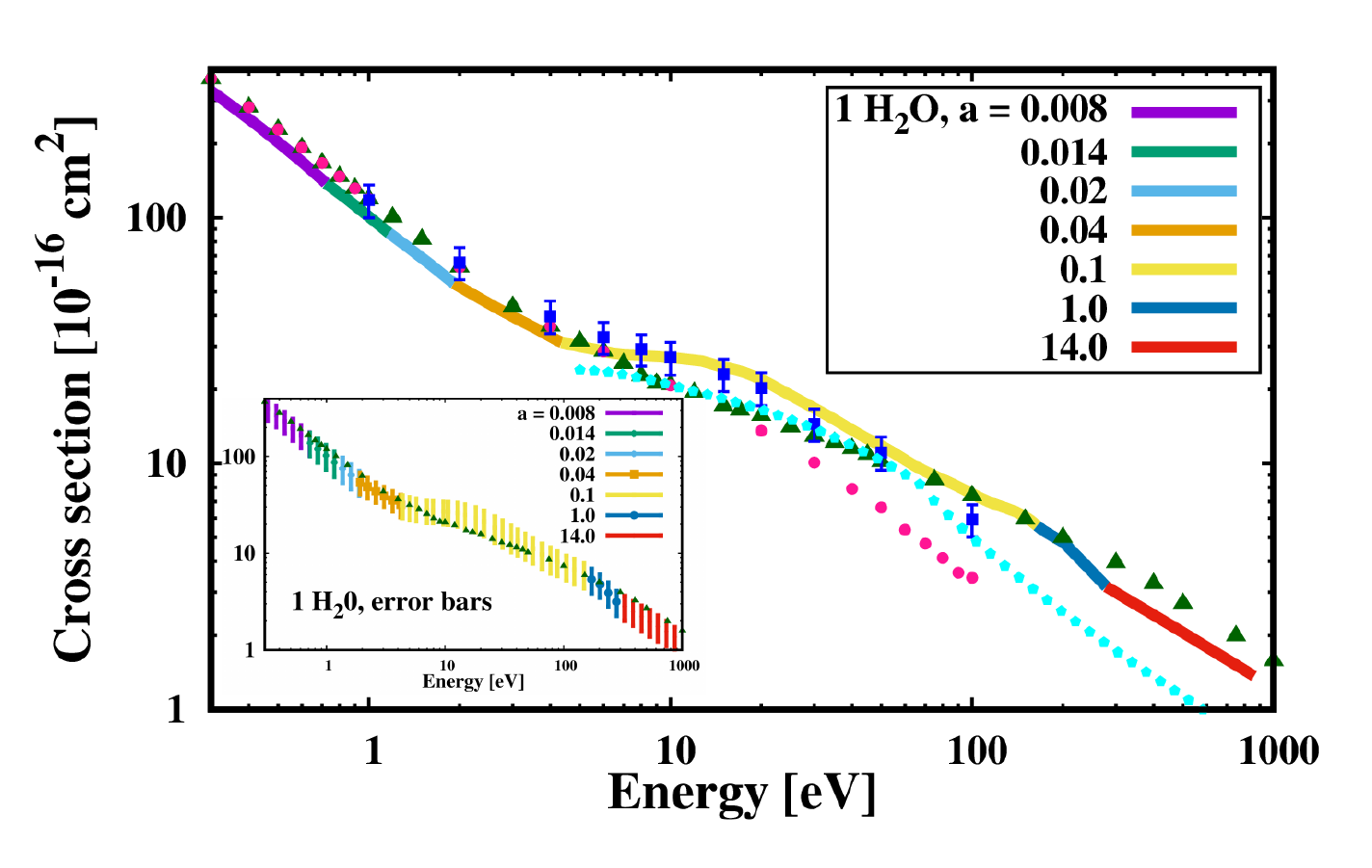}
\end{center}
\caption{Theoretical elastic scattering cross-section (continuous multicoloured line) of electrons hitting a single water molecule with different kinetic energies compared to mixed experimental/theoretical results from Ref. \cite{khakoo_2013} (full blue squares with error bars), with the recommended elastic (full purple circles) and total (green triangles) cross-sections from Ref. \cite{song_2021}. The recommended elastic cross-section from Ref. \cite{song_2021} is based on the results obtained in Ref. \cite{gorfinkiel_2002} up to 6 eV and by interpolation of the cross-sections obtained with different methods \cite{10.1063/1.1799251}. The recommended total cross-section for energies between 0.1 and 7 eV is determined by R-matrix calculations \cite{zhang_2009,faure_2004}, for 7–50 eV it is based on experiments from Refs. \cite{szmytkowski_1987,szmytkowski_2006,kadokura_2019} and above 50 eV on experimental measurements from Ref. \cite{munoz_2007}. The cyan-coloured circles show the results obtained with the ELSEPA code \cite{3ELSEPA}. In the inset, we give the statistical error bar of the cross-section compared to the total cross-section from Ref. \cite{song_2021}. The data in the $x$- and $y$-axes are given in $\log$-scale, $a$ is in $a_0^{-2}$ ($a_0$ is the Bohr radius). Source: Reprinted from \cite{triggiani2023elastic}.}\label{fig:single_molecule}
\end{figure}

\begin{figure}[htb!]
\begin{center}
\includegraphics[width=1.0\textwidth,trim={0cm 0.6cm 0cm 0cm},clip=true]{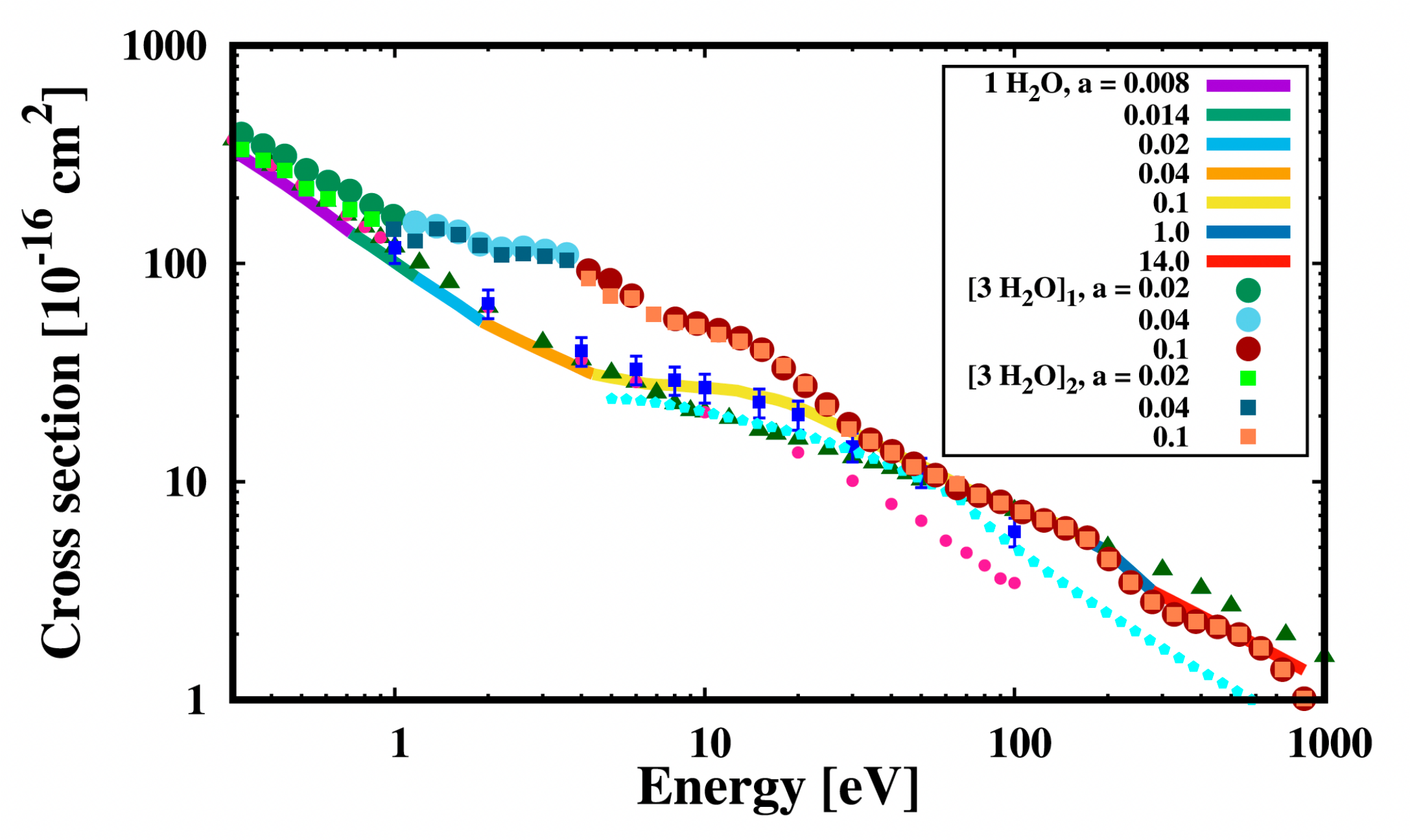}
\end{center}
\caption{ESCS of electrons hitting clusters of three water molecules in different molecular configurations (3 [H$_2$O]$_{1,2}$, multicoloured dots) with different kinetic energies compared to the results for the single water molecule (1 H$_2$O, multicoloured line). The cyan coloured dots show the values calculated with the ELSEPA code \cite{3ELSEPA}. The total cross-sections are normalised to the number of molecules in the clusters. The data on the $x$- and $y$-axis are given in $\log$-scale, $a$ is in $a_0^{-2}$ ($a_0$ is the Bohr radius). Source: Reprinted from \cite{triggiani2023elastic}.}\label{fig:cs_water_mol}
\end{figure}

\begin{figure}[htb!]
\begin{center}
\includegraphics[width=1.0\textwidth,clip=true]{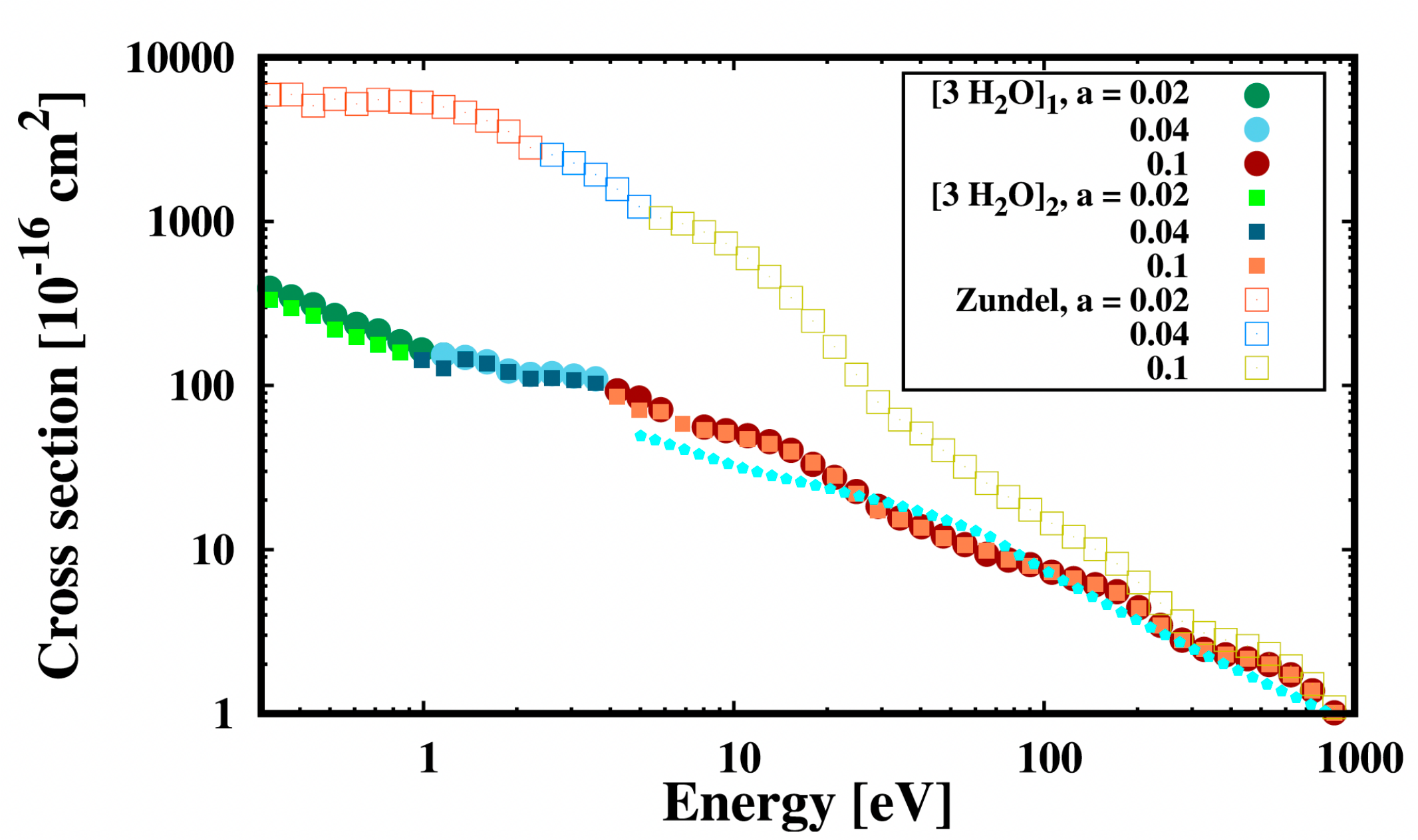}
\end{center}
\caption{ESCS of electrons colliding with the Zundel cation (multicoloured empty squares) at different kinetic energies compared to water clusters (3 [H$_2$O]$_{1,2}$, multicoloured dots). The cyan coloured dots show the values calculated with the ELSEPA code \cite{3ELSEPA} for the Zundel molecule, where the charge of the shared proton is compensated by an additional electron to achieve charge neutrality.
The total cross-section of the water clusters was divided by three to normalise with the number of molecules in the clusters and by two for the Zundel cation. The data in the $x$ and $y$ axes are in $\log$ scale, $a$ is in $a_0^{-2}$ ($a_0$ is the Bohr radius). Source: Reprinted from \cite{triggiani2023elastic}.}\label{fig:cs_zundel}
\end{figure}

We emphasise that these calculations are performed by freezing the roto-vibrational motion. We also stress that a single set of GBFs for projecting the interaction potential and wave functions unfortunately cannot accurately describe the elastic scattering cross-section in the whole spectrum, i.e. one has to include several GBFs of type $s$ with an energy-dependent width to accurately describe the oscillations of the continuum wave function in the whole energy range $[0,1000]$ eV. In Fig. \ref{fig:single_molecule}, each line colour actually represents a different Gaussian width (see the figure legend, where seven widths were needed to reproduce the experimental data).
The width of the Gaussian functions is set to capture the oscillations of the scattering wave functions at different energies. Especially at low energy, the GBFs must capture the behaviour of the long wavelengths; therefore, the width of the GBFs must be somehow small. At higher energies, larger values for the width are required to capture the narrow oscillations of the continuum wave function. 
To assess the robustness of the elastic scattering cross section calculations with respect to the change in the exponential factors $a$ (whose values are determined by using a cut-off radius $R$ for the scattering potential associated with the Gaussian variance by $R=1.25/\sqrt{(a)}$), we varied their value within a reasonable range relative to the wavelength of the scattering wave function. We then performed simulations to determine the statistical error. The inset in Fig. \ref{fig:single_molecule} shows the statistical error bars for different kinetic energies using seven different HGF basis sets.
Above 4 eV, the theoretical spectrum is in excellent agreement with the experimental data \cite{khakoo_2013}, showing that the use of a theoretical method based on the self-consistent calculation of the molecular orbitals, taking into account polarisation and the permanent dipole, is effective. At higher energies ($ > $ 10 eV), the ab initio simulations agree with the data indicating the total cross-section (green triangles in Fig. \ref{fig:single_molecule}), which also contain inelastic contributions, although they only take into account the elastic channel. This can be explained by the fact that the total scattering probability does not change by opening several channels, but is merely redistributed between them (see Refs. \cite{taioli2010electron,taioli2015computational}). In Fig. \ref{fig:single_molecule} ab initio calculations are compared with the Mott's cross-section using a semi-empirical treatment of the dipole moment \cite{3ELSEPA} (cyan coloured dots), which clearly shows the effect of the different treatment of the interaction potential between these two approaches.

Furthermore, the projected potential method can also be successfully applied to study polycentric systems in their interaction with an electron beam. In this context, we show in Fig. \ref{fig:cs_water_mol} the ESCS from ab initio simulations (upper multicoloured dots) for two clusters of three water molecules in different configurations with molecular dipole moments pointing in different directions compared to the calculations for a single water molecule (lower multicoloured line), ELSEPA (cyan dots) and experimental data recorded for water vapour under different conditions of relative humidity (symbols). This result shows the effect of the environment, as the ESCS of the water clusters are normalised to the number of molecules in the simulation cell. In particular, we notice the similarity between the ESCS of the two different water clusters. In the medium energy range ($1 < E < 30$ eV), the ESCS of the water clusters differ by almost 100\% with respect to the single molecule, showing a clear influence of multiple scattering and the need of a more accurate representation of the inter- and intramolecular potential provided by first-principles simulations.

Finally, we show in Fig. \ref{fig:cs_zundel} the ESCS of the Zundel cation (upper multicoloured empty squares) obtained with the projected potential relativistic ab initio method compared to the ESCS of the water clusters.
The influence of the environment is also clearly visible in this plot, which once again emphasises the importance of an accurate treatment of the interatomic potential.

\subsection{Inelastic scattering: ELF and IMFP}\label{applic_inel}

The inelastic scattering collisions can be treated with different accuracy within the dielectric theory by using either the DL/Mermin/Penn models (see section \ref{drudino}) or first-principles simulations (see section \ref{ELF_ab}) to determine the ELF.
We recall that the dielectric theory is based on the evaluation of the ELF in the entire Bethe surface. 

\begin{figure}[htb!]
\centering
\includegraphics[width=1.0\textwidth]{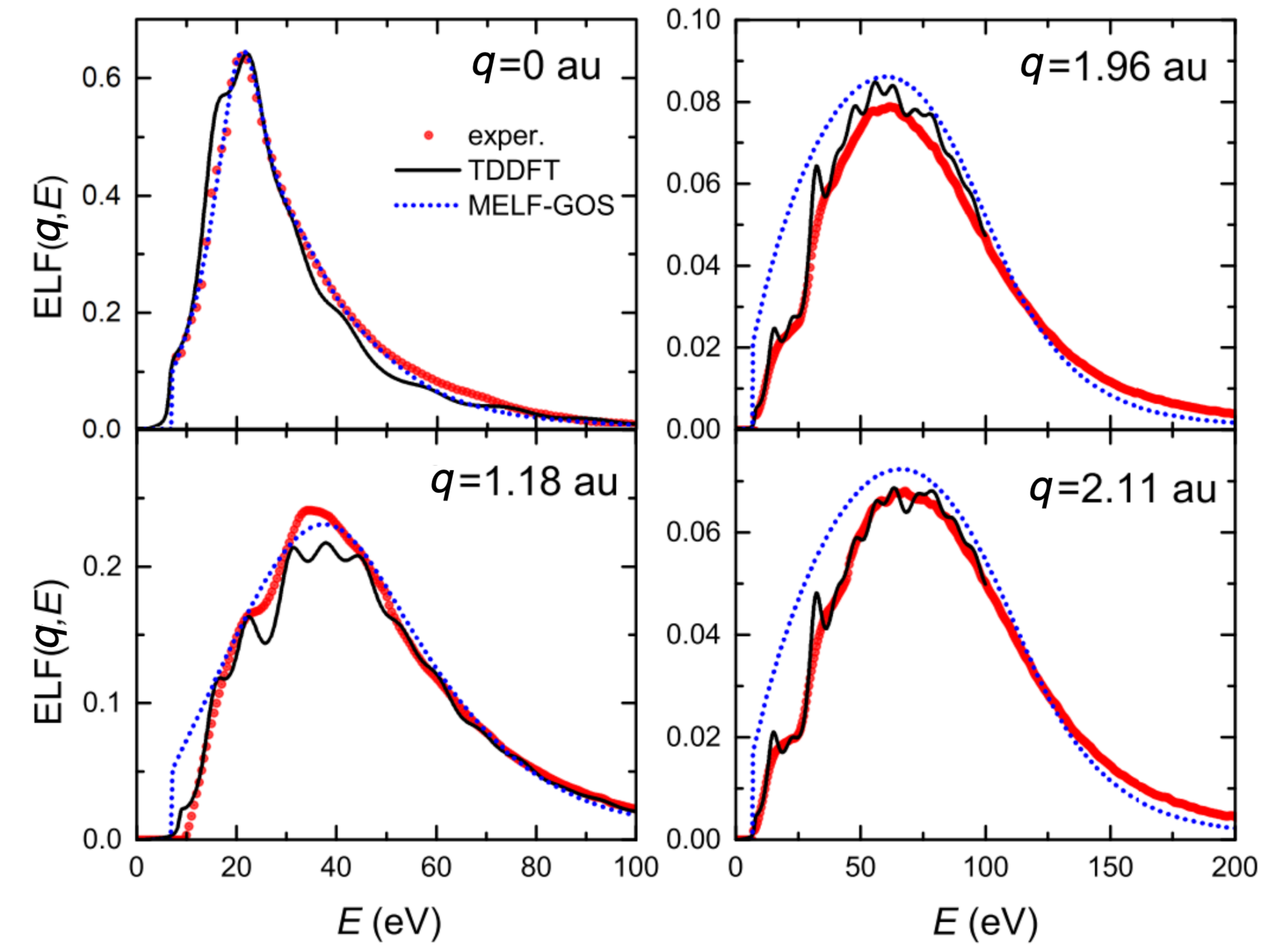}
\caption{Energy loss function (ELF) of liquid water as a function of energy loss $E$ at different momentum transfers $\hbar q$ from LR-TDDFT (black solid lines) and from MELF-GOS (blue dotted lines) approaches. The red circles correspond to the experimental data \cite{hayashi2000complete}. Source: Reprinted from \cite{taioli2020relative,de2022energy}.}
\label{fig:ELFWATER}
\end{figure}

In Fig. \ref{fig:ELFWATER} we report the ab initio (black liune) and semi-empirically (MELF-GOS, dotted blue line) calculated ELF of liquid water and compare it with measured data from inelastic X-ray scattering spectroscopy (red circles).

In particular, the upper left panel in Fig. \ref{fig:ELFWATER} shows in red circles the experimentally measured optical ELF (at $q=0$) and the fit obtained with the MELF-GOS method (dotted blue line). The following panels show the calculated ELFs for transferred momenta $q=1.18, 1.96$ and $2.11$ a.u., which were obtained from the analytical properties of the Mermin-type ELF. As momentum transfer increases, a broadening and reduction in the intensity of the ELF is observed, which is consistent with the theoretical expectation that individual excitations should gradually predominate over collective excitations at large momenta. The MELF-GOS results (dotted blue lines) agree quite well with the experimental data (red circles) over a wide range of energy transfers for $q=0$; however, this is not surprising as the MELF-GOS at $q=0$ is a fit of the experimental data. For higher values of momentum transfer, MELF-GOS assumes an analytical dispersion law (see Eqs. (\ref{merm}) and \ref{Lindhard1}). The corresponding data, while overall in good agreement, increasingly deviate from the experimental values, while the ab initio calculations better reproduce the ELF line shape. This is reflected in other quantities, such as the inelastic mean free path (see \cite{taioli2020relative,de2022energy}). We conclude that the use of an ab initio approach, especially at finite momentum transfer in the low energy range where the majority of transitions occurs and must be accurately represented, is critical to reproduce the experimental measurements.

\begin{figure}[htb!]
\centering
\includegraphics[width=1.0\textwidth]{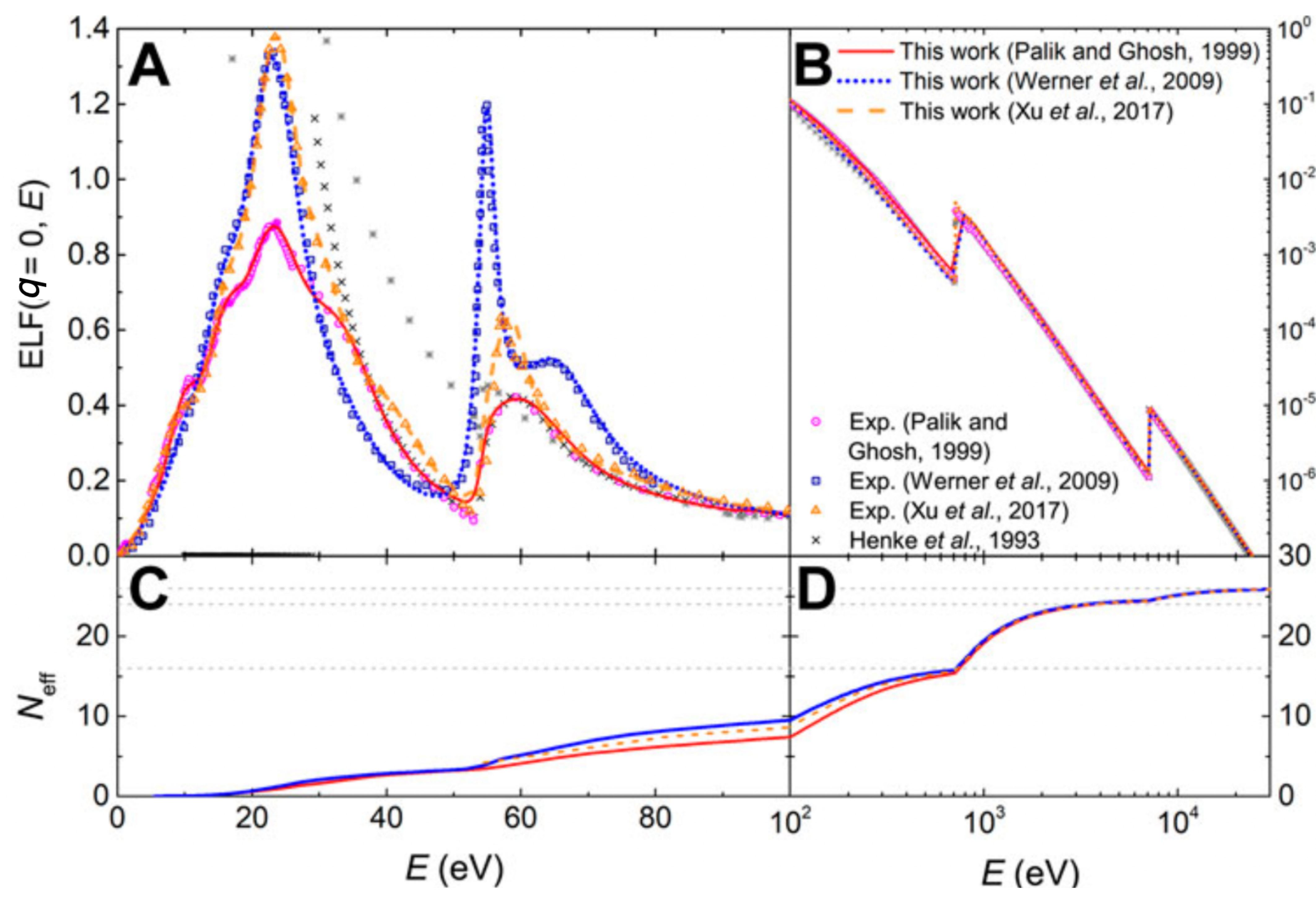}
\caption{(A) and (B): Optical ELF of iron. The experimental data from Refs. \cite{palik1999electronic} (magenta circles) \cite{werner2009optical} (square blue symbols) and \cite{PhysRevB.95.195417} (orange triangles) are presented together with the MELF-GOS fit (solid lines) and calculations from Henke et al. \cite{henke1993x} and the NIST compilation \cite{NIST}. (C) and (D): Effective number of electrons in Fe as a function of the excitation energy $E$, derived from the MELF-GOS model from different optical ELF. Source: reproduced from \cite{10.3389/fmats.2023.1249517}.
}
\label{fig:MELFdiff}
\end{figure}

In this context, semi-empirical methods, such as reverse Monte Carlo \cite{PhysRevB.95.195417}, can be used to derive the ELF from the REEL spectrum (or other experimental optical data), although this method is not free from uncertainties. In Fig. \ref{fig:MELFdiff}A,B (see Ref. \cite{10.3389/fmats.2023.1249517}), we give the optical ELF$(\omega,q=0)$ of iron, which consists of optical (magenta circles) \cite{palik1999electronic} and REELS \cite{werner2009optical,PhysRevB.95.195417} (blue squares and orange triangles) measurements, together with the MELF-GOS fits shown with a solid red curve, dotted blue and dashed orange lines. We note that the REELS-based measurements appear to provide sharper and brighter peaks with maxima approximately twice as high as the optical data \cite{werner2009optical}, which are broader and generally less intense. We note that all ELFs fulfil the $f$-sum rule (see Fig. \ref{fig:MELFdiff}C,D). However, the optical data \cite{palik1999electronic} indicate a slightly lower number of electrons for excitation energies up to several hundred eV. The effective electron numbers for each individual excitation channel for all optical ELFs are consistent with the expected orbital occupation, although the most recent figures, based on a reverse Monte Carlo algorithm using REELS measurements \cite{PhysRevB.95.195417,10.3389/fmats.2023.1249517,Li2023ImprovedRM}, appear to provide a smaller error in the individual sum rule for electrons described by Mermin functions. We conclude that different ELFs that fulfil the sum rules can be derived from different experimental data. 

In fact, summation rules are integrated quantities. As such, they can lose fine details in the description of electronic transitions, although they are acceptable on average. The fulfilment of the sum rules is per se a necessary but not a sufficient condition for obtaining the ELF. These three conflicting ELFs of iron, for example, lead to the determination of different stopping powers for proton and helium projectiles and to an underestimation or overestimation of experimental data in different energy ranges \cite{10.3389/fmats.2023.1249517}. %In this respect, the validity of using experimental data to reproduce other experimental data is questionable. In fact, according to this approach one should use a reverse MC to obtain the ELF from measured REEL spectra and then use MC to reproduce either the same REEL spectrum or other energy loss quantities associated with the ELF, such as stopping power.
Finally, all semi-empirical methods have the same flaw: the extension to finite momentum is done by analytic forms, which are typically polynomial with momentum transfer at all energies. This is not the correct behaviour even if known boundary conditions are met. This suggests that ab initio simulations offer the best alternative approach compared to all other methods mentioned above and also allow independent cross-checking of experimental measurements.

\begin{figure}[htb!]
\centering
\includegraphics[width=1.0\textwidth]{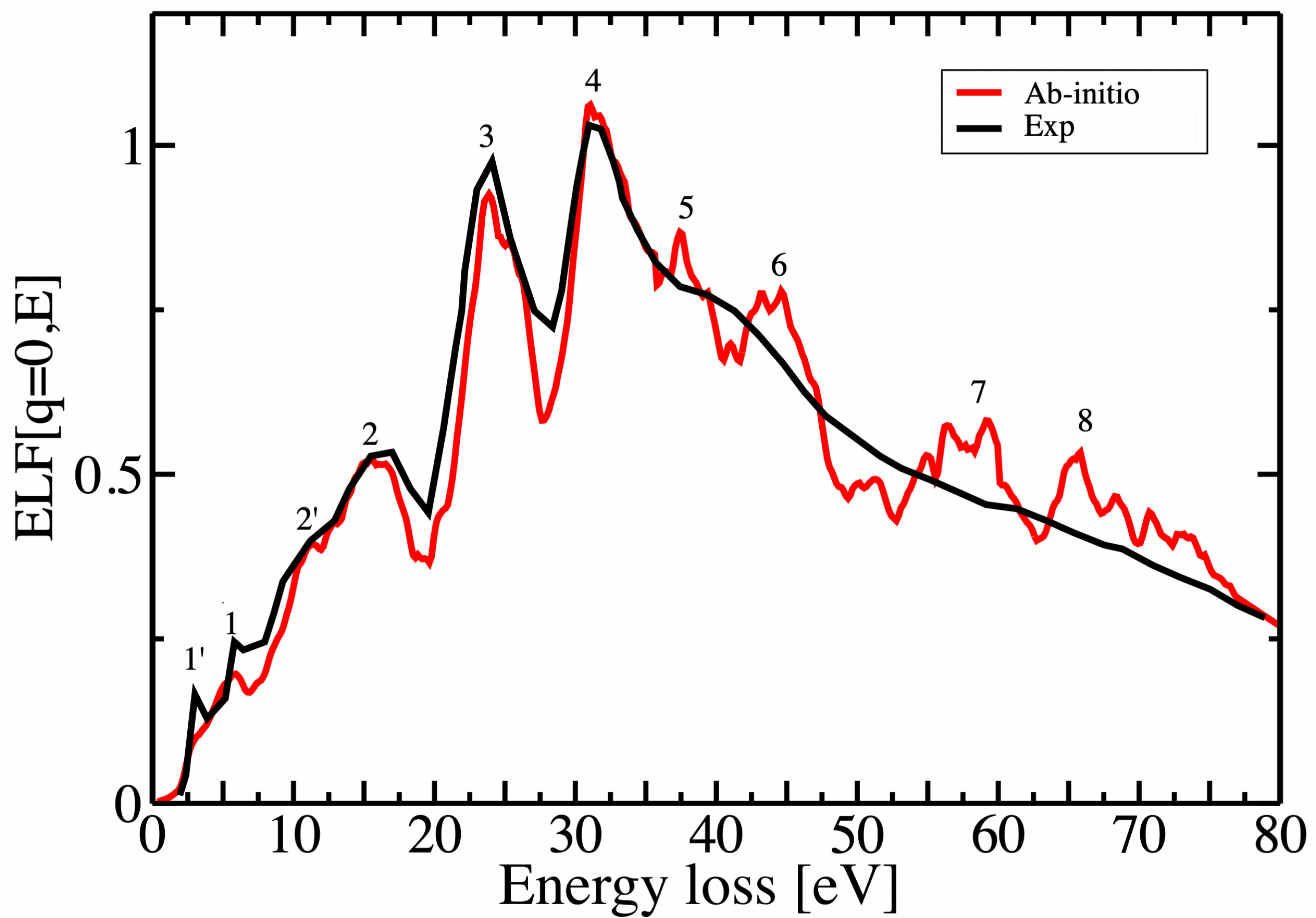}
\caption{Long-wavelength limit (${\bm q}\rightarrow 0$) of the ab initio ELF of Au (red curve) compared to experimental optical data from Ref. \cite{werner2009optical} (black curve).
The LSDA for the exchange correlation functional was used to include the SO coupling in addition to the ALDA approximation for the time-dependent exchange correlation kernel. Source: Reprinted from \cite{taioli2023role}.}
\label{fig:ELFPOLI1}
\end{figure}

\begin{table}[hbt!]
\begin{center}
\begin{tabular}{c|c|c|c|c|c}
peak label& $\omega^{a}$ (eV)  &$\omega^{b}$ (eV) & This work & Exp. & Origin\\
\hline
\hline
1'   &    2.2  &   &  2.66 &  2.5  & ME   \\
\hline
        1   &   5.1  &  5.5 &  5.94 &  5.9   & $5d$ plasmon    \\ 
        \hline
        2'   &  10.2   & 11.5 &  11.49 & 11.9  & Mainly-$6s$ plasmon     \\ 
        \hline
        2   &   15.5  &	 18 &    15.67 & 15.8  &  IT     	 \\
        \hline
        3   &   23.8  &	25 &   23.74 & 23.6 &  ME  \\
        \hline
        4   &   30.8  &	35 &    31.35 & 31.5 &  ME          	 \\
        \hline
        5  &    36.9 & 40.5	&    37.5 & 39.5  & IT              \\
        \hline
        6   &   43.5. &	43.5 &    44  & 44 & IT      \\
        \hline
         7  &   &	&    59.2  &  &	IT   \\
         \hline
          8   &  & 	&   65.62  & &  IT  \\
\end{tabular}
    \caption{Optical ELF (${\bm q} \rightarrow 0$) of bulk Au. First column: Peak designations according to Fig. \ref{fig:ELFPOLI1}. The second ($\omega^{a}$) and third ($\omega^{b}$) columns report the peak position in the ELF according to Refs. \cite{PhysRevB.102.035156} and \cite{GURTUBAY2001123}.
    In the fourth column we give the values obtained in this work with the number of digits representative of the numerical accuracy, while in the fifth column (Exp.) we give the experimental EELS data from \cite{werner2009optical}. In the last column (Origin), the theoretical interpretation of each energy loss peak is outlined. IT stands for an interband transition, while ME stands for a mixed excitation. Source: Reprinted from \cite{taioli2023role}.}
\label{tab:interpret}
\end{center}
\end{table}

\begin{figure}[htp!]
\centering
\includegraphics[width=0.75\textwidth]{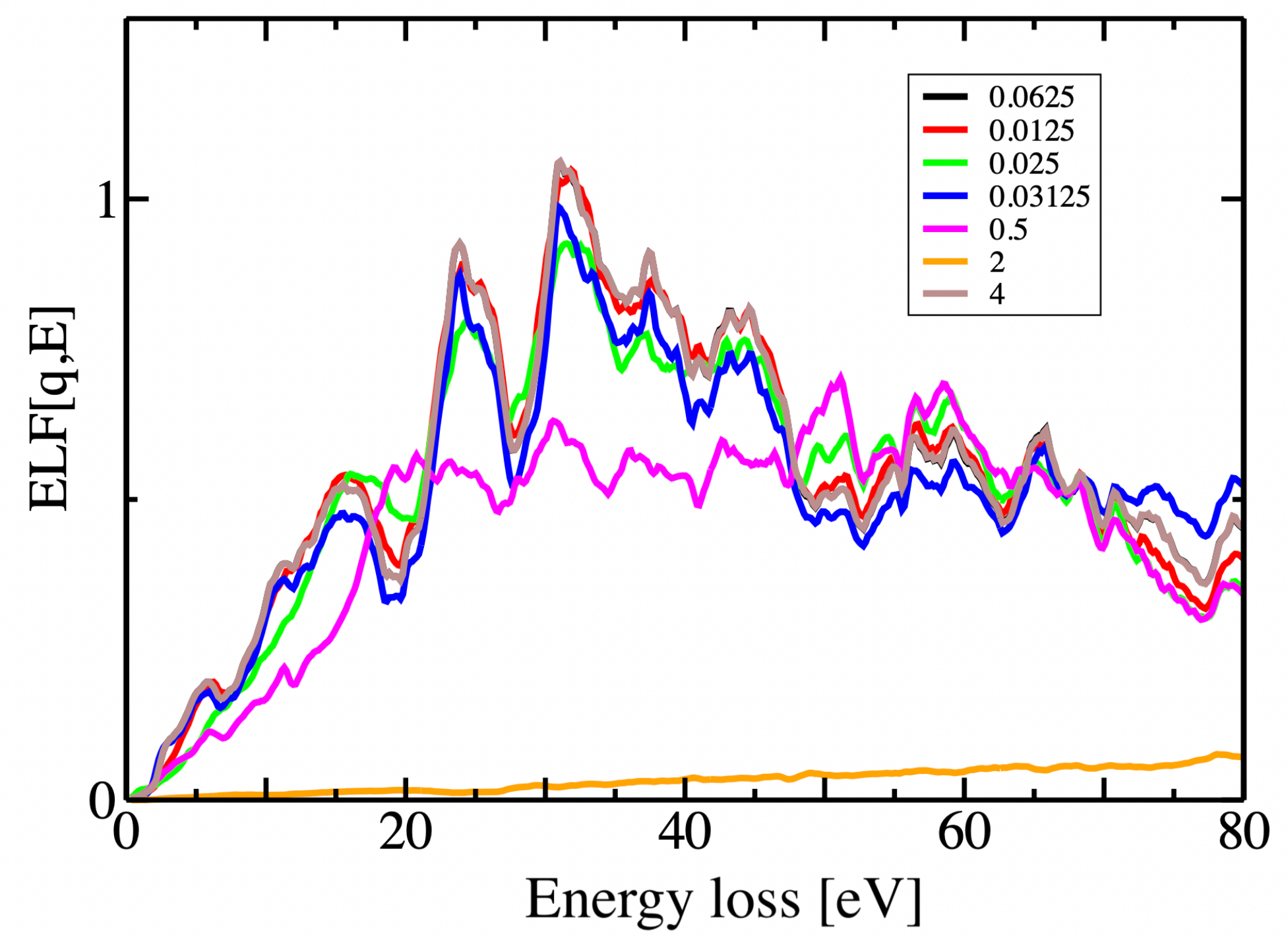}
\includegraphics[width=0.75\textwidth]{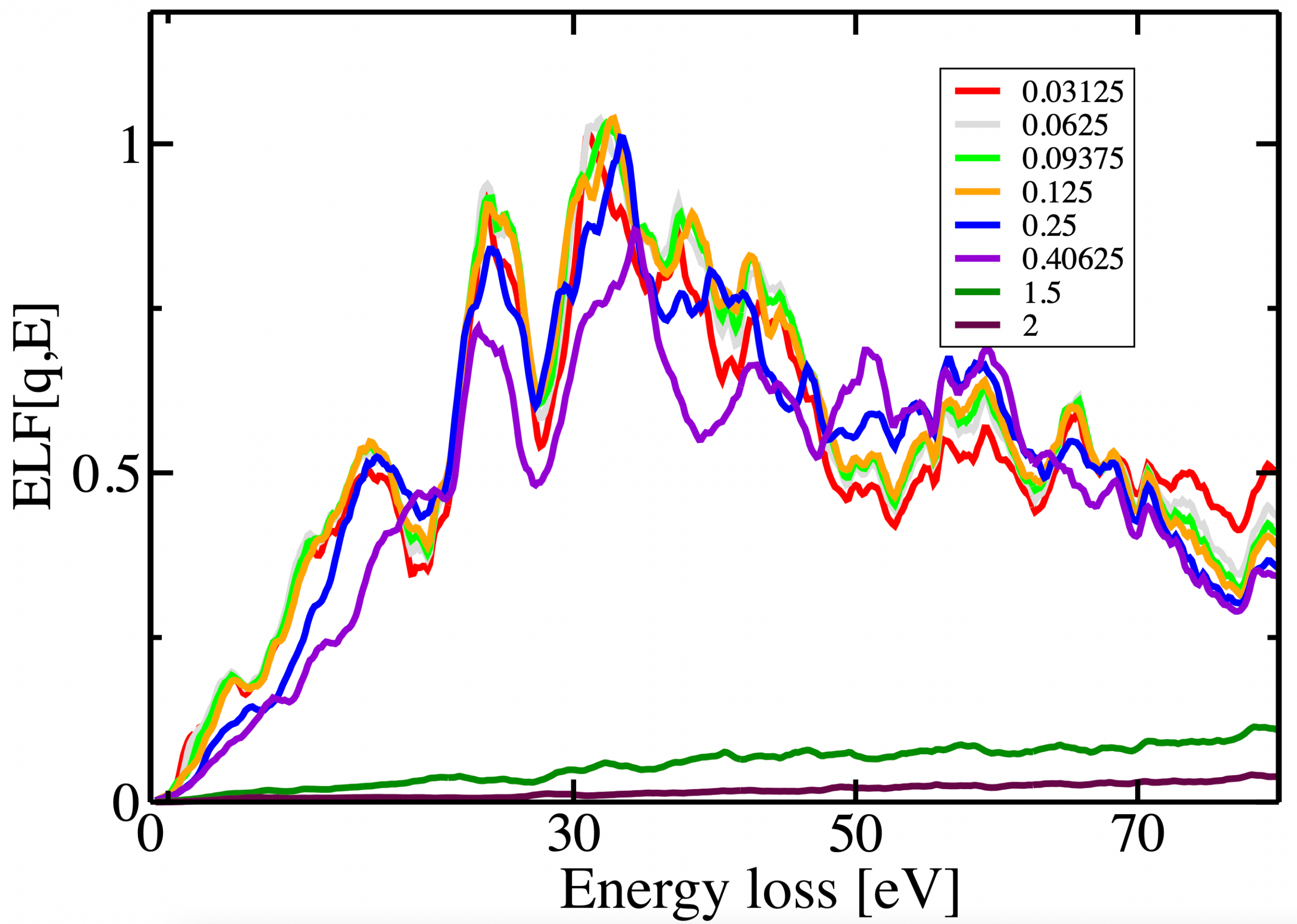}
\caption{Upper panel: ELF of Au for finite momentum transfers along the (1,1,1) direction. The calculated $q$-vectors are indicated in the legend in lattice coordinates (e.g. 0.0625 means the $q$-vector (0.0625, 0.0625, 0.0625)).
Bottom panel: ELF of Au for finite momentum transfers along the (1,1,0) direction.
The calculated $q$-vectors are indicated in the legend in lattice coordinates (e.g. 0.0625 means the $q$-vector (0.0625, 0.0625, 0)).
The LSDA approximation for the exchange correlation functional was used to account for the SO coupling in addition to the ALDA approximation for the time-dependent exchange correlation kernel. Source: Reprinted from \cite{taioli2023role}.}
\label{fig:ELFPOLI3}
\end{figure}

In Fig. \ref{fig:ELFPOLI1} we show the optical ELF of bulk gold (red line, see \cite{taioli2023role}) calculated with the ab initio approach described in section \ref{ELF_ab} compared to the experimental optical data \cite{werner2009optical,RIDZEL2020146824} (black line). This shows excellent agreement, although the surface plasmon excitations were neglected with respect to the experimental EELS data. We note that another approach to calculate the ELF is to use a Liouville-Lanczos method \cite{PhysRevB.88.064301,TIMROV2015460}, which avoids the computationally intensive summation over empty bands. The latter approach was described in Ref. \cite{PhysRevB.102.035156} and achieved comparable results.

An interpretation of the spectral features of the ELF lineshape can be found in Table \ref{tab:interpret} and a detailed discussion can be found in Ref. \cite{taioli2023role}. We only emphasise here that ab initio methods based on DFT are able to reproduce the well-resolved double-peak structure due to the localised $d-$bands that are a feature of other transition metals. In Au, the presence of these localised orbitals acts as a dielectric background and is responsible for both the lowering of the $5d$ plasmon energy from the free electron plasma frequency ($\simeq$9 eV) to 5.94 eV (peak 1 in Fig. \ref{fig:ELFPOLI1}) and the well-resolved $5d$ and $6s$ valence electron peaks. 

\begin{table}[hbt!]
\begin{center}
\begin{tabular}{c|c|c|c}
\textbf{$i$}& $A_i$ (eV$^2$)  &$\gamma_i$ (eV) & $\omega_i$ (eV)\\
\hline
\hline
        1   &   1.969      &   3.463  & 5.382   \\ 
        2   &   17.508      & 	6.121  &	  11.644 \\ 
         3   &   26.815     &   4.948  & 	 15.909 \\ 
         4   &   106.586 	&   5.784  & 24.449	 \\
        5   &   43.843 	&   3.086  & 	 30.993 \\
        6   &   98.402 	&   5.881 & 33.355	 \\
        7   &   90.344 	&   5.670 & 37.938	 \\
        8   &   177.047 	&   7.852 & 44.489	 \\
        9   &   551.321 	&   18.326 & 59.799	 \\
         10   &  29.773   	&   3.667 & 68.164 \\
        11   &  128.517   	&   7.449 & 74.020	 \\
        12   &  65.4861   	&   6.292 & 87.151  \\
        13   &    46.896 &   7.192 &  95.522 \\
        14   &  2500.0   &   350.0 & 330.0 \\
        15 & 2000.0  & 2000.0 & 2500.0 \\
         16 &  200.0 	&   5000.0 & 14000.0 \\
\end{tabular}
     \caption{Fitting parameters of Eq. (\ref{Drude2}) obtained for the optical ELF (${\bm q} \rightarrow 0$) in bulk Au. Source: Reprinted from \cite{taioli2023role}.}
\label{tab:fitELF1}
\end{center}
\end{table}

Fig. \ref{fig:ELFPOLI3} shows the peak dispersion for finite momentum transfers along two symmetry directions (1,1,1) (upper panel of Fig. \ref{fig:ELFPOLI3}) and (1,1,0) (lower panel of Fig. \ref{fig:ELFPOLI3}).
The dispersionless characteristic of the double peak indicates that it is formed by interband transitions and not by plasmon excitation. Other peaks at 2.66 eV (peak 1' in Fig. \ref{fig:ELFPOLI1}) and at 15.67 eV (peak 2 in Fig. \ref{fig:ELFPOLI1}) show the same dispersionless behaviour with finite momentum transfer, while the peaks at 5.94 and 11.49 eV (peaks 1 and 2' in Fig. \ref{fig:ELFPOLI1}), which are 5$d$ and 6$s$ plasmon signals, show a clear momentum dispersion. The peaks at higher energies (5, 6, 7 and 8 in Fig. \ref{fig:ELFPOLI1}) are also weakly dispersed, indicating their nature as interband transitions.
\begin{figure}[htb!]
\centering
\includegraphics[width=1.0\textwidth]{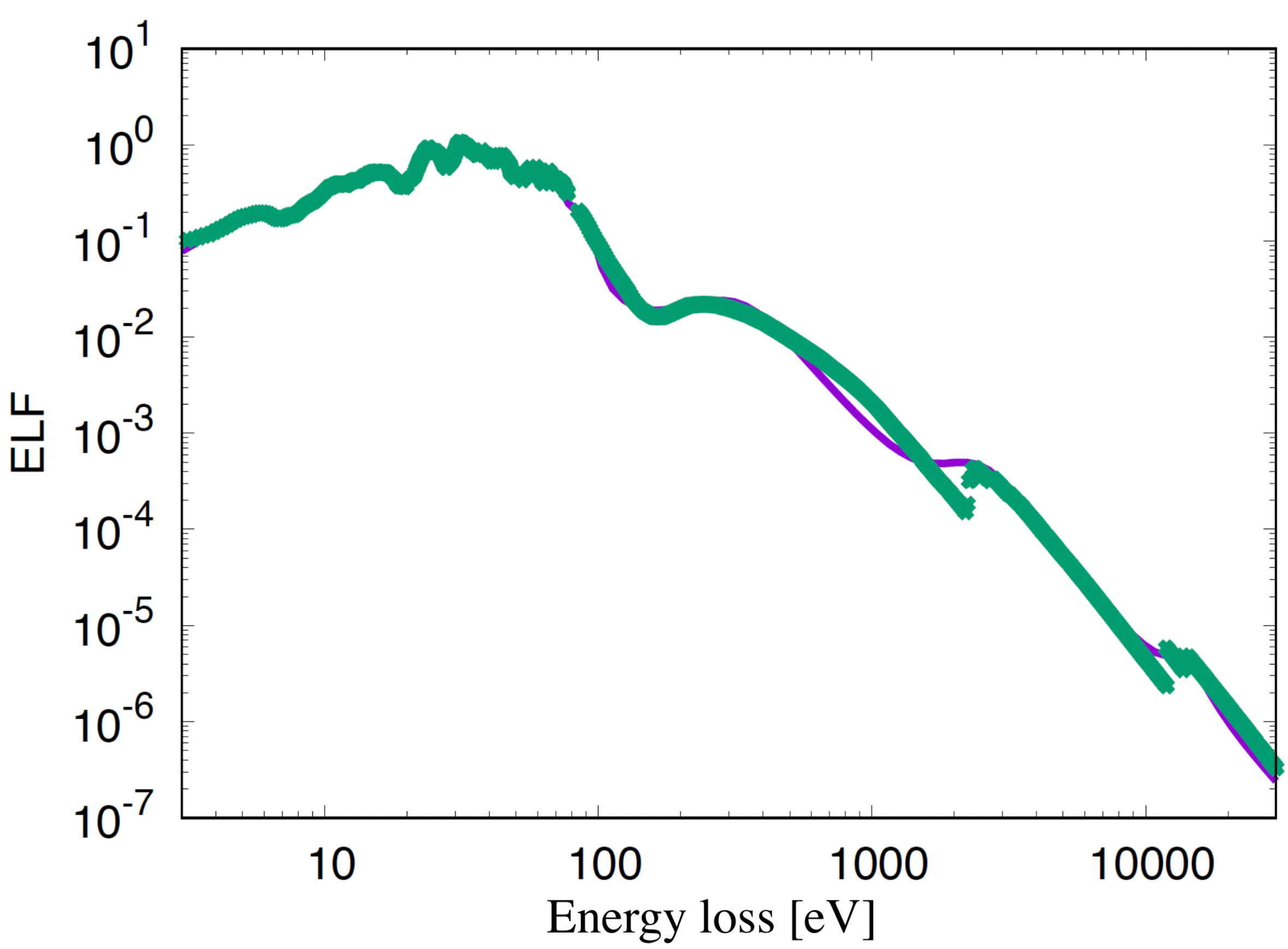}
\caption{Ab initio ELF extended to high excitation energies \cite{NIST} (green line) together with the DL fit (purple line). The matching point between ab initio and experimental data was set to the threshold energy of the semi-core transitions $E_{5p1/2} = 74.2 \ $ eV \cite{thompson2001x}. The fit parameters are shown in Table \ref{tab:fitELF1}. Source: Reprinted from \cite{taioli2023role}.}
\label{fig:ELFPOLI2}
\end{figure}

We also emphasise that (see section \ref{ELF_ab}) the first-principles simulations are very accurate, but at the price of a high computational cost that makes it impossible to extend the energy range and take into account the energy-momentum dispersion beyond a few tens of eV. However, in order to perform MC simulations for charged particle beams with high kinetic energy, it is necessary to determine the inelastic scattering cross-section far beyond these limits \cite{taioli2020relative,pedrielli2021electronic,azzolini2017monte,azzolini2018anisotropic,AZZOLINI2020109420}.

\begin{figure}[htb!]
\centering
\includegraphics[width=1.0\textwidth]{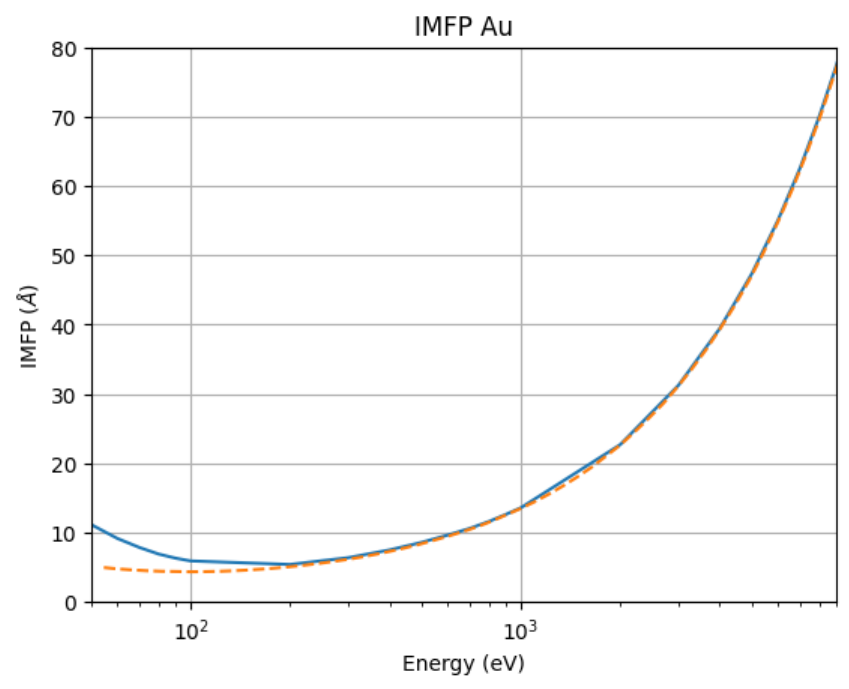}
\caption{IMFP (solid line \cite{taioli2023role}) as a function of incident electron energy in Au. Dashed line from Ref. \cite{tanuma2011calculations}.}
\label{fig:prova6}
\end{figure}

In this regard, the extension of the optical ELF along the excitation energy axis can be performed by matching the ab initio results with the experimental data, which typically cover a larger energy. In particular, in the NIST database \cite{NIST} one can find X-ray form factors, attenuations and scattering tables from which the ELF up to 30 keV can be derived.
In Fig. \ref{fig:ELFPOLI2} we show such an extension for Au, where the matching point between ab initio and experimental data was set to the threshold energy of the semi-core transitions $E_{5p1/2} = 74.2 \ $ eV \cite{thompson2001x}. In Table \ref{tab:fitELF1} we also show the DL parameters of Eq. (\ref{Drude2}), which were determined by a best-fit procedure. 

Finally, to determine the IMFP the ab initio data must also be extended beyond the optical limit for finite momentum transfers, e.g. by choosing one of the dispersion laws described in section \ref{drudino}.
The IMFP of Au can then be determined using the formula (\ref{exchange}) and is shown in Fig. \ref{fig:prova6} for a quadratic dispersion in $q$ (blue solid line) according to Eq. (\ref{fullRPA}).

\begin{figure}[hbt!]
\centering
\includegraphics[width=1.0\textwidth]{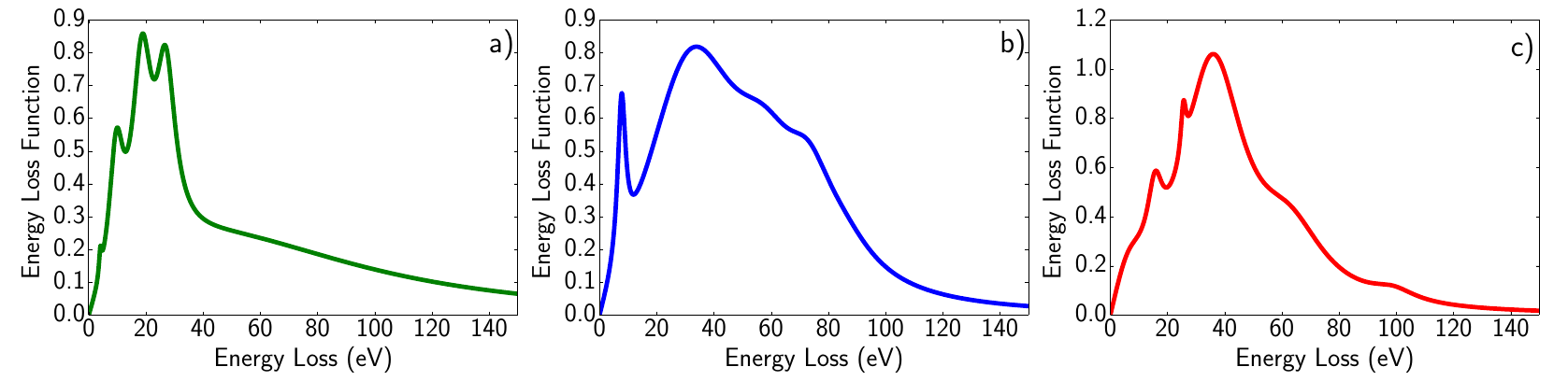}
\caption{Elf of bulk a) Cu, b) Ag and c) Au in the optical limit ($q\longrightarrow 0$), obtained by best-fitting the experimental data of zero momentum transfer with the DL model of Eq. (\ref{Drude2}). These data can be found in \cite{smith1985handbook} for Ag, in \cite{montanari2007calculation} and \cite{denton2008influence} for Cu and Au respectively. Source: Reprinted from \cite{AZZOLINI2020109420}.}\label{fig:ELF}
\end{figure}

\begin{figure}[htb!]
\centering
\includegraphics[width=1.0\textwidth]{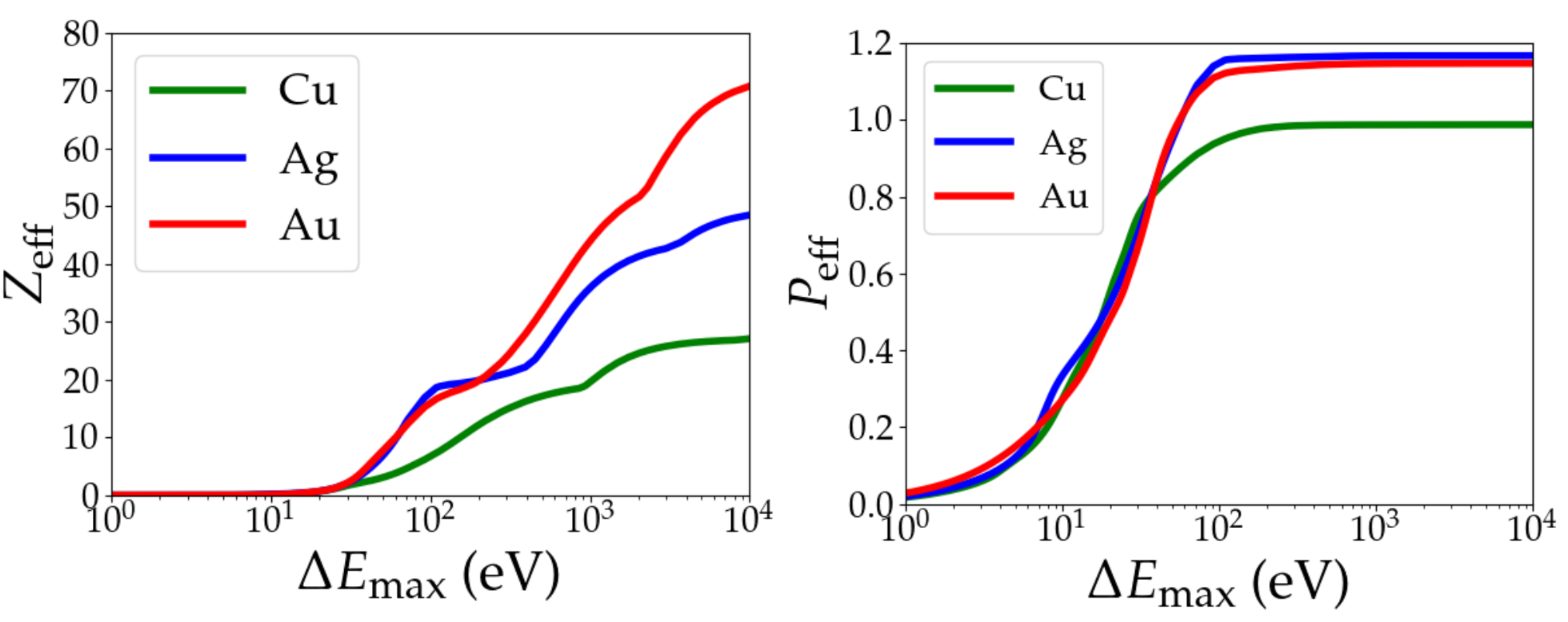}
\caption{$Z_{\mathrm{eff}}$ (left panel) and $P_{\mathrm{eff}}$ (right panel) versus the transferred energy ($\Delta E_{\mathrm{max}}$) for Cu, Ag and Au. The calculations were performed using the semi-empirical ELFs shown in Fig. \ref{fig:ELF}. Source: Reprinted from \cite{AZZOLINI2020109420}.}\label{fig:PZeff}
\end{figure}

As a further application of the DL model, we show in Fig. \ref{fig:ELF} the bulk ELFs in the optical limit of Cu, Ag and Au obtained by best-fitting the experimental data for zero momentum transfer from Ref. \cite{denton2008influence} for Au, Ref. \cite{montanari2007calculation} for Cu and Ref. \cite{smith1985handbook} for Ag. In Table \ref{tab:fit} we also give the best-fit parameters for fitting the experimental optical ELF by DL functions.

\begin{table}[hbt!]
    \centering
    \addtolength{\tabcolsep}{-1pt}
    \begin{tabular}{|cccc|cccc|cccc|}
        \hline
         && Cu &&  & & Ag & & && Au &\\
        \textit{i} & $\omega_i$ (eV) & $\gamma_i$ (eV) & $A_i$ (eV$^2$) & \textit{i} & $\omega_i$ (eV) & $\gamma_i$ (eV) & $A_i$ (eV$^2$) & \textit{i} & $\omega_i$ (eV) & $\gamma_i$ (eV) & $A_i$ (eV$^2$) \\ \hline 
             1 &  4.08 & 1.09 & 0.33 & 1 & 7.89 &	3.37 & 	12.80 & 1 & 9.52	& 14.97	& 18.49\\
             2 & 10.07 & 5.99	& 22.10 & 2 & 38.20 & 42.93 & 1109.46 & 2 & 15.92 & 6.26 & 25.85\\
             3 & 19.05	& 8.16	& 88.91  & 3 & 59.58	& 29.93	& 480.38 & 3 & 25.58	& 2.18	& 11.12\\
             4 & 27.21	& 8.16	& 112.54 & 4 & 73.81	& 20.12	& 300.6& 4 & 38.09	& 26.67	& 973.52 \\
             5 & 78.91	& 152.38 & 2216.74 & 5 & 85.70	& 27.70	&226.83  & 5 & 64.49	& 30.48	& 507.39\\
            & & & & & & & & 6 & 99.32	& 19.05	& 88.88 \\
            & & & & & & & & 7 & 402.71 & 612.23 & 337.32\\
            \hline
    \end{tabular}
     \caption{Fitting parameters of Eq. (\ref{Drude2}) for the optical ELF (${\bm q} \rightarrow 0$) in the bulk. The best-fitting parameters for Cu were provided by C.C. Montanari {et al.} \cite{montanari2007calculation}, while for Au by C.D. Denton {\it et al.} \cite{denton2008influence}. In the case of Ag, the parameters were determined by best-fitting the optical measurements of Smith et al. \cite{smith1985handbook}. Source: Reprinted from \cite{AZZOLINI2020109420}.}\label{tab:fit}
\end{table}

%\begin{table}[hbt!]
%    \centering
%    \addtolength{\tabcolsep}{-1pt}
%    \begin{tabular}{|cccc|ccc|ccc|}
%        \hline
%         && Au &  & &  Ag & &  & Cu &\\
%        \textit{i} & $\omega_i$ (eV) & $\gamma_i$ (eV) & $A_i$ (eV$^2$) &  $\omega_i$ (eV) & $\gamma_i$ (eV) & $A_i$ (eV$^2$) & $\omega_i$ (eV) & $\gamma_i$ (eV) & $A_i$ (eV$^2$) \\ \hline 
%        1 &  4.08	& 1.09	& 0.33  & 7.89 &	3.37 & 	12.80 \\     
%            \hline
%     &  & Cu & & \\
%     \textit{i} & $\omega_i$ (eV) & $\gamma_i$ (eV) & $A_i$ (eV$^2$) \\
%          1 & 9.52	& 14.97	& 18.49 \\  
%    \end{tabular}
%     \caption{Fitting parameters of Eq. (\ref{Drude2}) for the optical ELF (${\bm q} \rightarrow 0$) in the bulk. The best fitting parameters for Cu were provided by C.C. Montanari {et al.} \cite{montanari2007calculation}, while for Au by C.D. Denton {\it et al.} \cite{denton2008influence}. In the case of Ag, the parameters were determined by best-fitting the optical measurements of Smith et al. \cite{smith1985handbook}.}\label{tab:fit}
%\end{table}

\begin{table}[htb!]
    \centering
    \large
    \addtolength{\tabcolsep}{2pt}
    \begin{tabular}{|c|c c|c c|c c|}
        \hline
         & ~~~~~~~Cu & & ~~~~~~ Ag  & & ~~~~~ Au & \\
         $E$ (eV) &$Z_{\mathrm{eff}}$ &  $P_{\mathrm{eff}}$ & $Z_{\mathrm{eff}}$ &  $P_{\mathrm{eff}}$ & $Z_{\mathrm{eff}}$ &  $P_{\mathrm{eff}}$ \\ \hline 
         1 &    5.52e-05 & 0.017 & 9.24e-05 & 0.020 & 1.29e-04 & 0.028\\
          10&     0.09 & 0.275 & 0.17 & 0.336 & 0.1 & 0.27\\
          108.5 & 7.28 & 0.950 & 18.63  & 1.156 & 16.65 & 1.121\\
          1012   &   19.68 & 0.987 & 36.04 & 1.17 & 44.17 & 1.146\\
          9886.4 &   27.07 & 0.987 & 48.46 & 1.17 & 70.74 & 1.147\\
          29779 &   28.23 & 0.987 & 49.51 & 1.17 & 75.94 & 1.147\\
            \hline
    \end{tabular}
   \caption{$Z_{\mathrm{eff}}$, $P_{\mathrm{eff}}$ limit values of Cu, Ag and Au represented in Fig. (\ref{fig:PZeff}). These data were calculated using the semi-empirical ELFs in Fig. \ref{fig:ELF}. Source: Reprinted from \cite{AZZOLINI2020109420}.} \label{tab:val}
\end{table}

To test the accuracy of the DL model, we compared this data with the $f$-sum rule and perfect-screening (ps-sum or $P_{\mathrm{eff}}$) sum rule (see Eqs. (\ref{fsumrule}) and (\ref{psumrule}) respectively). In Fig. \ref{fig:PZeff} we show the $Z_{\mathrm{eff}}$ (left panel) for the ELF of Cu, Ag and Au, while in the right panel we show the $P_{\mathrm{eff}}$. The limit values $Z_{\mathrm{eff}}$, $P_{\mathrm{eff}}$ are also reported in the Table \ref{tab:val}.

\begin{figure}[htb!]
\centering
\includegraphics[width=1.0\textwidth]{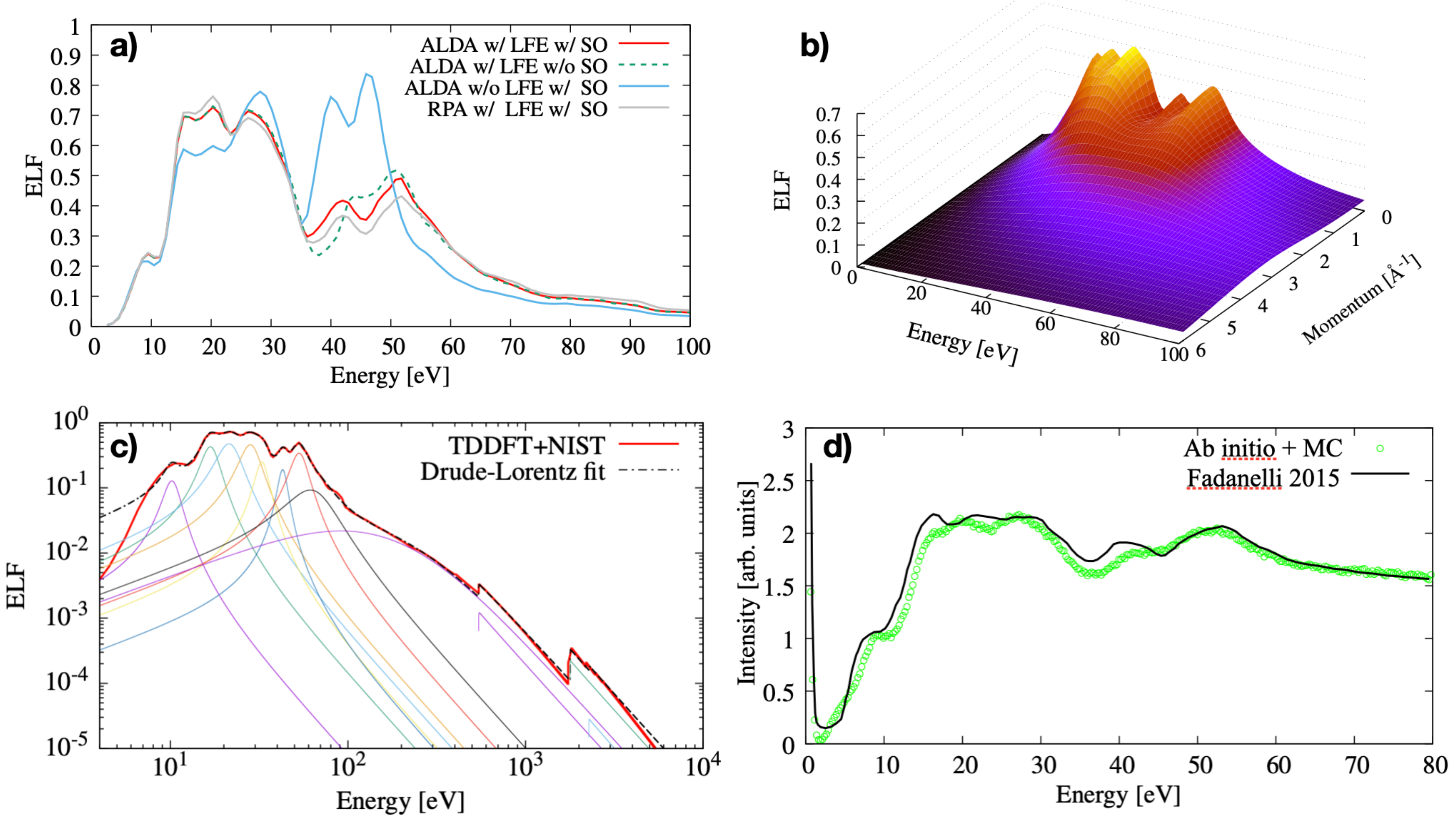}
\caption{a) ELF of $\gamma$-Ta$_2$O$_5$ compared to experimental data (dashed line) from references \cite{Franke2000,Fadanelli2015} at different approximation levels considering (or neglecting) the local field effects (LFE), spin-orbit coupling (SO), either in the framework of the adiabatic local density approximation (ALDA) or the random phase approximation (RPA). b) Ab-initio calculation of the ELF of $\gamma$-Ta$_2$O$_5$ with the extension to finite momentum (Bethe surface). c) Drude-Lorentz fit of the ab-initio ELF to include high-energy loss collisions. d) REEL spectrum (green line) compared to experimental data for the $\gamma$-Ta$_2$O$_5$ polymorph. Source: Adapted from \cite{pedrielli2022search}.}
\label{fig:ELFTa}
\end{figure}

\begin{figure}[htb!]
\centering
\begin{subfigure}{0.83\textwidth}
        \includegraphics[width=\linewidth]{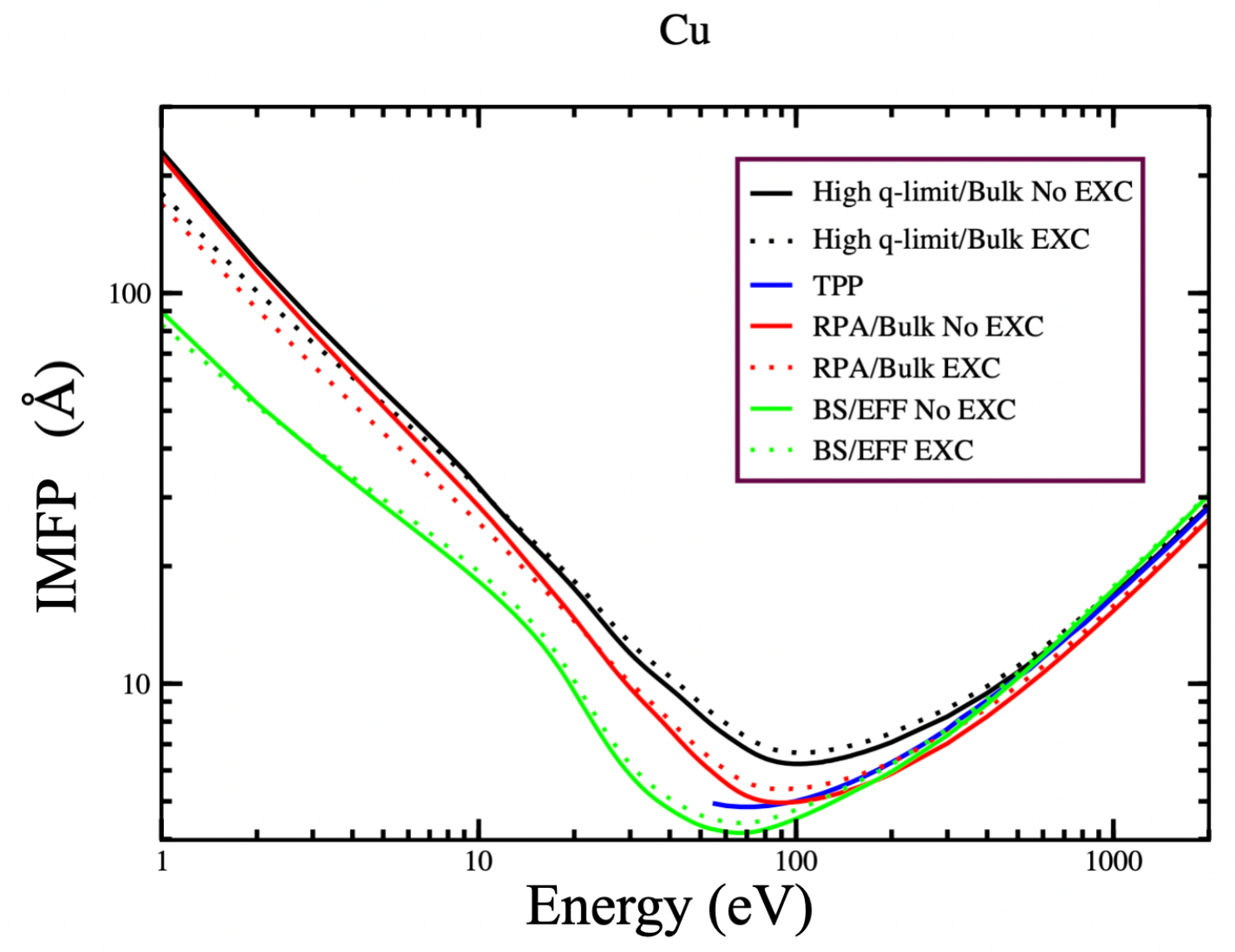}
        \subcaption{IMFP of Cu.}
        \label{fig:arm1}
    \end{subfigure}
    \begin{subfigure}{0.83\textwidth}
        \includegraphics[width=\linewidth]{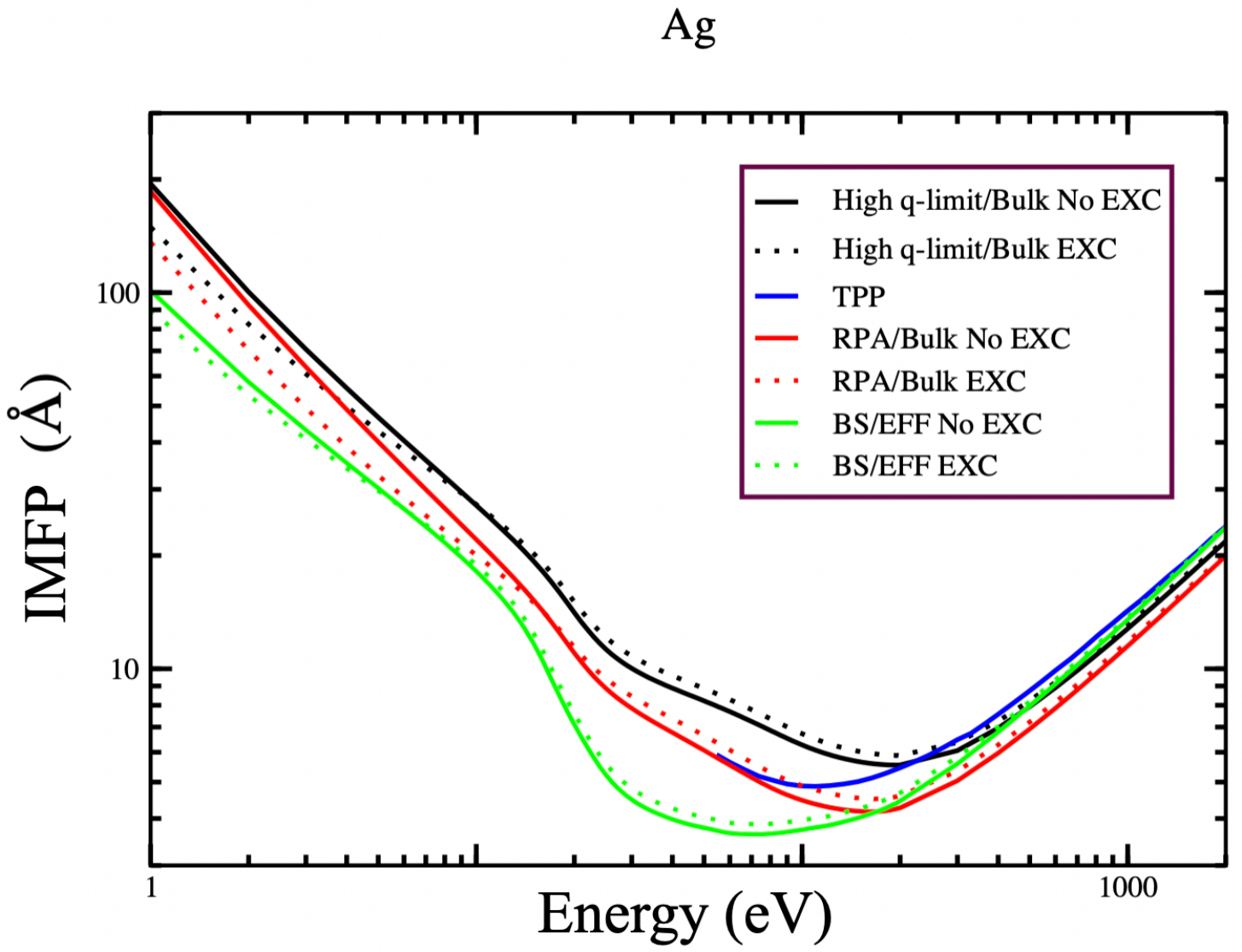}
        \subcaption{IMFP of Ag.}
        \label{fig:arm2}
    \end{subfigure}
%    \caption{$Q^{*}$ values for different arms}
\end{figure}%
\begin{figure}[htb!]\ContinuedFloat
    \centering
    \begin{subfigure}{0.83\textwidth}
    \includegraphics[width=1.0\linewidth, angle = 0]{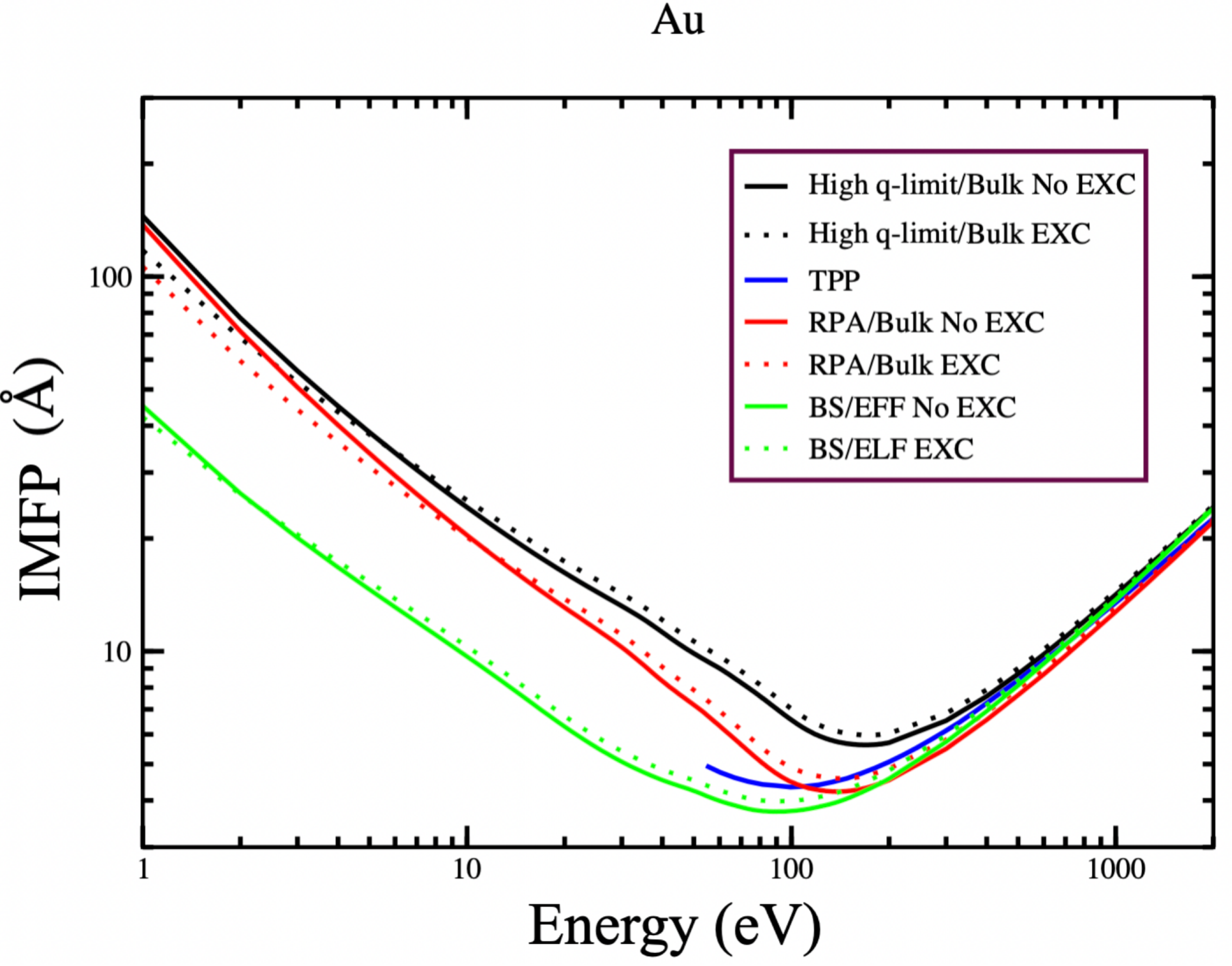}
        \subcaption{IMFP of Au.}
        \label{fig:arm3}
    \end{subfigure}
%\includegraphics[width=0.9\linewidth, angle = 0]{figures/IMFP_Cu_NoEXC.pdf}
%includegraphics[width=0.9\linewidth, angle = 0]{figures/IMFP_Ag_NoEXC.pdf}
%\includegraphics[width=0.9\linewidth, angle = 0]{figures/IMFP_Au_NOEXC.pdf}
\caption{IMFP of Cu, Ag and Au with (dashed curves) and without (solid curves) the Born--Ochkur electron exchange corrections (see Eq. (\ref{BornOchkur})) compared to calculations by Tanuma et al. \cite{tanuma2011calculations} (blue curve). The IMFPs are calculated using the bulk ELFs at high $q$ limit (black curves), the full 3D bulk ELF (red curves, RPA, see Eq. (\ref{fullRPA})) and the effective ELF with the bulk surface extension to finite momenta (green curves, BS, see Eq. (\ref{SB})), respectively. Abscissa and ordinate are given in logarithmic scale to emphasise the difference between the different treatments of the $q$-dispersion. Source: Reprinted from \cite{AZZOLINI2020109420}.}\label{fig:lambda_inel}
\end{figure}

We report another example dealing with the calculation of the ELF to show how different levels of theory, i.e. the adoption of different approximations in ab initio simulations, affect the evaluation of dielectric properties such as the IMFP and the REEL spectrum. In particular, we discuss how first-principles calculations can be used to determine the ground-state structure of thermally grown tantalum oxide by investigating the ELF in the optical limit of its different polymorphs.

Ab-initio methods can disentangle the respective spectral features of the ELF, starting from the different atomic arrangements of the polymorphs of Ta$_2$O$_5$, to determine the structure with the lowest energy. This task cannot be solved with semi-empirical approximations, as these assume the existence of the experimental ELF in order to calculate the REEL spectrum. In contrast, the atomic configuration of the different known polymorphs of Ta$_2$O$_5$ can be used to calculate their corresponding ELFs according to first principles and determine the REEL spectra. The results are shown in Fig. \ref{fig:ELFTa}, where the $\gamma$-phase of Ta$_2$O$_5$ is shown to be the ground state crystal structure by comparison with the experimental data of the different polymorphs. In panel a) of Fig. \ref{fig:ELFTa}, we plot the ELF for different levels of theory to show the effects of the different ab initio variants on the shape of the spectral lines. In particular, we use the adiabatic local density approximation (ALDA) for the time-dependent exchange correlation kernel, the local spin density approximation (LSDA) to include (or neglect) spin-orbit (SO) coupling in the spin-polarised mode, and the random phase approximation (RPA) with and without local field effects (LFE). We emphasise that the LFEs play an important role: They influence all features of the ELF and show the presence of strong spatial inhomogeneity in these systems.

To show how the use of different dispersion laws (see sections \ref{drudino} and \ref{pennino} for the mathematical details) and the addition of the electron exchange correction (see Eqs. (\ref{exchange}) and (\ref{BornOchkur})) affect the calculated inelastic scattering quantities, we plot the IMFPs of the three metals Cu, Ag and Au in Fig. \ref{fig:lambda_inel} \cite{AZZOLINI2020109420}. We find that the discrepancy between the dispersion laws is significant below 100 eV, while the inclusion of exchange in the calculation of the IMFP becomes significant below 10 eV.

\begin{figure}[htb!]
\centering
\includegraphics[width=1.0\textwidth]{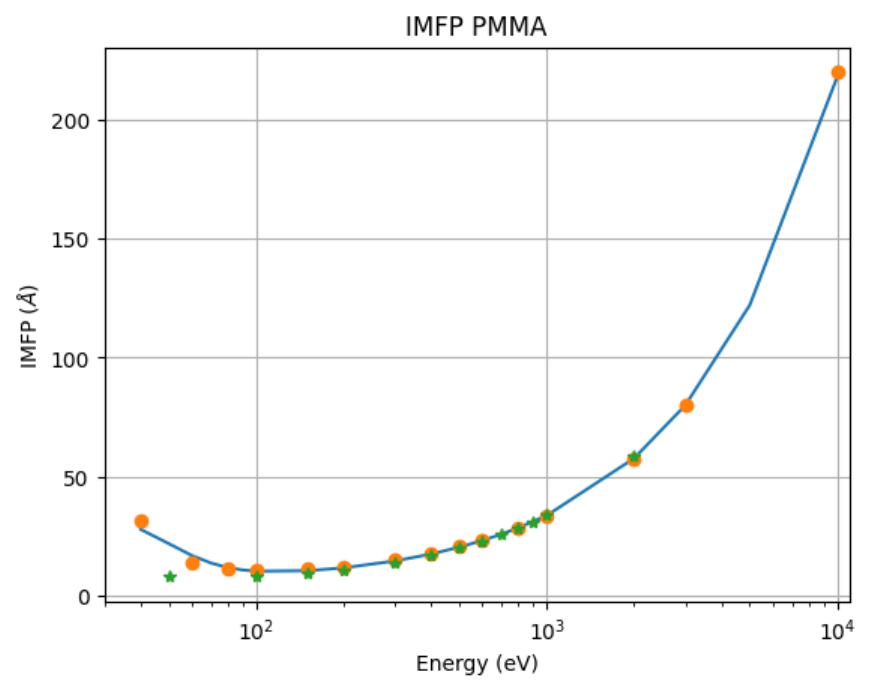}
\caption{IMFP as a function of incident electron energy in PMMA. Solid line from Ref. \cite{Dapor2015}, dots from Ref. \cite{Ashley1988} and stars from Ref. \cite{tanuma2011calculations}.}\label{fig:prova7}
\end{figure}

\begin{figure}[htb!]
\centering
\includegraphics[width=1.0\textwidth]{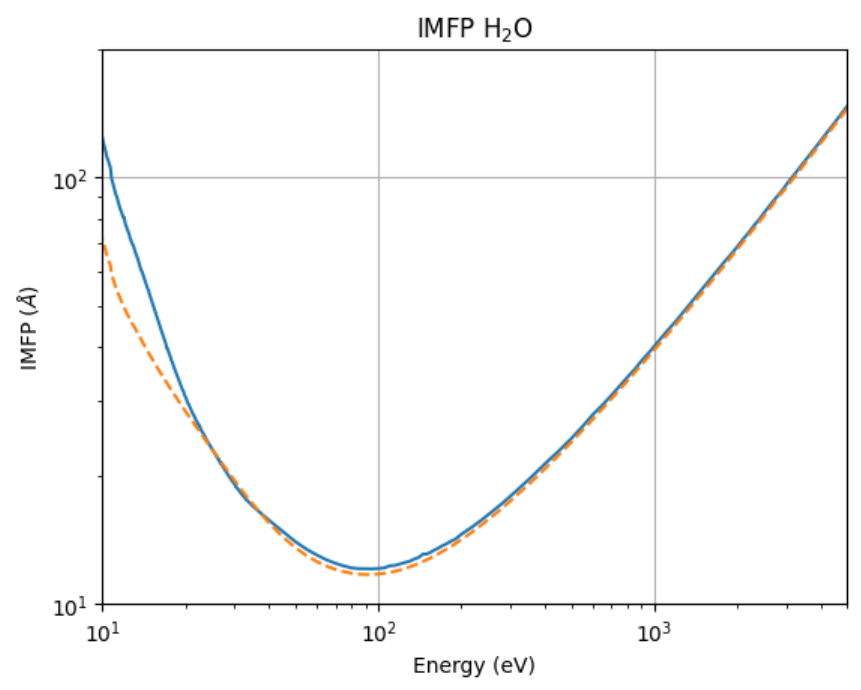}
\caption{IMFP as a function of incident electron energy in liquid water \cite{taioli2020relative}. Solid line: TDDFT calculations. Dashed line: MELF-GOS.}\label{fig:prova9}
\end{figure}

\begin{figure}[hbt!]
\centering
\includegraphics[width=1.0\textwidth]{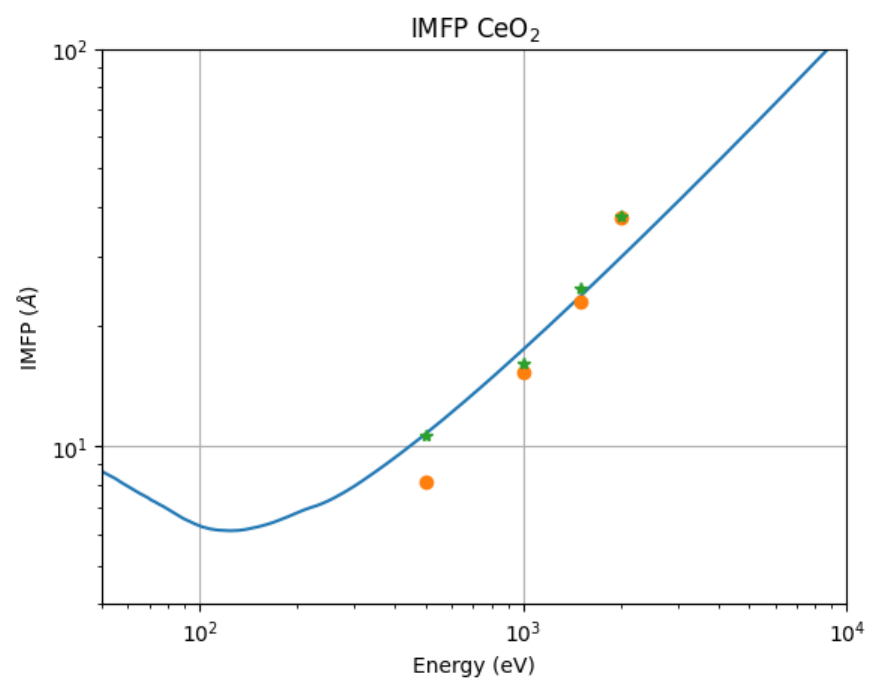}
\caption{IMFP as a function of the energy of the incident electrons in CeO$_2$. Solid line: TDDFT simulations \cite{pedrielli2021electronic}. Dots: experimental data using the Ni standard \cite{KRAWCZYK2015196}. Stars: experimental data using the Au standard \cite{KRAWCZYK2015196}.}\label{fig:prova8}
\end{figure}

Further examples of IMFPs of organic molecular materials, such as polymethyl methacrylate (PMMA) \cite{Dapor2017}, water \cite{taioli2020relative} and inorganic solids, such as cerium oxide \cite{pedrielli2021electronic}, using the DL, MELF-GOS and TDDFT methods are shown in the Figs. \ref{fig:prova7}, \ref{fig:prova9}, and \ref{fig:prova8}, respectively. A comprehensive collection of experimental and calculated data on IMFPs for different combinations of projectiles and materials can be found in the recently published Ref. \cite{10.3389/fmats.2023.1249517}.

\subsection{Auger electron spectra of nanostructures}

The ab initio method developed in section \ref{Auger_inles} can be used to calculate Auger spectra in molecules and solids. The spectrum of the non-radiative O $K-LL$ decay of silicon dioxide is shown in Fig. \ref{QMMC} \cite{taioli2010electron,taioli2009surprises,taioli2009mixed}.
We find that the theoretical spectrum  (orange line) accurately reflects the experimental transition energies (blue line), which are mainly distributed in three regions, while the intensities are lower than the measured values. In particular, the most intense peaks are found in the $[494-510]$ eV region, corresponding to $2p-2p$ double holes in the final states of the oxygen atom ($K-L_{23}L_{23}$ transitions), with the maximum intensity at $502.76$ eV, which is due to a singlet transition, in good agreement with the experimental value of $502.58$ eV; between $[475-492]$ eV a group of features corresponding to Auger emissions with final double-hole configurations $2s-2p$ ($K-L_1L_{23}$), with the main peak corresponding to the singlet transition at $481.1$ eV, in agreement with the experimental peak at $481.38$ eV; finally between $[460-470]$ eV, with the final double-hole configuration in the $2s-2s$ orbitals of oxygen ($K-L_1L_1$) and the strongest peak at 463.86 eV, which corresponds to the experimental value of 465.25 eV. In general, the ab initio method based on the theory of projected potentials to calculate the continuum wave functions shows that it is a valid tool to reproduce the transition energy of Auger decay.

Nevertheless, Auger electrons emitted from atomic centres in solids find an environment that can change both their spectral intensity and their distribution, e.g. due to inelastic scattering events. In this respect, the MC method can be used to account for the energy losses that Auger electrons suffer while travelling in the solid and before leaving the surface.

In this context, one can use the ab initio electron distribution (orange line)
as input for an MC routine to account for such changes in the spectrum caused by energy losses of the electrons on their way out of the solid.
Using this approach and assuming that Auger generation occurs at a depth of 40~\AA~\cite{VANRIESSEN2007150}, the Auger line profile that includes inelastic events (green line in Fig. \ref{QMMC}) can be determined. We find that such inclusion increases and broadens the Auger decay probability so that the measured experimental results are satisfactorily reproduced over the entire energy range.

\begin{figure}[hbt!]
\centerline{\includegraphics[width=14cm]{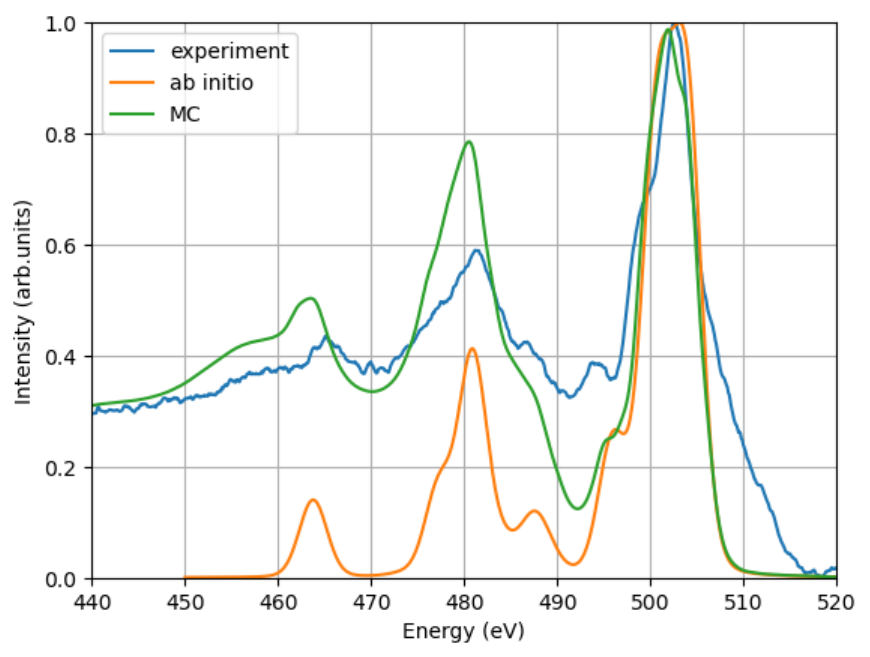}}
\caption{O $K-LL$ Auger spectrum in SiO$_2$. Comparison between the ab initio (orange line) and MC (green line) calculations with experimental data (blue line) \cite{taioli2010electron,taioli2009surprises,taioli2009mixed}.}
\label{QMMC}
\end{figure}

In particular, there is a strong broadening of the $K-L_1L_{23}$ peak, which is due to the SiO$_2$ plasmon excitation at $\approx$23~eV.
Finally, the Auger spectrum can also be determined in molecular systems by using the same approach of projected potentials \cite{taioli2010electron}, e.g. the Auger spectrum of ozone was recently calculated \cite{taioli2021resonant}.

\subsection{Spectra of backscattered electrons, secondary electrons and yields}

\begin{figure}[hbt!]
\centering
\includegraphics[width=1.0\textwidth]{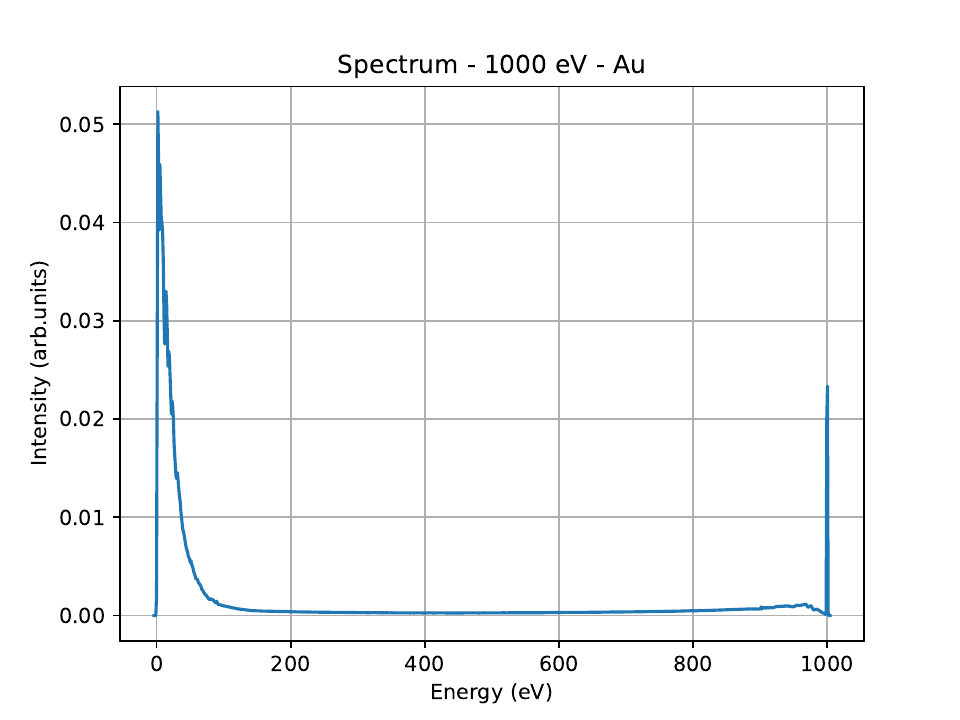}
\caption{Monte Carlo simulation of the energy loss spectrum of electrons from an Au sample for a kinetic energy of 1000 eV of the primary beam.}\label{fig:prova10}
\end{figure}

\begin{figure}[hbt!]
\centering
\includegraphics[width=1.0\textwidth]{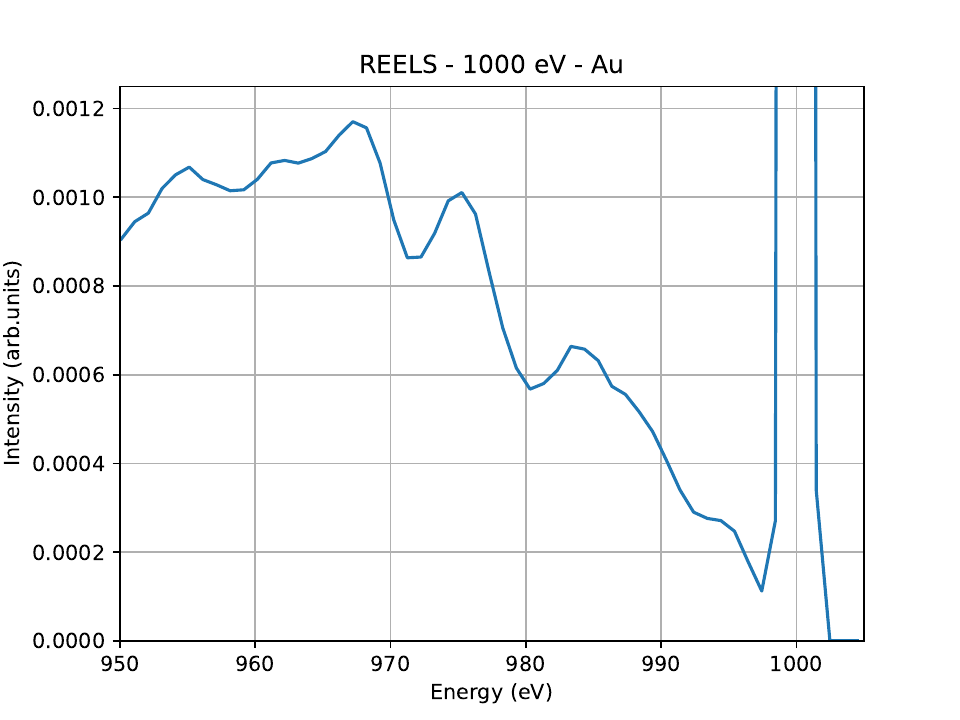}
\caption{Monte Carlo simulation of the REEL energy region of an Au sample for a kinetic energy of the primary beam of 1000 eV, showing the plasmon and part of the elastic peak.}\label{fig:prova11}
\end{figure}

Knowledge of the elastic and inelastic scattering cross-sections, which can be derived from the energy- and momentum-dependent ELF of a particular sample, enables the MC simulation in the entire energy spectrum, taking into account all processes such as single-electron and collective excitation, ionisation and elastic collisions. Each energy loss mechanism and each generation of secondary electrons leads to a well-resolved peak.
In Fig. \ref{fig:prova10} we show such a MC-simulated spectrum of both backscattered and secondary electrons emerging from Au. From right to left of the spectrum, one finds: the elastic -- or zero-loss -- peak, characterised by a width at half maximum of a few decimal eV to account for the intrinsic width of the beam (typically 0.8~eV); the bulk plasmon peaks, zoomed in the REEL spectrum of Fig. \ref{fig:prova11}, at $\approx$ 5.77 and 11.6 eV from the elastic peak, slightly shifted with respect to the maxima of the ELF (5.94 and 11.49 eV, see also Table \ref{tab:interpret}); finally, a series of clearly resolved spectral features (see also Fig. \ref{fig:prova11} for an enlarged view) in the range 15-50 eV from the elastic peak at $\approx$ 15.7, 24.7, 31.7 and 44 eV, which are due to $5d,4f$ $\rightarrow$ 6s interband electronic transitions mixed with plasmon-like features (Re$(\bar\varepsilon) \simeq 0$, Im$(\bar\varepsilon)$ small but $\neq 0$) at 24.7 and 31.7 eV.
To generate these spectra, 10$^{10}$ electron trajectories were used.

\begin{figure}[hbt!]
\centering
\includegraphics[width=1.0\textwidth]{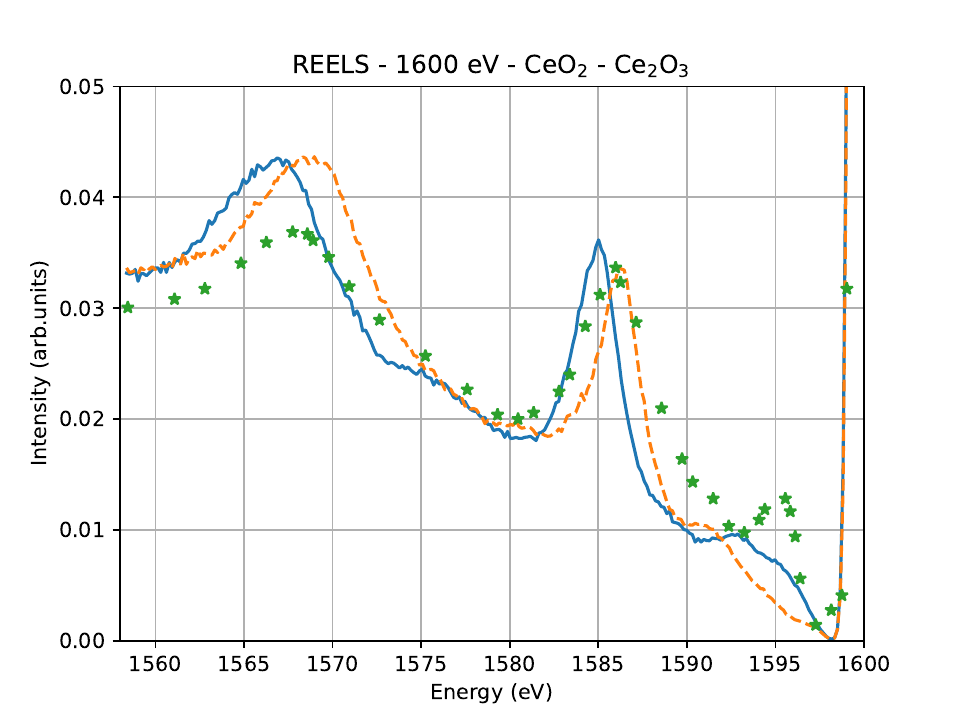}
\caption{Comparison between the theoretical REEL spectra of bulk CeO$_2$ (solid line) and Ce$_2$O$_3$ (dashed line), simulated with the Monte Carlo method \cite{pedrielli2021electronic}, and the measured values (stars) \cite{Pauly:17}. The spectra are normalised with respect to the area. The primary beam has a kinetic energy of 1600~eV and an angle of incidence of 60$^\circ$ with respect to the surface normal.}\label{fig:prova12}
\end{figure}

\begin{figure}[hbt!]
\centering
\includegraphics[width=1.0\textwidth]{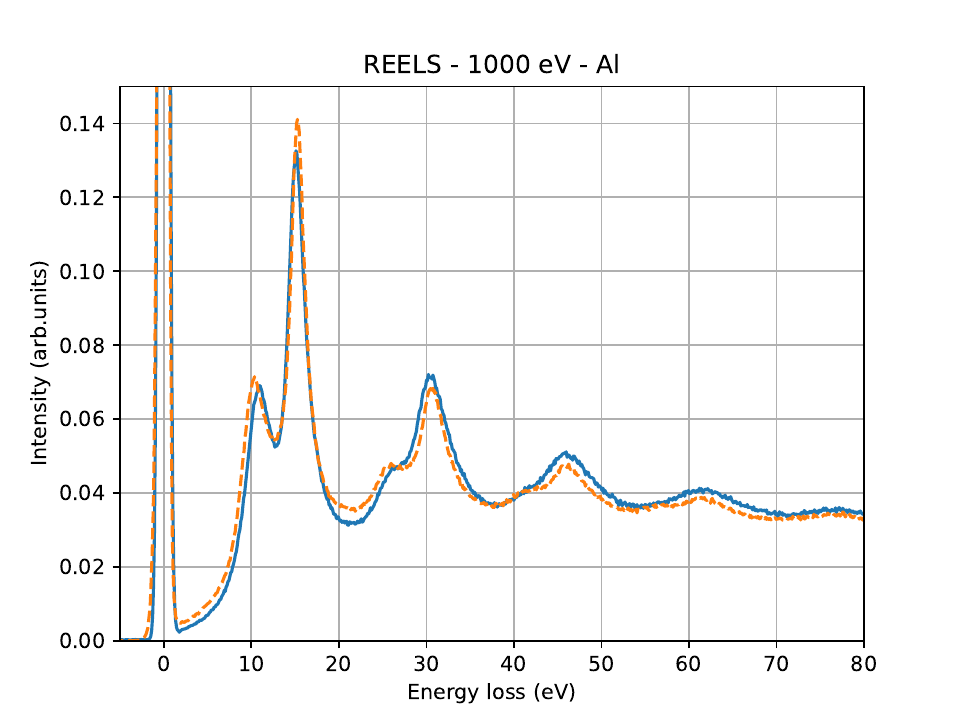}
\caption{REEL spectrum for 1000 eV electrons impinging on Al as a function of energy loss (the zero-loss peak is shown on the far left of the spectrum). The primary beam strikes the material with an angle of incidence of 30$^\circ$ with respect to the surface normal. The spectra were normalised to a common area of the zero-loss peak. Plasmon peaks up to the 5th order of scattering can be observed in both the experimental and simulated spectra. Please note the presence of surface plasmons and bulk plasmons in the spectra. Solid line: Monte Carlo simulations \cite{10.3389/fmats.2022.1068196}. Dashed line: experimental data from \cite{https://doi.org/10.1002/sia.4835}.}\label{fig:prova13}
\end{figure}

\begin{figure}[hbt!]
\centering
\includegraphics[width=1.0\textwidth]{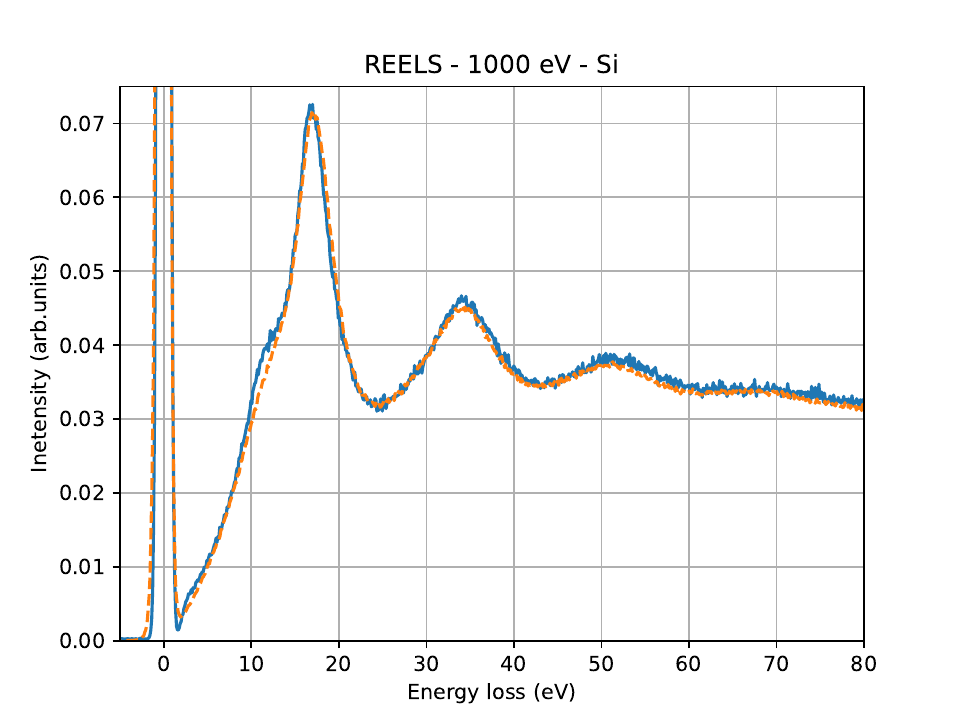}
\caption{REEL spectrum for 1000 eV electrons impinging on Si as a function of energy loss (the zero-loss peak is shown on the far left of the spectrum). The primary beam strikes the material with an angle of incidence of 30$^\circ$ with respect to the surface normal. The spectra are normalised to a common area of the zero-loss peak. Plasmon peaks up to the 4th order of scattering can be observed in both the experimental and simulated spectra. Please note the presence of surface plasmons and bulk plasmons in the spectra. Solid line: Monte Carlo simulations \cite{Dapor2023}. Dashed line: experimental data \cite{https://doi.org/10.1002/sia.4835}.}
\label{fig:prova14}
\end{figure}

To show the general applicability of the TMC approach to materials with different electronic properties, further examples of calculated REEL spectra for different kinetic energies of the incident electron beam are shown in Fig. \ref{fig:prova12}, \ref{fig:prova13} and \ref{fig:prova14} for cerium oxides \cite{pedrielli2021electronic}, which are insulating materials, metallic aluminium \cite{Dapor2023} and semiconductor silicon \cite{Dapor2023} in comparison to experimental results. The spectral behaviour reflects the presence of single and plasmon excitations for all samples.
To accurately reproduce the experimental results and reduce statistical noise, a number of 10$^9$ trajectories were used. In particular, we note the presence of surface and bulk plasmon peaks in both the theoretical and experimental spectra of Al and Si. In addition, several equally spaced plasmon peaks can be seen in Figs. \ref{fig:prova13} and \ref{fig:prova14}, the intensity of which decreases with increasing energy loss, in the case of Al up to the 5th energy level and in the case of Si up to the 4th energy level.

\begin{figure}[htb!]
\centering
\includegraphics[width=1.0\textwidth]{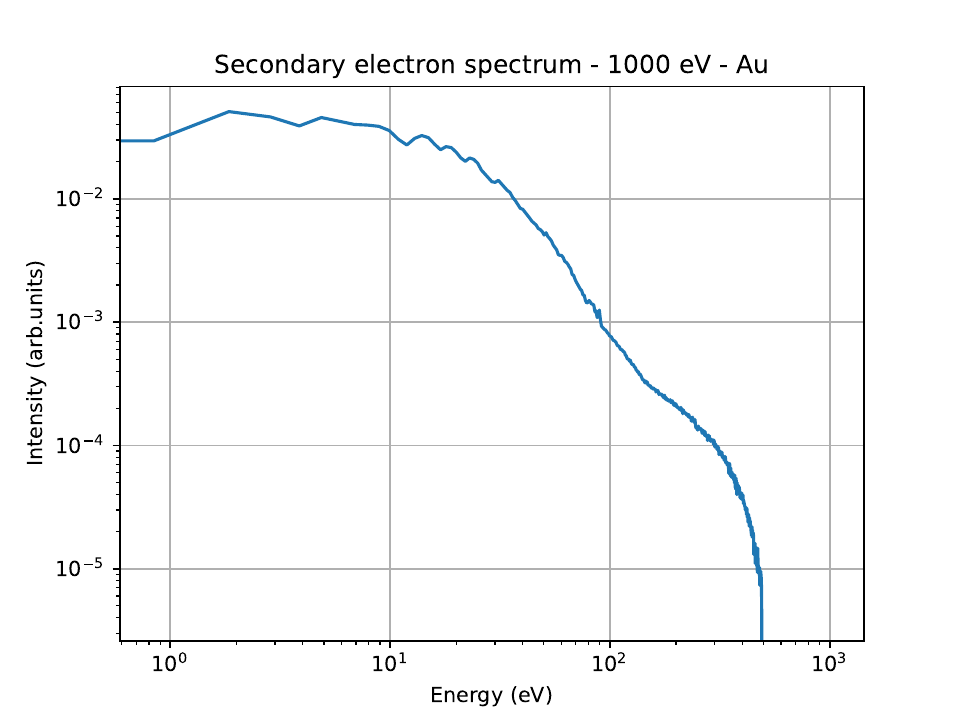}
\caption{Monte Carlo simulation of the spectrum of the secondary electron emission of an Au sample for a kinetic energy of the primary beam of 1000 eV, which shows a wiggling behaviour.}\label{fig:prova15}
\end{figure}

The energy straggling strategy can also be used in TMC for quantitative modelling of the secondary electron spectra and yields (SEY) of samples, such as Al \cite{KhurFitting1999}, Si \cite{KhurFitting1999,Dapor_2008,DAPOR20093055}, Cu \cite{azzolini2018secondary,DAPOR20093055}, Ag \cite{azzolini2018secondary,KhurFitting1999}, Au \cite{azzolini2018secondary,taioli2023role,KhurFitting1999}, PMMA \cite{Dapor2017}, P3HT \cite{DAPOR201895}, Al$_2$O$_3$ \cite{1Ganachaud} and SiO$_2$ \cite{SCHREIBER200225}. Essentially, this amounts to integrating the spectra of the secondary energy distribution (see e.g. Fig. \ref{fig:prova10}) between 0 and 50~eV by changing the incident kinetic energies of the electron beam. Recall that these so-called secondary electrons are, by definition, the particles produced by inelastic scattering events characterised by energy losses above the atomic ionisation threshold and ejected with an energy of less than 50 eV. Of course, they must also overcome the energy barrier at the solid-vacuum interface, i.e. the work function (see Eq. (\ref{eq:condsecond})). In this regard, there are also primary backscattered electrons that leave the target with a kinetic energy of less than 50~eV after suffering strong energy losses; however, their number is typically negligible compared to the number of secondary electrons that are emitted with an energy of less than 50~eV. The distinction between primary and secondary electrons, which is in principle meaningless due to the indistinguishability of electrons, can be used thanks to our classical TMC approach.

In Fig. \ref{fig:prova15} we show the MC simulation of the secondary electrons emitted by an Au sample for a primary beam with a kinetic energy of 1000 eV in the energy range where most of the secondary electrons are emitted. The wiggling behaviour of the lineshape reflects the ELF peaks. 
We also recall that the conditions for the emission of electrons at the sample surface are met if the transmission coefficient, which depends on the work function of the material, is greater than a random number (see Eq. (\ref{transmission})). The work function of Au depends strongly on the type of impurities present on the surface, the duration of exposure of the sample to air and the sputtering processes used to clean it \cite{turetta:hal-03240126,wass2019}: here we have assumed a value of $\Phi_{\rm W}= 4.7$ eV, which can be obtained after $\mathrm{Ar}^+$ sputter-cleaned Au. 

\begin{figure}[htb!]
\centering
\includegraphics[width=1.0\textwidth]{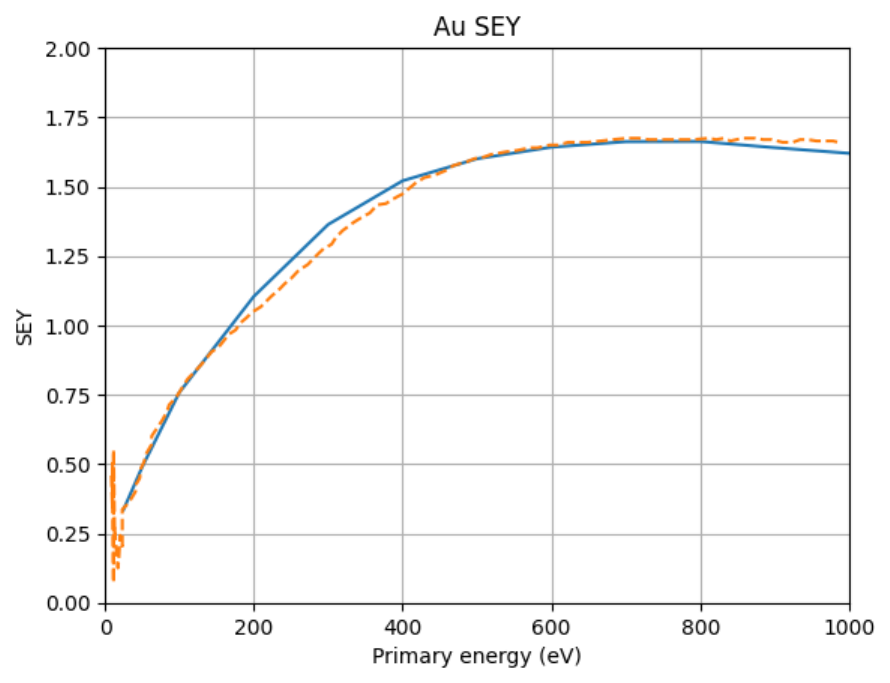}
\caption{Comparison between ab initio-fed MC calculations (solid blue line) of the SEY of Ar$^+$-sputtered Au samples for different kinetic energies of the primary beam and the experimental data (dashed orange line) \cite{gonzalez2017secondary}. The calculations were performed for normal incidence.}
\label{fig:prova16}
\end{figure}

\begin{figure}[htb!]
\centering
\includegraphics[width=1.0\textwidth]{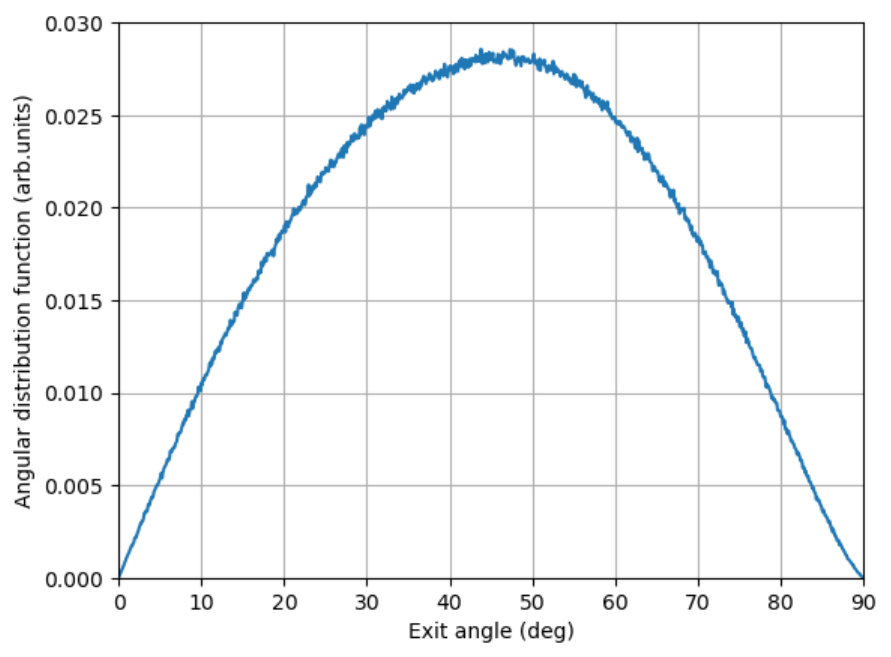}
\caption{Angular distribution of the SE emerging from the Au surface for a 1000 eV electron beam.}
\label{fig:angdist}
\end{figure}

In Fig. \ref{fig:prova16} we show a comparison between the SEY of Au simulated by TMC (blue line) \cite{taioli2023role} and the experimental data (orange dashed line) \cite{gonzalez2017secondary} as a function of the kinetic energy of the primary beam at perpendicular incidence with respect to the target surface. The agreement is remarkable.
We observe that the SEY increases with increasing primary energy up to a maximum and then decreases again. This behaviour can be explained by the fact that at a very low primary energy only a few secondary electrons can be generated and thus emitted from the surface; this number naturally increases with the primary energy together with the average depth at which the secondary electrons are generated. Finally, with increasing energy, the point at which the secondary electrons are generated is so deep inside the material that hardly any can escape from the solid surface.
 
The angular emission pattern of the secondary electrons is also shown in Fig. \ref{fig:angdist} for the same conditions (1 keV electron beam hitting an Au sample)
\cite{taioli2023role}. We note that the angular distribution closely follows a cosine function that is almost inversion symmetric around the maximum at $\approx$ 45\textdegree.
 
\begin{figure}[htb!]
\centering
\includegraphics[width=1.0\textwidth]{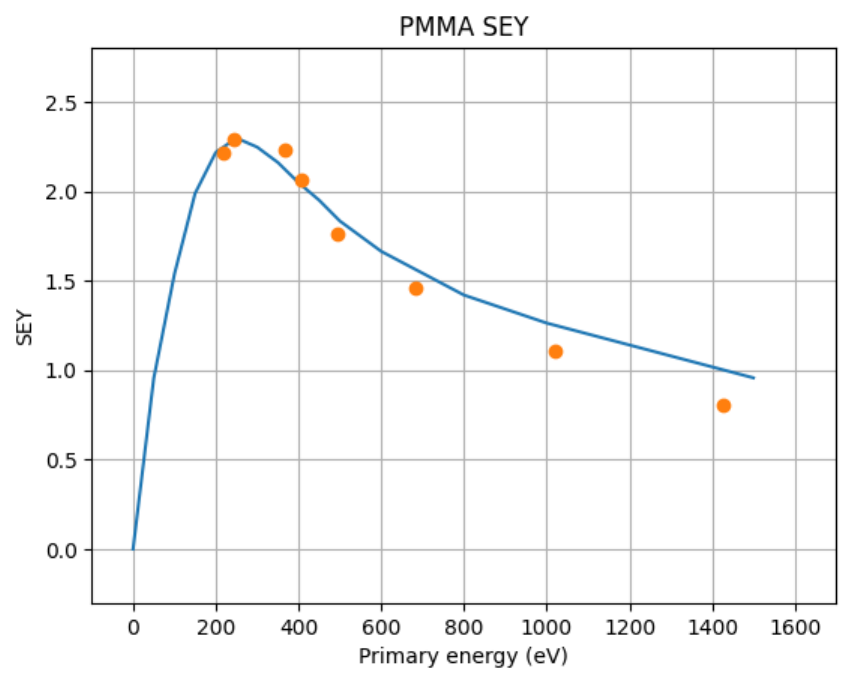}
\caption{Secondary electron yield of PMMA. The calculations were performed for normal incidence. Solid line: MC simulations \cite{Dapor2017}. Dots: experimental data \cite{Yasuda_2008}.}
\label{fig:prova17}
\end{figure}

\begin{figure}[htb!]
\centering
\includegraphics[width=1.0\textwidth]{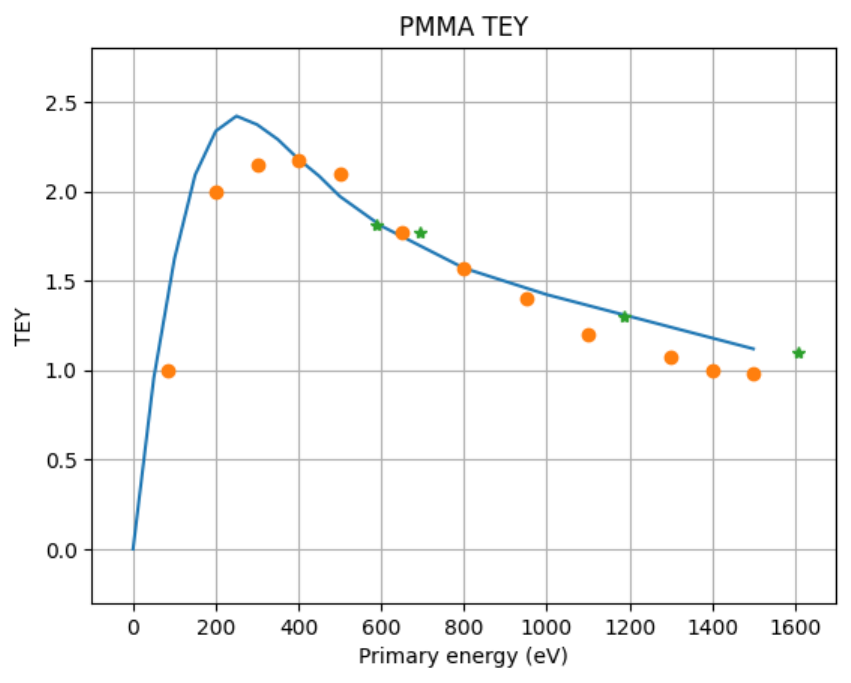}
\caption{Total electron yield (= secondary electron yield + backscattering coefficient) (TEY) of PMMA. Calculations were performed for normal incidence. Solid line: MC simulations \cite{Dapor2017}. Dots: Boubaya and Blaise experimental data \cite{Boubaya}. Stars: Rau et al. experimental data \cite{Rau}.}
\label{fig:prova19}
\end{figure}

\begin{figure}[htb!]
	\centering
\includegraphics[width=1.0\linewidth]{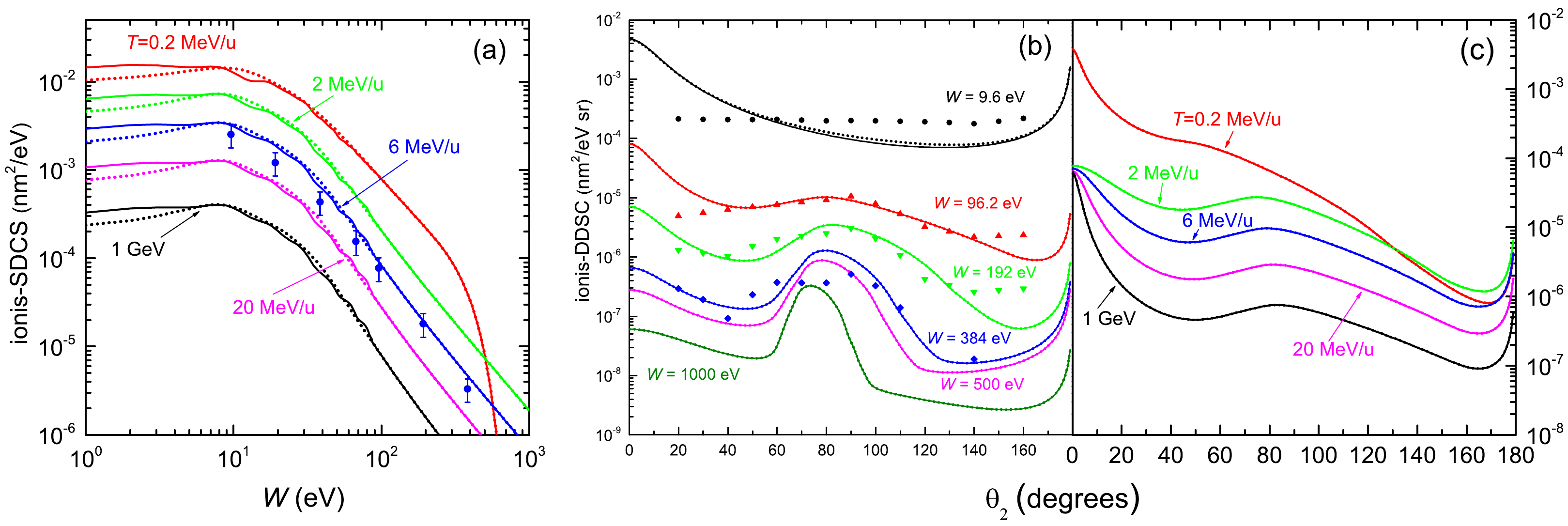}
\caption{Ionisation cross-sections of carbon ions in water. The solid and dashed lines correspond to the calculations performed with the LR-TDDFT (see section \ref{ELF_ab}) and MELF-GOS (see section \ref{drudino}) approaches. The symbols show the experimental data for water vapour \cite{DalCappello2009}.
(a) Energy distributions of electrons emitted during collisions of carbon ions with water molecules as a function of their kinetic energy $W$ for different carbon incident energies $T$.
(b) Angular distributions of the electrons emitted in the collision of 6 MeV/u C ions as a function of the ejection angle $\theta_2$ for different values of the emission energy $W$.
(c) Angular distribution of electrons emitted with a kinetic energy of 100 eV that are created in liquid water by the collision with C ions for different energies $T$. Source: Reprinted from  \cite{taioli2020relative,de2022energy}.  }\label{fig:ISDCS}
\end{figure}
\begin{figure}[htb!]
\centering
\includegraphics[width=1.0\linewidth]{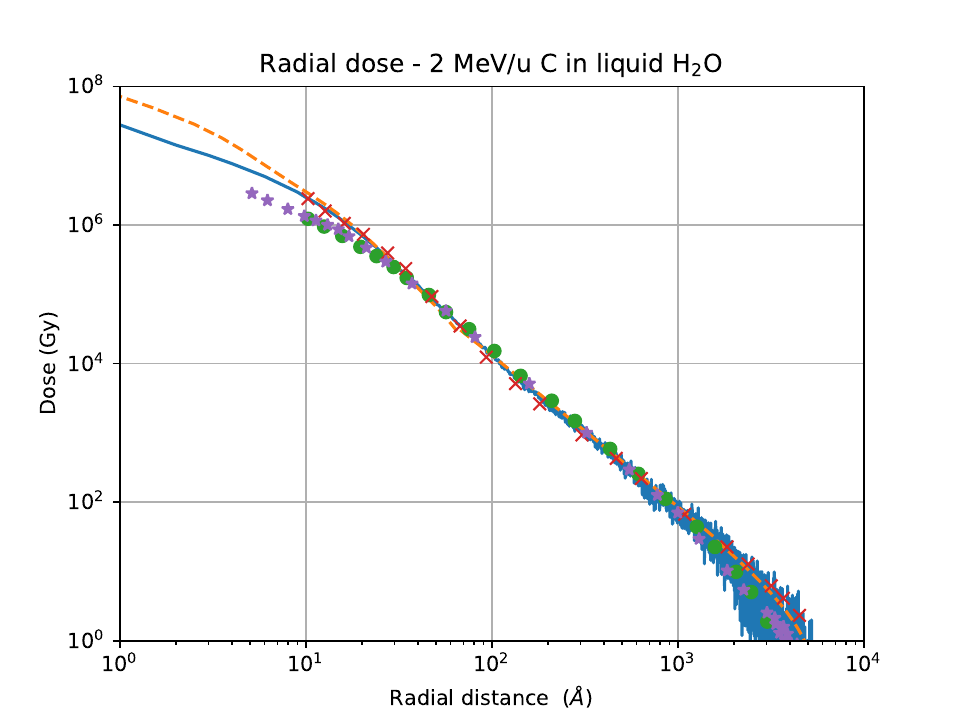}
\caption{Energy density radially deposited by secondary electrons from a 2~MeV/u carbon ion beam in H$_2$O. Solid line from Ref. \cite{de2022energy}. Dashed line from Ref. \cite{refId0}. Dots: Geant4-DNA \cite{INCERTI201492}. Symbols: $\times$ from Ref. \cite{WALIGORSKI1986309}, stars from Ref. \cite{Liamsuwan_2013}.}
\label{fig:prova18}
\end{figure}
\begin{figure}[hbt!]
\centering
\includegraphics[width=1.0\linewidth]{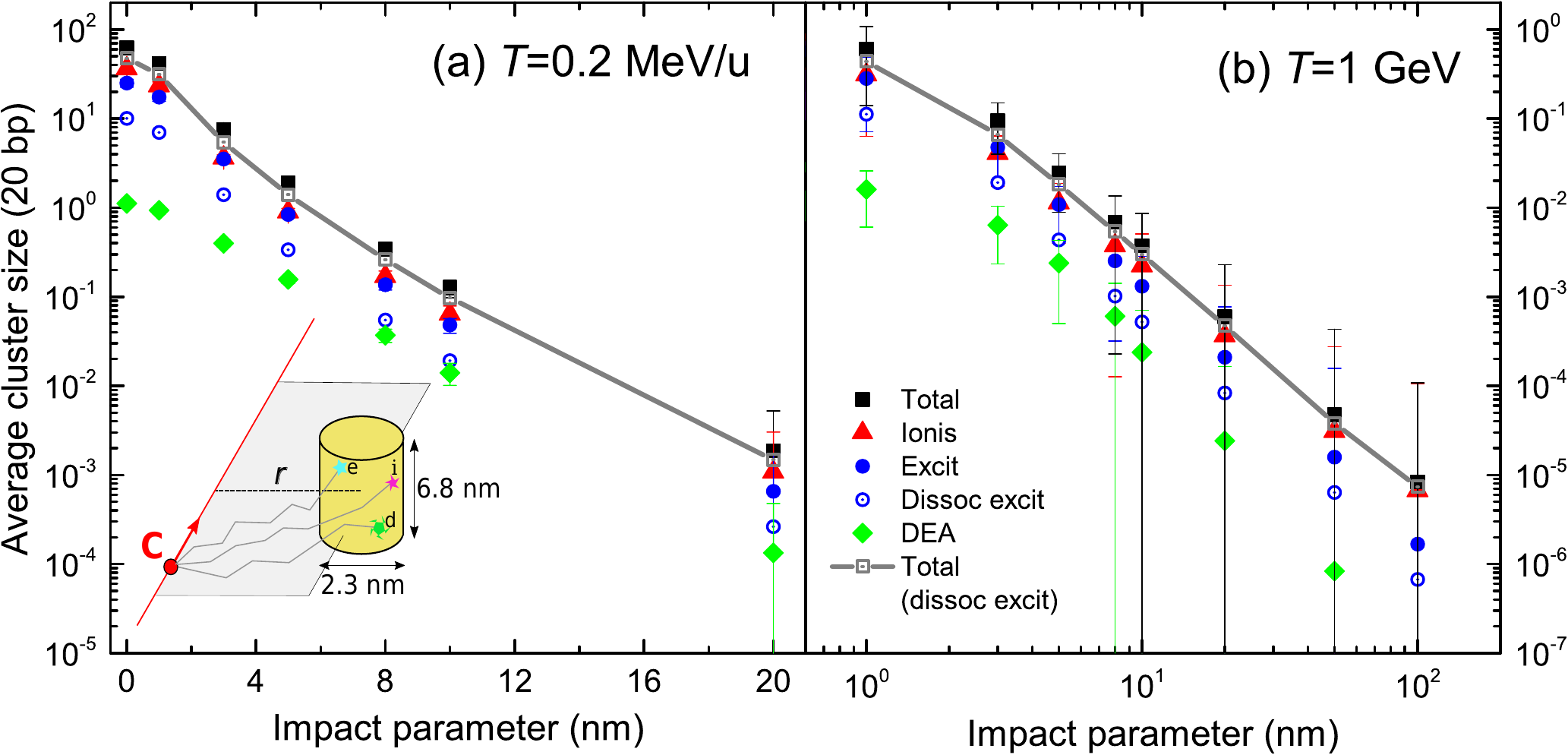}
\caption{Average cluster size in a sensitive volume of liquid water as a function of the impact parameter from the ionic pathway is shown for two kinetic energies \cite{taioli2020relative,de2022energy}: (a) $T$ = 0.2 MeV/u and (b) $T$= 1 GeV. The symbols correspond to the clusters produced by different types of events: Ionisations (red triangles), excitations (blue circles), dissociative excitations (blue hollow circles) and dissociative electron attachments (DEA) (green diamond). The total average cluster size, which includes all events, is represented by black squares, while the size that includes only the damaging events is represented by grey squares. The letters in the inset of panel (a) represent the sensitive cylindrical volume and refer to excitation (e), ionisation (i), and DEA (d). Source: Reprinted from  \cite{taioli2020relative,de2022energy}.}	\label{fig:AvClusterSize-Impact-2T}
\end{figure}
\begin{figure}[hbt!]
\centering
\includegraphics[width=0.9\linewidth]{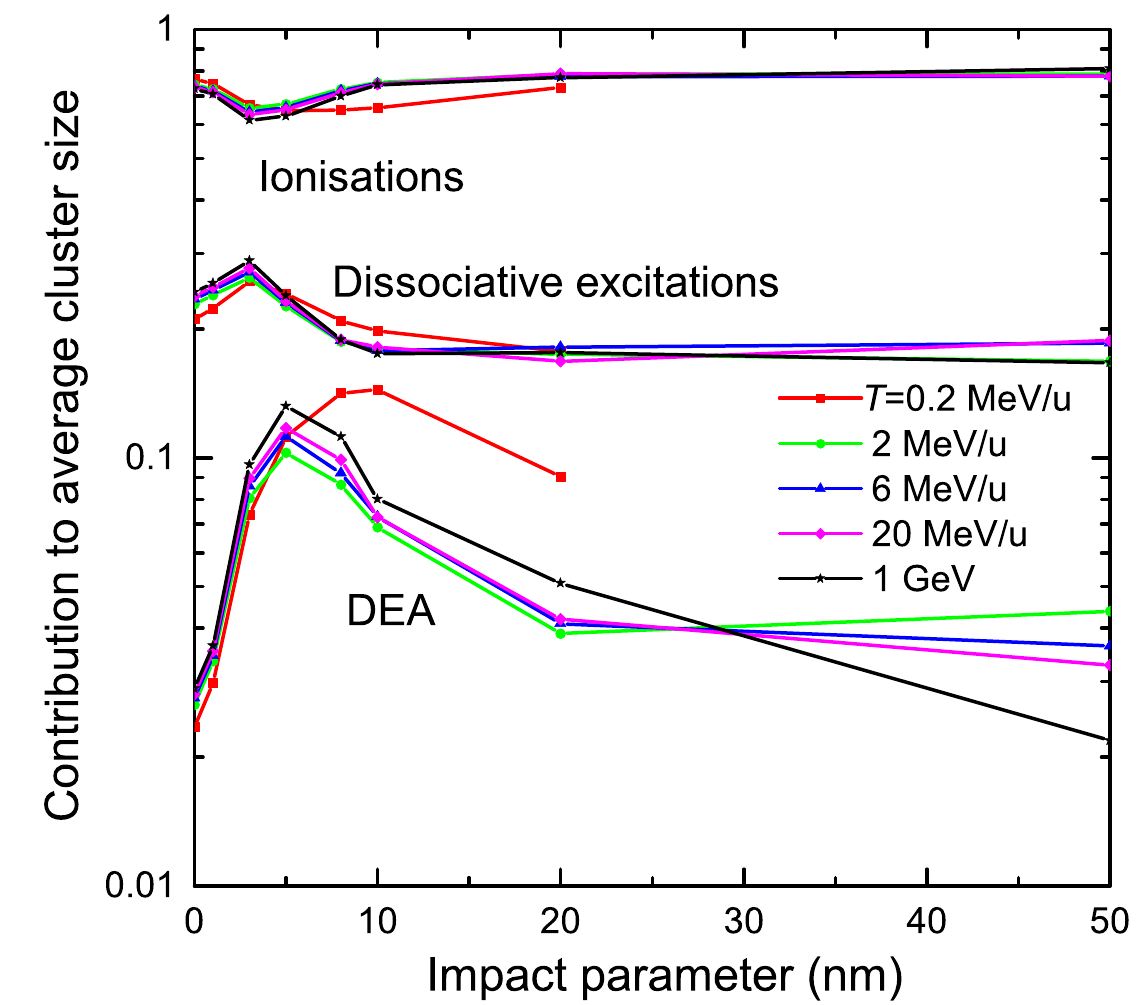}
\caption{Relative contribution of ionisation, dissociative excitation and DEA to the average cluster size for different kinetic energies of the carbon ions as a function of the radial distance of the test volume, represented by a cylinder with a dimension of two DNA turns. Source: Reprinted from  \cite{taioli2020relative,de2022energy}.}
\label{fig:M1-contrib}
\end{figure}
Further examples of the MC calculations of SEY and TEY for the organic material PMMA are shown in Figs. \ref{fig:prova17} and \ref{fig:prova19} respectively. The behaviour is similar to that previously observed for Au, i.e. the yield of secondary electron emission (which for TEY also includes the backscattered electrons of the primary beam that escape with a kinetic energy < 50 eV) increases until a maximum is reached with increasing primary energy; thereafter, the yield decreases with increasing primary energy.

\subsection{The radial dose in liquid water}

By accessing the elastic and inelastic scattering cross-sections, event-by-event MC simulations can also be performed to model secondary electron energy deposition \cite{taioli2020relative,de2022energy,PhysRevB.96.064113,Dapor2023}.
Especially for the development of novel and more effective radiotherapies to cure cancer, it is of utmost importance to investigate how charged particle beams such as protons and carbon ions interact with the biological material and deposit their energy. This study is also important for radiation protection and to prevent cosmic rays, which consist mainly of protons and nuclei, from damaging electronic equipment on board and harming personnel on manned space missions.
In the specific case of carbon ion beams, their successful use in hadron therapy is related to the low angular deflection they suffer on their path through the biological medium, as well as the well-defined penetration region characterised by a sharp peak of the depth-dose curve (known as Bragg peak), which means that a large part of their energy is deposited towards the end of their trajectory \cite{Schardt2010,Tsujii2008,Ebner2016}. In addition, carbon ions have an increased relative biological effectiveness (RBE) \cite{Amaldi2005}, i.e. the generation of numerous low-energy secondary electrons and chemically reactive species increases their ability to kill cells more effectively than photons, for example, for the same delivered dose \cite{Nikjoo2016,Schardt2010}.

This property, which depends on the stopping power, enables precise focussing on the tumour area, which can be treated without severely damaging the surrounding healthy tissues.
We note here that a relevant part of the damage inflicted on cancer cells is mainly due to the secondary electrons generated by the carbon ions on their way through the biological medium, which in computational biology is mimicked by liquid water, as it represents 60\% of the human body weight. The secondary electrons generated by the collisions between ions and atomic targets undergo elastic and inelastic processes that are actively involved in the generation of further secondary electrons, excitation of phonons, dissociative electron attachment (DEA) \cite{taioli2006waterwaves,taioli2006wave} and trapping.
The relative importance of these phenomena in terms of damage to the cells depends on the energy of the impacting ion. In order to evaluate the dose deposited radially in the material along the ion track, the first and most important mechanism to be simulated, which triggers the chemical-physical mechanisms behind the energy transfer into the material, is thus the generation of secondary electrons.

In this context, MC calculations for the specific kinetic energies of carbon ions \cite{taioli2020relative,de2022simulating} must first be performed. This calculation will be carried out a number of statistically significant times with different random seeds to obtain the structure of the trajectory, typically more than 1000. The length of the ion track depends on the energy. It must be long enough for virtually all secondary electrons generated along the carbon ion track to reach a sensitive volume (with the dimensions of a DNA-like target) located at a certain radial distance from the trajectory.
Using dielectric theory, each ion shot provides precise ionisation sites, the emission energy of the secondary electrons and their ejection angle.

To achieve an acceptable compromise between computational cost and low signal-to-noise ratio, the carbon ions should undergo a number of collisions with the target atoms that can generate on average $10^5$--$10^6$ electrons, which in turn produce an average 100 more electrons each. The avalanche of secondary electrons can easily lead to the simulation of billion trajectories per ion energy.
We emphasise that these values depend crucially on the charge state of the carbon ion, which can change dynamically during the motion to a fully ionised state \cite{de2022energy}. The charge state also has a strong influence on the stopping power, especially at energies around the maximum \cite{de2022energy}. The Bragg peak of the carbon ion is also influenced by fragmentation into nuclear fragments.
In Fig. \ref{fig:ISDCS} we show the results of MC simulations to determine the ionisation cross-section (panel (a)) of carbon ions in water as well as the energy and angular distributions of the electrons emitted after the collision
(panels (b) and (c)).
Once the energy and angular distributions of the secondary electrons generated by the impact of carbon ions on liquid water and the relevant elastic (see section \ref{elasmott} or \ref{abinito}) and inelastic (see section \ref{Ritchie_theory}) scattering cross-sections for electron transport in water are known, the simulation of the dose deposited by carbon ions and the relevant mechanisms triggered by their passage through the biological medium can be accurately performed.

In particular, we have access to: (i) the radial dose resulting from the energy deposition around the path of the ion, and (ii) the clustering of damaging events generated in DNA-sized volumes located at different radial distances from the ion's path. While the former is a macroscopic quantity that can be useful in semi-empirical radiobiological models \cite{Friedrich2018}, the latter provides fundamental information for the assessment of radiobiological effectiveness (RBE) for ion-induced radiation damage \cite{Surdutovich2014,Solovyov2017}.

In general, the absorbed dose represents the average energy deposited per unit mass by ionising radiation. The radial dose is an approximation of this. It is estimated microscopically using concentric rings centred around the ion trajectory with increasing radius. Using an MC algorithm \cite{PhysRevB.96.064113} with the previously calculated elastic and inelastic scattering cross sections, one can calculate the amount of energy deposited by secondary electrons in cylinders with different radii divided by the mass of such an element filled with water at standard concentration. In Fig. \ref{fig:prova18} we show the radial dose around carbon ions impinging on liquid water with a kinetic energy of 2 MeV/u \cite{de2022energy}.

The average cluster size of damaging events on the nanoscale takes into account processes such as ionisation, electronic excitation and DEA.
To account for DEA, the cross-section for liquid water was taken from experimental data \cite{10.1063/1.1799251}.
The average cluster size is commonly used to represent the distribution of damage events generated in nanovolumes. These are modelled by a cylinder of 2.3 nm (diameter) and 6.8 nm (height) positioned with the axis perpendicular to the ion path at different impact parameters to mimic sensitive biological targets that resemble the dimensions of two turns of DNA molecules (a strand of 20 base pairs in length).
In Fig. \ref{fig:AvClusterSize-Impact-2T}(a) and (b) we show the average cluster size for impact parameters from 0 to 100 nm for 0.2 MeV/u and 1 GeV carbon ions, respectively.

First, we find that at shorter radial distances the total damage is larger for 0.2 MeV/u carbon ions, a tendency that weakens at larger distances.
Second, we find that ionisation processes (triangles in Fig. \ref{fig:AvClusterSize-Impact-2T}) dominate for all impact parameters, with electronic excitation (full circles) having a significant impact on causing severe damage, although only $\approx$40\% of electronic excitations are effective in dissociating water. Conversely, the DEA apparently plays a subordinate role as it is almost two orders of magnitude smaller than the ionisation clusters. However, this view assumes a linear energy transfer (LET), whereas non-linear effects should also be taken into account. 

In Fig. \ref{fig:M1-contrib} finally, we plot the relative contribution to the average size of damage clusters produced by ionisation events, dissociative electronic excitations and DEA, respectively, for different carbon ion energies as a function of the impact parameter. We find that, in general, ionisation processes are responsible for $\approx$80\% of the cluster size, followed by dissociative excitations ($\approx$15-17\%) and DEA $\approx$5\%. Thus, ionisation makes the largest contribution to the clusters of damaging events triggered by carbon ions in liquid water, while DEA, which is normally considered a very relevant biological damage mechanism \cite{Boudaiffa2000}, surprisingly plays a minor role at least in LET.

\section{Summary, conclusions and outlook }\label{Conclusions}

This article examines the mechanisms of electron scattering, including elastic and inelastic interactions with target atoms, molecules and solids, by reviewing in detail the theoretical and computational methods that can be used to interpret the experimental results. In particular, the application of quantum mechanics techniques to the analysis of the interaction processes of charged particles with various forms of matter is thoroughly explored. The text also introduces the Monte Carlo method and provides a large number of examples and extensive comparisons between simulated results and experimental data from the literature.
Modern derivations are discussed and ab initio methods are explained in detail, providing a rigorous approach to the calculation of elastic and inelastic scattering cross sections. Aspects such as the energy and angular distributions of secondary electrons, the backscattering coefficient, the yield of secondary electrons and their role in hadron therapy are discussed.

In particular, Monte Carlo results on the radial distribution of energy deposited in the medium by secondary electrons generated by high-energy ion beams are presented. The macroscopic nature of the radial dose resulting from the energy deposition around the ion's trajectory is emphasised and its usefulness in semi-empirical radiobiological models is demonstrated. Furthermore, the presented Monte Carlo results on the clustering of damage events within DNA-sized volumes at different radial distances from the ion's trajectory provide fundamental information for the evaluation of radiobiological effectiveness in the assessment of ion-induced radiation damage.
%\FloatBarrier

In this context, we point out that despite the maturity of the field of energy loss spectroscopy and backscattered electron analysis and the thorough understanding of electron transport and emission mechanisms, there are still approximations, assumptions and challenges to overcome.

We believe, in particular, that future experimental research in this area should focus on key aspects of fabrication, e.g. developing instruments such as detectors that can collect more secondary electrons. This will enable high-resolution imaging and improve the signal-to-noise ratio for better visualisation of surface morphology and fine topographical details.

In addition, advances in the miniaturisation of devices at the nanoscale and the ability to non-destructively analyse materials to extract energy and momentum information from low-energy electrons in SEM and helium ion scanning microscopes (HIM) are essential for improved characterisation of nanomaterials. In particular, efforts should be made to develop non-destructive microscopy techniques for imaging and mapping quantum electronic properties and band structure information in nanostructures. Indeed, obtaining electronic band structure information from SEM devices is still in its infancy compared to other electron spectroscopy techniques such as angle-resolved photoemission electron spectroscopy (ARPES).

On the theoretical side, there is a clear need to improve the integration of ab initio methods into Monte Carlo simulations for a spectrally resolved study of the generation and emission of secondary electrons in response to excitation by an electron/ion primary beam, as can be observed in SEM or HIM. Specifically, the aim is to improve the calculations of differential and total elastic scattering cross sections for solids using first principles methods in order to replace Mott's theory. In particular, current methods are inadequate to account for solid-state effects such as multiple scattering from neighbouring atomic centres, especially in the low energy range (< 10 eV). In addition, a relativistic dielectric theory that goes beyond the limitations of the Ritchie approach is needed to determine the inelastic mean free path, the differential inverse inelastic mean free path and the cross sections to describe the interaction between electrons and ions with matter.

Theoretical models should also aim to establish a quantitative and robust link between the band structure, the density of states (DOS) and the information obtained from the energy spectra of secondary electrons in solids, which is still lacking. Furthermore, extending the energy range of secondary electrons to a few tens of electron volts could reveal additional features related to plasmon and core-level energy loss and deliver information on conduction bands that may help to interpret experimental results. These advances would contribute enormously to the development of a comprehensive theory for the generation of secondary electrons produced by the interaction between electron and ion beams in solids.

\section{Declaration of competing interest}

The authors declare that they have no known competing financial interests or personal relationships that could have appeared to influence the work reported in this paper. 

\section{Acknowledgments}

This action has received funding from the European Union under grant agreement No 101046651. The authors gratefully acknowledge discussions with I. Abril, P. de Vera, G. Garberoglio, R. Garcia-Molina, T. Morresi, S. Simonucci and P. Trevisanutto. 

%\section*{References}


\begin{thebibliography}{242}
\expandafter\ifx\csname natexlab\endcsname\relax\def\natexlab#1{#1}\fi
\providecommand{\url}[1]{\texttt{#1}}
\providecommand{\href}[2]{#2}
\providecommand{\path}[1]{#1}
\providecommand{\DOIprefix}{doi:}
\providecommand{\ArXivprefix}{arXiv:}
\providecommand{\URLprefix}{URL: }
\providecommand{\Pubmedprefix}{pmid:}
\providecommand{\doi}[1]{\href{http://dx.doi.org/#1}{\path{#1}}}
\providecommand{\Pubmed}[1]{\href{pmid:#1}{\path{#1}}}
\providecommand{\bibinfo}[2]{#2}
\ifx\xfnm\relax \def\xfnm[#1]{\unskip,\space#1}\fi
%Type = Article
\bibitem[{Werner(2023)}]{werner2023}
\bibinfo{author}{W.~Werner},
\newblock \bibinfo{title}{Electron beams near surfaces: The concept of partial
  intensities for surface analysis and perspective on the low energy regime},
\newblock \bibinfo{journal}{Front. Mater.} \bibinfo{volume}{10}
  (\bibinfo{year}{2023}) \bibinfo{pages}{1202456}.
  \DOIprefix\doi{https://doi.org/10.3389/fmats.2023.1202456}.
%Type = Article
\bibitem[{Li et~al.(2024)Li, Gong, Harada, Da, Zeng, and ZJ}]{li2024}
\bibinfo{author}{Z.~Li}, \bibinfo{author}{J.~Gong},
  \bibinfo{author}{Y.~Harada}, \bibinfo{author}{B.~Da},
  \bibinfo{author}{R.~Zeng}, \bibinfo{author}{D.~ZJ},
\newblock \bibinfo{title}{Determination of the energy loss function of tungsten
  from reflection electron energy loss spectroscopy spectra},
\newblock \bibinfo{journal}{Results Phys.} \bibinfo{volume}{56}
  (\bibinfo{year}{2024}) \bibinfo{pages}{107247}.
  \DOIprefix\doi{https://doi.org/10.1016/j.rinp.2023.107247}.
%Type = Article
\bibitem[{Taioli et~al.(2009)Taioli, Umari, and
  De~Souza}]{taioli2009electronic}
\bibinfo{author}{S.~Taioli}, \bibinfo{author}{P.~Umari},
  \bibinfo{author}{M.~De~Souza},
\newblock \bibinfo{title}{Electronic properties of extended graphene
  nanomaterials from gw calculations},
\newblock \bibinfo{journal}{Phys. Status Solidi (B)} \bibinfo{volume}{246}
  (\bibinfo{year}{2009}) \bibinfo{pages}{2572--2576}.
  \DOIprefix\doi{https://doi.org/10.1002/pssb.200982339}.
%Type = Article
\bibitem[{Umari et~al.(2012)Umari, Petrenko, Taioli, and
  De~Souza}]{umari2012communication}
\bibinfo{author}{P.~Umari}, \bibinfo{author}{O.~Petrenko},
  \bibinfo{author}{S.~Taioli}, \bibinfo{author}{M.~De~Souza},
\newblock \bibinfo{title}{Communication: Electronic band gaps of semiconducting
  zig-zag carbon nanotubes from many-body perturbation theory calculations},
\newblock \bibinfo{journal}{J. Chem. Phys.} \bibinfo{volume}{136}
  (\bibinfo{year}{2012}). \DOIprefix\doi{https://doi.org/10.1063/1.4716178}.
%Type = Article
\bibitem[{Pedrielli et~al.(2022)Pedrielli, Pugno, Dapor, and
  Taioli}]{pedrielli2022search}
\bibinfo{author}{A.~Pedrielli}, \bibinfo{author}{N.~M. Pugno},
  \bibinfo{author}{M.~Dapor}, \bibinfo{author}{S.~Taioli},
\newblock \bibinfo{title}{In search of the ground state crystal structure of
  $\rm{Ta_2O_5}$ from ab-initio and $\rm{M}$onte $\rm{C}$arlo simulations},
\newblock \bibinfo{journal}{Comput. Mater. Sci.} \bibinfo{volume}{216}
  (\bibinfo{year}{2022}) \bibinfo{pages}{111828}.
  \DOIprefix\doi{https://doi.org/10.1016/j.commatsci.2022.111828}.
%Type = Article
\bibitem[{Ding and Shimizu(1996)}]{ding1996}
\bibinfo{author}{Z.~Ding}, \bibinfo{author}{R.~Shimizu},
\newblock \bibinfo{title}{A $\rm{M}$onte $\rm{C}$arlo modeling of electron
  interaction with solids including cascade secondary electron production},
\newblock \bibinfo{journal}{Scanning} \bibinfo{volume}{18}
  (\bibinfo{year}{1996}) \bibinfo{pages}{92--113}.
  \DOIprefix\doi{https://doi.org/10.1002/sca.1996.4950180204}.
%Type = Article
\bibitem[{Joy and Joy(1996)}]{joy1996}
\bibinfo{author}{D.~Joy}, \bibinfo{author}{C.~Joy},
\newblock \bibinfo{title}{Low voltage scanning electron microscopy},
\newblock \bibinfo{journal}{Micron} \bibinfo{volume}{27} (\bibinfo{year}{1996})
  \bibinfo{pages}{247--263}.
  \DOIprefix\doi{https://doi.org/10.1016/0968-4328(96)00023-6}.
%Type = Article
\bibitem[{Ding et~al.(2001)Ding, Tang, and Shimizu}]{ding2001}
\bibinfo{author}{Z.~Ding}, \bibinfo{author}{X.~Tang},
  \bibinfo{author}{R.~Shimizu},
\newblock \bibinfo{title}{$\rm{M}$onte $\rm{C}$arlo study of secondary electron
  emission},
\newblock \bibinfo{journal}{J. Appl. Phys.} \bibinfo{volume}{89}
  (\bibinfo{year}{2001}) \bibinfo{pages}{718--726}.
  \DOIprefix\doi{https://doi.org/10.1063/1.1331645}.
%Type = Article
\bibitem[{Vos(2001)}]{Vos2001}
\bibinfo{author}{M.~Vos},
\newblock \bibinfo{title}{Observing atom motion by electron-atom $\rm{C}$ompton
  scattering},
\newblock \bibinfo{journal}{Phys. Rev. A} \bibinfo{volume}{65}
  (\bibinfo{year}{2001}) \bibinfo{pages}{012703}.
  \DOIprefix\doi{https://doi.org/10.1103/PhysRevA.65.012703}.
%Type = Article
\bibitem[{Orosz et~al.(2004)Orosz, Gergely, Menyhard, T{\'o}th, Varga~D, and
  A}]{Orosz2004}
\bibinfo{author}{G.~Orosz}, \bibinfo{author}{G.~Gergely},
  \bibinfo{author}{M.~Menyhard}, \bibinfo{author}{J.~T{\'o}th},
  \bibinfo{author}{B.~Varga~D, Lesiak}, \bibinfo{author}{J.~A},
\newblock \bibinfo{title}{Hydrogen and surface excitation in electron spectra
  of polyethylene},
\newblock \bibinfo{journal}{Surf. Sci.} \bibinfo{volume}{566–568}
  (\bibinfo{year}{2004}) \bibinfo{pages}{544–548}.
  \DOIprefix\doi{https://doi.org/10.1016/j.susc.2004.05.106}.
%Type = Article
\bibitem[{Yubero et~al.(2008)Yubero, Pauly, Dubus, and Tougaard}]{Yubero2008}
\bibinfo{author}{F.~Yubero}, \bibinfo{author}{N.~Pauly},
  \bibinfo{author}{A.~Dubus}, \bibinfo{author}{S.~Tougaard},
\newblock \bibinfo{title}{Test of validity of the $\rm{V}$-type approach for
  electron trajectories in reflection electron energy loss spectroscopy},
\newblock \bibinfo{journal}{Phys. Rev. B} \bibinfo{volume}{77}
  (\bibinfo{year}{2008}) \bibinfo{pages}{245405}.
  \DOIprefix\doi{https://doi.org/10.1103/PhysRevB.77.245405}.
%Type = Article
\bibitem[{Mao et~al.(2008)Mao, Li, Zeng, and Ding}]{ding2008}
\bibinfo{author}{S.~Mao}, \bibinfo{author}{Y.~Li}, \bibinfo{author}{R.~Zeng},
  \bibinfo{author}{Z.~Ding},
\newblock \bibinfo{title}{Electron inelastic scattering and secondary electron
  emission calculated without the single pole approximation},
\newblock \bibinfo{journal}{J. Appl. Phys.} \bibinfo{volume}{104}
  (\bibinfo{year}{2008}) \bibinfo{pages}{114907}.
  \DOIprefix\doi{https://doi.org/10.1063/1.3033564}.
%Type = Article
\bibitem[{Larciprete et~al.(2013)Larciprete, Grosso, Commisso, Flammini, and
  Cimino}]{larciprete2013}
\bibinfo{author}{R.~Larciprete}, \bibinfo{author}{D.~Grosso},
  \bibinfo{author}{M.~Commisso}, \bibinfo{author}{R.~Flammini},
  \bibinfo{author}{R.~Cimino},
\newblock \bibinfo{title}{Secondary electron yield of $\rm{C}$u technical
  surfaces: dependence on electron irradiation},
\newblock \bibinfo{journal}{Phys. Rev. ST Accel. Beams.} \bibinfo{volume}{16}
  (\bibinfo{year}{2013}) \bibinfo{pages}{011002}.
  \DOIprefix\doi{https://doi.org/10.1103/PhysRevSTAB.16.011002}.
%Type = Article
\bibitem[{Salvat‐Pujol and Werner(2013)}]{salvatpuyol2013}
\bibinfo{author}{F.~Salvat‐Pujol}, \bibinfo{author}{W.~Werner},
\newblock \bibinfo{title}{Surface excitations in electron spectroscopy.
  $\rm{P}$art $\rm{I}$: dielectric formalism and $\rm{M}$onte $\rm{C}$arlo
  algorithm},
\newblock \bibinfo{journal}{Surf. Interface Anal.} \bibinfo{volume}{45}
  (\bibinfo{year}{2013}) \bibinfo{pages}{873--894}.
  \DOIprefix\doi{https://doi.org/10.1002/sia.5175}.
%Type = Article
\bibitem[{Bellissimo et~al.(2020)Bellissimo, Pierantozzi, Ruocco, Stefani,
  Ridzel, Vytautas, Werner, and Taborelli}]{bellissimo2020}
\bibinfo{author}{A.~Bellissimo}, \bibinfo{author}{G.~Pierantozzi},
  \bibinfo{author}{A.~Ruocco}, \bibinfo{author}{G.~Stefani},
  \bibinfo{author}{O.~Ridzel}, \bibinfo{author}{A.~Vytautas},
  \bibinfo{author}{W.~Werner}, \bibinfo{author}{M.~Taborelli},
\newblock \bibinfo{title}{Secondary electron generation mechanisms in carbon
  allotropes at low impact electron energies},
\newblock \bibinfo{journal}{J. Electron Spectrosc. Relat. Phenom.}
  \bibinfo{volume}{241} (\bibinfo{year}{2020}) \bibinfo{pages}{146883}.
  \DOIprefix\doi{https://doi.org/10.1016/j.elspec.2019.07.004}.
%Type = Article
\bibitem[{Tougaard et~al.(2022)Tougaard, Pauly, and Yubero}]{tougaard2022}
\bibinfo{author}{S.~Tougaard}, \bibinfo{author}{N.~Pauly},
  \bibinfo{author}{F.~Yubero},
\newblock \bibinfo{title}{$\rm{QUEELS}$: Software to calculate the energy loss
  processes in $\rm{TEELS}$, $\rm{REELS}$, $\rm{XPS}$ and $\rm{AES}$ including
  effects of the core hole},
\newblock \bibinfo{journal}{Surf. Interface Anal.} \bibinfo{volume}{54}
  (\bibinfo{year}{2022}) \bibinfo{pages}{820--833}.
  \DOIprefix\doi{https://doi.org/10.1002/sia.7095}.
%Type = Article
\bibitem[{Polak and Morgan(2021)}]{polak2021}
\bibinfo{author}{M.~Polak}, \bibinfo{author}{D.~Morgan},
\newblock \bibinfo{title}{$\rm{MAST-SEY}$: Material simulation toolkit for
  secondary electron yield. $\rm{A}$ $\rm{M}$onte $\rm{C}$arlo approach to
  secondary electron emission based on complex dielectric functions},
\newblock \bibinfo{journal}{Comput. Mater. Sci.} \bibinfo{volume}{193}
  (\bibinfo{year}{2021}) \bibinfo{pages}{110281}.
  \DOIprefix\doi{https://doi.org/10.48550/arXiv.2108.11582}.
%Type = Article
\bibitem[{Khan et~al.(2023)Khan, Mao, Zou, Lu, Da, Li, and Ding}]{khan2023}
\bibinfo{author}{M.~Khan}, \bibinfo{author}{S.~Mao}, \bibinfo{author}{Y.~Zou},
  \bibinfo{author}{D.~Lu}, \bibinfo{author}{B.~Da}, \bibinfo{author}{Y.~Li},
  \bibinfo{author}{Z.~Ding},
\newblock \bibinfo{title}{An extensive theoretical quantification of secondary
  electron emission from silicon},
\newblock \bibinfo{journal}{Vacuum} \bibinfo{volume}{215}
  (\bibinfo{year}{2023}) \bibinfo{pages}{112257}.
  \DOIprefix\doi{10.1016/j.vacuum.2023.112257}.
%Type = Article
\bibitem[{Pauly et~al.(2023)Pauly, Yubero, and Tougaard}]{pauly2023}
\bibinfo{author}{N.~Pauly}, \bibinfo{author}{F.~Yubero},
  \bibinfo{author}{S.~Tougaard},
\newblock \bibinfo{title}{Determination of the primary excitation spectra in
  $\rm{XPS}$ and $\rm{AES}$},
\newblock \bibinfo{journal}{Nanomaterials} \bibinfo{volume}{13}
  (\bibinfo{year}{2023}) \bibinfo{pages}{339}.
  \DOIprefix\doi{https://doi.org/10.3390/nano13020339}.
%Type = Article
\bibitem[{Dapor et~al.(2023)Dapor, Garcia-Molina, Trevisanutto, Ding, and
  Taioli}]{dapor2023editorial}
\bibinfo{author}{M.~Dapor}, \bibinfo{author}{R.~Garcia-Molina},
  \bibinfo{author}{P.~E. Trevisanutto}, \bibinfo{author}{Z.~Ding},
  \bibinfo{author}{S.~Taioli},
\newblock \bibinfo{title}{Editorial: Methodological and computational
  developments for modeling the transport of particles within materials},
\newblock \bibinfo{journal}{Front. Mater.} \bibinfo{volume}{10}
  (\bibinfo{year}{2023}) \bibinfo{pages}{1302000}.
  \DOIprefix\doi{https://doi.org/10.3389/fmats.2023.1302000}.
%Type = Article
\bibitem[{Durante and Cucinotta(2011)}]{durante2011}
\bibinfo{author}{M.~Durante}, \bibinfo{author}{F.~Cucinotta},
\newblock \bibinfo{title}{Physical basis of radiation protection in space
  travel},
\newblock \bibinfo{journal}{Rev. Mod. Phys.} \bibinfo{volume}{83}
  (\bibinfo{year}{2011}) \bibinfo{pages}{1245--1281}.
  \DOIprefix\doi{https://doi.org/10.1103/RevModPhys.83.1245}.
%Type = Article
\bibitem[{Taioli et~al.(2023)Taioli, Dapor, Dimiccoli, Fabi, Ferroni, Grimani,
  Villani, and Weber}]{taioli2023role}
\bibinfo{author}{S.~Taioli}, \bibinfo{author}{M.~Dapor},
  \bibinfo{author}{F.~Dimiccoli}, \bibinfo{author}{M.~Fabi},
  \bibinfo{author}{V.~Ferroni}, \bibinfo{author}{C.~Grimani},
  \bibinfo{author}{M.~Villani}, \bibinfo{author}{W.~J. Weber},
\newblock \bibinfo{title}{The role of low-energy electrons in the charging
  process of $\rm{LISA}$ test masses},
\newblock \bibinfo{journal}{Class. Quantum Gravity} \bibinfo{volume}{40}
  (\bibinfo{year}{2023}) \bibinfo{pages}{075001}.
  \DOIprefix\doi{https://doi.org/110.1088/1361-6382/acbadd}.
%Type = Article
\bibitem[{Solov’yov et~al.(2009)Solov’yov, Surdutovich, Scifoni, Mishustin,
  and Greiner}]{Solovyov2009}
\bibinfo{author}{A.~Solov’yov}, \bibinfo{author}{E.~Surdutovich},
  \bibinfo{author}{E.~Scifoni}, \bibinfo{author}{I.~Mishustin},
  \bibinfo{author}{W.~Greiner},
\newblock \bibinfo{title}{Physics of ion beam cancer therapy: A multiscale
  approach},
\newblock \bibinfo{journal}{Phys. Rev. E} \bibinfo{volume}{79}
  (\bibinfo{year}{2009}) \bibinfo{pages}{011909}.
  \DOIprefix\doi{https://doi.org/10.1103/PhysRevE.79.011909}.
%Type = Article
\bibitem[{Surdutovich and Solov’yov(2014)}]{Surdutovich2014}
\bibinfo{author}{E.~Surdutovich}, \bibinfo{author}{A.~V. Solov’yov},
\newblock \bibinfo{title}{Multiscale approach to the physics of radiation
  damage with ions},
\newblock \bibinfo{journal}{Eur. Phys. J. D} \bibinfo{volume}{68}
  (\bibinfo{year}{2014}) \bibinfo{pages}{353}.
  \DOIprefix\doi{:https://doi.org/110.1063/1.4802412}.
%Type = Article
\bibitem[{Surdutovich and Solov’yov(2015)}]{Surdutovich2015}
\bibinfo{author}{E.~Surdutovich}, \bibinfo{author}{A.~Solov’yov},
\newblock \bibinfo{title}{Transport of secondary electrons and reactive species
  in ion tracks},
\newblock \bibinfo{journal}{Eur. Phys. J. D} \bibinfo{volume}{69}
  (\bibinfo{year}{2015}) \bibinfo{pages}{193}.
  \DOIprefix\doi{https://doi.org/10.1140/epjd/e2015-60242-1}.
%Type = Article
\bibitem[{Nikjoo et~al.(2016)Nikjoo, Emfietzoglou, Liamsuwan, Taleei, and
  Uehara}]{Nikjoo2016}
\bibinfo{author}{H.~Nikjoo}, \bibinfo{author}{D.~Emfietzoglou},
  \bibinfo{author}{T.~Liamsuwan}, \bibinfo{author}{D.~Taleei, R~andt},
  \bibinfo{author}{S.~Uehara},
\newblock \bibinfo{title}{Radiation track, $\rm{DNA}$ damage and response:
  $\rm{A}$ review},
\newblock \bibinfo{journal}{Rep. Prog. Phys.} \bibinfo{volume}{79}
  (\bibinfo{year}{2016}) \bibinfo{pages}{116601}.
  \DOIprefix\doi{https://doi.org/10.1088/0034-4885/79/11/116601}.
%Type = Book
\bibitem[{Solov’yov(2017)}]{Solovyov2017}
\bibinfo{editor}{A.~V. Solov’yov} (Ed.), \bibinfo{title}{Nanoscale Insights
  into Ion-Beam Cancer Therapy}, \bibinfo{publisher}{Springer International
  Publishing Switzerland}, \bibinfo{year}{2017}. \URLprefix
  \url{https://link.springer.com/book/10.1007/978-3-319-43030-0}.
%Type = Article
\bibitem[{Conte et~al.(2017)Conte, Selva, Colautti, Hilgers, and
  Rabus}]{conte2017}
\bibinfo{author}{V.~Conte}, \bibinfo{author}{A.~Selva},
  \bibinfo{author}{P.~Colautti}, \bibinfo{author}{G.~Hilgers},
  \bibinfo{author}{H.~Rabus},
\newblock \bibinfo{title}{Track structure characterization and its link to
  radiobiology},
\newblock \bibinfo{journal}{Radiat. Meas.} \bibinfo{volume}{83}
  (\bibinfo{year}{2017}) \bibinfo{pages}{506--511}.
  \DOIprefix\doi{https://doi.org/10.1016/j.radmeas.2017.06.010}.
%Type = Article
\bibitem[{Conte et~al.(2018)Conte, Selva, Colautti, Hilgers, Rabus, Bantsar,
  Pietrzak, and Pszona}]{conte2018}
\bibinfo{author}{V.~Conte}, \bibinfo{author}{A.~Selva},
  \bibinfo{author}{P.~Colautti}, \bibinfo{author}{G.~Hilgers},
  \bibinfo{author}{H.~Rabus}, \bibinfo{author}{A.~Bantsar},
  \bibinfo{author}{M.~Pietrzak}, \bibinfo{author}{S.~Pszona},
\newblock \bibinfo{title}{Nanodosimetry: Towards a new concept of radiation
  quality},
\newblock \bibinfo{journal}{Radiat. Prot. Dosim.} \bibinfo{volume}{180}
  (\bibinfo{year}{2018}) \bibinfo{pages}{150--156}.
  \DOIprefix\doi{https://doi.org/10.1093/rpd/ncx175}.
%Type = Article
\bibitem[{Friedrich et~al.(2018)Friedrich, Ilicic, Greubel, Girst, Reindl,
  Sammer, Schwarz, Siebenwirth, Walsh, Schmid, Scholz, and
  Dollinger}]{Friedrich2018}
\bibinfo{author}{H.~Friedrich}, \bibinfo{author}{K.~Ilicic},
  \bibinfo{author}{C.~Greubel}, \bibinfo{author}{S.~Girst},
  \bibinfo{author}{J.~Reindl}, \bibinfo{author}{M.~Sammer},
  \bibinfo{author}{B.~Schwarz}, \bibinfo{author}{C.~Siebenwirth},
  \bibinfo{author}{D.~W.~M. Walsh}, \bibinfo{author}{T.~E. Schmid},
  \bibinfo{author}{M.~Scholz}, \bibinfo{author}{G.~Dollinger},
\newblock \bibinfo{title}{{DNA damage interactions on both nanometer and
  micrometer scale determine overall cellular damage}},
\newblock \bibinfo{journal}{Sci. Rep.} \bibinfo{volume}{8}
  (\bibinfo{year}{2018}) \bibinfo{pages}{16063}.
  \DOIprefix\doi{https://doi.org/10.1038/s41598-018-34323-9}.
%Type = Article
\bibitem[{Surdutovich and Solov’yov(2019)}]{Surdutovich2019}
\bibinfo{author}{E.~Surdutovich}, \bibinfo{author}{A.~Solov’yov},
\newblock \bibinfo{title}{Multiscale modeling for cancer radiotherapies},
\newblock \bibinfo{journal}{Cancer Nanotechnol.} \bibinfo{volume}{10}
  (\bibinfo{year}{2019}) \bibinfo{pages}{6}.
  \DOIprefix\doi{https://doi.org/10.1186/s12645-019-0051-2}.
%Type = Article
\bibitem[{Taioli et~al.(2020)Taioli, Trevisanutto, de~Vera, Simonucci, Abril,
  Garcia-Molina, and Dapor}]{taioli2020relative}
\bibinfo{author}{S.~Taioli}, \bibinfo{author}{P.~E. Trevisanutto},
  \bibinfo{author}{P.~de~Vera}, \bibinfo{author}{S.~Simonucci},
  \bibinfo{author}{I.~Abril}, \bibinfo{author}{R.~Garcia-Molina},
  \bibinfo{author}{M.~Dapor},
\newblock \bibinfo{title}{Relative role of physical mechanisms on complex
  biodamage induced by carbon irradiation},
\newblock \bibinfo{journal}{J. Phys. Chem. Lett.} \bibinfo{volume}{12}
  (\bibinfo{year}{2020}) \bibinfo{pages}{487--493}.
  \DOIprefix\doi{https://doi.org/10.1021/acs.jpclett.0c03250}.
%Type = Article
\bibitem[{de~Vera et~al.(2022{\natexlab{a}})de~Vera, Taioli, Trevisanutto,
  Dapor, Abril, Simonucci, and Garcia-Molina}]{de2022energy}
\bibinfo{author}{P.~de~Vera}, \bibinfo{author}{S.~Taioli},
  \bibinfo{author}{P.~E. Trevisanutto}, \bibinfo{author}{M.~Dapor},
  \bibinfo{author}{I.~Abril}, \bibinfo{author}{S.~Simonucci},
  \bibinfo{author}{R.~Garcia-Molina},
\newblock \bibinfo{title}{Energy deposition around swift carbon-ion tracks in
  liquid water},
\newblock \bibinfo{journal}{Int. J. Mol. Sci.} \bibinfo{volume}{23}
  (\bibinfo{year}{2022}{\natexlab{a}}) \bibinfo{pages}{6121}.
  \DOIprefix\doi{https://doi.org/10.3390/ijms23116121}.
%Type = Article
\bibitem[{de~Vera et~al.(2022{\natexlab{b}})de~Vera, Simonucci, Trevisanutto,
  Abril, Dapor, Taioli, and Garcia-Molina}]{de2022simulating}
\bibinfo{author}{P.~de~Vera}, \bibinfo{author}{S.~Simonucci},
  \bibinfo{author}{P.~E. Trevisanutto}, \bibinfo{author}{I.~Abril},
  \bibinfo{author}{M.~Dapor}, \bibinfo{author}{S.~Taioli},
  \bibinfo{author}{R.~Garcia-Molina},
\newblock \bibinfo{title}{Simulating the nanometric track-structure of carbon
  ion beams in liquid water at energies relevant for hadrontherapy},
\newblock \bibinfo{journal}{J. Phys. Conf. Ser.} \bibinfo{volume}{2326}
  (\bibinfo{year}{2022}{\natexlab{b}}) \bibinfo{pages}{012017}.
  \DOIprefix\doi{https://doi.org/10.1088/1742-6596/2326/1/012017}.
%Type = Article
\bibitem[{Dapor et~al.(2015)Dapor, Abril, de~Vera, and
  Garcia-Molina}]{daporEPJD2015}
\bibinfo{author}{M.~Dapor}, \bibinfo{author}{I.~Abril},
  \bibinfo{author}{P.~de~Vera}, \bibinfo{author}{R.~Garcia-Molina},
\newblock \bibinfo{title}{Simulation of the secondary electrons energy
  deposition produced by proton beams in pmma: influence of the target
  electronic excitation description},
\newblock \bibinfo{journal}{Eur. Phys. J. D} \bibinfo{volume}{69}
  (\bibinfo{year}{2015}) \bibinfo{pages}{165}.
  \DOIprefix\doi{https://doi.org/10.1140/epjd/e2015-60123-7}.
%Type = Article
\bibitem[{Dapor et~al.(2017)Dapor, Abril, de~Vera, and
  Garcia-Molina}]{PhysRevB.96.064113}
\bibinfo{author}{M.~Dapor}, \bibinfo{author}{I.~Abril},
  \bibinfo{author}{P.~de~Vera}, \bibinfo{author}{R.~Garcia-Molina},
\newblock \bibinfo{title}{Energy deposition around swift proton tracks in
  polymethylmethacrylate: How much and how far},
\newblock \bibinfo{journal}{Phys. Rev. B} \bibinfo{volume}{96}
  (\bibinfo{year}{2017}) \bibinfo{pages}{064113}.
  \DOIprefix\doi{https://doi.org/10.1103/PhysRevB.96.064113}.
%Type = Article
\bibitem[{Gorfinkiel et~al.(2005)Gorfinkiel, Faure, Taioli, Piccarreta,
  Halmova, and Tennyson}]{gorfinkiel2005electron}
\bibinfo{author}{J.~Gorfinkiel}, \bibinfo{author}{A.~Faure},
  \bibinfo{author}{S.~Taioli}, \bibinfo{author}{C.~Piccarreta},
  \bibinfo{author}{G.~Halmova}, \bibinfo{author}{J.~Tennyson},
\newblock \bibinfo{title}{Electron-molecule collisions at low and intermediate
  energies using the r-matrix method},
\newblock \bibinfo{journal}{Eur. Phys. J. D} \bibinfo{volume}{35}
  (\bibinfo{year}{2005}) \bibinfo{pages}{231--237}.
  \DOIprefix\doi{https://doi.org/10.1140/epjd/e2005-00179-4}.
%Type = Article
\bibitem[{Taioli and Tennyson(2006)}]{taioli2006waterwaves}
\bibinfo{author}{S.~Taioli}, \bibinfo{author}{J.~Tennyson},
\newblock \bibinfo{title}{Waterwaves: wave particles dynamics on a complex
  triatomic potential},
\newblock \bibinfo{journal}{Comput. Phys. Commun.} \bibinfo{volume}{175}
  (\bibinfo{year}{2006}) \bibinfo{pages}{41--51}.
  \DOIprefix\doi{https://doi.org/10.1016/j.cpc.2006.01.008}.
%Type = Article
\bibitem[{Taioli et~al.(2010)Taioli, Simonucci, Calliari, and
  Dapor}]{taioli2010electron}
\bibinfo{author}{S.~Taioli}, \bibinfo{author}{S.~Simonucci},
  \bibinfo{author}{L.~Calliari}, \bibinfo{author}{M.~Dapor},
\newblock \bibinfo{title}{Electron spectroscopies and inelastic processes in
  nanoclusters and solids: Theory and experiment},
\newblock \bibinfo{journal}{Phys. Rep.} \bibinfo{volume}{493}
  (\bibinfo{year}{2010}) \bibinfo{pages}{237--319}.
  \DOIprefix\doi{https://doi.org/10.1016/j.physrep.2010.04.003}.
%Type = Incollection
\bibitem[{Taioli and Simonucci(2015)}]{taioli2015computational}
\bibinfo{author}{S.~Taioli}, \bibinfo{author}{S.~Simonucci},
\newblock \bibinfo{title}{A computational perspective on multichannel
  scattering theory with applications to physical and nuclear chemistry},
\newblock in: \bibinfo{booktitle}{Annual Reports in Computational Chemistry},
  volume~\bibinfo{volume}{11}, \bibinfo{publisher}{Elsevier},
  \bibinfo{year}{2015}, pp. \bibinfo{pages}{191--310}.
  \DOIprefix\doi{https://doi.org/10.1016/bs.arcc.2015.09.005}.
%Type = Article
\bibitem[{Shimizu and Ding(1992)}]{Shimizu_1992}
\bibinfo{author}{R.~Shimizu}, \bibinfo{author}{Z.-J. Ding},
\newblock \bibinfo{title}{Monte $\rm{C}$arlo modelling of electron-solid
  interactions},
\newblock \bibinfo{journal}{Rep. Prog. Phys.} \bibinfo{volume}{55}
  (\bibinfo{year}{1992}) \bibinfo{pages}{487--531}.
  \DOIprefix\doi{https://doi.org/10.1088/0034-4885/55/4/002}.
%Type = Book
\bibitem[{Joy(1995)}]{joy1995monte}
\bibinfo{author}{D.~C. Joy}, \bibinfo{title}{Monte Carlo modeling for electron
  microscopy and microanalysis}, volume~\bibinfo{volume}{9},
  \bibinfo{publisher}{Oxford University Press}, \bibinfo{year}{1995}.
  \URLprefix \url{https://academic.oup.com/book/54583}.
%Type = Book
\bibitem[{Dapor(2003)}]{Dapor2003book}
\bibinfo{author}{M.~Dapor}, \bibinfo{title}{Electron-Beam Interactions with
  Solids. Application of the Monte Carlo Method to Electron Scattering
  Problems}, volume \bibinfo{volume}{186} of \textit{\bibinfo{series}{Springer
  Tracts in Modern Physics}}, \bibinfo{publisher}{Springer-Verlag Berlin
  Heidelberg}, \bibinfo{year}{2003}.
  \DOIprefix\doi{https://doi.org/10.1007/3-540-36507-9}.
%Type = Book
\bibitem[{Dapor(2023)}]{Dapor2023}
\bibinfo{author}{M.~Dapor}, \bibinfo{title}{Transport of Energetic Electrons in
  Solids. Computer Simulation with Applications to Materials Analysis and
  Characterization}, volume \bibinfo{volume}{290} of
  \textit{\bibinfo{series}{Springer Tracts in Modern Physics}},
  \bibinfo{edition}{4$^{\rm{th}}$} ed., \bibinfo{publisher}{Springer Nature
  Switzerland AG}, \bibinfo{year}{2023}.
  \DOIprefix\doi{https://doi.org/10.1007/978-3-031-37242-1}.
%Type = Article
\bibitem[{Liljequist(2008)}]{5Liljequist2008}
\bibinfo{author}{D.~Liljequist},
\newblock \bibinfo{title}{A study of errors in trajectory simulation with
  relevance for 0.2–50 e$\rm{V}$ electrons in liquid water},
\newblock \bibinfo{journal}{Radiat. Phys. Chem.} \bibinfo{volume}{77}
  (\bibinfo{year}{2008}) \bibinfo{pages}{835--853}.
  \DOIprefix\doi{10.1016/j.radphyschem.2008.03.004}.
%Type = Article
\bibitem[{Liljequist(2013)}]{5Liljequist2013}
\bibinfo{author}{D.~Liljequist},
\newblock \bibinfo{title}{Contribution from inelastic scattering to the
  validity of trajectory methods},
\newblock \bibinfo{journal}{J. Electron Spectrosc. Relat. Phenom.}
  \bibinfo{volume}{189} (\bibinfo{year}{2013}) \bibinfo{pages}{5--11}.
  \DOIprefix\doi{https://doi.org/10.1016/j.elspec.2013.04.011}.
%Type = Article
\bibitem[{Liljequist and Nikjoo(2014)}]{LILJEQUIST201445}
\bibinfo{author}{D.~Liljequist}, \bibinfo{author}{H.~Nikjoo},
\newblock \bibinfo{title}{On the validity of trajectory methods for calculating
  the transport of very low energy (<1kev) electrons in liquids and amorphous
  media},
\newblock \bibinfo{journal}{Radiation Physics and Chemistry}
  \bibinfo{volume}{99} (\bibinfo{year}{2014}) \bibinfo{pages}{45--52}.
  \URLprefix
  \url{https://www.sciencedirect.com/science/article/pii/S0969806X14000607}.
  \DOIprefix\doi{https://doi.org/10.1016/j.radphyschem.2014.02.015}.
%Type = Article
\bibitem[{Ritchie(1957)}]{1Ritchie57}
\bibinfo{author}{R.~H. Ritchie},
\newblock \bibinfo{title}{Plasma losses by fast electrons in thin films},
\newblock \bibinfo{journal}{Phys. Rev.} \bibinfo{volume}{106}
  (\bibinfo{year}{1957}) \bibinfo{pages}{874--881}.
  \DOIprefix\doi{https://doi.org/10.1103/PhysRev.106.874}.
%Type = Article
\bibitem[{Pedrielli et~al.(2021)Pedrielli, de~Vera, Trevisanutto, Pugno,
  Garcia-Molina, Abril, Taioli, and Dapor}]{pedrielli2021electronic}
\bibinfo{author}{A.~Pedrielli}, \bibinfo{author}{P.~de~Vera},
  \bibinfo{author}{P.~E. Trevisanutto}, \bibinfo{author}{N.~M. Pugno},
  \bibinfo{author}{R.~Garcia-Molina}, \bibinfo{author}{I.~Abril},
  \bibinfo{author}{S.~Taioli}, \bibinfo{author}{M.~Dapor},
\newblock \bibinfo{title}{Electronic excitation spectra of cerium oxides: from
  ab initio dielectric response functions to $\rm{M}$onte $\rm{C}$arlo electron
  transport simulations},
\newblock \bibinfo{journal}{Phys. Chem. Chem. Phys.} \bibinfo{volume}{23}
  (\bibinfo{year}{2021}) \bibinfo{pages}{19173--19187}.
  \DOIprefix\doi{https://doi.org/10.1039/D1CP01810H}.
%Type = Article
\bibitem[{Azzolini et~al.(2017)Azzolini, Morresi, Garberoglio, Calliari, Pugno,
  Taioli, and Dapor}]{azzolini2017monte}
\bibinfo{author}{M.~Azzolini}, \bibinfo{author}{T.~Morresi},
  \bibinfo{author}{G.~Garberoglio}, \bibinfo{author}{L.~Calliari},
  \bibinfo{author}{N.~M. Pugno}, \bibinfo{author}{S.~Taioli},
  \bibinfo{author}{M.~Dapor},
\newblock \bibinfo{title}{Monte $\rm{C}$arlo simulations of measured electron
  energy-loss spectra of diamond and graphite: role of dielectric-response
  models},
\newblock \bibinfo{journal}{Carbon} \bibinfo{volume}{118}
  (\bibinfo{year}{2017}) \bibinfo{pages}{299--309}.
  \DOIprefix\doi{http://dx.doi.org/10.1016/j.carbon.2017.03.041}.
%Type = Article
\bibitem[{Azzolini et~al.(2018)Azzolini, Morresi, Abrams, Masters, Stehling,
  Rodenburg, Pugno, Taioli, and Dapor}]{azzolini2018anisotropic}
\bibinfo{author}{M.~Azzolini}, \bibinfo{author}{T.~Morresi},
  \bibinfo{author}{K.~Abrams}, \bibinfo{author}{R.~Masters},
  \bibinfo{author}{N.~Stehling}, \bibinfo{author}{C.~Rodenburg},
  \bibinfo{author}{N.~M. Pugno}, \bibinfo{author}{S.~Taioli},
  \bibinfo{author}{M.~Dapor},
\newblock \bibinfo{title}{Anisotropic approach for simulating electron
  transport in layered materials: computational and experimental study of
  highly oriented pyrolitic graphite},
\newblock \bibinfo{journal}{J. Phys. Chem. C} \bibinfo{volume}{122}
  (\bibinfo{year}{2018}) \bibinfo{pages}{10159--10166}.
  \DOIprefix\doi{https://doi.org/10.1021/acs.jpcc.8b02256}.
%Type = Article
\bibitem[{Triggiani et~al.(2023)Triggiani, Morresi, Taioli, and
  Simonucci}]{triggiani2023elastic}
\bibinfo{author}{F.~Triggiani}, \bibinfo{author}{T.~Morresi},
  \bibinfo{author}{S.~Taioli}, \bibinfo{author}{S.~Simonucci},
\newblock \bibinfo{title}{Elastic scattering of electrons by water: an ab
  initio study},
\newblock \bibinfo{journal}{Front. Mater.} \bibinfo{volume}{10}
  (\bibinfo{year}{2023}) \bibinfo{pages}{1145261}.
  \DOIprefix\doi{https://doi.org/10.3389/fmats.2023.1145261}.
%Type = Book
\bibitem[{Newton(1982)}]{Newton}
\bibinfo{author}{R.~Newton}, \bibinfo{title}{Scattering Theory of Waves and
  Particles}, \bibinfo{publisher}{Springer Berlin, Heidelberg},
  \bibinfo{year}{1982}.
  \DOIprefix\doi{https://doi.org/10.1007/978-3-642-88128-2}.
%Type = Article
\bibitem[{Tennyson(2010)}]{TENNYSON201029}
\bibinfo{author}{J.~Tennyson},
\newblock \bibinfo{title}{{Electron–molecule collision calculations using the
  R-matrix method}},
\newblock \bibinfo{journal}{Phys. Rep.} \bibinfo{volume}{491}
  (\bibinfo{year}{2010}) \bibinfo{pages}{29--76}.
  \DOIprefix\doi{https://doi.org/10.1016/j.physrep.2010.02.001}.
%Type = Article
\bibitem[{Mott(1929)}]{1Mott}
\bibinfo{author}{N.~F. Mott},
\newblock \bibinfo{title}{The scattering of fast electrons by atomic nuclei},
\newblock \bibinfo{journal}{Proc. R. soc. Lond. Ser. A-Contain. Pap. Math.
  Phys. Character} \bibinfo{volume}{124} (\bibinfo{year}{1929})
  \bibinfo{pages}{425--442}.
  \DOIprefix\doi{https://doi.org/10.1098/rspa.1929.0127}.
%Type = Article
\bibitem[{Salvat et~al.(1987)Salvat, Martinez, Mayol, and
  Parellada}]{Salvat1987}
\bibinfo{author}{F.~Salvat}, \bibinfo{author}{J.~D. Martinez},
  \bibinfo{author}{R.~Mayol}, \bibinfo{author}{J.~Parellada},
\newblock \bibinfo{title}{$\rm{A}$nalytical
  $\rm{D}$irac-$\rm{H}$artree-$\rm{F}$ock-$\rm{S}$later screening function for
  atoms ($\rm{Z}$=1--92)},
\newblock \bibinfo{journal}{Phys. Rev. A} \bibinfo{volume}{36}
  (\bibinfo{year}{1987}) \bibinfo{pages}{467--474}.
  \DOIprefix\doi{https://doi.org/10.1103/PhysRevA.36.467}.
%Type = Article
\bibitem[{Onida et~al.(2002)Onida, Reining, and Rubio}]{RevModPhys.74.601}
\bibinfo{author}{G.~Onida}, \bibinfo{author}{L.~Reining},
  \bibinfo{author}{A.~Rubio},
\newblock \bibinfo{title}{Electronic excitations: density-functional versus
  many-body green's-function approaches},
\newblock \bibinfo{journal}{Rev. Mod. Phys.} \bibinfo{volume}{74}
  (\bibinfo{year}{2002}) \bibinfo{pages}{601--659}.
  \DOIprefix\doi{https://doi.org/10.1103/RevModPhys.74.601}.
%Type = Article
\bibitem[{Ashley(1988)}]{Ashley1988}
\bibinfo{author}{J.~C. Ashley},
\newblock \bibinfo{title}{Interaction of low-energy electrons with condensed
  matter: stopping powers and inelastic mean free paths from optical data},
\newblock \bibinfo{journal}{J. Electron Spectrosc. Relat. Phenom.}
  \bibinfo{volume}{46} (\bibinfo{year}{1988}) \bibinfo{pages}{199--214}.
  \DOIprefix\doi{https://doi.org/10.1016/0368-2048(88)80019-7}.
%Type = Article
\bibitem[{Ashley(1990)}]{Ashley1990}
\bibinfo{author}{J.~C. Ashley},
\newblock \bibinfo{title}{Energy loss rate and inelastic mean free path of
  low-energy electrons and positrons in condensed matter},
\newblock \bibinfo{journal}{J. Electron Spectrosc. Relat. Phenom.}
  \bibinfo{volume}{50} (\bibinfo{year}{1990}) \bibinfo{pages}{323--334}.
  \DOIprefix\doi{https://doi.org/10.1016/0368-2048(90)87075-Y}.
%Type = Article
\bibitem[{Tanuma et~al.(1993)Tanuma, Powell, and Penn}]{TPP1993}
\bibinfo{author}{S.~Tanuma}, \bibinfo{author}{C.~J. Powell},
  \bibinfo{author}{D.~R. Penn},
\newblock \bibinfo{title}{Calculation of electron inelastic mean free paths},
\newblock \bibinfo{journal}{Surf. Interface Anal.} \bibinfo{volume}{21}
  (\bibinfo{year}{1993}) \bibinfo{pages}{165--176}.
  \DOIprefix\doi{https://doi.org/110.1002/sia.740210302}.
%Type = Inproceedings
\bibitem[{Garcia-Molina et~al.(2012)Garcia-Molina, Abril, Kyriakou, and
  Emfietzoglou}]{GarciaMolina2012}
\bibinfo{author}{R.~Garcia-Molina}, \bibinfo{author}{I.~Abril},
  \bibinfo{author}{I.~Kyriakou}, \bibinfo{author}{D.~Emfietzoglou},
\newblock \bibinfo{title}{Energy loss of swift protons in liquid water: Role of
  optical data input and extension algorithms},
\newblock in: \bibinfo{booktitle}{Radiation Damage in Biomolecular Systems},
  \bibinfo{year}{2012}, pp. \bibinfo{pages}{239--261}.
  \DOIprefix\doi{https://doi.org/10.1007/978-94-007-2564-5 15}.
%Type = Article
\bibitem[{Abril et~al.(1998)Abril, Garcia-Molina, Denton, P\'erez-P\'erez, and
  Arista}]{EAbril}
\bibinfo{author}{I.~Abril}, \bibinfo{author}{R.~Garcia-Molina},
  \bibinfo{author}{C.~D. Denton}, \bibinfo{author}{F.~J. P\'erez-P\'erez},
  \bibinfo{author}{N.~R. Arista},
\newblock \bibinfo{title}{Dielectric description of wakes and stopping powers
  in solids},
\newblock \bibinfo{journal}{Phys. Rev. A} \bibinfo{volume}{58}
  (\bibinfo{year}{1998}) \bibinfo{pages}{357--366}.
  \DOIprefix\doi{https://doi.org/10.1103/PhysRevA.58.357}.
%Type = Book
\bibitem[{Egerton(2011)}]{1Egerton}
\bibinfo{author}{R.~F. Egerton}, \bibinfo{title}{Electron Energy-Loss
  Spectroscopy in the Electron Microscope}, \bibinfo{publisher}{Springer, New
  York, Dordrecht, Heidelberg, London}, \bibinfo{year}{2011}.
  \DOIprefix\doi{https://doi.org/10.1007/978-1-4419-9583-4}.
%Type = Article
\bibitem[{{Fr{\"o}hlich}(1954)}]{1Frohlich}
\bibinfo{author}{H.~{Fr{\"o}hlich}},
\newblock \bibinfo{title}{{Electrons in lattice fields}},
\newblock \bibinfo{journal}{Adv. Phys.} \bibinfo{volume}{3}
  (\bibinfo{year}{1954}) \bibinfo{pages}{325--361}.
  \DOIprefix\doi{https://doi.org/10.1080/00018735400101213}.
%Type = Article
\bibitem[{Zhou et~al.(2021)Zhou, Park, Lu, Maliyov, Tong, and
  Bernardi}]{ZHOU2021107970}
\bibinfo{author}{J.-J. Zhou}, \bibinfo{author}{J.~Park}, \bibinfo{author}{I.-T.
  Lu}, \bibinfo{author}{I.~Maliyov}, \bibinfo{author}{X.~Tong},
  \bibinfo{author}{M.~Bernardi},
\newblock \bibinfo{title}{Perturbo: A software package for ab initio
  electron–phonon interactions, charge transport and ultrafast dynamics},
\newblock \bibinfo{journal}{Comput. Phys. Commun.} \bibinfo{volume}{264}
  (\bibinfo{year}{2021}) \bibinfo{pages}{107970}.
  \DOIprefix\doi{https://doi.org/10.1016/j.cpc.2021.107970}.
%Type = Article
\bibitem[{Ganachaud and Mokrani(1995)}]{1Ganachaud}
\bibinfo{author}{J.~Ganachaud}, \bibinfo{author}{A.~Mokrani},
\newblock \bibinfo{title}{Theoretical study of the secondary electron emission
  of insulating targets},
\newblock \bibinfo{journal}{Surface Science} \bibinfo{volume}{334}
  (\bibinfo{year}{1995}) \bibinfo{pages}{329--341}.
  \DOIprefix\doi{https://doi.org/10.1016/0039-6028(95)00474-2}.
%Type = Article
\bibitem[{Franchini et~al.(2021)Franchini, Reticcioli, Setvin, and
  Diebold}]{Franchini2021}
\bibinfo{author}{C.~Franchini}, \bibinfo{author}{M.~Reticcioli},
  \bibinfo{author}{M.~Setvin}, \bibinfo{author}{U.~Diebold},
\newblock \bibinfo{title}{{Polarons in materials}},
\newblock \bibinfo{journal}{Nat. Rev. Mater.} \bibinfo{volume}{6}
  (\bibinfo{year}{2021}) \bibinfo{pages}{560--586}.
  \DOIprefix\doi{10.1038/s41578-021-00289-w}.
%Type = Article
\bibitem[{Azzolini et~al.(2020)Azzolini, Ridzel, Kaplya, Afanas’ev, Pugno,
  Taioli, and Dapor}]{AZZOLINI2020109420}
\bibinfo{author}{M.~Azzolini}, \bibinfo{author}{O.~Y. Ridzel},
  \bibinfo{author}{P.~S. Kaplya}, \bibinfo{author}{V.~Afanas’ev},
  \bibinfo{author}{N.~M. Pugno}, \bibinfo{author}{S.~Taioli},
  \bibinfo{author}{M.~Dapor},
\newblock \bibinfo{title}{A comparison between $\rm{M}$onte $\rm{C}$arlo method
  and the numerical solution of the $\rm{A}$mbartsumian-$\rm{C}$handrasekhar
  equations to unravel the dielectric response of metals},
\newblock \bibinfo{journal}{Comput. Mater. Sci.} \bibinfo{volume}{173}
  (\bibinfo{year}{2020}) \bibinfo{pages}{109420}.
  \DOIprefix\doi{https://doi.org/10.1016/j.commatsci.2019.109420}.
%Type = Book
\bibitem[{Ibach(1977)}]{ibach1977}
\bibinfo{author}{H.~Ibach}, \bibinfo{title}{Electron Spectroscopy for Surface
  Analysis}, \bibinfo{publisher}{Springer, Berlin, Heidelberg},
  \bibinfo{year}{1977}. \URLprefix
  \url{https://link.springer.com/book/10.1007/978-3-642-81099-2}.
  \DOIprefix\doi{https://doi.org/10.1007/978-3-642-81099-2}.
%Type = Article
\bibitem[{Gergely(2002)}]{1Gergely}
\bibinfo{author}{G.~Gergely},
\newblock \bibinfo{title}{Elastic backscattering of electrons: determination of
  physical parameters of electron transport processes by elastic peak electron
  spectroscopy},
\newblock \bibinfo{journal}{Progr. Surf. Sci.} \bibinfo{volume}{71}
  (\bibinfo{year}{2002}) \bibinfo{pages}{31--88}.
  \DOIprefix\doi{https://doi.org/10.1016/S0079-6816(02)00019-9}.
%Type = Article
\bibitem[{Jablonski(2003)}]{1JablonskiII}
\bibinfo{author}{A.~Jablonski},
\newblock \bibinfo{title}{Analytical applications of elastic electron
  backscattering from surfaces},
\newblock \bibinfo{journal}{Progr. Surf. Sci.} \bibinfo{volume}{74}
  (\bibinfo{year}{2003}) \bibinfo{pages}{357--374}.
  \DOIprefix\doi{https://doi.org/10.1016/j.progsurf.2003.08.028}.
%Type = Article
\bibitem[{Taioli et~al.(2009)Taioli, Simonucci, and
  Dapor}]{taioli2009surprises}
\bibinfo{author}{S.~Taioli}, \bibinfo{author}{S.~Simonucci},
  \bibinfo{author}{M.~Dapor},
\newblock \bibinfo{title}{Surprises: when ab initio meets statistics in
  extended systems},
\newblock \bibinfo{journal}{Comput. Sci. Discov.} \bibinfo{volume}{2}
  (\bibinfo{year}{2009}) \bibinfo{pages}{015002}.
  \DOIprefix\doi{https://doi.org/10.1088/1749-4699/2/1/015002}.
%Type = Article
\bibitem[{Taioli and Simonucci(2021)}]{taioli2021resonant}
\bibinfo{author}{S.~Taioli}, \bibinfo{author}{S.~Simonucci},
\newblock \bibinfo{title}{The resonant and normal auger spectra of ozone},
\newblock \bibinfo{journal}{Symmetry} \bibinfo{volume}{13}
  (\bibinfo{year}{2021}) \bibinfo{pages}{516}.
  \DOIprefix\doi{https://doi.org/10.3390/sym13030516}.
%Type = Article
\bibitem[{Colle et~al.(2004{\natexlab{a}})Colle, Embriaco, Massini, Simonucci,
  and Taioli}]{colle2004abc}
\bibinfo{author}{R.~Colle}, \bibinfo{author}{D.~Embriaco},
  \bibinfo{author}{M.~Massini}, \bibinfo{author}{S.~Simonucci},
  \bibinfo{author}{S.~Taioli},
\newblock \bibinfo{title}{Ab initio calculation of the normal $\rm{A}$uger
  spectrum of $\rm{C}_2\rm{H}_2$},
\newblock \bibinfo{journal}{J. Phys. B} \bibinfo{volume}{37}
  (\bibinfo{year}{2004}{\natexlab{a}}) \bibinfo{pages}{1237}.
  \DOIprefix\doi{https://doi.org/10.1088/0953-4075/37/6/008}.
%Type = Article
\bibitem[{Colle et~al.(2004{\natexlab{b}})Colle, Embriaco, Massini, Simonucci,
  and Taioli}]{colle2004auger}
\bibinfo{author}{R.~Colle}, \bibinfo{author}{D.~Embriaco},
  \bibinfo{author}{M.~Massini}, \bibinfo{author}{S.~Simonucci},
  \bibinfo{author}{S.~Taioli},
\newblock \bibinfo{title}{Auger-electron angular distributions calculated
  without the two-step approximation: Calculation of angle-resolved resonant
  $\rm{A}$uger spectra of $\rm{C}_2\rm{H}_2$},
\newblock \bibinfo{journal}{Phys. Rev. A} \bibinfo{volume}{70}
  (\bibinfo{year}{2004}{\natexlab{b}}) \bibinfo{pages}{042708}.
  \DOIprefix\doi{https://doi.org/10.1103/PhysRevA.70.042708}.
%Type = Article
\bibitem[{Colle et~al.(2004{\natexlab{c}})Colle, Embriaco, Massini, Simonucci,
  and Taioli}]{colle2004ab}
\bibinfo{author}{R.~Colle}, \bibinfo{author}{D.~Embriaco},
  \bibinfo{author}{M.~Massini}, \bibinfo{author}{S.~Simonucci},
  \bibinfo{author}{S.~Taioli},
\newblock \bibinfo{title}{Ab initio calculation of the c1s photoelectron
  spectrum of $\rm{C}_2\rm{H}_2$},
\newblock \bibinfo{journal}{Nucl. Instrum. Methods Phys. Res. B}
  \bibinfo{volume}{213} (\bibinfo{year}{2004}{\natexlab{c}})
  \bibinfo{pages}{65--70}.
  \DOIprefix\doi{https://doi.org/10.1016/S0168-583X(03)01535-0}.
%Type = Article
\bibitem[{Dapor et~al.(2010)Dapor, Ciappa, and Fichtner}]{1DaporetalSE2010}
\bibinfo{author}{M.~Dapor}, \bibinfo{author}{M.~Ciappa},
  \bibinfo{author}{W.~Fichtner},
\newblock \bibinfo{title}{Monte carlo modeling in the low-energy domain of the
  secondary electron emission of polymethylmethacrylate for critical-dimension
  scanning electron microscopy},
\newblock \bibinfo{journal}{Journal of Micro/Nanolithography, MEMS, and MOEMS}
  \bibinfo{volume}{9} (\bibinfo{year}{2010}) \bibinfo{pages}{023001}.
  \DOIprefix\doi{https://doi.org/10.1117/1.3373517}.
%Type = Article
\bibitem[{Garcia-Molina et~al.(2006)Garcia-Molina, Abril, Denton, and
  Heredia-Avalos}]{garciamolina2006}
\bibinfo{author}{R.~Garcia-Molina}, \bibinfo{author}{I.~Abril},
  \bibinfo{author}{C.~D. Denton}, \bibinfo{author}{S.~Heredia-Avalos},
\newblock \bibinfo{title}{Allotropic effects on the energy loss of swift
  $\rm{H}$+ and $\rm{H}$e+ ion beams through thin foils},
\newblock \bibinfo{journal}{Nucl. Instrum. Methods Phys. Res. B}
  \bibinfo{volume}{249} (\bibinfo{year}{2006}) \bibinfo{pages}{6--12}.
  \DOIprefix\doi{https://doi.org/10.1016/j.nimb.2006.03.011}.
%Type = Article
\bibitem[{Everhart(1960)}]{6Everhart}
\bibinfo{author}{T.~E. Everhart},
\newblock \bibinfo{title}{{Simple Theory Concerning the Reflection of Electrons
  from Solids}},
\newblock \bibinfo{journal}{J. Appl. Phys.} \bibinfo{volume}{31}
  (\bibinfo{year}{1960}) \bibinfo{pages}{1483--1490}.
  \DOIprefix\doi{https://doi.org/10.1063/1.1735868}.
%Type = Article
\bibitem[{Archard(2004)}]{Archard}
\bibinfo{author}{G.~D. Archard},
\newblock \bibinfo{title}{{Back Scattering of Electrons}},
\newblock \bibinfo{journal}{J. Appl. Phys.} \bibinfo{volume}{32}
  (\bibinfo{year}{2004}) \bibinfo{pages}{1505--1509}.
  \DOIprefix\doi{https://doi.org/10.1063/1.1728385}.
%Type = Article
\bibitem[{Dapor(1992)}]{PhysRevB.46.618}
\bibinfo{author}{M.~Dapor},
\newblock \bibinfo{title}{Monte carlo simulation of backscattered electrons and
  energy from thick targets and surface films},
\newblock \bibinfo{journal}{Phys. Rev. B} \bibinfo{volume}{46}
  (\bibinfo{year}{1992}) \bibinfo{pages}{618--625}. \URLprefix
  \url{https://link.aps.org/doi/10.1103/PhysRevB.46.618}.
  \DOIprefix\doi{10.1103/PhysRevB.46.618}.
%Type = Article
\bibitem[{Dapor et~al.(2011)Dapor, Bazzanella, Toniutti, Miotello, and
  Gialanella}]{1DaporetalBSE2011}
\bibinfo{author}{M.~Dapor}, \bibinfo{author}{N.~Bazzanella},
  \bibinfo{author}{L.~Toniutti}, \bibinfo{author}{A.~Miotello},
  \bibinfo{author}{S.~Gialanella},
\newblock \bibinfo{title}{Backscattered electrons from surface films deposited
  on bulk targets: A comparison between computational and experimental
  results},
\newblock \bibinfo{journal}{Nucl. Instrum. Methods Phys. Res. B}
  \bibinfo{volume}{269} (\bibinfo{year}{2011}) \bibinfo{pages}{1672--1674}.
  \DOIprefix\doi{https://doi.org/10.1016/j.nimb.2010.11.016}.
%Type = Book
\bibitem[{Thomson(1906)}]{6Thomson}
\bibinfo{author}{J.~J. Thomson}, \bibinfo{title}{Conduction of Electricity
  Through Gases}, \bibinfo{edition}{2$^{nd}$} ed.,
  \bibinfo{publisher}{Cambridge University Press, Cambridge, England},
  \bibinfo{year}{1906}.
%Type = Article
\bibitem[{Whiddington(1914)}]{Whiddington}
\bibinfo{author}{R.~Whiddington},
\newblock \bibinfo{title}{The transmission of cathode rays through matter},
\newblock \bibinfo{journal}{Proc. R. Soc. Lond.} \bibinfo{volume}{89}
  (\bibinfo{year}{1914}) \bibinfo{pages}{554--560}.
  \DOIprefix\doi{https://doi.org/10.1098/rspa.1912.0028}.
%Type = Article
\bibitem[{Niedrig(1982)}]{Niedrig2020AnalyticalMI}
\bibinfo{author}{H.~Niedrig},
\newblock \bibinfo{title}{Analytical models in electron backscattering},
\newblock \bibinfo{journal}{Scanning Electron Microscopy}
  \bibinfo{volume}{1982} (\bibinfo{year}{1982}) \bibinfo{pages}{51--68}.
  \URLprefix \url{https://digitalcommons.usu.edu/electron/vol1982/iss1/5}.
%Type = Article
\bibitem[{Neubert and Rogaschewski(1980)}]{6NeubertRogaschewski}
\bibinfo{author}{G.~Neubert}, \bibinfo{author}{S.~Rogaschewski},
\newblock \bibinfo{title}{Backscattering coefficient measurements of 15 to 60
  kev electrons for solids at various angles of incidence},
\newblock \bibinfo{journal}{Phys. Status Solidi A} \bibinfo{volume}{59}
  (\bibinfo{year}{1980}) \bibinfo{pages}{35--41}.
  \DOIprefix\doi{https://doi.org/10.1515/9783112501726-005}.
%Type = Article
\bibitem[{Vicanek and Urbassek(1991)}]{6VicanekUrbassek1991}
\bibinfo{author}{M.~Vicanek}, \bibinfo{author}{H.~M. Urbassek},
\newblock \bibinfo{title}{Reflection coefficient of low-energy light ions},
\newblock \bibinfo{journal}{Phys. Rev. B} \bibinfo{volume}{44}
  (\bibinfo{year}{1991}) \bibinfo{pages}{7234--7242}.
  \DOIprefix\doi{https://doi.org/10.1103/PhysRevB.44.7234}.
%Type = Article
\bibitem[{Coleman et~al.(1992)Coleman, Albrecht, Jensen, and
  Walker}]{6Colemanetal1992}
\bibinfo{author}{P.~G. Coleman}, \bibinfo{author}{L.~Albrecht},
  \bibinfo{author}{K.~O. Jensen}, \bibinfo{author}{A.~B. Walker},
\newblock \bibinfo{title}{Positron backscattering from elemental solids},
\newblock \bibinfo{journal}{J. Phys. Condens. Matter} \bibinfo{volume}{4}
  (\bibinfo{year}{1992}) \bibinfo{pages}{10311--10322}.
  \DOIprefix\doi{https://doi.org/10.1088/0953-8984/4/50/018}.
%Type = Article
\bibitem[{Dapor(1996)}]{6DaporJAP1996}
\bibinfo{author}{M.~Dapor},
\newblock \bibinfo{title}{{Elastic scattering calculations for electrons and
  positrons in solid targets}},
\newblock \bibinfo{journal}{J. Appl. Phys.} \bibinfo{volume}{79}
  (\bibinfo{year}{1996}) \bibinfo{pages}{8406--8411}.
  \DOIprefix\doi{https://doi.org/10.1063/1.362514}.
%Type = Article
\bibitem[{Wolff(1954{\natexlab{a}})}]{Wolff}
\bibinfo{author}{P.~A. Wolff},
\newblock \bibinfo{title}{Theory of secondary electron cascade in metals},
\newblock \bibinfo{journal}{Phys. Rev.} \bibinfo{volume}{95}
  (\bibinfo{year}{1954}{\natexlab{a}}) \bibinfo{pages}{56--66}.
  \DOIprefix\doi{https://doi.org/10.1103/PhysRev.95.56}.
%Type = Article
\bibitem[{Wolff(1954{\natexlab{b}})}]{PhysRev.95.1415}
\bibinfo{author}{P.~A. Wolff},
\newblock \bibinfo{title}{Theory of electron multiplication in silicon and
  germanium},
\newblock \bibinfo{journal}{Phys. Rev.} \bibinfo{volume}{95}
  (\bibinfo{year}{1954}{\natexlab{b}}) \bibinfo{pages}{1415--1420}.
  \DOIprefix\doi{https://doi.org/10.1103/PhysRev.95.1415}.
%Type = Article
\bibitem[{Amelio(1970)}]{10.1116/1.1315884}
\bibinfo{author}{G.~F. Amelio},
\newblock \bibinfo{title}{{Theory for the Energy Distribution of Secondary
  Electrons}},
\newblock \bibinfo{journal}{J. Vac. Sci. Technol.} \bibinfo{volume}{7}
  (\bibinfo{year}{1970}) \bibinfo{pages}{593--604}.
  \DOIprefix\doi{https://doi.org/10.1116/1.1315884}.
%Type = Article
\bibitem[{Streitwolf(1959)}]{https://doi.org/10.1002/andp.19594580308}
\bibinfo{author}{H.~Streitwolf},
\newblock \bibinfo{title}{On the theory of secondary electron emission from
  metals the excitation process},
\newblock \bibinfo{journal}{Ann. Phys.} \bibinfo{volume}{458}
  (\bibinfo{year}{1959}) \bibinfo{pages}{183--196}.
  \DOIprefix\doi{https://doi.org/10.1002/andp.19594580308}.
%Type = Article
\bibitem[{Chung and Everhart(1974)}]{10.1063/1.1663306}
\bibinfo{author}{M.~S. Chung}, \bibinfo{author}{T.~E. Everhart},
\newblock \bibinfo{title}{{Simple calculation of energy distribution of
  low‐energy secondary electrons emitted from metals under electron
  bombardment}},
\newblock \bibinfo{journal}{J. Appl. Phys.} \bibinfo{volume}{45}
  (\bibinfo{year}{1974}) \bibinfo{pages}{707--709}.
  \DOIprefix\doi{https://doi.org/10.1063/1.1663306}.
%Type = Article
\bibitem[{Ceperley and Alder(1986)}]{Ceperley}
\bibinfo{author}{D.~Ceperley}, \bibinfo{author}{B.~Alder},
\newblock \bibinfo{title}{Quantum $\rm{M}$onte $\rm{C}$arlo},
\newblock \bibinfo{journal}{Science} \bibinfo{volume}{231}
  (\bibinfo{year}{1986}) \bibinfo{pages}{555--560}.
  \DOIprefix\doi{https://doi.org/10.1126/science.231.4738.555}.
%Type = Book
\bibitem[{Becca and Sorella(2017)}]{becca2017quantum}
\bibinfo{author}{F.~Becca}, \bibinfo{author}{S.~Sorella},
  \bibinfo{title}{Quantum Monte Carlo approaches for correlated systems},
  \bibinfo{publisher}{Cambridge University Press}, \bibinfo{year}{2017}.
  \DOIprefix\doi{https://doi.org/10.1017/9781316417041}.
%Type = Article
\bibitem[{Joy and Luo(1989)}]{joyluo}
\bibinfo{author}{D.~C. Joy}, \bibinfo{author}{S.~Luo},
\newblock \bibinfo{title}{An empirical stopping power relationship for
  low-energy electrons},
\newblock \bibinfo{journal}{Scanning} \bibinfo{volume}{11}
  (\bibinfo{year}{1989}) \bibinfo{pages}{176--180}. \URLprefix
  \url{https://onlinelibrary.wiley.com/doi/abs/10.1002/sca.4950110404}.
  \DOIprefix\doi{https://doi.org/10.1002/sca.4950110404}.
  \href{http://arxiv.org/abs/https://onlinelibrary.wiley.com/doi/pdf/10.1002/sca.4950110404}{{\tt
  arXiv:https://onlinelibrary.wiley.com/doi/pdf/10.1002/sca.4950110404}}.
%Type = Article
\bibitem[{Tung et~al.(1979)Tung, Ashley, and Ritchie}]{TUNG1979427}
\bibinfo{author}{C.~Tung}, \bibinfo{author}{J.~Ashley},
  \bibinfo{author}{R.~Ritchie},
\newblock \bibinfo{title}{Electron inelastic mean free paths and energy losses
  in solids ii: Electron gas statistical model},
\newblock \bibinfo{journal}{Surface Science} \bibinfo{volume}{81}
  (\bibinfo{year}{1979}) \bibinfo{pages}{427--439}. \URLprefix
  \url{https://www.sciencedirect.com/science/article/pii/0039602879901109}.
  \DOIprefix\doi{https://doi.org/10.1016/0039-6028(79)90110-9}.
%Type = Article
\bibitem[{Yubero and Tougaard(2 II)}]{yuberotougaard}
\bibinfo{author}{F.~Yubero}, \bibinfo{author}{S.~Tougaard},
\newblock \bibinfo{title}{Model for quantitative analysis of
  reflection-electron-energy-loss spectra},
\newblock \bibinfo{journal}{Phys. Rev. B} \bibinfo{volume}{46}
  (\bibinfo{year}{1992-II}) \bibinfo{pages}{2489--2497}.
  \DOIprefix\doi{https://doi.org/10.1103/PhysRevB.46.2486}.
%Type = Article
\bibitem[{Giannozzi et~al.(2009)}]{qe}
\bibinfo{author}{P.~Giannozzi}, et~al.,
\newblock \bibinfo{title}{$\rm{QUANTUM}$ $\rm{ESPRESSO}$: a modular and
  open-source software project for quantum simulations of materials},
\newblock \bibinfo{journal}{J. Phys. Condens. Matter} \bibinfo{volume}{21}
  (\bibinfo{year}{2009}) \bibinfo{pages}{395502}.
  \DOIprefix\doi{https://doi.org/10.1088/0953-8984/21/39/395502}.
%Type = Misc
\bibitem[{The Elk Code(2024)}]{elk}
The Elk Code, \bibinfo{year}{2024}.
  \bibinfo{note}{\url{http://elk.sourceforge.net/}}.
%Type = Inbook
\bibitem[{Giannozzi and Baroni(2005)}]{baroni_book}
\bibinfo{author}{P.~Giannozzi}, \bibinfo{author}{S.~Baroni},
  \bibinfo{title}{Density-Functional Perturbation Theory},
  \bibinfo{publisher}{Springer Science+Business Media B.V., 2005, p. 195},
  \bibinfo{year}{2005}, pp. \bibinfo{pages}{195--214}.
  \DOIprefix\doi{https://doi.org/110.1007/978-1-4020-3286-8 11}.
%Type = Article
\bibitem[{Egerton(2017)}]{EGERTON2017115}
\bibinfo{author}{R.~Egerton},
\newblock \bibinfo{title}{Scattering delocalization and radiation damage in
  stem-eels},
\newblock \bibinfo{journal}{Ultramicroscopy} \bibinfo{volume}{180}
  (\bibinfo{year}{2017}) \bibinfo{pages}{115--124}. \URLprefix
  \url{https://www.sciencedirect.com/science/article/pii/S0304399117300815}.
  \DOIprefix\doi{https://doi.org/10.1016/j.ultramic.2017.02.007},
  \bibinfo{note}{ondrej Krivanek: A research life in EELS and aberration
  corrected STEM}.
%Type = Article
\bibitem[{Khan et~al.(2023)Khan, Mao, Zou, Lu, Da, Li, and
  Ding}]{KHAN2023112257}
\bibinfo{author}{M.~Khan}, \bibinfo{author}{S.~Mao}, \bibinfo{author}{Y.~Zou},
  \bibinfo{author}{D.~Lu}, \bibinfo{author}{B.~Da}, \bibinfo{author}{Y.~Li},
  \bibinfo{author}{Z.~Ding},
\newblock \bibinfo{title}{An extensive theoretical quantification of secondary
  electron emission from silicon},
\newblock \bibinfo{journal}{Vacuum} \bibinfo{volume}{215}
  (\bibinfo{year}{2023}) \bibinfo{pages}{112257}. \URLprefix
  \url{https://www.sciencedirect.com/science/article/pii/S0042207X23004542}.
  \DOIprefix\doi{https://doi.org/10.1016/j.vacuum.2023.112257}.
%Type = Book
\bibitem[{Kessler(1985)}]{3KesslerBook}
\bibinfo{author}{J.~Kessler}, \bibinfo{title}{Polarized Electrons},
  \bibinfo{publisher}{Springer-Verlag, Berlin}, \bibinfo{year}{1985}.
  \DOIprefix\doi{https://doi.org/10.1007/978-3-662-02434-8}.
%Type = Book
\bibitem[{Burke and Joachain(1995)}]{3BurkeJoachainBook}
\bibinfo{author}{P.~G. Burke}, \bibinfo{author}{C.~J. Joachain},
  \bibinfo{title}{Theory of Electron-Atom Collisions},
  \bibinfo{publisher}{Plenum Press, New York}, \bibinfo{year}{1995}.
  \DOIprefix\doi{https://doi.org/10.1007/978-1-4899-1567-2}.
%Type = Article
\bibitem[{Salvat et~al.(2005)Salvat, Jablonski, and Powell}]{3ELSEPA}
\bibinfo{author}{F.~Salvat}, \bibinfo{author}{A.~Jablonski},
  \bibinfo{author}{C.~J. Powell},
\newblock \bibinfo{title}{$\rm{ELSEPA}$—$\rm{D}$irac partial-wave calculation
  of elastic scattering of electrons and positrons by atoms, positive ions and
  molecules},
\newblock \bibinfo{journal}{Comp. Phys. Comm.} \bibinfo{volume}{165}
  (\bibinfo{year}{2005}) \bibinfo{pages}{157--190}.
  \DOIprefix\doi{https://doi.org/10.1016/j.cpc.2004.09.006}.
%Type = Book
\bibitem[{Dapor(2022)}]{Dapor2022}
\bibinfo{author}{M.~Dapor}, \bibinfo{title}{Electron-Atom Collisions.
  Quantum-Relativistic Theory and Exercises}, \bibinfo{publisher}{De Gruyter,
  Berlin, Boston}, \bibinfo{year}{2022}.
  \DOIprefix\doi{https://doi.org/10.1515/9783110675375}.
%Type = Article
\bibitem[{Riley et~al.(1975)Riley, MacCallum, and Biggs}]{3Rileyetal}
\bibinfo{author}{M.~E. Riley}, \bibinfo{author}{C.~J. MacCallum},
  \bibinfo{author}{F.~Biggs},
\newblock \bibinfo{title}{Theoretical electron-atom elastic scattering cross
  sections: Selected elements, 1 ke$\rm{V}$ to 256 ke$\rm{V}$},
\newblock \bibinfo{journal}{At. Data Nucl. Data Tables} \bibinfo{volume}{15}
  (\bibinfo{year}{1975}) \bibinfo{pages}{443--476}.
  \DOIprefix\doi{https://doi.org/10.1016/0092-640X(75)90012-1}.
%Type = Book
\bibitem[{Sigmund(2006)}]{3Sigmund}
\bibinfo{author}{P.~Sigmund}, \bibinfo{title}{Particle Penetration and
  Radiation Effects}, \bibinfo{publisher}{Springer-Verlag, Berlin},
  \bibinfo{year}{2006}. \DOIprefix\doi{https://doi.org/10.1007/3-540-31718-X}.
%Type = Article
\bibitem[{Egerton(2009)}]{3EgertonII}
\bibinfo{author}{R.~F. Egerton},
\newblock \bibinfo{title}{Electron energy-loss spectroscopy in the $\rm{TEM}$},
\newblock \bibinfo{journal}{Rep. Prog. Phys.} \bibinfo{volume}{72}
  (\bibinfo{year}{2009}) \bibinfo{pages}{016502}.
  \DOIprefix\doi{https://doi.org/110.1088/0034-4885/72/1/016502}.
%Type = Article
\bibitem[{Jablonski et~al.(2004)Jablonski, Salvat, and Powell}]{3Jablonski}
\bibinfo{author}{A.~Jablonski}, \bibinfo{author}{F.~Salvat},
  \bibinfo{author}{C.~J. Powell},
\newblock \bibinfo{title}{Comparison of electron elastic-scattering cross
  sections calculated from two commonly used atomic potentials},
\newblock \bibinfo{journal}{J. Phys. Chem. Data} \bibinfo{volume}{33}
  (\bibinfo{year}{2004}) \bibinfo{pages}{409--451}.
  \DOIprefix\doi{https://doi.org/10.1063/1.159565}.
%Type = Article
\bibitem[{Mayol and Salvat(1997)}]{MayolSalvat1997}
\bibinfo{author}{R.~Mayol}, \bibinfo{author}{M.~Salvat},
\newblock \bibinfo{title}{Total and transport cross sections for elastic
  scattering of electrons by atoms},
\newblock \bibinfo{journal}{At. Data Nucl. Data Tables} \bibinfo{volume}{65}
  (\bibinfo{year}{1997}) \bibinfo{pages}{55--154}.
  \DOIprefix\doi{https://doi.org/10.1006/adnd.1997.0734}.
%Type = Article
\bibitem[{Taioli and Simonucci(2021)}]{taioli2021relativistic}
\bibinfo{author}{S.~Taioli}, \bibinfo{author}{S.~Simonucci},
\newblock \bibinfo{title}{Relativistic quantum theory and algorithms: a toolbox
  for modeling many-fermion systems in different scenarios},
\newblock \bibinfo{journal}{Annual Reports in Computational Chemistry 17}
  (\bibinfo{year}{2021}) \bibinfo{pages}{55--111}.
  \DOIprefix\doi{https://doi.org/10.1016/bs.arcc.2021.08.003}.
%Type = Article
\bibitem[{Colle et~al.(1988)Colle, Fortunelli, and
  Simonucci}]{Colle1988HermiteGF}
\bibinfo{author}{R.~Colle}, \bibinfo{author}{A.~Fortunelli},
  \bibinfo{author}{S.~Simonucci},
\newblock \bibinfo{title}{Hermite gaussian functions modulated by plane waves:
  a general basis set for bound and continuum states},
\newblock \bibinfo{journal}{Il Nuovo Cimento D} \bibinfo{volume}{10}
  (\bibinfo{year}{1988}) \bibinfo{pages}{805--818}.
  \DOIprefix\doi{https://doi.org/10.1007/BF02450141}.
%Type = Article
\bibitem[{Bethe(1930)}]{3Bethe}
\bibinfo{author}{H.~Bethe},
\newblock \bibinfo{title}{Zur theorie des durchgangs schneller
  korpuskularstrahlen durch materie},
\newblock \bibinfo{journal}{Ann. Phys. (Leipzig)} \bibinfo{volume}{397}
  (\bibinfo{year}{1930}) \bibinfo{pages}{325--400}.
  \DOIprefix\doi{https://doi.org/10.1002/andp.19303970303}.
%Type = Article
\bibitem[{Lane and Zaffarano(1954)}]{3Lane}
\bibinfo{author}{R.~O. Lane}, \bibinfo{author}{D.~J. Zaffarano},
\newblock \bibinfo{title}{Transmission of 0-40 kev electrons by thin films with
  application to beta-ray spectroscopy},
\newblock \bibinfo{journal}{Phys. Rev.} \bibinfo{volume}{94}
  (\bibinfo{year}{1954}) \bibinfo{pages}{960--964}.
  \DOIprefix\doi{https://doi.org/10.1103/PhysRev.94.960}.
%Type = Article
\bibitem[{Kanaya and Okayama(1972)}]{3Kanaya}
\bibinfo{author}{K.~Kanaya}, \bibinfo{author}{S.~Okayama},
\newblock \bibinfo{title}{Penetration and energy-loss theory of electrons in
  solid targets},
\newblock \bibinfo{journal}{J. Phys. D: Appl. Phys.} \bibinfo{volume}{5}
  (\bibinfo{year}{1972}) \bibinfo{pages}{43--58}.
  \DOIprefix\doi{10.1088/0022-3727/5/1/308}.
%Type = Book
\bibitem[{Bromley and Greiner(2013)}]{Greiner}
\bibinfo{author}{D.~Bromley}, \bibinfo{author}{W.~Greiner},
  \bibinfo{title}{Relativistic Quantum Mechanics. Wave Equations},
  \bibinfo{publisher}{Springer Berlin Heidelberg}, \bibinfo{year}{2013}.
  \DOIprefix\doi{https://doi.org/10.1007/978-3-662-04275-5}.
%Type = Article
\bibitem[{Slater(1951)}]{Slater1951}
\bibinfo{author}{J.~C. Slater},
\newblock \bibinfo{title}{A simplification of the $\rm{H}$artree-$\rm{F}$ock
  method},
\newblock \bibinfo{journal}{Phys. Rev.} \bibinfo{volume}{81}
  (\bibinfo{year}{1951}) \bibinfo{pages}{385--390}.
  \DOIprefix\doi{https://doi.org/10.1103/PhysRev.81.385}.
%Type = Article
\bibitem[{Schwierz et~al.(2007)Schwierz, Wiedenhover, and Volya}]{Woodsaxon}
\bibinfo{author}{N.~Schwierz}, \bibinfo{author}{I.~Wiedenhover},
  \bibinfo{author}{A.~Volya},
\newblock \bibinfo{title}{Parameterization of the woods-saxon potential for
  shell-model calculations},
\newblock \bibinfo{journal}{arXiv:0709.3525 [nucl-th]}  (\bibinfo{year}{2007}).
  \DOIprefix\doi{https://doi.org/10.48550/arXiv.0709.3525}.
%Type = Article
\bibitem[{Morresi et~al.(2018)Morresi, Taioli, and
  Simonucci}]{morresi2018nuclear}
\bibinfo{author}{T.~Morresi}, \bibinfo{author}{S.~Taioli},
  \bibinfo{author}{S.~Simonucci},
\newblock \bibinfo{title}{Nuclear beta decay: Relativistic theory and ab initio
  simulations of electroweak decay spectra in medium-heavy nuclei and of atomic
  and molecular electronic structure},
\newblock \bibinfo{journal}{Adv. Theory Simul.} \bibinfo{volume}{1}
  (\bibinfo{year}{2018}) \bibinfo{pages}{1870030}.
  \DOIprefix\doi{https://doi.org/10.1002/adts.201870030}.
%Type = Book
\bibitem[{Bohm(1951)}]{3Bohm}
\bibinfo{author}{D.~Bohm}, \bibinfo{title}{Quantum Theory}, Dover Books on
  Physics Series, \bibinfo{publisher}{Dover Publications},
  \bibinfo{year}{1951}. \URLprefix
  \url{https://store.doverpublications.com/products/9780486659695}.
%Type = Article
\bibitem[{Dapor et~al.(2018)Dapor, Masters, Ross, Lidzey, Pearson, Abril,
  Garcia-Molina, Sharp, Unčovský, Vystavel, Mika, and
  Rodenburg}]{DAPOR201895}
\bibinfo{author}{M.~Dapor}, \bibinfo{author}{R.~C. Masters},
  \bibinfo{author}{I.~Ross}, \bibinfo{author}{D.~G. Lidzey},
  \bibinfo{author}{A.~Pearson}, \bibinfo{author}{I.~Abril},
  \bibinfo{author}{R.~Garcia-Molina}, \bibinfo{author}{J.~Sharp},
  \bibinfo{author}{M.~Unčovský}, \bibinfo{author}{T.~Vystavel},
  \bibinfo{author}{F.~Mika}, \bibinfo{author}{C.~Rodenburg},
\newblock \bibinfo{title}{{Secondary electron spectra of semi-crystalline
  polymers – A novel polymer characterisation tool?}},
\newblock \bibinfo{journal}{J. Electron Spectrosc. Relat. Phenom.}
  \bibinfo{volume}{222} (\bibinfo{year}{2018}) \bibinfo{pages}{95--105}.
  \DOIprefix\doi{https://doi.org/10.1016/j.elspec.2017.08.001}.
%Type = Article
\bibitem[{{McEachran} and {Elford}(2003)}]{McEachran2003}
\bibinfo{author}{R.~P. {McEachran}}, \bibinfo{author}{M.~T. {Elford}},
\newblock \bibinfo{title}{{The momentum transfer cross section and transport
  coefficients for low energy electrons in mercury}},
\newblock \bibinfo{journal}{J. Phys. B At. Mol. Opt. Phys.}
  \bibinfo{volume}{36} (\bibinfo{year}{2003}) \bibinfo{pages}{427--441}.
  \DOIprefix\doi{https://doi.org/10.1088/0953-4075/36/3/303}.
%Type = Article
\bibitem[{Palmerini et~al.(2016)Palmerini, Busso, Simonucci, and
  Taioli}]{palmerini2016lithium}
\bibinfo{author}{S.~Palmerini}, \bibinfo{author}{M.~Busso},
  \bibinfo{author}{S.~Simonucci}, \bibinfo{author}{S.~Taioli},
\newblock \bibinfo{title}{Lithium abundances in $\rm{AGB}$ stars and a new
  estimate for the $^7$$\rm{B}$e life-time},
\newblock \bibinfo{journal}{J. Phys. Conf. Ser.} \bibinfo{volume}{665}
  (\bibinfo{year}{2016}) \bibinfo{pages}{012014}.
  \DOIprefix\doi{https://doi.org/10.1088/1742-6596/665/1/012014}.
%Type = Article
\bibitem[{Vescovi et~al.(2019)Vescovi, Piersanti, Cristallo, Busso, Vissani,
  Palmerini, Simonucci, and Taioli}]{vescovi2019effects}
\bibinfo{author}{D.~Vescovi}, \bibinfo{author}{L.~Piersanti},
  \bibinfo{author}{S.~Cristallo}, \bibinfo{author}{M.~Busso},
  \bibinfo{author}{F.~Vissani}, \bibinfo{author}{S.~Palmerini},
  \bibinfo{author}{S.~Simonucci}, \bibinfo{author}{S.~Taioli},
\newblock \bibinfo{title}{The effects of a revised $^7\rm{B}$e e$^-$-capture
  rate on solar neutrino fluxes},
\newblock \bibinfo{journal}{Astron. Astrophys.} \bibinfo{volume}{623}
  (\bibinfo{year}{2019}) \bibinfo{pages}{A126}.
  \DOIprefix\doi{https://doi.org/10.1051/0004-6361/201834993}.
%Type = Article
\bibitem[{Taioli et~al.(2022)Taioli, Vescovi, Busso, Palmerini, Cristallo,
  Mengoni, and Simonucci}]{taioli2022theoretical}
\bibinfo{author}{S.~Taioli}, \bibinfo{author}{D.~Vescovi},
  \bibinfo{author}{M.~Busso}, \bibinfo{author}{S.~Palmerini},
  \bibinfo{author}{S.~Cristallo}, \bibinfo{author}{A.~Mengoni},
  \bibinfo{author}{S.~Simonucci},
\newblock \bibinfo{title}{Theoretical estimate of the half-life for the
  radioactive $^{134}\rm{C}$s and $^{135}\rm{C}$s in astrophysical scenarios},
\newblock \bibinfo{journal}{Astrophys. J.} \bibinfo{volume}{933}
  (\bibinfo{year}{2022}) \bibinfo{pages}{158}.
  \DOIprefix\doi{https://doi.org/10.3847/1538-4357/ac74b3}.
%Type = Article
\bibitem[{Palmerini et~al.(2021)Palmerini, Busso, Vescovi, Naselli, Pidatella,
  Mucciola, Cristallo, Mascali, Mengoni, Simonucci
  et~al.}]{palmerini2021presolar}
\bibinfo{author}{S.~Palmerini}, \bibinfo{author}{M.~Busso},
  \bibinfo{author}{D.~Vescovi}, \bibinfo{author}{E.~Naselli},
  \bibinfo{author}{A.~Pidatella}, \bibinfo{author}{R.~Mucciola},
  \bibinfo{author}{S.~Cristallo}, \bibinfo{author}{D.~Mascali},
  \bibinfo{author}{A.~Mengoni}, \bibinfo{author}{S.~Simonucci}, et~al.,
\newblock \bibinfo{title}{Presolar grain isotopic ratios as constraints to
  nuclear and stellar parameters of asymptotic giant branch star
  nucleosynthesis},
\newblock \bibinfo{journal}{Astrophys. J.} \bibinfo{volume}{921}
  (\bibinfo{year}{2021}) \bibinfo{pages}{7}.
  \DOIprefix\doi{https://doi.org/10.3847/1538-4357/ac1786}.
%Type = Article
\bibitem[{Mascali et~al.(2022)Mascali, Santonocito, Amaducci, And{\`o},
  Antonuccio, Biri, Bonanno, Bonanno, Briefi, Busso et~al.}]{mascali2022novel}
\bibinfo{author}{D.~Mascali}, \bibinfo{author}{D.~Santonocito},
  \bibinfo{author}{S.~Amaducci}, \bibinfo{author}{L.~And{\`o}},
  \bibinfo{author}{V.~Antonuccio}, \bibinfo{author}{S.~Biri},
  \bibinfo{author}{A.~Bonanno}, \bibinfo{author}{V.~P. Bonanno},
  \bibinfo{author}{S.~Briefi}, \bibinfo{author}{M.~Busso}, et~al.,
\newblock \bibinfo{title}{A novel approach to $\beta$-decay: Pandora, a new
  experimental setup for future in-plasma measurements},
\newblock \bibinfo{journal}{Universe} \bibinfo{volume}{8}
  (\bibinfo{year}{2022}) \bibinfo{pages}{80}.
  \DOIprefix\doi{https://doi.org/10.3390/universe8020080}.
%Type = Article
\bibitem[{{Mascali, D.} et~al.(2023){Mascali, D.}, {Santonocito, D.}, {Busso,
  M.}, {Celona, L.}, {Galatà, A.}, {La Cognata, M.}, {Mauro, G. S.}, {Mengoni,
  A.}, {Naselli, E.}, {Odorici, F.}, {Palmerini, S.}, {Pidatella, A.}, {Ràcz,
  R.}, {Taioli, S.}, and {Torrisi, G.}}]{mascali2023new}
\bibinfo{author}{{Mascali, D.}}, \bibinfo{author}{{Santonocito, D.}},
  \bibinfo{author}{{Busso, M.}}, \bibinfo{author}{{Celona, L.}},
  \bibinfo{author}{{Galatà, A.}}, \bibinfo{author}{{La Cognata, M.}},
  \bibinfo{author}{{Mauro, G. S.}}, \bibinfo{author}{{Mengoni, A.}},
  \bibinfo{author}{{Naselli, E.}}, \bibinfo{author}{{Odorici, F.}},
  \bibinfo{author}{{Palmerini, S.}}, \bibinfo{author}{{Pidatella, A.}},
  \bibinfo{author}{{Ràcz, R.}}, \bibinfo{author}{{Taioli, S.}},
  \bibinfo{author}{{Torrisi, G.}},
\newblock \bibinfo{title}{A new approach to $\beta$-decays studies impacting
  nuclear physics and astrophysics: The pandora setup},
\newblock \bibinfo{journal}{EPJ Web Conf.} \bibinfo{volume}{279}
  (\bibinfo{year}{2023}) \bibinfo{pages}{06007}.
  \DOIprefix\doi{https://doi.org/10.1051/epjconf/202327906007}.
%Type = Article
\bibitem[{Palmerini et~al.(2023)Palmerini, Busso, Vescovi, Cristallo, Mengoni,
  Simonucci, and Taioli}]{palmerini2023presolar}
\bibinfo{author}{S.~Palmerini}, \bibinfo{author}{M.~Busso},
  \bibinfo{author}{D.~Vescovi}, \bibinfo{author}{S.~Cristallo},
  \bibinfo{author}{A.~Mengoni}, \bibinfo{author}{S.~Simonucci},
  \bibinfo{author}{S.~Taioli},
\newblock \bibinfo{title}{Presolar grain isotopic ratios as constraints to
  nuclear physics inputs for s-process calculations},
\newblock \bibinfo{journal}{EPJ Web of Conferences} \bibinfo{volume}{279}
  (\bibinfo{year}{2023}) \bibinfo{pages}{06006}.
  \DOIprefix\doi{https://doi.org/10.1051/epjconf/202327906006}.
%Type = Article
\bibitem[{Agodi et~al.(2023)Agodi, Cappuzzello, Cardella, Cirrone, De~Filippo,
  Di~Pietro, Gargano, La~Cognata, Mascali, Milluzzo et~al.}]{agodi2023nuclear}
\bibinfo{author}{C.~Agodi}, \bibinfo{author}{F.~Cappuzzello},
  \bibinfo{author}{G.~Cardella}, \bibinfo{author}{G.~Cirrone},
  \bibinfo{author}{E.~De~Filippo}, \bibinfo{author}{A.~Di~Pietro},
  \bibinfo{author}{A.~Gargano}, \bibinfo{author}{M.~La~Cognata},
  \bibinfo{author}{D.~Mascali}, \bibinfo{author}{G.~Milluzzo}, et~al.,
\newblock \bibinfo{title}{Nuclear physics midterm plan at $\rm{LNS}$},
\newblock \bibinfo{journal}{Eur. Phys. J. Plus} \bibinfo{volume}{138}
  (\bibinfo{year}{2023}) \bibinfo{pages}{1038}.
  \DOIprefix\doi{https://doi.org/10.1140/epjp/s13360-023-04358-7}.
%Type = Article
\bibitem[{Savukov(2006)}]{PhysRevLett.96.073202}
\bibinfo{author}{I.~M. Savukov},
\newblock \bibinfo{title}{Simple method for obtaining electron scattering phase
  shifts from energies of an atom in a cavity},
\newblock \bibinfo{journal}{Phys. Rev. Lett.} \bibinfo{volume}{96}
  (\bibinfo{year}{2006}) \bibinfo{pages}{073202}.
  \DOIprefix\doi{https://doi.org/10.1103/PhysRevLett.96.073202}.
%Type = Article
\bibitem[{J{\"o}nsson et~al.(2022)J{\"o}nsson, Godefroid, Gaigalas, Ekman,
  Grumer, Li, Li, Brage, Grant, Biero{\'n} et~al.}]{jonsson2022introduction}
\bibinfo{author}{P.~J{\"o}nsson}, \bibinfo{author}{M.~Godefroid},
  \bibinfo{author}{G.~Gaigalas}, \bibinfo{author}{J.~Ekman},
  \bibinfo{author}{J.~Grumer}, \bibinfo{author}{W.~Li},
  \bibinfo{author}{J.~Li}, \bibinfo{author}{T.~Brage}, \bibinfo{author}{I.~P.
  Grant}, \bibinfo{author}{J.~Biero{\'n}}, et~al.,
\newblock \bibinfo{title}{An introduction to relativistic theory as implemented
  in grasp},
\newblock \bibinfo{journal}{Atoms} \bibinfo{volume}{11} (\bibinfo{year}{2022})
  \bibinfo{pages}{7}. \DOIprefix\doi{https://doi.org/10.3390/atoms11010007}.
%Type = Article
\bibitem[{van Faassen et~al.(2007)van Faassen, Wasserman, Engel, Zhang, and
  Burke}]{PhysRevLett.99.043005}
\bibinfo{author}{M.~van Faassen}, \bibinfo{author}{A.~Wasserman},
  \bibinfo{author}{E.~Engel}, \bibinfo{author}{F.~Zhang},
  \bibinfo{author}{K.~Burke},
\newblock \bibinfo{title}{Time-dependent density functional calculation of
  e-$\rm{H}$ scattering},
\newblock \bibinfo{journal}{Phys. Rev. Lett.} \bibinfo{volume}{99}
  (\bibinfo{year}{2007}) \bibinfo{pages}{043005}.
  \DOIprefix\doi{https://doi.org/10.1103/PhysRevLett.99.043005}.
%Type = Book
\bibitem[{Nikjoo et~al.(2012)Nikjoo, Uehara, and Emfietzoglou}]{dimitris2012}
\bibinfo{author}{H.~Nikjoo}, \bibinfo{author}{S.~Uehara},
  \bibinfo{author}{D.~Emfietzoglou}, \bibinfo{title}{Interaction of Radiation
  with Matter}, \bibinfo{publisher}{CRC Press}, \bibinfo{year}{2012}.
  \DOIprefix\doi{https://doi.org/10.1201/b12109}.
%Type = Book
\bibitem[{Raether(1982)}]{CRaether}
\bibinfo{author}{H.~Raether}, \bibinfo{title}{Excitation of Plasmons and
  Interband Transitions by Electrons}, \bibinfo{publisher}{Springer-Verlag,
  Berlin}, \bibinfo{year}{1982}.
  \DOIprefix\doi{https://doi.org/10.1007/BFb0045955}.
%Type = Book
\bibitem[{Jackson(2003)}]{jackson}
\bibinfo{author}{J.~D. Jackson}, \bibinfo{title}{Electrodynamics, Classical},
  \bibinfo{publisher}{John Wiley \& Sons, Ltd}, \bibinfo{year}{2003}.
  \DOIprefix\doi{https://doi.org/10.1002/3527600434.eap109}.
%Type = Article
\bibitem[{Wiser(1963)}]{Wiser}
\bibinfo{author}{N.~Wiser},
\newblock \bibinfo{title}{Dielectric constant with local field effects
  included},
\newblock \bibinfo{journal}{Phys. Rev.} \bibinfo{volume}{129}
  (\bibinfo{year}{1963}) \bibinfo{pages}{62--69}.
  \DOIprefix\doi{https://doi.org/10.1103/PhysRev.129.62}.
%Type = Article
\bibitem[{Adler(1962)}]{Adler}
\bibinfo{author}{S.~L. Adler},
\newblock \bibinfo{title}{Quantum theory of the dielectric constant in real
  solids},
\newblock \bibinfo{journal}{Phys. Rev.} \bibinfo{volume}{126}
  (\bibinfo{year}{1962}) \bibinfo{pages}{413--420}.
  \DOIprefix\doi{https://doi.org/10.1103/PhysRev.126.413}.
%Type = Article
\bibitem[{Aspnes(1982)}]{doi:10.1119/1.12734}
\bibinfo{author}{D.~E. Aspnes},
\newblock \bibinfo{title}{Local‐field effects and effective‐medium theory:
  A microscopic perspective},
\newblock \bibinfo{journal}{Am. J. Phys.} \bibinfo{volume}{50}
  (\bibinfo{year}{1982}) \bibinfo{pages}{704--709}.
  \DOIprefix\doi{https://doi.org/10.1119/1.12734}.
%Type = Article
\bibitem[{Gurtubay et~al.(2001)}]{GURTUBAY2001123}
\bibinfo{author}{I.~G. Gurtubay}, et~al.,
\newblock \bibinfo{title}{Dynamic structure factor of gold},
\newblock \bibinfo{journal}{Comput. Mater. Sci.} \bibinfo{volume}{22}
  (\bibinfo{year}{2001}) \bibinfo{pages}{123--128}.
  \DOIprefix\doi{https://doi.org/10.1016/S0927-0256(01)00178-1}.
%Type = Article
\bibitem[{Alkauskas et~al.(2013)}]{PhysRevB.88.195124}
\bibinfo{author}{A.~Alkauskas}, et~al.,
\newblock \bibinfo{title}{Dynamic structure factors of $\rm{C}$u, $\rm{A}$g,
  and $\rm{A}$u: Comparative study from first principles},
\newblock \bibinfo{journal}{Phys. Rev. B} \bibinfo{volume}{88}
  (\bibinfo{year}{2013}) \bibinfo{pages}{195124}.
  \DOIprefix\doi{https://doi.org/10.1103/PhysRevB.88.195124}.
%Type = Article
\bibitem[{Mermin(1970)}]{EMermin}
\bibinfo{author}{N.~D. Mermin},
\newblock \bibinfo{title}{Lindhard dielectric function in the relaxation-time
  approximation},
\newblock \bibinfo{journal}{Phys. Rev. B} \bibinfo{volume}{1}
  (\bibinfo{year}{1970}) \bibinfo{pages}{2362--2363}.
  \DOIprefix\doi{https://doi.org/10.1103/PhysRevB.1.2362}.
%Type = Article
\bibitem[{Planes et~al.(1996)Planes, Garcia-Molina, Abril, and
  Arista}]{EPlanes}
\bibinfo{author}{D.~J. Planes}, \bibinfo{author}{R.~Garcia-Molina},
  \bibinfo{author}{I.~Abril}, \bibinfo{author}{N.~R. Arista},
\newblock \bibinfo{title}{Wavenumber dependence of the energy loss function of
  graphite and aluminium},
\newblock \bibinfo{journal}{J. Electron Spectrosc. Relat. Phenom.}
  \bibinfo{volume}{82} (\bibinfo{year}{1996}) \bibinfo{pages}{23--29}.
  \DOIprefix\doi{https://doi.org/10.1016/S0368-2048(96)03043-5}.
%Type = Article
\bibitem[{Penn(1987)}]{PhysRevB.35.482}
\bibinfo{author}{D.~R. Penn},
\newblock \bibinfo{title}{Electron mean-free-path calculations using a model
  dielectric function},
\newblock \bibinfo{journal}{Phys. Rev. B} \bibinfo{volume}{35}
  (\bibinfo{year}{1987}) \bibinfo{pages}{482--486}.
  \DOIprefix\doi{https://doi.org/10.1103/PhysRevB.35.482}.
%Type = Article
\bibitem[{Lindhard(1954)}]{ELindhard}
\bibinfo{author}{J.~Lindhard},
\newblock \bibinfo{title}{On the properties of a gas of charged particles},
\newblock \bibinfo{journal}{Kgl. Danske Videnskab. Selskab Mat.-fys. Medd.}
  \bibinfo{volume}{28} (\bibinfo{year}{1954}).
%Type = Article
\bibitem[{Heredia-Avalos et~al.(2005)Heredia-Avalos, Garcia-Molina,
  Fern\'andez-Varea, and Abril}]{PhysRevA.72.052902}
\bibinfo{author}{S.~Heredia-Avalos}, \bibinfo{author}{R.~Garcia-Molina},
  \bibinfo{author}{J.~M. Fern\'andez-Varea}, \bibinfo{author}{I.~Abril},
\newblock \bibinfo{title}{Calculated energy loss of swift $\rm{H}$e, $\rm{L}$i,
  $\rm{B}$, and $\rm{N}$ ions in $\rm{SiO_2}$, $\rm{Al_2O_3}$, and
  $\rm{ZrO_2}$},
\newblock \bibinfo{journal}{Phys. Rev. A} \bibinfo{volume}{72}
  (\bibinfo{year}{2005}) \bibinfo{pages}{052902}.
  \DOIprefix\doi{https://doi.org/10.1103/PhysRevA.72.052902}.
%Type = Article
\bibitem[{Garcia-Molina et~al.(2017)Garcia-Molina, Abril, Kyriakou, and
  Emfietzoglou}]{https://doi.org/10.1002/sia.5947}
\bibinfo{author}{R.~Garcia-Molina}, \bibinfo{author}{I.~Abril},
  \bibinfo{author}{I.~Kyriakou}, \bibinfo{author}{D.~Emfietzoglou},
\newblock \bibinfo{title}{Inelastic scattering and energy loss of swift
  electron beams in biologically relevant materials},
\newblock \bibinfo{journal}{Surf. Interface Anal.} \bibinfo{volume}{49}
  (\bibinfo{year}{2017}) \bibinfo{pages}{11--17}.
  \DOIprefix\doi{https://doi.org/10.1002/sia.5947}.
%Type = Article
\bibitem[{de~la Cruz and Yubero(2007)}]{EdelaCruzandYubero}
\bibinfo{author}{W.~de~la Cruz}, \bibinfo{author}{F.~Yubero},
\newblock \bibinfo{title}{Electron inelastic mean free paths: influence of the
  modelling energy-loss function},
\newblock \bibinfo{journal}{Surf. Interface Anal.} \bibinfo{volume}{39}
  (\bibinfo{year}{2007}) \bibinfo{pages}{460--463}.
  \DOIprefix\doi{https://doi.org/10.1002/sia.2545}.
%Type = Article
\bibitem[{Kyriakou et~al.(2011)Kyriakou, Emfietzoglou, Garcia-Molina, Abril,
  and Kostarelos}]{10.1063/1.3626460}
\bibinfo{author}{I.~Kyriakou}, \bibinfo{author}{D.~Emfietzoglou},
  \bibinfo{author}{R.~Garcia-Molina}, \bibinfo{author}{I.~Abril},
  \bibinfo{author}{K.~Kostarelos},
\newblock \bibinfo{title}{{Simple model of bulk and surface excitation effects
  to inelastic scattering in low-energy electron beam irradiation of
  multi-walled carbon nanotubes}},
\newblock \bibinfo{journal}{J. Appl. Phys.} \bibinfo{volume}{110}
  (\bibinfo{year}{2011}) \bibinfo{pages}{054304}.
  \DOIprefix\doi{https://doi.org/10.1063/1.3626460}.
%Type = Article
\bibitem[{Da et~al.(2019)Da, Shinotsuka, Yoshikawa, and Tanuma}]{DaBo}
\bibinfo{author}{B.~Da}, \bibinfo{author}{H.~Shinotsuka},
  \bibinfo{author}{H.~Yoshikawa}, \bibinfo{author}{S.~Tanuma},
\newblock \bibinfo{title}{Comparison of the mermin and penn models for
  inelastic mean-free path calculations for electrons based on a model using
  optical energy-loss functions},
\newblock \bibinfo{journal}{Surface and Interface Analysis}
  \bibinfo{volume}{51} (\bibinfo{year}{2019}) \bibinfo{pages}{627--640}.
  \URLprefix
  \url{https://analyticalsciencejournals.onlinelibrary.wiley.com/doi/abs/10.1002/sia.6628}.
  \DOIprefix\doi{https://doi.org/10.1002/sia.6628}.
  \href{http://arxiv.org/abs/https://analyticalsciencejournals.onlinelibrary.wiley.com/doi/pdf/10.1002/sia.6628}{{\tt
  arXiv:https://analyticalsciencejournals.onlinelibrary.wiley.com/doi/pdf/10.1002/sia.6628}}.
%Type = Article
\bibitem[{Shinotsuka~H. and R.(2017)}]{d504rm44w}
\bibinfo{author}{P.~C.~J. Shinotsuka~H., Tanuma~S.}, \bibinfo{author}{P.~D.
  R.},
\newblock \bibinfo{title}{Calculations of electron inelastic mean free paths.
  x. data for 41 elemental solids over the 50 ev to 200 kev range with the
  relativistic full penn algorithm},
\newblock \bibinfo{journal}{Surf. Interface Anal.} \bibinfo{volume}{49}
  (\bibinfo{year}{2017}) \bibinfo{pages}{238--252}.
  \DOIprefix\doi{https://doi.org/10.1002/sia.6123}.
%Type = Article
\bibitem[{Chiarello et~al.(1984)Chiarello, Colavita, De~Crescenzi, and
  Nannarone}]{PhysRevB.29.4878}
\bibinfo{author}{G.~Chiarello}, \bibinfo{author}{E.~Colavita},
  \bibinfo{author}{M.~De~Crescenzi}, \bibinfo{author}{S.~Nannarone},
\newblock \bibinfo{title}{Reflection electron-energy-loss investigation of the
  electronic and structural properties of palladium},
\newblock \bibinfo{journal}{Phys. Rev. B} \bibinfo{volume}{29}
  (\bibinfo{year}{1984}) \bibinfo{pages}{4878--4889}. \URLprefix
  \url{https://link.aps.org/doi/10.1103/PhysRevB.29.4878}.
  \DOIprefix\doi{10.1103/PhysRevB.29.4878}.
%Type = Article
\bibitem[{Ohno(1989)}]{PhysRevB.39.8209}
\bibinfo{author}{Y.~Ohno},
\newblock \bibinfo{title}{Kramers-kronig analysis of reflection
  electron-energy-loss spectra measured with a cylindrical mirror analyzer},
\newblock \bibinfo{journal}{Phys. Rev. B} \bibinfo{volume}{39}
  (\bibinfo{year}{1989}) \bibinfo{pages}{8209--8219}. \URLprefix
  \url{https://link.aps.org/doi/10.1103/PhysRevB.39.8209}.
  \DOIprefix\doi{10.1103/PhysRevB.39.8209}.
%Type = Article
\bibitem[{Yubero et~al.(1990)Yubero, Sanz, Elizalde, and
  Galán}]{YUBERO1990173}
\bibinfo{author}{F.~Yubero}, \bibinfo{author}{J.~Sanz},
  \bibinfo{author}{E.~Elizalde}, \bibinfo{author}{L.~Galán},
\newblock \bibinfo{title}{Kramers-krönig analysis of reflection electron
  energy loss spectra (reels) of zr and zro2},
\newblock \bibinfo{journal}{Surface Science} \bibinfo{volume}{237}
  (\bibinfo{year}{1990}) \bibinfo{pages}{173--180}. \URLprefix
  \url{https://www.sciencedirect.com/science/article/pii/003960289090529H}.
  \DOIprefix\doi{https://doi.org/10.1016/0039-6028(90)90529-H}.
%Type = Article
\bibitem[{Ding et~al.(2002)Ding, Li, Pu, Zhang, and
  Shimizu}]{PhysRevB.66.085411}
\bibinfo{author}{Z.~J. Ding}, \bibinfo{author}{H.~M. Li},
  \bibinfo{author}{Q.~R. Pu}, \bibinfo{author}{Z.~M. Zhang},
  \bibinfo{author}{R.~Shimizu},
\newblock \bibinfo{title}{Reflection electron energy loss spectrum of surface
  plasmon excitation of ag: A monte carlo study},
\newblock \bibinfo{journal}{Phys. Rev. B} \bibinfo{volume}{66}
  (\bibinfo{year}{2002}) \bibinfo{pages}{085411}. \URLprefix
  \url{https://link.aps.org/doi/10.1103/PhysRevB.66.085411}.
  \DOIprefix\doi{10.1103/PhysRevB.66.085411}.
%Type = Article
\bibitem[{Werner(2006)}]{PhysRevB.74.075421}
\bibinfo{author}{W.~S.~M. Werner},
\newblock \bibinfo{title}{Differential surface and volume excitation
  probability of medium-energy electrons in solids},
\newblock \bibinfo{journal}{Phys. Rev. B} \bibinfo{volume}{74}
  (\bibinfo{year}{2006}) \bibinfo{pages}{075421}. \URLprefix
  \url{https://link.aps.org/doi/10.1103/PhysRevB.74.075421}.
  \DOIprefix\doi{10.1103/PhysRevB.74.075421}.
%Type = Article
\bibitem[{Werner et~al.(2001)Werner, Smekal, Tomastik, and
  Störi}]{WERNER2001L461}
\bibinfo{author}{W.~S. Werner}, \bibinfo{author}{W.~Smekal},
  \bibinfo{author}{C.~Tomastik}, \bibinfo{author}{H.~Störi},
\newblock \bibinfo{title}{Surface excitation probability of medium energy
  electrons in metals and semiconductors},
\newblock \bibinfo{journal}{Surface Science} \bibinfo{volume}{486}
  (\bibinfo{year}{2001}) \bibinfo{pages}{L461--L466}. \URLprefix
  \url{https://www.sciencedirect.com/science/article/pii/S0039602801010913}.
  \DOIprefix\doi{https://doi.org/10.1016/S0039-6028(01)01091-3}.
%Type = Article
\bibitem[{Chen and Kwei(1996)}]{CHEN1996131}
\bibinfo{author}{Y.~Chen}, \bibinfo{author}{C.~Kwei},
\newblock \bibinfo{title}{Electron differential inverse mean free path for
  surface electron spectroscopy},
\newblock \bibinfo{journal}{Surface Science} \bibinfo{volume}{364}
  (\bibinfo{year}{1996}) \bibinfo{pages}{131--140}. \URLprefix
  \url{https://www.sciencedirect.com/science/article/pii/0039602896006164}.
  \DOIprefix\doi{https://doi.org/10.1016/0039-6028(96)00616-4}.
%Type = Article
\bibitem[{Emfietzoglou et~al.(2017)Emfietzoglou, Kyriakou, Garcia-Molina, and
  Abril}]{doi:10.1002/sia.5878}
\bibinfo{author}{D.~Emfietzoglou}, \bibinfo{author}{I.~Kyriakou},
  \bibinfo{author}{R.~Garcia-Molina}, \bibinfo{author}{I.~Abril},
\newblock \bibinfo{title}{Inelastic mean free path of low-energy electrons in
  condensed media: beyond the standard models},
\newblock \bibinfo{journal}{Surf. Interface Anal.} \bibinfo{volume}{49}
  (\bibinfo{year}{2017}) \bibinfo{pages}{4--10}.
  \DOIprefix\doi{https://doi.org/10.1002/sia.5878}.
%Type = Article
\bibitem[{Azzolini et~al.(2018)Azzolini, Angelucci, Cimino, Larciprete, Pugno,
  Taioli, and Dapor}]{azzolini2018secondary}
\bibinfo{author}{M.~Azzolini}, \bibinfo{author}{M.~Angelucci},
  \bibinfo{author}{R.~Cimino}, \bibinfo{author}{R.~Larciprete},
  \bibinfo{author}{N.~M. Pugno}, \bibinfo{author}{S.~Taioli},
  \bibinfo{author}{M.~Dapor},
\newblock \bibinfo{title}{Secondary electron emission and yield spectra of
  metals from monte carlo simulations and experiments},
\newblock \bibinfo{journal}{J. Phys. Condens. Matter} \bibinfo{volume}{31}
  (\bibinfo{year}{2018}) \bibinfo{pages}{055901}.
  \DOIprefix\doi{https://doi.org/10.1021/acs.jpcc.8b02256}.
%Type = Article
\bibitem[{Ochkur(1964)}]{3BornOchkur}
\bibinfo{author}{V.~I. Ochkur},
\newblock \bibinfo{title}{The born-oppenheimer method in the theory of atomic
  collisions},
\newblock \bibinfo{journal}{Soviet Phys. J.E.T.P.} \bibinfo{volume}{18}
  (\bibinfo{year}{1964}) \bibinfo{pages}{503--508}.
%Type = Article
\bibitem[{Fern\'andez-Varea et~al.(1993)Fern\'andez-Varea, Mayol, Liljequist,
  and Salvat}]{3FernandezVarea}
\bibinfo{author}{J.~M. Fern\'andez-Varea}, \bibinfo{author}{R.~Mayol},
  \bibinfo{author}{D.~Liljequist}, \bibinfo{author}{F.~Salvat},
\newblock \bibinfo{title}{Inelastic scattering of electrons in solids from a
  generalized oscillator strength model using optical and photoelectric data},
\newblock \bibinfo{journal}{J. Phys.: Condens. Matter} \bibinfo{volume}{5}
  (\bibinfo{year}{1993}) \bibinfo{pages}{3593--3610}.
  \DOIprefix\doi{https://doi.org/10.1088/0953-8984/5/22/011}.
%Type = Article
\bibitem[{de~Vera et~al.(2011)de~Vera, Abril, and Garcia-Molina}]{3deVera}
\bibinfo{author}{P.~de~Vera}, \bibinfo{author}{I.~Abril},
  \bibinfo{author}{R.~Garcia-Molina},
\newblock \bibinfo{title}{Inelastic scattering of electron and light ion beams
  in organic polymers},
\newblock \bibinfo{journal}{J. Appl. Phys.} \bibinfo{volume}{109}
  (\bibinfo{year}{2011}) \bibinfo{pages}{094901}.
  \DOIprefix\doi{https://doi.org/10.1063/1.3581120}.
%Type = Article
\bibitem[{Bourke(2019)}]{3Bourke}
\bibinfo{author}{J.~D. Bourke},
\newblock \bibinfo{title}{Exchange corrections for inelastic electron
  scattering rates in condensed matter},
\newblock \bibinfo{journal}{Phys. Rev. B} \bibinfo{volume}{100}
  (\bibinfo{year}{2019}) \bibinfo{pages}{184311}.
  \DOIprefix\doi{https://doi.org/10.1103/PhysRevB.100.184311}.
%Type = Article
\bibitem[{Segatta et~al.(2017)Segatta, Cupellini, Jurinovich, Mukamel, Dapor,
  Taioli, Garavelli, and Mennucci}]{segatta2017quantum}
\bibinfo{author}{F.~Segatta}, \bibinfo{author}{L.~Cupellini},
  \bibinfo{author}{S.~Jurinovich}, \bibinfo{author}{S.~Mukamel},
  \bibinfo{author}{M.~Dapor}, \bibinfo{author}{S.~Taioli},
  \bibinfo{author}{M.~Garavelli}, \bibinfo{author}{B.~Mennucci},
\newblock \bibinfo{title}{A quantum chemical interpretation of two-dimensional
  electronic spectroscopy of light-harvesting complexes},
\newblock \bibinfo{journal}{J. Am. Chem. Soc.} \bibinfo{volume}{139}
  (\bibinfo{year}{2017}) \bibinfo{pages}{7558}.
  \DOIprefix\doi{https://doi.org/10.1021/jacs.7b02130}.
%Type = Article
\bibitem[{Medvedev et~al.(2017)Medvedev, Bushmarinov, Sun, Perdew, and
  Lyssenko}]{medvedev2017density}
\bibinfo{author}{M.~G. Medvedev}, \bibinfo{author}{I.~S. Bushmarinov},
  \bibinfo{author}{J.~Sun}, \bibinfo{author}{J.~P. Perdew},
  \bibinfo{author}{K.~A. Lyssenko},
\newblock \bibinfo{title}{Density functional theory is straying from the path
  toward the exact functional},
\newblock \bibinfo{journal}{Science} \bibinfo{volume}{355}
  (\bibinfo{year}{2017}) \bibinfo{pages}{49--52}.
  \DOIprefix\doi{https://doi.org/10.1126/science.aah5975}.
%Type = Article
\bibitem[{Borlido et~al.(2020)Borlido, Schmidt, Huran et~al.}]{Borlido}
\bibinfo{author}{P.~Borlido}, \bibinfo{author}{J.~Schmidt},
  \bibinfo{author}{A.~W. Huran}, et~al.,
\newblock \bibinfo{title}{Exchange-correlation functionals for band gaps of
  solids: benchmark, reparametrization and machine learning},
\newblock \bibinfo{journal}{Npj Comput. Mater.} \bibinfo{volume}{6}
  (\bibinfo{year}{2020}) \bibinfo{pages}{96}.
  \DOIprefix\doi{https://doi.org/10.1038/s41524-020-00360-0}.
%Type = Article
\bibitem[{Ryabov et~al.(2022)Ryabov, Akhatov, and Zhilyaev}]{Ryabov}
\bibinfo{author}{A.~Ryabov}, \bibinfo{author}{I.~Akhatov},
  \bibinfo{author}{P.~Zhilyaev},
\newblock \bibinfo{title}{Application of two-component neural network for
  exchange-correlation functional interpolation},
\newblock \bibinfo{journal}{Sci. Rep.} \bibinfo{volume}{12}
  (\bibinfo{year}{2022}) \bibinfo{pages}{14133}.
  \DOIprefix\doi{https://doi.org/10.1038/s41598-022-18083-1}.
%Type = Article
\bibitem[{Emfietzoglou and Nikjoo(2005)}]{10.1667RR3281}
\bibinfo{author}{D.~Emfietzoglou}, \bibinfo{author}{H.~Nikjoo},
\newblock \bibinfo{title}{{The Effect of Model Approximations on
  Single-Collision Distributions of Low-Energy Electrons in Liquid Water}},
\newblock \bibinfo{journal}{Radiation Research} \bibinfo{volume}{163}
  (\bibinfo{year}{2005}) \bibinfo{pages}{98 -- 111}.
  \DOIprefix\doi{https://doi.org/10.1667/RR3281}.
%Type = Article
\bibitem[{Emfietzoglou et~al.(2013)Emfietzoglou, Kyriakou, Garcia-Molina, and
  Abril}]{doi:10.1063/1.4824541}
\bibinfo{author}{D.~Emfietzoglou}, \bibinfo{author}{I.~Kyriakou},
  \bibinfo{author}{R.~Garcia-Molina}, \bibinfo{author}{I.~Abril},
\newblock \bibinfo{title}{The effect of static many-body local-field
  corrections to inelastic electron scattering in condensed media},
\newblock \bibinfo{journal}{J. Appl. Phys.} \bibinfo{volume}{114}
  (\bibinfo{year}{2013}) \bibinfo{pages}{144907}.
  \DOIprefix\doi{https://doi.org/10.1063/1.4824541}.
%Type = Article
\bibitem[{Weissker et~al.(2010)}]{Weissker2010}
\bibinfo{author}{H.~C. Weissker}, et~al.,
\newblock \bibinfo{title}{Dynamic structure factor and dielectric function of
  silicon for finite momentum transfer: Inelastic {X}-ray scattering
  experiments and ab initio calculations},
\newblock \bibinfo{journal}{Phys. Rev. B} \bibinfo{volume}{81}
  (\bibinfo{year}{2010}) \bibinfo{pages}{085104}.
  \DOIprefix\doi{https://doi.org/10.1103/PhysRevB.81.085104}.
%Type = Article
\bibitem[{Marques and
  Gross(2004)}]{doi:10.1146/annurev.physchem.55.091602.094449}
\bibinfo{author}{M.~Marques}, \bibinfo{author}{E.~Gross},
\newblock \bibinfo{title}{Time-dependent density functional theory},
\newblock \bibinfo{journal}{Annu. Rev. Phys. Chem.} \bibinfo{volume}{55}
  (\bibinfo{year}{2004}) \bibinfo{pages}{427--455}.
  \DOIprefix\doi{https://doi.org/10.1146/annurev.physchem.55.091602.094449}.
%Type = Article
\bibitem[{Dobson et~al.(1997)Dobson, B\"unner, and Gross}]{PhysRevLett.79.1905}
\bibinfo{author}{J.~F. Dobson}, \bibinfo{author}{M.~J. B\"unner},
  \bibinfo{author}{E.~K.~U. Gross},
\newblock \bibinfo{title}{Time-dependent density functional theory beyond
  linear response: An exchange-correlation potential with memory},
\newblock \bibinfo{journal}{Phys. Rev. Lett.} \bibinfo{volume}{79}
  (\bibinfo{year}{1997}) \bibinfo{pages}{1905--1908}.
  \DOIprefix\doi{https://doi.org/10.1103/PhysRevLett.79.1905}.
%Type = Article
\bibitem[{Bloch(1933{\natexlab{a}})}]{Bloch1}
\bibinfo{author}{F.~Bloch},
\newblock \bibinfo{title}{Bremsverm{\"o}gen von atomen mit mehreren
  elektronen},
\newblock \bibinfo{journal}{Z. Phys.} \bibinfo{volume}{81}
  (\bibinfo{year}{1933}{\natexlab{a}}) \bibinfo{pages}{363--376}.
  \DOIprefix\doi{https://doi.org/10.1007/BF01344553}.
%Type = Article
\bibitem[{Bloch(1933{\natexlab{b}})}]{Bloch2}
\bibinfo{author}{F.~Bloch},
\newblock \bibinfo{title}{For braking rapidly moving particles as they pass
  through matter},
\newblock \bibinfo{journal}{Ann. Phys.} \bibinfo{volume}{408}
  (\bibinfo{year}{1933}{\natexlab{b}}) \bibinfo{pages}{285--320}.
  \DOIprefix\doi{https://doi.org/10.1002/andp.19334080303}.
%Type = Article
\bibitem[{Fano(1961)}]{PhysRev.124.1866}
\bibinfo{author}{U.~Fano},
\newblock \bibinfo{title}{Effects of configuration interaction on intensities
  and phase shifts},
\newblock \bibinfo{journal}{Phys. Rev.} \bibinfo{volume}{124}
  (\bibinfo{year}{1961}) \bibinfo{pages}{1866--1878}.
  \DOIprefix\doi{https://doi.org/10.1103/PhysRev.124.1866}.
%Type = Article
\bibitem[{{{\r{A}}berg} and {Howat}(1982)}]{1982HD469A}
\bibinfo{author}{T.~{{\r{A}}berg}}, \bibinfo{author}{G.~{Howat}},
\newblock \bibinfo{title}{{Theory of the Auger Effect}},
\newblock \bibinfo{journal}{Handb. Phys.} \bibinfo{volume}{6}
  (\bibinfo{year}{1982}) \bibinfo{pages}{469--619}.
%Type = Article
\bibitem[{Taioli et~al.(2009)Taioli, Simonucci, Calliari, Filippi, and
  Dapor}]{taioli2009mixed}
\bibinfo{author}{S.~Taioli}, \bibinfo{author}{S.~Simonucci},
  \bibinfo{author}{L.~Calliari}, \bibinfo{author}{M.~Filippi},
  \bibinfo{author}{M.~Dapor},
\newblock \bibinfo{title}{Mixed ab initio quantum mechanical and monte carlo
  calculations of secondary emission from $\rm{SiO_2}$ nanoclusters},
\newblock \bibinfo{journal}{Physical Review B} \bibinfo{volume}{79}
  (\bibinfo{year}{2009}) \bibinfo{pages}{085432}.
  \DOIprefix\doi{https://doi.org/10.1103/PhysRevB.79.085432}.
%Type = Article
\bibitem[{Carvalho et~al.(2021)Carvalho, Trevisanutto, Taioli, and
  Neto}]{carvalho2021computational}
\bibinfo{author}{A.~Carvalho}, \bibinfo{author}{P.~Trevisanutto},
  \bibinfo{author}{S.~Taioli}, \bibinfo{author}{A.~Neto},
\newblock \bibinfo{title}{Computational methods for 2d materials modelling},
\newblock \bibinfo{journal}{Rep. Prog. Phys.} \bibinfo{volume}{84}
  (\bibinfo{year}{2021}) \bibinfo{pages}{106501}.
  \DOIprefix\doi{https://doi.org/10.1088/1361-6633/ac2356}.
%Type = Article
\bibitem[{Bromberg(1969)}]{10.1063/1.1672634}
\bibinfo{author}{J.~P. Bromberg},
\newblock \bibinfo{title}{{Absolute Differential Cross Sections of Elastically
  Scattered Electrons. II. A Polarization Scattering Potential for Hg at 500,
  400, and 300 eV}},
\newblock \bibinfo{journal}{J. Chem. Phys.} \bibinfo{volume}{51}
  (\bibinfo{year}{1969}) \bibinfo{pages}{4117--4122}.
  \DOIprefix\doi{https://doi.org/10.1063/1.1672634}.
%Type = Article
\bibitem[{Holtkamp et~al.(1987)Holtkamp, Jost, Peitzmann, and
  Kessler}]{Holtkamp_1987}
\bibinfo{author}{G.~Holtkamp}, \bibinfo{author}{K.~Jost},
  \bibinfo{author}{F.~J. Peitzmann}, \bibinfo{author}{J.~Kessler},
\newblock \bibinfo{title}{Absolute differential cross sections for elastic
  electron scattering from mercury},
\newblock \bibinfo{journal}{J. Phys. B} \bibinfo{volume}{20}
  (\bibinfo{year}{1987}) \bibinfo{pages}{4543--4569}.
  \DOIprefix\doi{https://doi.org/10.1088/0022-3700/20/17/030}.
%Type = Article
\bibitem[{Katase et~al.(1986)Katase, Ishibashi, Matsumoto, Sakae, Maezono,
  Murakami, Watanabe, and Maki}]{Katase_1986}
\bibinfo{author}{A.~Katase}, \bibinfo{author}{K.~Ishibashi},
  \bibinfo{author}{Y.~Matsumoto}, \bibinfo{author}{T.~Sakae},
  \bibinfo{author}{S.~Maezono}, \bibinfo{author}{E.~Murakami},
  \bibinfo{author}{K.~Watanabe}, \bibinfo{author}{H.~Maki},
\newblock \bibinfo{title}{Elastic scattering of electrons by water molecules
  over the range 100-1000 e$\rm{V}$},
\newblock \bibinfo{journal}{J. Phys. B} \bibinfo{volume}{19}
  (\bibinfo{year}{1986}) \bibinfo{pages}{2715--2734}.
  \DOIprefix\doi{https://doi.org/10.1088/0022-3700/19/17/020}.
%Type = Article
\bibitem[{Itikawa and Mason(2005)}]{10.1063/1.1799251}
\bibinfo{author}{Y.~Itikawa}, \bibinfo{author}{N.~Mason},
\newblock \bibinfo{title}{Cross sections for electron collisions with water
  molecules},
\newblock \bibinfo{journal}{J. Phys. Chem. Ref. Data} \bibinfo{volume}{34}
  (\bibinfo{year}{2005}) \bibinfo{pages}{1--22}.
  \DOIprefix\doi{https://doi.org/10.1063/1.1799251}.
%Type = Article
\bibitem[{Khakoo et~al.(2013)Khakoo, Silva, Muse, Lopes, Winstead, and
  McKoy}]{khakoo_2013}
\bibinfo{author}{M.~A. Khakoo}, \bibinfo{author}{H.~Silva},
  \bibinfo{author}{J.~Muse}, \bibinfo{author}{M.~C.~A. Lopes},
  \bibinfo{author}{C.~Winstead}, \bibinfo{author}{V.~McKoy},
\newblock \bibinfo{title}{Erratum: Electron scattering from $\rm{H_2O}$:
  Elastic scattering [$\rm{P}$hys. $\rm{R}$ev. $\rm{A}$ 78, 052710 (2008)]},
\newblock \bibinfo{journal}{Phys. Rev. A} \bibinfo{volume}{87}
  (\bibinfo{year}{2013}) \bibinfo{pages}{049902}.
  \DOIprefix\doi{https://doi.org/10.1103/PhysRevA.87.049902}.
%Type = Article
\bibitem[{Song et~al.(2021)Song, Cho, Karwasz, Kokoouline, Nakamura, Tennyson,
  Faure, Mason, and Itikawa}]{song_2021}
\bibinfo{author}{M.-Y. Song}, \bibinfo{author}{H.~Cho}, \bibinfo{author}{G.~P.
  Karwasz}, \bibinfo{author}{V.~Kokoouline}, \bibinfo{author}{Y.~Nakamura},
  \bibinfo{author}{J.~Tennyson}, \bibinfo{author}{A.~Faure},
  \bibinfo{author}{N.~J. Mason}, \bibinfo{author}{Y.~Itikawa},
\newblock \bibinfo{title}{Cross sections for electron collisions with
  $\rm{H_2O}$},
\newblock \bibinfo{journal}{J. Phys. Chem. Ref. Data} \bibinfo{volume}{50}
  (\bibinfo{year}{2021}) \bibinfo{pages}{023103}.
  \DOIprefix\doi{https://doi.org/10.1063/5.0035315}.
%Type = Article
\bibitem[{Gorfinkiel et~al.(2002)Gorfinkiel, Morgan, and
  Tennyson}]{gorfinkiel_2002}
\bibinfo{author}{J.~D. Gorfinkiel}, \bibinfo{author}{L.~A. Morgan},
  \bibinfo{author}{J.~Tennyson},
\newblock \bibinfo{title}{Electron impact dissociative excitation of water
  within the adiabatic nuclei approximation},
\newblock \bibinfo{journal}{J. Phys. B} \bibinfo{volume}{35}
  (\bibinfo{year}{2002}) \bibinfo{pages}{543--555}.
  \DOIprefix\doi{https://doi.org/10.1088/0953-4075/35/3/309}.
%Type = Article
\bibitem[{Zhang et~al.(2009)Zhang, Faure, and Tennyson}]{zhang_2009}
\bibinfo{author}{R.~Zhang}, \bibinfo{author}{A.~Faure},
  \bibinfo{author}{J.~Tennyson},
\newblock \bibinfo{title}{Electron and positron collisions with polar
  molecules: studies with the benchmark water molecule},
\newblock \bibinfo{journal}{Phys. Scr.} \bibinfo{volume}{80}
  (\bibinfo{year}{2009}) \bibinfo{pages}{015301}.
  \DOIprefix\doi{https://doi.org/10.1088/0031-8949/80/01/015301}.
%Type = Article
\bibitem[{Faure et~al.(2004)Faure, Gorfinkiel, and Tennyson}]{faure_2004}
\bibinfo{author}{A.~Faure}, \bibinfo{author}{J.~D. Gorfinkiel},
  \bibinfo{author}{J.~Tennyson},
\newblock \bibinfo{title}{Low-energy electron collisions with water: elastic
  and rotationally inelastic scattering},
\newblock \bibinfo{journal}{J. Phys. B} \bibinfo{volume}{37}
  (\bibinfo{year}{2004}) \bibinfo{pages}{801--807}.
  \DOIprefix\doi{10.1088/0953-4075/37/4/007}.
%Type = Article
\bibitem[{Szmytkowski(1987)}]{szmytkowski_1987}
\bibinfo{author}{C.~Szmytkowski},
\newblock \bibinfo{title}{Absolute total cross sections for electron-water
  vapour scattering},
\newblock \bibinfo{journal}{Chem. Phys. Lett.} \bibinfo{volume}{136}
  (\bibinfo{year}{1987}) \bibinfo{pages}{363--367}.
  \DOIprefix\doi{https://doi.org/10.1016/0009-2614(87)80267-1}.
%Type = Article
\bibitem[{Szmytkowski and Mozejko(2006)}]{szmytkowski_2006}
\bibinfo{author}{C.~Szmytkowski}, \bibinfo{author}{P.~Mozejko},
\newblock \bibinfo{title}{Electron-scattering total cross sections for
  triatomic molecules: $\rm{NO_2}$ and $\rm{H_2O}$},
\newblock \bibinfo{journal}{Opt. Appl.} \bibinfo{volume}{36}
  (\bibinfo{year}{2006}) \bibinfo{pages}{543--550}.
%Type = Article
\bibitem[{Kadokura et~al.(2019)Kadokura, Loreti, K\"ov\'er, Faure, Tennyson,
  and Laricchia}]{kadokura_2019}
\bibinfo{author}{R.~Kadokura}, \bibinfo{author}{A.~Loreti},
  \bibinfo{author}{A.~K\"ov\'er}, \bibinfo{author}{A.~Faure},
  \bibinfo{author}{J.~Tennyson}, \bibinfo{author}{G.~Laricchia},
\newblock \bibinfo{title}{Angle-resolved electron scattering from $\rm{H_2O}$
  near 0\ifmmode^\circ\else\textdegree\fi{}},
\newblock \bibinfo{journal}{Phys. Rev. Lett.} \bibinfo{volume}{123}
  (\bibinfo{year}{2019}) \bibinfo{pages}{033401}.
  \DOIprefix\doi{https://doi.org/10.1103/PhysRevLett.123.033401}.
%Type = Article
\bibitem[{Mu\~noz et~al.(2007)Mu\~noz, Oller, Blanco, Gorfinkiel, Lim\~ao
  Vieira, and Garc\'{\i}a}]{munoz_2007}
\bibinfo{author}{A.~Mu\~noz}, \bibinfo{author}{J.~C. Oller},
  \bibinfo{author}{F.~Blanco}, \bibinfo{author}{J.~D. Gorfinkiel},
  \bibinfo{author}{P.~Lim\~ao Vieira}, \bibinfo{author}{G.~Garc\'{\i}a},
\newblock \bibinfo{title}{Electron-scattering cross sections and stopping
  powers in $\rm{H_2O}$},
\newblock \bibinfo{journal}{Phys. Rev. A} \bibinfo{volume}{76}
  (\bibinfo{year}{2007}) \bibinfo{pages}{052707}.
  \DOIprefix\doi{https://doi.org/10.1103/PhysRevA.76.052707}.
%Type = Article
\bibitem[{Hayashi et~al.(2000)Hayashi, Watanabe, Udagawa, and
  Kao}]{hayashi2000complete}
\bibinfo{author}{H.~Hayashi}, \bibinfo{author}{N.~Watanabe},
  \bibinfo{author}{Y.~Udagawa}, \bibinfo{author}{C.-C. Kao},
\newblock \bibinfo{title}{The complete optical spectrum of liquid water
  measured by inelastic x-ray scattering},
\newblock \bibinfo{journal}{Proceedings of the National Academy of Sciences}
  \bibinfo{volume}{97} (\bibinfo{year}{2000}) \bibinfo{pages}{6264--6266}.
%Type = Book
\bibitem[{Palik and Gosh(1999)}]{palik1999electronic}
\bibinfo{author}{E.~D. Palik}, \bibinfo{author}{G.~Gosh}, \bibinfo{title}{The
  electronic handbook of optical constants of solids},
  \bibinfo{publisher}{Academic Press}, \bibinfo{year}{1999}.
%Type = Article
\bibitem[{Werner et~al.(2009)Werner, Glantschnig, and
  Ambrosch-Draxl}]{werner2009optical}
\bibinfo{author}{W.~S.~M. Werner}, \bibinfo{author}{K.~Glantschnig},
  \bibinfo{author}{C.~Ambrosch-Draxl},
\newblock \bibinfo{title}{Optical constants and inelastic electron-scattering
  data for 17 elemental metals},
\newblock \bibinfo{journal}{J. Phys. Chem. Ref. Data} \bibinfo{volume}{38}
  (\bibinfo{year}{2009}) \bibinfo{pages}{1013--1092}.
  \DOIprefix\doi{https://doi.org/10.1063/1.3243762}.
%Type = Article
\bibitem[{Xu et~al.(2017)Xu, Da, T\'oth, T\ifmmode~\mbox{\H{o}}\else
  \H{o}\fi{}k\'esi, and Ding}]{PhysRevB.95.195417}
\bibinfo{author}{H.~Xu}, \bibinfo{author}{B.~Da}, \bibinfo{author}{J.~T\'oth},
  \bibinfo{author}{K.~T\ifmmode~\mbox{\H{o}}\else \H{o}\fi{}k\'esi},
  \bibinfo{author}{Z.~J. Ding},
\newblock \bibinfo{title}{Absolute determination of optical constants by
  reflection electron energy loss spectroscopy},
\newblock \bibinfo{journal}{Phys. Rev. B} \bibinfo{volume}{95}
  (\bibinfo{year}{2017}) \bibinfo{pages}{195417}. \URLprefix
  \url{https://link.aps.org/doi/10.1103/PhysRevB.95.195417}.
  \DOIprefix\doi{10.1103/PhysRevB.95.195417}.
%Type = Article
\bibitem[{Henke et~al.(1993)Henke, Gullikson, and Davis}]{henke1993x}
\bibinfo{author}{B.~L. Henke}, \bibinfo{author}{E.~M. Gullikson},
  \bibinfo{author}{J.~C. Davis},
\newblock \bibinfo{title}{X-ray interactions: photoabsorption, scattering,
  transmission, and reflection at e= 50-30,000 ev, z= 1-92},
\newblock \bibinfo{journal}{Atomic data and nuclear data tables}
  \bibinfo{volume}{54} (\bibinfo{year}{1993}) \bibinfo{pages}{181--342}.
%Type = Misc
\bibitem[{Seltzer(1995)}]{NIST}
\bibinfo{author}{S.~Seltzer}, \bibinfo{title}{{X-ray form factor, attenuation
  and scattering tables, NIST standard reference database 66}},
  \bibinfo{year}{1995}.
%Type = Article
\bibitem[{de~Vera et~al.(2023)de~Vera, Abril, and
  Garcia-Molina}]{10.3389/fmats.2023.1249517}
\bibinfo{author}{P.~de~Vera}, \bibinfo{author}{I.~Abril},
  \bibinfo{author}{R.~Garcia-Molina},
\newblock \bibinfo{title}{Electronic cross section, stopping power and
  energy-loss straggling of metals for swift protons, alpha particles and
  electrons},
\newblock \bibinfo{journal}{Front. Mater.} \bibinfo{volume}{10}
  (\bibinfo{year}{2023}) \bibinfo{pages}{1249517}.
  \DOIprefix\doi{https://doi.org/10.3389/fmats.2023.1249517}.
%Type = Article
\bibitem[{Li et~al.(2023)Li, Gong, Da, T{\'o}th, Tők{\'e}si, Zeng, and
  Ding}]{Li2023ImprovedRM}
\bibinfo{author}{Z.~Li}, \bibinfo{author}{J.~Gong}, \bibinfo{author}{B.~Da},
  \bibinfo{author}{J.~T{\'o}th}, \bibinfo{author}{K.~Tők{\'e}si},
  \bibinfo{author}{R.~Zeng}, \bibinfo{author}{Z.~J. Ding},
\newblock \bibinfo{title}{Improved reverse monte carlo analysis of optical
  property of fe and ni from reflection electron energy loss spectroscopy
  spectra},
\newblock \bibinfo{journal}{Scientific Reports} \bibinfo{volume}{13}
  (\bibinfo{year}{2023}). \URLprefix
  \url{https://api.semanticscholar.org/CorpusID:260377825}.
%Type = Article
\bibitem[{Motornyi et~al.(2020)Motornyi, Vast, Timrov, Baseggio, Baroni, and
  Dal~Corso}]{PhysRevB.102.035156}
\bibinfo{author}{O.~Motornyi}, \bibinfo{author}{N.~Vast},
  \bibinfo{author}{I.~Timrov}, \bibinfo{author}{O.~Baseggio},
  \bibinfo{author}{S.~Baroni}, \bibinfo{author}{A.~Dal~Corso},
\newblock \bibinfo{title}{Electron energy loss spectroscopy of bulk gold with
  ultrasoft pseudopotentials and the $\rm{L}$iouville-$\rm{L}$anczos method},
\newblock \bibinfo{journal}{Phys. Rev. B} \bibinfo{volume}{102}
  (\bibinfo{year}{2020}) \bibinfo{pages}{035156}.
  \DOIprefix\doi{https://doi.org/10.1103/PhysRevB.102.035156}.
%Type = Article
\bibitem[{Ridzel et~al.(2020)Ridzel, Astašauskas, and
  Werner}]{RIDZEL2020146824}
\bibinfo{author}{O.~Y. Ridzel}, \bibinfo{author}{V.~Astašauskas},
  \bibinfo{author}{W.~S. Werner},
\newblock \bibinfo{title}{{Low energy ($1–100$ eV) electron inelastic mean
  free path (IMFP) values determined from analysis of secondary electron yields
  (SEY) in the incident energy range of $0.1–10$ keV}},
\newblock \bibinfo{journal}{J. Electron Spectrosc. Relat. Phenom.}
  \bibinfo{volume}{241} (\bibinfo{year}{2020}) \bibinfo{pages}{146824}.
  \DOIprefix\doi{https://doi.org/10.1016/j.elspec.2019.02.003}.
%Type = Article
\bibitem[{Timrov et~al.(2013)Timrov, Vast, Gebauer, and
  Baroni}]{PhysRevB.88.064301}
\bibinfo{author}{I.~Timrov}, \bibinfo{author}{N.~Vast},
  \bibinfo{author}{R.~Gebauer}, \bibinfo{author}{S.~Baroni},
\newblock \bibinfo{title}{Electron energy loss and inelastic x-ray scattering
  cross sections from time-dependent density-functional perturbation theory},
\newblock \bibinfo{journal}{Phys. Rev. B} \bibinfo{volume}{88}
  (\bibinfo{year}{2013}) \bibinfo{pages}{064301}.
  \DOIprefix\doi{https://doi.org/10.1103/PhysRevB.88.064301}.
%Type = Article
\bibitem[{Timrov et~al.(2015)Timrov, Vast, Gebauer, and
  Stefano}]{TIMROV2015460}
\bibinfo{author}{I.~Timrov}, \bibinfo{author}{N.~Vast},
  \bibinfo{author}{R.~Gebauer}, \bibinfo{author}{B.~Stefano},
\newblock \bibinfo{title}{{turboEELS—A code for the simulation of the
  electron energy loss and inelastic X-ray scattering spectra using the
  Liouville–Lanczos approach to time-dependent density-functional
  perturbation theory}},
\newblock \bibinfo{journal}{Comput. Phys. Commun.} \bibinfo{volume}{196}
  (\bibinfo{year}{2015}) \bibinfo{pages}{460--469}.
  \DOIprefix\doi{https://doi.org/10.1016/j.cpc.2015.05.021}.
%Type = Book
\bibitem[{Thompson(2001)}]{thompson2001x}
\bibinfo{author}{A.~Thompson}, \bibinfo{title}{X-ray Data Booklet},
  \bibinfo{publisher}{Lawrence Berkeley National Laboratory, University of
  California}, \bibinfo{year}{2001}.
%Type = Article
\bibitem[{Tanuma et~al.(2011)Tanuma, Powell, and Penn}]{tanuma2011calculations}
\bibinfo{author}{S.~Tanuma}, \bibinfo{author}{C.~Powell},
  \bibinfo{author}{D.~Penn},
\newblock \bibinfo{title}{Calculations of electron inelastic mean free paths.
  {IX}. data for 41 elemental solids over the 50 e{V} to 30 ke{V} range},
\newblock \bibinfo{journal}{Surf. Interface Anal.} \bibinfo{volume}{43}
  (\bibinfo{year}{2011}) \bibinfo{pages}{689--713}.
  \DOIprefix\doi{https://doi.org/10.1002/sia.3522}.
%Type = Article
\bibitem[{Smith et~al.(1985)Smith, Shiles, Inokuti, and
  Palik}]{smith1985handbook}
\bibinfo{author}{D.~Smith}, \bibinfo{author}{E.~Shiles},
  \bibinfo{author}{M.~Inokuti}, \bibinfo{author}{E.~Palik},
\newblock \bibinfo{title}{Handbook of optical constants of solids},
\newblock \bibinfo{journal}{Handbook of Optical Constants of Solids}
  \bibinfo{volume}{1} (\bibinfo{year}{1985}) \bibinfo{pages}{369--408}.
%Type = Article
\bibitem[{Montanari et~al.(2007)Montanari, Miraglia, Heredia-Avalos,
  Garcia-Molina, and Abril}]{montanari2007calculation}
\bibinfo{author}{C.~Montanari}, \bibinfo{author}{J.~Miraglia},
  \bibinfo{author}{S.~Heredia-Avalos}, \bibinfo{author}{R.~Garcia-Molina},
  \bibinfo{author}{I.~Abril},
\newblock \bibinfo{title}{Calculation of energy-loss straggling of {C}, {Al},
  {Si}, and {Cu} for fast {H}, {He}, and {L}i ions},
\newblock \bibinfo{journal}{Phys. Rev. A} \bibinfo{volume}{75}
  (\bibinfo{year}{2007}) \bibinfo{pages}{022903}.
  \DOIprefix\doi{https://doi.org/10.1103/PhysRevA.75.022903}.
%Type = Article
\bibitem[{Denton et~al.(2008)Denton, Abril, Garcia-Molina, and
  Heredia-Avalos}]{denton2008influence}
\bibinfo{author}{C.~Denton}, \bibinfo{author}{I.~Abril},
  \bibinfo{author}{J.~Garcia-Molina, Rand Moreno-Mar{\'\i}n},
  \bibinfo{author}{S.~Heredia-Avalos},
\newblock \bibinfo{title}{Influence of the description of the target
  energy-loss function on the energy loss of swift projectiles},
\newblock \bibinfo{journal}{Surf. Interface Anal.} \bibinfo{volume}{40}
  (\bibinfo{year}{2008}) \bibinfo{pages}{1481--1487}.
  \DOIprefix\doi{https://doi.org/10.1002/sia.2936}.
%Type = Article
\bibitem[{Franke et~al.(2000)Franke, Trimble, DeVries, Woollam, Schubert, and
  Frost}]{Franke2000}
\bibinfo{author}{E.~Franke}, \bibinfo{author}{C.~L. Trimble},
  \bibinfo{author}{M.~J. DeVries}, \bibinfo{author}{J.~A. Woollam},
  \bibinfo{author}{M.~Schubert}, \bibinfo{author}{F.~Frost},
\newblock \bibinfo{title}{Dielectric function of amorphous tantalum oxide from
  the far infrared to the deep ultraviolet spectral region measured by
  spectroscopic ellipsometry},
\newblock \bibinfo{journal}{Journal of Applied Physics} \bibinfo{volume}{88}
  (\bibinfo{year}{2000}) \bibinfo{pages}{5166--5174}.
%Type = Article
\bibitem[{Fadanelli et~al.(2015)Fadanelli, Behar, Nagamine, Vos, Arista,
  Nascimento, Garcia-Molina, and Abril}]{Fadanelli2015}
\bibinfo{author}{R.~C. Fadanelli}, \bibinfo{author}{M.~Behar},
  \bibinfo{author}{L.~C. C.~M. Nagamine}, \bibinfo{author}{M.~Vos},
  \bibinfo{author}{N.~R. Arista}, \bibinfo{author}{C.~D. Nascimento},
  \bibinfo{author}{R.~Garcia-Molina}, \bibinfo{author}{I.~Abril},
\newblock \bibinfo{title}{Energy loss function of solids assessed by ion beam
  energy-loss measurements: Practical application to {T}a$_2${O}$_5$},
\newblock \bibinfo{journal}{The Journal of Physical Chemistry C}
  \bibinfo{volume}{119} (\bibinfo{year}{2015}) \bibinfo{pages}{20561--20570}.
%Type = Article
\bibitem[{Dapor(2015)}]{Dapor2015}
\bibinfo{author}{M.~Dapor},
\newblock \bibinfo{title}{Energy loss of fast electrons impinging upon
  polymethylmethacrylate},
\newblock \bibinfo{journal}{Nucl. Instrum. Methods Phys. Res. B}
  \bibinfo{volume}{352} (\bibinfo{year}{2015}) \bibinfo{pages}{190--194}.
  \DOIprefix\doi{https://doi.org/10.1016/j.nimb.2014.11.101}.
%Type = Article
\bibitem[{Krawczyk et~al.(2015)Krawczyk, Holdynski, Lisowski, Sobczak, and
  Jablonski}]{KRAWCZYK2015196}
\bibinfo{author}{M.~Krawczyk}, \bibinfo{author}{M.~Holdynski},
  \bibinfo{author}{W.~Lisowski}, \bibinfo{author}{J.~Sobczak},
  \bibinfo{author}{A.~Jablonski},
\newblock \bibinfo{title}{Electron inelastic mean free paths in cerium
  dioxide},
\newblock \bibinfo{journal}{Appl. Surf. Sci.} \bibinfo{volume}{341}
  (\bibinfo{year}{2015}) \bibinfo{pages}{196--202}.
  \DOIprefix\doi{https://doi.org/10.1016/j.apsusc.2015.02.177}.
%Type = Article
\bibitem[{Dapor(2017)}]{Dapor2017}
\bibinfo{author}{M.~Dapor},
\newblock \bibinfo{title}{Role of the tail of high-energy secondary electrons
  in the monte carlo evaluation of the fraction of electrons backscattered from
  polymethylmethacrylate},
\newblock \bibinfo{journal}{Appl. Surf. Sci.} \bibinfo{volume}{391}
  (\bibinfo{year}{2017}) \bibinfo{pages}{3--11}.
  \DOIprefix\doi{https://doi.org/10.1016/j.apsusc.2015.12.043}.
%Type = Article
\bibitem[{{van Riessen} et~al.(2007){van Riessen}, Thurgate, and
  Ramaker}]{VANRIESSEN2007150}
\bibinfo{author}{G.~{van Riessen}}, \bibinfo{author}{S.~Thurgate},
  \bibinfo{author}{D.~Ramaker},
\newblock \bibinfo{title}{{Auger-photoelectron coincidence spectroscopy of
  SiO2}},
\newblock \bibinfo{journal}{J. Electron Spectrosc. Relat. Phenom.}
  \bibinfo{volume}{161} (\bibinfo{year}{2007}) \bibinfo{pages}{150--159}.
  \DOIprefix\doi{https://doi.org/10.1016/j.elspec.2007.02.028}.
%Type = Article
\bibitem[{Pauly et~al.(2017)Pauly, Yubero, Espin\'{o}s, and
  Tougaard}]{Pauly:17}
\bibinfo{author}{N.~Pauly}, \bibinfo{author}{F.~Yubero}, \bibinfo{author}{J.~P.
  Espin\'{o}s}, \bibinfo{author}{S.~Tougaard},
\newblock \bibinfo{title}{Optical properties and electronic transitions of zinc
  oxide, ferric oxide, cerium oxide, and samarium oxide in the ultraviolet and
  extreme ultraviolet},
\newblock \bibinfo{journal}{Appl. Opt.} \bibinfo{volume}{56}
  (\bibinfo{year}{2017}) \bibinfo{pages}{6611--6621}.
  \DOIprefix\doi{https://doi.org/10.1364/AO.56.006611}.
%Type = Article
\bibitem[{Dapor(2022)}]{10.3389/fmats.2022.1068196}
\bibinfo{author}{M.~Dapor},
\newblock \bibinfo{title}{Aluminum electron energy loss spectra. a comparison
  between monte carlo and experimental data},
\newblock \bibinfo{journal}{Front. Mater.} \bibinfo{volume}{9}
  (\bibinfo{year}{2022}) \bibinfo{pages}{1068196}.
  \DOIprefix\doi{https://doi.org/10.3389/fmats.2022.1068196}.
%Type = Article
\bibitem[{Dapor et~al.(2012)Dapor, Calliari, and
  Fanchenko}]{https://doi.org/10.1002/sia.4835}
\bibinfo{author}{M.~Dapor}, \bibinfo{author}{L.~Calliari},
  \bibinfo{author}{S.~Fanchenko},
\newblock \bibinfo{title}{{Energy loss of electrons backscattered from solids:
  measured and calculated spectra for Al and Si}},
\newblock \bibinfo{journal}{Surf. Interface Anal.} \bibinfo{volume}{44}
  (\bibinfo{year}{2012}) \bibinfo{pages}{1110--1113}.
  \DOIprefix\doi{https://doi.org/10.1002/sia.4835}.
%Type = Article
\bibitem[{Kuhr and Fitting(1999)}]{KhurFitting1999}
\bibinfo{author}{J.-C. Kuhr}, \bibinfo{author}{H.-J. Fitting},
\newblock \bibinfo{title}{{Monte Carlo simulation of electron emission from
  solids}},
\newblock \bibinfo{journal}{J. Electron Spectrosc. Relat. Phenom.}
  \bibinfo{volume}{105} (\bibinfo{year}{1999}) \bibinfo{pages}{257--273}.
  \DOIprefix\doi{https://doi.org/10.1016/S0368-2048(99)00082-1}.
%Type = Article
\bibitem[{Dapor et~al.(2008)Dapor, Inkson, Rodenburg, and
  Rodenburg}]{Dapor_2008}
\bibinfo{author}{M.~Dapor}, \bibinfo{author}{B.~J. Inkson},
  \bibinfo{author}{C.~Rodenburg}, \bibinfo{author}{J.~M. Rodenburg},
\newblock \bibinfo{title}{{A comprehensive Monte Carlo calculation of dopant
  contrast in secondary-electron imaging}},
\newblock \bibinfo{journal}{EPL} \bibinfo{volume}{82} (\bibinfo{year}{2008})
  \bibinfo{pages}{30006}.
  \DOIprefix\doi{https://doi.org/10.1209/0295-5075/82/30006 - Erratum:
  https://doi.org/10.1209/0295-5075/82/49901}.
%Type = Article
\bibitem[{Dapor(2009)}]{DAPOR20093055}
\bibinfo{author}{M.~Dapor},
\newblock \bibinfo{title}{{A Monte Carlo investigation of secondary electron
  emission from solid targets: Spherical symmetry versus momentum conservation
  within the classical binary collision model}},
\newblock \bibinfo{journal}{Nucl. Instrum. Methods Phys. Res. B}
  \bibinfo{volume}{267} (\bibinfo{year}{2009}) \bibinfo{pages}{3055--3058}.
  \DOIprefix\doi{https://doi.org/10.1016/j.nimb.2009.06.025}.
%Type = Article
\bibitem[{Schreiber and Fitting(2002)}]{SCHREIBER200225}
\bibinfo{author}{E.~Schreiber}, \bibinfo{author}{H.-J. Fitting},
\newblock \bibinfo{title}{{Monte Carlo simulation of secondary electron
  emission from the insulator SiO2}},
\newblock \bibinfo{journal}{J. Electron Spectrosc. Relat. Phenom.}
  \bibinfo{volume}{124} (\bibinfo{year}{2002}) \bibinfo{pages}{25--37}.
  \DOIprefix\doi{https://doi.org/10.1016/S0368-2048(01)00368-1}.
%Type = Article
\bibitem[{Turetta et~al.(2021)Turetta, Sedona, Liscio, Sambi, and
  Paolo}]{turetta:hal-03240126}
\bibinfo{author}{N.~Turetta}, \bibinfo{author}{F.~Sedona},
  \bibinfo{author}{A.~Liscio}, \bibinfo{author}{M.~Sambi},
  \bibinfo{author}{S.~Paolo},
\newblock \bibinfo{title}{Au(111) surface contamination in ambient conditions:
  Unravelling the dynamics of the work function in air},
\newblock \bibinfo{journal}{Adv. Mater. Interfaces} \bibinfo{volume}{8}
  (\bibinfo{year}{2021}) \bibinfo{pages}{2100068}.
  \DOIprefix\doi{https://doi.org/10.1002/admi.202100068}.
%Type = Article
\bibitem[{Wass et~al.(2019)Wass, Hollington, Sumner, Yang, and
  Pfeil}]{wass2019}
\bibinfo{author}{P.~J. Wass}, \bibinfo{author}{D.~Hollington},
  \bibinfo{author}{T.~J. Sumner}, \bibinfo{author}{F.~Yang},
  \bibinfo{author}{M.~Pfeil},
\newblock \bibinfo{title}{Effective decrease of photoelectric emission
  threshold from gold plated surfaces},
\newblock \bibinfo{journal}{Rev. Sci. Instrum.} \bibinfo{volume}{90}
  (\bibinfo{year}{2019}) \bibinfo{pages}{064501}.
  \DOIprefix\doi{https://doi.org/10.1063/1.5088135}.
%Type = Article
\bibitem[{Gonzalez et~al.(2017)Gonzalez, Angelucci, Larciprete, and
  Cimino}]{gonzalez2017secondary}
\bibinfo{author}{L.~A. Gonzalez}, \bibinfo{author}{M.~Angelucci},
  \bibinfo{author}{R.~Larciprete}, \bibinfo{author}{R.~Cimino},
\newblock \bibinfo{title}{The secondary electron yield of noble metal
  surfaces},
\newblock \bibinfo{journal}{AIP Advances} \bibinfo{volume}{7}
  (\bibinfo{year}{2017}) \bibinfo{pages}{115203}.
  \DOIprefix\doi{https://doi.org/10.1063/1.5000118}.
%Type = Article
\bibitem[{Yasuda et~al.(2008)Yasuda, Morimoto, Kainuma, Kawata, and
  Hirai}]{Yasuda_2008}
\bibinfo{author}{M.~Yasuda}, \bibinfo{author}{K.~Morimoto},
  \bibinfo{author}{Y.~Kainuma}, \bibinfo{author}{H.~Kawata},
  \bibinfo{author}{Y.~Hirai},
\newblock \bibinfo{title}{Analysis of charging phenomena of polymer films on
  silicon substrates under electron beam irradiation},
\newblock \bibinfo{journal}{Jpn. J. Appl. Phys.} \bibinfo{volume}{47}
  (\bibinfo{year}{2008}) \bibinfo{pages}{4890}.
  \DOIprefix\doi{https://doi.org/10.1143/JJAP.47.4890}.
%Type = Article
\bibitem[{Boubaya and Blaise(2007)}]{Boubaya}
\bibinfo{author}{M.~Boubaya}, \bibinfo{author}{G.~Blaise},
\newblock \bibinfo{title}{{Charging regime of PMMA studied by secondary
  electron emission}},
\newblock \bibinfo{journal}{Eur. Phys. J.: Appl. Phys.} \bibinfo{volume}{37}
  (\bibinfo{year}{2007}) \bibinfo{pages}{79}.
  \DOIprefix\doi{https://doi.org/10.1051/epjap:2006128}.
%Type = Article
\bibitem[{Rau et~al.(2008)Rau, Evstaf'eva, and Adrianov}]{Rau}
\bibinfo{author}{E.~I. Rau}, \bibinfo{author}{E.~N. Evstaf'eva},
  \bibinfo{author}{M.~V. Adrianov},
\newblock \bibinfo{title}{Mechanisms of charging of insulators under
  irradiation with medium-energy electron beams},
\newblock \bibinfo{journal}{Phys. Solid State} \bibinfo{volume}{50}
  (\bibinfo{year}{2008}) \bibinfo{pages}{621--630}.
  \DOIprefix\doi{https://doi.org/10.1134/S1063783408040057}.
%Type = Article
\bibitem[{Dal~Cappello et~al.(2009)Dal~Cappello, Champion, Boudrioua, Lekadir,
  Sato, and Ohsawa}]{DalCappello2009}
\bibinfo{author}{C.~Dal~Cappello}, \bibinfo{author}{C.~Champion},
  \bibinfo{author}{O.~Boudrioua}, \bibinfo{author}{H.~Lekadir},
  \bibinfo{author}{Y.~Sato}, \bibinfo{author}{D.~Ohsawa},
\newblock \bibinfo{title}{Theoretical and experimental investigations of
  electron emission in {C}$^{6+}$ + {H}$_2${O} collisions},
\newblock \bibinfo{journal}{Nucl. Instrum. Methods Phys. Res. B}
  \bibinfo{volume}{267} (\bibinfo{year}{2009}) \bibinfo{pages}{781--790}.
  \DOIprefix\doi{https://doi.org/10.1016/j.nimb.2008.12.010}.
%Type = Article
\bibitem[{de~Vera et~al.(2017)de~Vera, Surdutovich, Mason, and
  Solov’yov}]{refId0}
\bibinfo{author}{P.~de~Vera}, \bibinfo{author}{E.~Surdutovich},
  \bibinfo{author}{N.~J. Mason}, \bibinfo{author}{A.~V. Solov’yov},
\newblock \bibinfo{title}{Radial doses around energetic ion tracks and the
  onset of shock waves on the nanoscale},
\newblock \bibinfo{journal}{Eur. Phys. J. D} \bibinfo{volume}{71}
  (\bibinfo{year}{2017}) \bibinfo{pages}{281}.
  \DOIprefix\doi{https://doi.org/10.1140/epjd/e2017-80176-8}.
%Type = Article
\bibitem[{Incerti et~al.(2014)}]{INCERTI201492}
\bibinfo{author}{S.~Incerti}, et~al.,
\newblock \bibinfo{title}{{Simulating radial dose of ion tracks in liquid water
  simulated with Geant4-DNA: A comparative study}},
\newblock \bibinfo{journal}{Nucl. Instrum. Methods Phys. Res. B}
  \bibinfo{volume}{333} (\bibinfo{year}{2014}) \bibinfo{pages}{92--98}.
  \DOIprefix\doi{https://doi.org/10.1016/j.nimb.2014.04.025}.
%Type = Article
\bibitem[{Waligórski et~al.(1986)Waligórski, Hamm, and
  Katz}]{WALIGORSKI1986309}
\bibinfo{author}{M.~Waligórski}, \bibinfo{author}{R.~Hamm},
  \bibinfo{author}{R.~Katz},
\newblock \bibinfo{title}{The radial distribution of dose around the path of a
  heavy ion in liquid water},
\newblock \bibinfo{journal}{International Journal of Radiation Applications and
  Instrumentation. Part D. Nuclear Tracks and Radiation Measurements}
  \bibinfo{volume}{11} (\bibinfo{year}{1986}) \bibinfo{pages}{309--319}.
  \DOIprefix\doi{https://doi.org/10.1016/1359-0189(86)90057-9}.
%Type = Article
\bibitem[{Liamsuwan and Nikjoo(2013)}]{Liamsuwan_2013}
\bibinfo{author}{T.~Liamsuwan}, \bibinfo{author}{H.~Nikjoo},
\newblock \bibinfo{title}{{A Monte Carlo track structure simulation code for
  the full-slowing-down carbon projectiles of energies 1 keV
  $\rm{u}^{–1}$–10 MeV $\rm{u}^{–1}$ in water}},
\newblock \bibinfo{journal}{Phys. Med. Biol.} \bibinfo{volume}{58}
  (\bibinfo{year}{2013}) \bibinfo{pages}{673}.
  \DOIprefix\doi{https://doi.org/10.1088/0031-9155/58/3/673}.
%Type = Article
\bibitem[{Schardt et~al.(2010)Schardt, Els\"asser, and
  Schulz-Ertner}]{Schardt2010}
\bibinfo{author}{D.~Schardt}, \bibinfo{author}{T.~Els\"asser},
  \bibinfo{author}{D.~Schulz-Ertner},
\newblock \bibinfo{title}{Heavy-ion tumor therapy: Physical and radiobiological
  benefits},
\newblock \bibinfo{journal}{Rev. Mod. Phys.} \bibinfo{volume}{82}
  (\bibinfo{year}{2010}) \bibinfo{pages}{383--425}.
  \DOIprefix\doi{https://doi.org/10.1103/RevModPhys.82.383}.
%Type = Article
\bibitem[{Tsujii et~al.(2008)Tsujii, Kamada, Baba, Tsuji, Kato, Kato, Yamada,
  Yasuda, Yanagi, Kato, Hara, Yamamoto, and Mizoe}]{Tsujii2008}
\bibinfo{author}{H.~Tsujii}, \bibinfo{author}{T.~Kamada},
  \bibinfo{author}{M.~Baba}, \bibinfo{author}{H.~Tsuji},
  \bibinfo{author}{H.~Kato}, \bibinfo{author}{S.~Kato},
  \bibinfo{author}{S.~Yamada}, \bibinfo{author}{S.~Yasuda},
  \bibinfo{author}{T.~Yanagi}, \bibinfo{author}{H.~Kato},
  \bibinfo{author}{R.~Hara}, \bibinfo{author}{N.~Yamamoto},
  \bibinfo{author}{J.~Mizoe},
\newblock \bibinfo{title}{{Clinical advantages of carbon-ion radiotherapy}},
\newblock \bibinfo{journal}{New J. Phys.} \bibinfo{volume}{10}
  (\bibinfo{year}{2008}) \bibinfo{pages}{075009}.
  \DOIprefix\doi{https://doi.org/10.1088/1367-2630/10/7/075009}.
%Type = Article
\bibitem[{Ebner and Kamada(2016)}]{Ebner2016}
\bibinfo{author}{D.~K. Ebner}, \bibinfo{author}{T.~Kamada},
\newblock \bibinfo{title}{The emerging role of carbon-ion radiotherapy},
\newblock \bibinfo{journal}{Front. Oncol.} \bibinfo{volume}{6}
  (\bibinfo{year}{2016}) \bibinfo{pages}{140}.
  \DOIprefix\doi{https://doi.org/10.3389/fonc.2016.00140}.
%Type = Article
\bibitem[{Amaldi and Kraft(2005)}]{Amaldi2005}
\bibinfo{author}{U.~Amaldi}, \bibinfo{author}{G.~Kraft},
\newblock \bibinfo{title}{Radiotherapy with beams of carbon ions},
\newblock \bibinfo{journal}{Rep. Progr. Phys.} \bibinfo{volume}{68}
  (\bibinfo{year}{2005}) \bibinfo{pages}{1861--1882}.
  \DOIprefix\doi{10.1088/0034-4885/68/8/r04}.
%Type = Article
\bibitem[{Taioli and Tennyson(2006)}]{taioli2006wave}
\bibinfo{author}{S.~Taioli}, \bibinfo{author}{J.~Tennyson},
\newblock \bibinfo{title}{A wave packet method for treating nuclear dynamics on
  complex potentials},
\newblock \bibinfo{journal}{J. Phys. B} \bibinfo{volume}{39}
  (\bibinfo{year}{2006}) \bibinfo{pages}{4379}.
  \DOIprefix\doi{https://doi.org/10.1088/0953-4075/39/21/004}.
%Type = Article
\bibitem[{Bouda\"{\i}ffa et~al.(2000)Bouda\"{\i}ffa, Cloutier, Hunting, Huels,
  and Sanche}]{Boudaiffa2000}
\bibinfo{author}{B.~Bouda\"{\i}ffa}, \bibinfo{author}{P.~Cloutier},
  \bibinfo{author}{D.~Hunting}, \bibinfo{author}{M.~A. Huels},
  \bibinfo{author}{L.~Sanche},
\newblock \bibinfo{title}{{Resonant formation of DNA strand breaks by
  low-energy (3 to 20 eV) electrons}},
\newblock \bibinfo{journal}{Science} \bibinfo{volume}{287}
  (\bibinfo{year}{2000}) \bibinfo{pages}{1658--1660}.
  \DOIprefix\doi{https://doi.org/10.1126/science.287.5458.165}.

\end{thebibliography}
\end{document}